\documentclass[12pt,curves,epsf]{article}

\usepackage{epsf,amsfonts,hyperref}
\usepackage{makeidx}
\usepackage{graphicx}

\makeindex

\renewcommand{\appendix}[1]{
    \addtocounter{section}{1}
    \setcounter{equation}{0}
    \renewcommand{\thesection}{\Alph{section}}
    \section*{Appendix \thesection\protect\indent #1}
    \addcontentsline{toc}{section}{Appendix \thesection\ \ \ #1}
}
\newcommand\encadremath[1]{\vbox{\hrule\hbox{\vrule\kern8pt
\vbox{\kern8pt \hbox{$\displaystyle #1$}\kern8pt}
\kern8pt\vrule}\hrule}}
\def\enca#1{\vbox{\hrule\hbox{
\vrule\kern8pt\vbox{\kern8pt \hbox{$\displaystyle #1$}
\kern8pt} \kern8pt\vrule}\hrule}}

\newcommand\figureframex[3]{
\begin{figure}[bth]
\hrule\hbox{\vrule\kern8pt
\vbox{\kern8pt \vbox{
\begin{center}
{\mbox{\epsfxsize=#1.truecm\epsfbox{#2}}}
\end{center}
\caption{#3}
}\kern8pt}
\kern8pt\vrule}\hrule
\end{figure}
}
\newcommand\figureframey[3]{
\begin{figure}[bth]
\hrule\hbox{\vrule\kern8pt
\vbox{\kern8pt \vbox{
\begin{center}
{\mbox{\epsfysize=#1.truecm\epsfbox{#2}}}
\end{center}
\caption{#3}
}\kern8pt}
\kern8pt\vrule}\hrule
\end{figure}
}

\newcommand{\eq}[1]{eq.(\ref{#1})}
\newcommand{\Eq}[1]{Eq.(\ref{#1})}

\newcommand{\beq}{\begin{equation}}
\newcommand{\eeq}{\end{equation}}
\newcommand{\bea}{\begin{eqnarray}}
\newcommand{\eea}{\end{eqnarray}}

\newcommand{\vs}{\vspace{0.7cm}}
\renewcommand{\thesection}{\arabic{section}}

\makeatletter
\@addtoreset{equation}{section}
\makeatother
\newtheorem{theorem}{Theorem}[section]

\newtheorem{remark}{Remark}[section]
\newtheorem{proposition}{Proposition}[section]
\newtheorem{lemma}{Lemma}[section]
\newtheorem{corollary}{Corollary}[section]
\newtheorem{definition}{Definition}[section]
\def\br{\begin{remark}\rm\small}
\def\er{\end{remark}}
\def\bt{\begin{theorem}}
\def\et{\end{theorem}}
\def\bd{\begin{definition}}
\def\ed{\end{definition}}
\def\bp{\begin{proposition}}
\def\ep{\end{proposition}}
\def\bl{\begin{lemma}}
\def\el{\end{lemma}}
\def\bc{\begin{corollary}}
\def\ec{\end{corollary}}
\def\beaq{\begin{eqnarray}}
\def\eeaq{\end{eqnarray}}
\newcommand{\proof}[1]{{\noindent \bf proof:}\par
{#1} $\square$}

\renewcommand{\and}{{\qquad {\rm and} \qquad}}

\newcommand{\virg}{{\qquad , \qquad}}

 \newcommand{\Tr}{{\,\rm Tr}\:}
\newcommand{\tr}{{\,\rm tr}\:}

\newcommand{\Res}{\mathop{\,\rm Res\,}}

\newcommand{\td}[1]{{\tilde{#1}}}

\renewcommand{\l}{\lambda}
\newcommand{\om}{\omega}

\newcommand{\ee}[1]{{{\rm e}^{#1}}}

\renewcommand{\d}{{{\partial}}}

\newcommand{\Pint}{{\int\kern -1.em -\kern-.25em}}

\renewcommand{\Re}{{\mathrm{Re}}}
\renewcommand{\Im}{{\mathrm{Im}}}

\newcommand{\genus}{\bar{g}}
\newcommand{\spcurve}{{\cal L}}
\newcommand{\acycle}{{\cal A}}
\newcommand{\bcycle}{{\cal B}}

\newcommand{\TT}{{\cal T}}
\renewcommand{\l}{\lambda}
\renewcommand{\L}{\Lambda}

\newcommand{\ovl}{\overline}

\begin{document}
%=============================Page de titre===============
\begin{titlepage}
{}~
\hfill\vbox{
\hbox{IPhT-T08/189}

\hbox{CERN-PH-TH-2008-222}
}\break

\vskip .6cm

\centerline{\Large \bf
Algebraic methods in random matrices}
\centerline{\Large \bf
and enumerative geometry}
\vspace*{1.0ex}

\medskip

\vspace*{4.0ex}

\centerline{\large \rm Bertrand Eynard$^a$ and Nicolas Orantin$^b$}

\vspace*{4.0ex}

\centerline{\rm ~$^a$  Institut de Physique Th\'eorique, CEA, IPhT}
\centerline{ F-91191 Gif-sur-Yvette, France}
\centerline{CNRS, URA 2306, F-91191 Gif-sur-Yvette, France }

\centerline{{\tt bertrand.eynard@cea.fr}}

\vspace*{1.8ex}

\centerline{ \rm ~$^b$  Theory department, CERN}
\centerline{ Geneva 23, CH-1211 Switzerland}

\centerline{\tt
nicolas.orantin@cern.ch}

\vspace*{6ex}

\centerline{\bf Abstract}
\medskip

We review the method of symplectic invariants recently introduced to solve matrix models loop equations, and further extended beyond the context of matrix models.
For any given spectral curve, one defines a sequence of differential forms, and a sequence of complex numbers $F_g$.
We recall the definition of the invariants $F_g$, and we explain their main properties, in particular symplectic invariance, integrability, modularity,...
Then, we give several examples of applications, in particular matrix models, enumeration of discrete surfaces (maps), algebraic geometry and topological strings, non-intersecting brownian motions,...

\end{titlepage}
\vfill
\eject

\pagestyle{plain}
\setcounter{page}{1}

%*********************************************************************
%==================== ARTICLE ========================================
%*********************************************************************

\tableofcontents

\section{Introduction}

Recently, it was understood how to solve, order by order in the so-called "topological expansion", the loop equations (Schwinger-Dyson equations) for matrix integrals \cite{eynloop1mat}. The solution brought an unexpectedly rich structure \cite{EOinvariants}, which, did not only solve the 1-matrix model, but which also solved multi-matrix models, as well as their limits. Later, it was understood that this structure also appears in other matrix models, and in problems of enumerative geometry, not directly related to matrix models.

Thus, there is an underlying structure which can be defined beyond the context of matrix models, and relies only on the intrinsic algebro-geometric properties of a plane curve, called the spectral curve.

\smallskip
In other words, for any regular (to be defined below) complex plane curve ${\cal E}=\{y(x)\}$ (whether it is related to a matrix model or not), we can define a sequence of numbers $F_g({\cal E})$, $g=0,1,2,\dots,\infty$.
Those numbers $F_g$ are called the {\bf symplectic invariants} of the spectral curve ${\cal E}$ (first introduced in \cite{EOinvariants}).
The reason is because two spectral curves ${\cal E}$ and $\td{\cal E}$ which can be deduced from one another by a symplectic transformation (i.e. they have the same wedge product $dx\wedge dy = d\td{x}\wedge d\td{y}$), have the same $F_g$'s. $F_g$ is called the symplectic invariant of degree $2-2g$ , because under a rescaling $y\to \l y$, $F_g$ scales as $F_g\to \l^{2-2g} F_g$ (except $F_1$ which is logarithmic).

\medskip
Moreover, for a spectral curve ${\cal E}=\{y(x)\}$, we define not only its symplectic invariants $F_g$'s, we also define a doubly infinite sequence of symmetric meromorphic forms $\om_n^{(g)}(x_1,\dots,x_n)[{\cal E}]$, $n\in {\mathbb N}, g\in {\mathbb N}$, and such that $F_g=\om_0^{(g)}$.
For $n\geq 1$, those forms are not symplectic invariants, but they have many nice properties. They allow to compute the derivatives of the $F_g$'s with respect to any parameter on which ${\cal E}$ could depend.

\bigskip
Those geometric objects are interesting, in particular  for their applications to various problems of enumerative geometry (each problem corresponding to a given spectral curve ${\cal E}$), but also on their own. Indeed they have remarkable properties for arbitrary spectral curves, i.e. even for spectral curves not known to correspond to any enumerative geometry problem.

In particular, they are related to the Kodaira-Spencer field theory, to Frobenius manifolds, to the WDVV special geometry, topological strings and Dijkgraaf-Vafa conjecture. They are expected to be the B-model partition function, and through mirror symmetry, the $F_g$'s are thus expected to be the generating functions of Gromov-Witten invariants of genus $g$ for some toric geometries.

As another example of interesting properties, the $F_g$'s have nice modular behaviors, and, for instance, they provide a solution to holomorphic anomaly equations.

They also contain an integrable structure, related to "multicomponent KP" hierarchy, e.g. they satisfy determinantal formulae, Hirota equations,...

Another nice property, is that they can be computed by a simple diagrammatic method, which makes them really easy to use. For instance the holomorphic anomaly equations can be proved only by drawing diagrams.

\bigskip
Regarding the applications, we will consider the following examples:

- Enumeration of discrete surfaces, possibly carrying colors on their faces (Ising model), as well as the asymptotics of large discrete surfaces.

- For the curve $y={1\over 2\pi}\,\sin{(2\pi\sqrt{x})}$, the $F_g$'s compute the Weyl-Petersson volumes.

- We will consider also the Kontsevich spectral curve, related to the Kontsevich integral, for which the $F_g$'s are generating functions for intersection numbers of Chern classes of cotangent bundles at marked points, and the $W_n^{(g)}$'s are generating functions of Mumford $\kappa$ classes (in some sense by "forgetting" some marked points).

- For the curve $y={\rm Argcosh}{(x)}$, and deformations of that curve, the $F_g$'s are generating functions for counting partitions with the Plancherel measure, related to the computation of Hurwitz numbers.

- $q$-deformed versions of Plancherel measure sums of partitions can also be computed with symplectic invariants of some appropriate spectral curve, which, not so surprisingly, is the (singular locus of the) mirror of a toric Calabi-Yau manifold. This is consistent with the conjecture that the $F_g$'s are related to Gromov-Witten invariants. Indeed, Gromov-Witten invariants of toric Calabi-Yau 3-folds can be computed, using the topological vertex, as sums of partitions, typically q-deformed Plancherel sums, for the simplest examples of toric Calabi-Yau 3-folds.

%%%%%%%%%%%%%%%%%%%%%%%%%%%%%%%%%%%%%%%%%%%%%%%%%%%%%%%%%%%%%%%%%%%%%%%%%%%%%%%%%%%%%%%%%%%%%%%%%%%%%%%%%%%%

\section{Symplectic invariants of spectral curves}\label{secsymplectic}

Symplectic invariants were introduced in \cite{EOinvariants}, as a common framework for the solution of loop equations of several matrix models: 1-matrix, 2-matrix, matrix with external fields,..., as well as their double scaling limits. Then it was discovered that they have many nice properties, in particular symplectic invariance, and that they appear in other problems of enumerative geometry, not necessarily related to random matrices.

Here we only briefly summarize the construction of \cite{EOinvariants}, without proofs, and we refer the reader to the original article for more details.

\subsection{Spectral curves}

In this article, we define a spectral curve as follows\footnote{This definition is not exactly the one usually encountered in integrable systems \cite{BBT}, in fact it turns out that the plane curve we are considering here, is the "classical limit" of the full spectral curve. We call it spectral curve by abuse of language, and because it has become customary to do so.}:
\bd
A spectral curve ${\cal E}=(\spcurve, x,y)$, is the data of a compact Riemann surface $\spcurve$, and two analytical functions $x$ and $y$ on some open domain in $\spcurve$.

\ed

In some sense, we consider a parametric representation of the spectral curve $y(x)$, where the space of the parameter $z$ is a Riemann surface $\spcurve$.

\bd
If $\spcurve$ is a compact Riemann surface of genus $\genus$, and $x$ and $y$ are meromorphic functions on $\spcurve$, we say that the spectral curve is algebraic.
If in addition, $\spcurve$ is the Riemann sphere ($\spcurve={\mathbb P}^1={\mathbb C}\cup\{\infty\}$, i.e. of genus $\genus=0$), we say that the spectral curve is rational.
\ed

Indeed, for an algebraic spectral curve, it is always possible to find a polynomial relationship between $x$ and $y$:
\beq
{\rm Pol}(x,y)=0.
\eeq
For a rational spectral curve, the polynomial equation ${\rm Pol}(x,y)=0$, can be parameterized with two rational functions $x(z)$ and $y(z)$ of a complex variable $z$.

\bd\label{defregular}
A spectral curve $(\spcurve, x,y)$ is called regular if:

$\bullet$ The differential form $dx$ has a finite number of zeroes $dx(a_i)=0$, and all zeroes of $dx$ are simple zeroes.

$\bullet$ The differential $dy$ does not vanish  at the zeroes of $dx$, i.e. $dy(a_i)\neq 0$.\\

This means that near $x(a_i)$, $y$ behaves locally like a square-root $y(z) \sim y(a_i) + C\,\sqrt{x(z)-x(a_i)}$, or in other words, that the curve $y(x)$ has a vertical tangent at $a_i$.
\ed

From now on, we assume that we are considering only regular spectral curves.
Symplectic invariants are defined only for regular spectral curves, and they diverge when the spectral curve becomes singular.
Examples of singular spectral curves are considered in section \ref{secsingFg}, they play a central role in the double scaling limit in chapter \ref{secdsl}.

\bd
We say that two spectral curves ${\cal E}=(\spcurve,x,y)$ and  $\td{\cal E}=(\td\spcurve,\td{x},\td{y})$ are symplectically equivalent if there is a conformal mapping $\spcurve\to\td\spcurve$, and
if under this mapping $dx \wedge dy\to d\td{x}\wedge d\td{y}  $.
%$\td{y}\td{dx}-ydx$ is an exact form, i.e. it is the differential of a function on $\spcurve$.
The group of symplectomorphisms is generated by:
\begin{itemize}
\item $\td{x}=x$, $\td{y}= y+R(x)$, $R(x)=$ rational function of $x$.
\item $\td{x}= {ax+b\over cx+d}$, $\td{y}= {(cx+d)^2\over ad-bc}\,y$.
\item $\td{x}= f(x)$, $\td{y}= {1\over f'(x)}\,y$, where $f$ is analytical and injective in the image of $x$.
\item $\td{x}= y$, $\td{y}= -x$.
\end{itemize}

All those transformations conserve the symplectic form on $\spcurve$, whence the name:
\beq
d\td{x}\wedge d\td{y} = dx\wedge dy .
\eeq

%We say that the spectral curve is the equivalence class of equations $E(x,y)=0$ modulo symplectomorphisms.

\ed

The main property of the $F_g$'s we are going to define, is that they are symplectic invariants, i.e. two curves which are symplecticaly equivalent, have the same $F_g$'s.

\subsubsection{Examples of spectral curves}
\label{secexspcurves}

Interesting examples of spectral curves may come from several areas of physics or mathematics, and are related to some problems of enumerative geometry.
We will study in details some examples between section \ref{secMM} and \ref{sectopostring}. Here, in order to illustrate our notion of spectral curve, we give some examples of spectral curves of interest extracted from those applications.

\smallskip

For the readers familiar with matrix models, the spectral curve under consideration here, can be thought of, as the "equilibrium density of eigenvalues of the random matrix". It is not to be confused with the large $N$ density of eigenvalues, although, for many simple cases the two may coincide\footnote{The two notions coincide for example for matrix integrals with a polynomial potential. They do not coincide for example when the potential has an explicit dependence on $N$.}.
In the most simple matrix models, the spectral curve is algebraic.
For formal random matrix models, designed as combinatorics generating funcions for counting discrete surfaces, the spectral curve is shown to be rational (see section \ref{secmaps}).

In the context of string theory, the spectral curve is often given by a transcendental equation of the form $H(e^x,e^y)=0$, where $H$ is a polynomial. It is not an algebraic spectral curve, but is closely related to an algebraic curve.
In that case, $dx$ and $dy$ are abelian meromorphic differentials on the compact Riemann surface $\spcurve$ corresponding to $H$ (see section \ref{sectopostring}).

\bigskip
The origin of all the examples below are described with more details in sections \ref{secMM} to \ref{sectopostring}.

\medskip

$\bullet$ The following curve is a rational spectral curve:
\beq
\spcurve={\mathbb P}^1={\mathbb C}\cup\{\infty\}
\virg
x(z) = z^2-2 \virg y(z) =z^3-3z .
\eeq
It satisfies the algebraic equation $y^2-2 = x^3-3x$. It is an hyperelliptical curve of genus $\genus=0$.
This spectral curve is related to the so-called "pure gravity Liouville field theory".
It will often be called the "pure gravity" spectral curve, or also the $(3,2)$ spectral curve, because pure gravity is the $(3,2)$ minimal conformal field theory, it has central charge $c=0$. See sections \ref{secsingFg} and \ref{secdsl}.

\bigskip

$\bullet$ The curve $y=\sqrt{x}$, is also a rational spectral curve which satisfies $y^2=x$, and which can be parameterized by:
\beq\label{Airy}
\spcurve={\mathbb P}^1
\virg
x(z) = z^2 \virg y(z) =z .
\eeq
This spectral curve arises in the study of the extreme eigenvalues statistics of a random matrix, i.e. in the study of the Tracy-Widom law and of the Airy kernel \cite{TWlaw}.
It will often be called the "Airy" spectral curve.
It will also be called the $(1,2)$ spectral curve in order to match the classification of minimal conformal field theories. The minimal model $(1,2)$ has central charge $c=-2$.
See section \ref{secdsl}.

\bigskip

$\bullet$ The following spectral curve is also a rational spectral curve:
\beq
\spcurve={\mathbb P}^1
\virg
x(z) = \gamma\left(z+{1\over z}\right) \virg y(z) = -{t\over \gamma z} + {t_4\gamma^3\over z^3}
\eeq
where $\gamma^2 = {1-\sqrt{1-12 t t_4}\over 6  t_4}$.
This spectral curve arises in the enumeration of quadrangulated surfaces, i.e. in the formal quartic matrix model.
See section \ref{secquadr}.

\bigskip

$\bullet$ The following spectral curve
\beq
\spcurve={\mathbb P}^1
\virg
x(z) = z^2 \virg y(z) = {1\over 2\pi}\sin{(2\pi z)}
\eeq
is related to the computation of Weyl-Petersson volumes. Notice that it is not algebraic, but it can be parameterized by a complex variable, i.e. by a $\genus=0$  Riemann surface.
See section \ref{secWP}.

\bigskip

$\bullet$ The following rational spectral curve
\beq
\spcurve={\mathbb P}^1
\virg
x(z) = z^2 \virg y(z) = z^3-3tz
\eeq
is singular at $t=0$. Indeed at $t=0$, the differential $dy=3(z^2-t)dz$ vanishes at $z=0$ which is the zero of $dx$.
We will see below that the $F_g$'s diverge for singular curves, and thus the function $F_g(t)$ has a singularity at $t=0$. Just from homogeneity, and by considering the change of variable $z\to\sqrt{t}\, z$, we see that:
\beq
F_g(t) = t^{5(1-g)}\,F_g(1)
\eeq
which indeed diverges at $t=0$.
See section \ref{secsingFg}.

\bigskip

$\bullet$ The following spectral curve depends on two parameters $p\in {\mathbb Z}$ and $z_0\in {\mathbb C}^*$:
\beq
\spcurve={\mathbb P}^1
\virg
\left\{\begin{array}{l}
\displaystyle
x(z) = {(1-{z\over z_0})(1-{1\over z z_0})\over (1+{1\over z_0})^2} \cr
\cr
\displaystyle
y(z) = {1\over x(z)}\left(-\ln{z}+{p\over 2}\ln{\left(1-z/z_0\over 1-1/z z_0\right)}\right)
\end{array}\right. .
\eeq
It appears in the enumeration of $q-$deformed Plancherel sums of partitions, i.e. in the computation of the Gromov-Witten invariants of the toric Calabi-Yau manifold $X_p = O(-p)\oplus O(p-2) \to {\mathbb P}^1$.
See section \ref{secqdeformpart}.

This spectral curve is symplectically equivalent to (compute $dx\wedge dy$ in both cases):
\beq
\spcurve={\mathbb P}^1
\virg
\left\{\begin{array}{l}
\displaystyle
x(z) = \ln{\left((1-{z\over z_0})(1-{1\over z z_0})\right)}
\cr \cr
\displaystyle
y(z) = \ln{\left({1\over z}\,\,\left( 1-z/z_0\over 1-1/z z_0\right)^{p\over 2}\right)}
\end{array}\right. .
\eeq
This last spectral curve is such that $\ee{x}$ and $\ee{y}$ are rational functions of $z$, and thus by eliminating $z$, there exists a polynomial $H(\ee{x},\ee{y})$ such that:
\beq
H(\ee{x},\ee{y})=0.
\eeq
This equation is precisely the singular locus of the mirror manifold of $X_p$.
The full mirror manifold (not only its singular locus) is the 3 dimensional submanifold of ${\mathbb C}^4$ locally given by $\{(x,y,\om_+,\om_-)\in {\mathbb C}^4\, / \,\,\, H(\ee{x},\ee{y})=\om_+\om_- \}$.
See section \ref{sectopostring}.

\bigskip

$\bullet$ The following spectral curve is of genus $\genus=1$, it is algebraic but not rational:
\beq
\spcurve={\mathbb C}/({\mathbb Z}+\tau {\mathbb Z})
\virg
x(z) = \wp(z,\tau)
\virg y(z) = \wp'(z,\tau)
\eeq
where $\wp$ is the Weierstrass function, and $\spcurve$ is the torus of modulus $\tau$.
It is algebraic because the Weierstrass function obeys the differential equation:
\beq
\wp'^2 = 4\wp^3 - g_2 \wp - g_3 .
\eeq
This spectral curve is called the Seiberg-Witten curve since it first appeared in a solution
to ${\cal{N}}=2$ Supersymmetric Yang-Mills theory proposed by Seiberg and Witten in \cite{SeibWit}.

\subsection{Geometry of the spectral curve}

\subsubsection{Genus and cycles}
\label{defgenuscycle}

The only compact Riemann surface of genus $\genus=0$ is the Riemann sphere ${\mathbb P}^1={\mathbb C}\cup\{\infty\}$. It is simply connected.

\bigskip

A compact Riemann surface $\spcurve$ of genus $\genus\geq 1$, can be equipped with a symplectic basis (not unique) of $2\genus$ non-contractible cycles such that:
\beq
\acycle_{i}\cap\bcycle_j = \delta_{i,j}
\virg
\acycle_{i}\cap\acycle_j = 0
\virg
\bcycle_{i}\cap\bcycle_j = 0 .
\eeq
They are such that $\spcurve \backslash \left( \cup_i \acycle_i \cup_i \bcycle_i\right)$ is a simply connected domain of $\spcurve$, which we shall call the fundamental domain.

$${\epsfxsize 9cm\epsffile{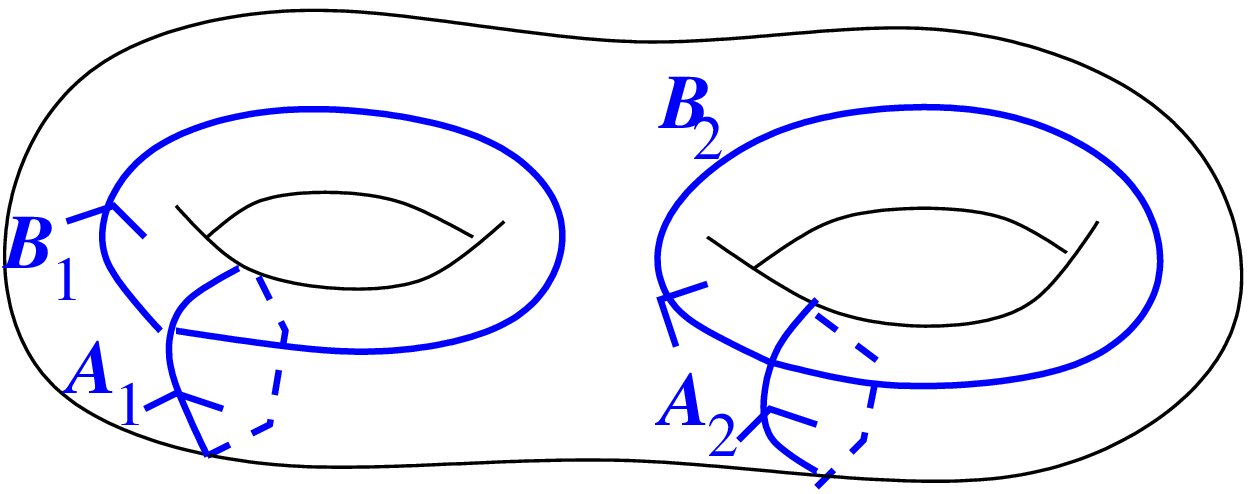}}$$

The choice of cycles and of a fundamental domain is rather arbitrary, and is not unique.
Many of the quantities we are going to consider depend on that choice.

The quantities which do not depend on that choice are called modular invariant.

\medskip

On a compact Riemann surface of genus $\genus\geq 1$, there exist holomorphic differential forms (analytical everywhere on $\spcurve$, in particular with no pole). Those holomorphic forms clearly form a vector space (linear combinations are also holomorphic) over ${\mathbb C}$, and this vector space has dimension $\genus$.

When we have a choice of cycles $\acycle_i,\bcycle_j$, it is possible to choose a basis (which is unique), which we call $du_1,\dots, du_{\genus}$, and normalized such that:
\beq
\oint_{\acycle_i} du_j =\delta_{i,j} .
\eeq
Once we have defined those $du_j$'s, we can compute the following Riemann matrix of periods:
\beq
\tau_{i,j} =  \oint_{\bcycle_i} du_j .
\eeq
This matrix $\tau_{i,j}$ is symmetric, and its imaginary part is positive definite:
\beq
\tau_{i,j}=\tau_{j,i}
\virg
\Im \tau >0 .
\eeq

\subsubsection{Abel map}

Consider an arbitrary origin $o$ in the fundamental domain, and fixed throughout all this article.

For any point $z$ in the fundamental domain, the vector $(u_1(z),\dots,u_{\genus}(z))$:
\beq
u_i(z) = \int^z_o du_i
\eeq
where the integration path is in the fundamental domain, is called the Abel map of $z$.
It is a vector in ${\mathbb C}^{\genus}$.
It depends on the choice of $o$ by an additive constant, and it depends on the choice of the fundamental domain, by a vector in the lattice ${\mathbb Z}^{\genus}+\tau {\mathbb Z}^{\genus}$.
The quotient ${\mathbb C}^{\genus}/({\mathbb Z}^{\genus}+\tau {\mathbb Z}^{\genus})$ is called the Jacobian.

The Abel map, sends points of $\spcurve$ to points in the Jacobian.

\subsubsection{Bergmann kernel}
\label{secBergmann}

Given a choice of cycles, we define the Bergmann kernel:
\beq
B(z_1,z_2)
\eeq
as the unique bilinear differential having one double pole at $z_1=z_2$ (it is called  "2nd kind") and no other pole, and such that, in any local parameter $z$:
\beq
B(z_1,z_2) \mathop{\sim}_{z_1\to z_2} {dz_1 dz_2\over (z_1-z_2)^2}+{\rm reg}
\virg
\forall i=1,\dots,\genus,\,\,\,
\oint_{\acycle_i} B(z_1,z_2)=0 .
\eeq

One should keep in mind that the Bergmann kernel depends only on $\spcurve$, and not on the functions $x$ and $y$.

\medskip

The Bermann kernel can be seen as the derivative of the Green function, i.e. the solution of the heat kernel equation on $\spcurve$.

\medskip

The Bergmann kernel is clearly unique because the difference of two Bergmann kernels would have no pole, and vanishing $\acycle$-cycle integrals, therefore it would vanish. It is also interesting to note that it is symmetric in its variables $z_1$ and $z_2$.

\medskip
{\bf Examples:}
\begin{itemize}
\item if $\spcurve={\mathbb P}^1={\mathbb C}\cup\{\infty\}=$the Riemann Sphere, the Bergmann kernel is a rational expression:
\beq
B(z_1,z_2) = {dz_1 dz_2\over (z_1-z_2)^2} .
\eeq
Most of the applications between section \ref{secMM} and section \ref{sectopostring}, will be on $\spcurve=\mathbb P^1$, and will use this rational Bergmann kernel.

\item if $\spcurve={\mathbb C}/({\mathbb Z}+\tau{\mathbb Z})=$Torus of modulus $\tau$, the Bergmann kernel is
\beq\label{Bergmanntorus}
B(z_1,z_2) = \left(\wp(z_1-z_2,\tau)+{\pi\over \Im\tau}\right)\,dz_1 dz_2
\eeq
where $\wp$ is the Weierstrass elliptical function.

\item if $\spcurve$ is a compact Riemann surface of genus $\genus\geq 1$, of Riemann matrix of periods $\tau_{i,j}$,
the Bergmann kernel is
\beq
B(z_1,z_2) = d_{z_1} d_{z_2} \,\ln{(\theta(u(z_1)-u(z_2)-c,\tau))}
\eeq
where $u(z)$ is the Abel map, $c$ is an odd characteristic, and $\theta$ is the Riemann theta function of genus $\genus$ (cf \cite{Fay, Farkas} for theta-functions).

\end{itemize}

\subsubsection{Generalized Bergmann kernel}
\label{secBkappa}

Given an arbitrary symmetric matrix $\kappa$ of size $\genus\times \genus$, we consider a "deformed" Bergmann kernel:
\beq
B_\kappa(z_1,z_2) = B(z_1,z_2) + 2i\pi \sum_{i,j=1}^{\genus} \kappa_{i,j}\,\, du_i(z_1) du_j(z_2).
\eeq
If $\kappa=0$ we recover the usual Bergmann kernel $B_0=B$.

\smallskip
The reason for introducing this $\kappa$, is that a change of basis of cycles and fundamental domain, can be rewritten as a change of $\kappa$.

Indeed, perform a $Sp_{2\genus}({\mathbb Z})$ change of symplectic basis of cycles ($C,D,\td{C},\td{D}$ have coefficients in ${\mathbb Z}$ and $CD^t=DC^t$, $\td{C}\td{D}^t=\td{D}\td{C}^t$, $C\td{D}^t-D\td{C}^t=\mathbf{1}$):
\beq
\acycle_i=\sum_j C_{i,j} \acycle'_j + \sum_j D_{i,j} \bcycle'_j
\virg
\bcycle_i=\sum_j \td{C}_{i,j} \acycle'_j + \sum_j \td{D}_{i,j} \bcycle'_j \, .
\eeq
The Riemann matrix of periods $\tau'$ in the new basis $\acycle',\bcycle'$, is related to the old one by the modular transformation:
\beq
\tau' = (\td{D}-\tau D)^{-1} (\tau C-\td{C})
\eeq
%\beq
%A'=\td{D}^t A - D^t B
%\virg
%B'=C^t B - \td{C}^t A
%\eeq
%\beq
%du = (\td{D}-\tau D) du'
%\eeq
%\beq
%B-\tau A = (\td{D}-\tau D).(B'-\tau' A')
%\eeq
and the Bergmann kernel changes as:
\beq
B_0 \to B'_{0}= B_{0} + 2i\pi \sum_{i,j=1}^{\genus} \kappa_{i,j}\,\, du_i(z_1) du_j(z_2)
\virg \kappa = (\td{D} D^{-1} -\tau )^{-1}
\eeq
in other words, the change of cycles can be reabsorbed as a change of $\kappa$.

More generally, the kernel $B_{\kappa}$ in a basis ${\acycle,\bcycle}$ is equal to $B'_{\kappa'}$ in the basis ${\acycle',\bcycle'}$, where:
\beq
\kappa' =  (\td{D}^t-D^t \tau)\kappa (\td{D}-\tau D) - (\td{D}^t-D^t \tau) D.
\eeq

%\medskip
%By choosing $\kappa=0$ we recover the usual Bergmann kernel.

\medskip
From now on, we will always consider $B_\kappa$, and we will write $B$ instead of $B_\kappa$, i.e. we will omit the $\kappa$ subscript, unless ambiguity.
However, for most of the practical applications, one often chooses $\kappa=0$.

\subsubsection{Schiffer kernel}\label{Schiffer}

In particular if we choose $\kappa$ to be the Zamolodchikov K\"ahler metric:
\beq
\kappa = (\ovl\tau - \tau)^{-1} = {i\over 2}\,\, (\Im\,\tau)^{-1} ,
\eeq
we see that in the new basis $\acycle',\bcycle'$, the matrix $\kappa$ becomes
\beq
\kappa' = {i\over 2}\,\, (\Im\tau')^{-1} ,
\eeq
i.e. it takes the same form as in the initial basis.
Therefore, with this special value of $\kappa$, the Bergmann kernel $B_\kappa$ is called the Schiffer kernel \cite{BergSchif} and it is modular invariant: it does not depend on a choice of cycles.
However, the price to pay to have modular invariance, is to have a non analytical dependence in $\tau$, and thus a non  analytical dependence in the spectral curve.
This incompatibility between analyticity and modular invariance is the origin of the so-called "holomorphic anomaly equation", see section \ref{secholoano}.

\medskip

{\bf Example:} if $\spcurve={\mathbb C}/({\mathbb Z}+\tau{\mathbb Z})=$Torus of modulus $\tau$, the Schiffer kernel is
\beq
B(z_1,z_2) = \wp(z_1-z_2,\tau)\,dz_1 dz_2
\eeq
where $\wp$ is the Weierstrass elliptical function. Compare with \eq{Bergmanntorus}.

\subsubsection{Branchpoints}

Branchpoints are the points with a vertical tangent, they are the {\bf zeroes of ${dx}$}.
Let us write them $a_i$, $i=1,\dots,\#$bp:
\beq
\forall i,\quad dx(a_i)=0 .
\eeq
Since we consider a regular spectral curve, all branchpoints are simple zeroes of $dx$, the curve $y(x)$ behaves locally like a square root $y(z)\sim y(a_i)+C_i\sqrt{x(z)-x(a_i)}$, near a branchpoint $a_i$, and thus, for any $z$ close to $a_i$, there is exactly one point $\bar{z}\neq z$ in the vicinity of $a_i$ such that:
\beq
x(\bar{z})=x(z).
\eeq
$\bar{z}$ is called the {\bf conjugated point} of $z$. It is defined locally near each branchpoint $a_i$, and it is not necessarily defined globally.

\medskip

{\bf Examples:}
\begin{itemize}
\item {\bf enumeration of maps $\mathbf \approx$ 1-matrix model in the 1-cut case}:

in this case we have $\spcurve=$Riemann sphere, and (see section \ref{sec1MMrational}):
\beq
x(z)=\alpha+\gamma(z+1/z)
\virg
dx(z) = x'(z) dz = \gamma(1-z^{-2})\,dz .
\eeq
The zeroes of $dx(z)$ are $z=\pm 1$, and we clearly have $\bar{z}=1/z$:
\beq
a_1=1,\,\,\, a_2=-1 \virg \bar{z}=1/z.
\eeq
In this case $\bar{z}$ is defined globally.

\item {\bf pure gravity $(3,2)$}:

in that case we have $\spcurve=$Riemann sphere, and
\beq
x(z)= z^2-2
\virg
dx(z) =  2z\, dz.
\eeq
The only zeroe of $dx(z)$ is $z=0$, and we have $\bar{z}=-z$:
\beq
a=0 \virg \bar{z}=-z .
\eeq
In this case $\bar{z}$ is defined globally.

\item {\bf Ising model $(4,3)$}:

in that case we have $\spcurve=$Riemann sphere, and
\beq
x(z)= z^3-3z
\virg
dx(z) =  3(z^2-1)\, dz.
\eeq
The zeroes of $dx(z)$ are $a_i=\pm 1$, and near $a_i=\pm 1$ we have:
\beq
a_i=\pm 1 \virg \bar{z}=-{1\over 2}(z-a_i \sqrt{12-3z^2}).
\eeq
In this case $\bar{z}$ is not defined globally, and it depends on $a_i$.

\end{itemize}

\subsubsection{Recursion Kernel}

For any $z_0\in \spcurve$, and any $z$ close to a branchpoint, we define the recursion kernel:
\beq\label{defkernelK}
\encadremath{
K(z_0,z) = {-1\over 2}\,\,{\int_{z'=\bar{z}}^z B(z_0,z') \over (y(z)-y(\bar{z}))\, dx(z)}
}\eeq
where the integral is taken in a small domain in the vicinity of the concerned branchpoint.
\smallskip

$K(z_0,z)$ is a meromorphic 1-form in the variable $z_0$, it is defined globally for all $ z_0\in \spcurve$, it has simple poles at $z_0=z$ and $z_0=\bar z$.
\smallskip

On the contrary, in the variable $z$, the kernel $K(z_0,z)$ is defined only locally near branchpoints $z\sim a_i$, and it is the inverse of a differential. As we shall see below, $K(z_0,z)$ will always be used only in the vicinity of branchpoints, and it will always be multiplied by a quadratic differential in $z$, so that the product will be a differential form.
\smallskip

Let us notice that $K(z_0,z)=K(z_0,\bar{z})$, and that $K(z_0,z)$ has a simple pole when $z$ approaches the branchpoint.
Using De L'Hopital's rule, the leading behavior near the branchpoint is:
\beq
K(z_0,z) \mathop{\sim}_{z\to a_i} - {B(z,z_0)\over 2\,dy(z) dx(z)} + {\rm regular}\,\dots
\eeq

\subsection{Correlation functions}

We start by defining a sequence of meromorphic $n$-forms $\om_n^{(g)}$ with $n=1,2,\dots$ and $g=0,1,2,\dots$,
called correlators or correlation functions, by the following recursion:

\bd\label{defCorrnspinv}

Given a spectral curve ${\cal E} = (\spcurve,x,y)$, and a matrix $\kappa$ (see section \ref{secBkappa}),
we define recursively the following meromorphic forms:
\beq
\om_1^{(0)}(z)=-y(z)dx(z)
\eeq
\beq
\om_2^{(0)}(z_1,z_2) = B(z_1,z_2)
\eeq
and if $2g-2+n\geq 0$, and $J$ is a collective notation for $n$ variables $J=\{z_1,\dots,z_n\}$:
\beq\label{defWnginv}
\encadremath{
\om_{n+1}^{(g)}(z_0,J)
= \sum_i \Res_{z\to a_i}\, K(z_0,z)\,\Big[
\om_{n+2}^{(g-1)}(z,\bar{z},J)
+ \sum_{h=0}^g\sum'_{I\subset J} \om_{1+|I|}^{(h)}(z,I) \om_{1+n-|I|}^{(g-h)}(\bar{z},J\backslash I) \Big]
}\eeq
where ${\displaystyle \sum'}$ in the RHS means that we exclude the terms with $(h,I)=(0,\emptyset)$, and $(g,J)$.
\smallskip

This definition is indeed a recursive one, because all the terms in the RHS have a strictly smaller $2g-2+n$ than the LHS.

\smallskip
The functions $\om_n^{(g)}$ with $2-2g-n<0$ are called stable, the others are unstable (the only unstable ones are thus $\om_1^{(0)}$ and $\om_2^{(0)}$).
\ed

$\om_n^{(g)}(z_1,\dots,z_n)$ is a meromorphic 1-form on $\spcurve$ in each variables $z_i$.
It can be proved by recursion, that it is in fact a symmetric form.
Moreover, if $2-2g-n<0$, its only poles are at branchpoints $z_i\to a_j$, and have no residues:
\beq\label{omngnores}
\Res_{z_1\to a_i} \om_n^{(g)}(z_1,z_2,\dots,z_n) = 0.
\eeq
Those properties can be proved by recursion, and we refer the reader to \cite{EOinvariants}.

\subsection{Free energies}

The previous definition, defines $\om_n^{(g)}$ only if $n\geq 1$.
Now, we define $F_g=\om_0^{(g)}$, called ``Free energies'' or ``symplectic invariant of degree $2-2g$'', by the following:

\bd\label{defsympinvFg}
 {\bf Symplectic invariants}

We define for $g\geq 2$:
\beq\label{defFginv}
\encadremath{
F_g({\cal E},\kappa) = \om_0^{(g)} = {1\over 2-2g}\,\sum_i \Res_{z\to a_i}\, \Phi(z)\,\om_1^{(g)}(z)
}\eeq
where $\Phi$ is any function defined locally near branchpoints, such that $d\Phi=ydx$ ($\Phi$ is defined up to an additive constant, but thanks to \eq{omngnores}, $F_g$ does not depend on the choice of that constant).
\ed

The unstable cases $g=0$ and $g=1$ are special. We have:

\bd
For $g=1$ we define
\beq
\encadremath{
F_1 ={1\over 24} \ln{\left(\tau_B(\{x(a_i)\}) \,\, \prod_i y'(a_i)\right)}
}
\eeq
where we define:
\beq
y'(a_i)  = \mathop{{{\rm lim}}}_{z\to a_i}\, {y(z)-y(a_i)\over \sqrt{x(z)-x(a_i)}}
\eeq
and $\tau_B$ is the Bergmann $\tau$-function of Kokotov-Korotkin \cite{KoKo}.
%, it depends only and $\spcurve$ and on $\kappa$. It is independent of $x$ and $y$.
If $x(z)$ is a meromorphic function on $\spcurve$,  $\tau_B$ depends only on the values of $x$ at its branch points, i.e. $X_i=x(a_i)$. It is defined by:
\beq
{\partial \ln{\tau_B(\{X_i)\})}\over \partial X_i} = \Res_{z\to a_i} {B(z,\bar{z})\over dx(z)} .
\eeq
\ed
Notice that $B$ here stands for $B_\kappa$ with arbitrary $\kappa$.

\medskip

The definition of $F_0$ is more involved, and we refer the reader to \cite{EOinvariants}.
A convenient way to define $F_0$, is through its 3rd derivatives, using theorem \ref{thvariat} below.
In fact, all the $F_g$'s with $g\geq 1$ are obtained in terms of local behaviors around branchpoints, but $F_0$ depends on the whole spectral curve, not only on the vicinity of branchpoints.
In the context of topological strings, $F_0$ is called the prepotential.

%%%%%%%%%%%%%%%%%%%%%%%%%%%%%%%%%%%%%%%%%%%%%%%%%%%%%%%%%%%%%%%%%%%%%%%%%%%%%%%%%%%%%%%%%%%%%%%%%%%%%%%%%%%%%

\section{Diagrammatic representation} \label{sectiondiagrepresent}

The recursive  definitions of $\om_k^{(g)}$ and $F^{(g)}$ can be represented {\bf graphically}.

We represent the $k-$form $\om_k^{(g)}(p_1,\dots,p_k)$ as a ``blob-like surface'' with $g$ holes and
$k$ legs (or punctures) labeled with the variables $p_1,\dots, p_k$, and $F^{(g)}=\om_0^{(g)}$ with $0$ legs and
$g$ holes.
\beq
\om_{k+1}^{(g)}(p,p_1,\dots,p_k):=\begin{array}{r}
{\epsfxsize 4.5cm\epsffile{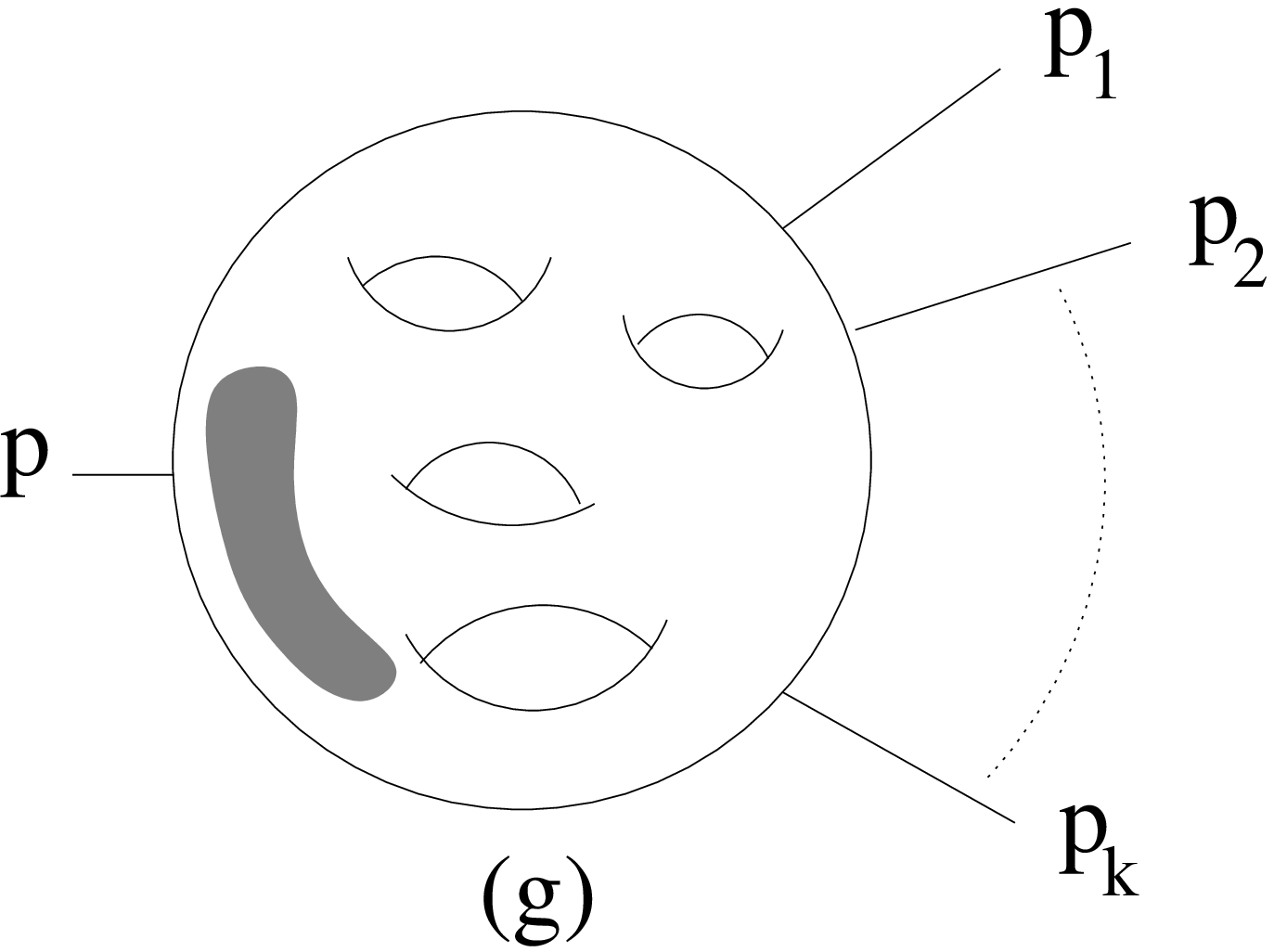}}
\end{array}
\virg
F^{(g)}:= \begin{array}{r}
{\epsfxsize 2.5cm\epsffile{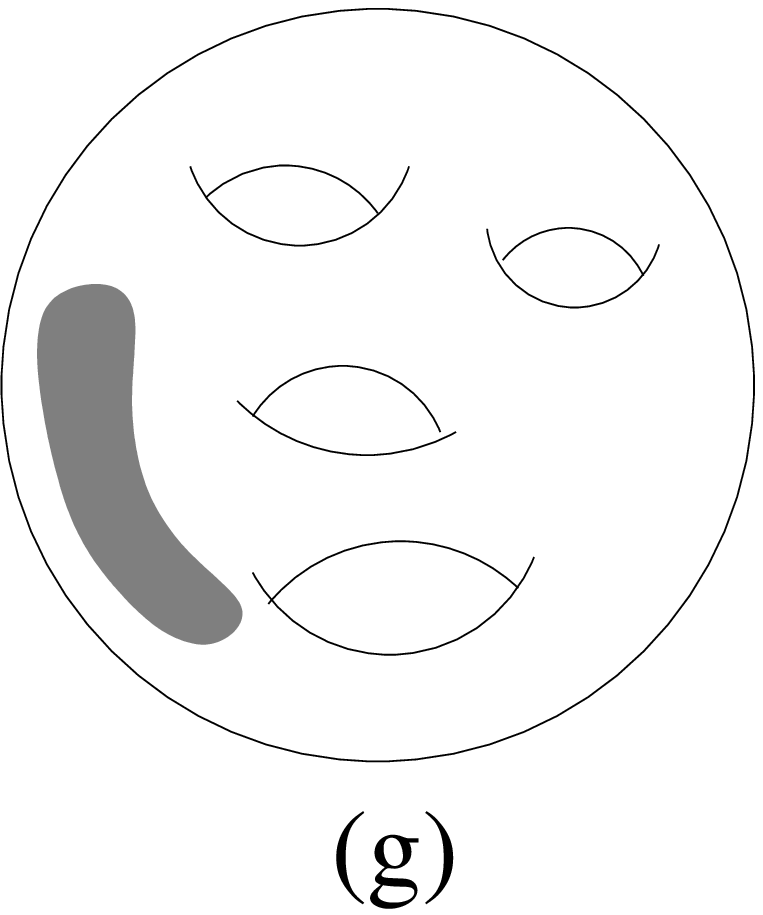}}
\end{array} .
\eeq

We represent the Bergmann kernel $B(p,q)$ (which is also $\om_2^{(0)}$, i.e. a blob with 2 legs and no hole) as a straight non-oriented line between $p$ and $q$
\beq
B(p,q):= \begin{array}{r}
{\epsfxsize 2cm\epsffile{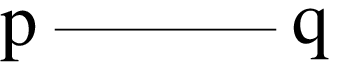}}
\end{array}.
\eeq

We represent $K(p,q)$ as a straight arrowed line with the arrow from $p$ towards $q$, and with a tri-valent vertex whose left leg is $q$ and right leg is $\overline{q}$
\beq
K(p,q):= \begin{array}{r}
{\epsfxsize 3cm\epsffile{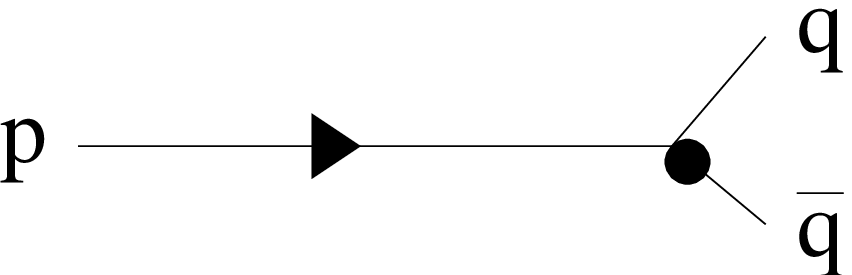}}
\end{array}.
\eeq

\medskip

\subsubsection*{Graphs}

\bd\label{defgraphs}
For any $k\geq 0$ and $g\geq 0$ such that $k+2g\geq 3$, we define:

Let ${\cal G}_{k+1}^{(g)}(p,p_1,\dots,p_k)$ be the set of connected trivalent graphs defined as follows:
\begin{enumerate}

\item there are $2g+k-1$ trivalent vertices called vertices.
\item there is one 1-valent vertex labelled by $p$, called the root.
\item there are $k$ 1-valent vertices labelled with $p_1,\dots,p_k$ called the leaves.
\item There are $3g+2k-1$ edges.
\item Edges can be arrowed or non-arrowed. There are $k+g$ non-arrowed edges and $2g+k-1$ arrowed edges.
\item The edge starting at $p$ has an arrow leaving from the root $p$.
\item The $k$ edges ending at the leaves $p_1,\dots, p_k$ are non-arrowed.
\item The arrowed edges form a "spanning\footnote{It goes through all vertices.} planar\footnote{Planar tree means that the left child and right child are not equivalent. The right child is marked by a black disk on the outgoing edge.} binary skeleton\footnote{A binary skeleton tree is a binary tree from which we have removed the leaves, i.e. a tree with vertices of valence 1, 2 or 3.} tree" with root $p$. The arrows are oriented from root towards leaves. In particular, this induces a partial ordering of all vertices.
\item There are $k$ non-arrowed edges going from a vertex to a leaf, and $g$ non arrowed edges joining two inner vertices. Two inner vertices can be connected by a non arrowed edge only if one is the parent of the other following the arrows along the tree.
\item If an arrowed edge and a non-arrowed inner edge come out of a vertex, then the arrowed edge is the left child. This rule
only applies when the non-arrowed edge links this vertex to one of its descendants (not one of its parents).

\end{enumerate}

\ed

%We have the following useful lemma:
%\bl\label{lemgraphaddleg}
%There is a $1$ to $3g+2k-1$ map from ${\cal G}_{k+1}^{(g)}(p,{\bf p_K})$ to ${\cal G}_{k+2}^{(g)}(p,{\bf p_K},p_{k+1})$.
%\el
%\proof{If $G$ is  a graph in ${\cal G}_{k+2}^{(g)}(p,p_1,\dots,p_k,p_{k+1})$, remove the non-arrowed edge attached to the leaf $p_{k+1}$ and remove the corresponding vertex, and merge the incoming and the other outgoing edges of that vertex. You clearly get a graph $G'\in{\cal G}_{k+1}^{(g)}(p,p_1,\dots,p_k)$. It is clear that the same graph is obtained $3g+2k-1$ times (the number of edges of $G'$.
%And it is clear that from any $G'\in{\cal G}_{k+1}^{(g)}(p,p_1,\dots,p_k)$, you can obtain $3g+2k-1$ graphs $G\in{\cal G}_{k+2}^{(g)}(p,p_1,\dots,p_k,p_{k+1})$ by adding a new vertex on any edge, and linking this new vertex to the leaf $p_{k+1}$.}

\vs

\noindent {\bf Example of ${\cal G}_{1}^{(2)}(p)$}

As an example, let us build step by step all the graphs of ${\cal G}_{1}^{(2)}(p)$, i.e. $g=2$ and $k=0$.

We first draw all planar binary skeleton trees with one root $p$ and $2g+k-1=3$ arrowed edges:
\beq
\begin{array}{r}
{\epsfxsize 2cm\epsffile{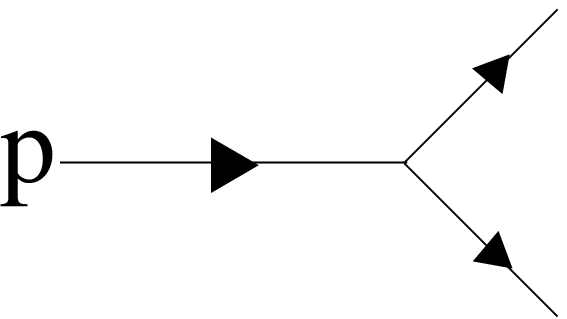}}
\end{array}
\virg
\begin{array}{r}
{\epsfxsize 2.5cm\epsffile{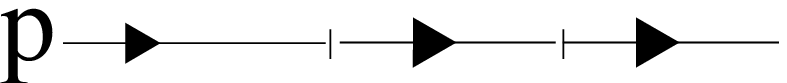}}
\end{array}.
\eeq
Then, we draw $g+k=2$ non-arrowed edges in all possible ways such that every vertex is trivalent, also satisfying rule 9) of definition \ref{defgraphs}. There is only
one possibility for the first tree, and two for the second one:
\beq
\begin{array}{r}
\begin{array}{r}
{\epsfxsize 2.5cm\epsffile{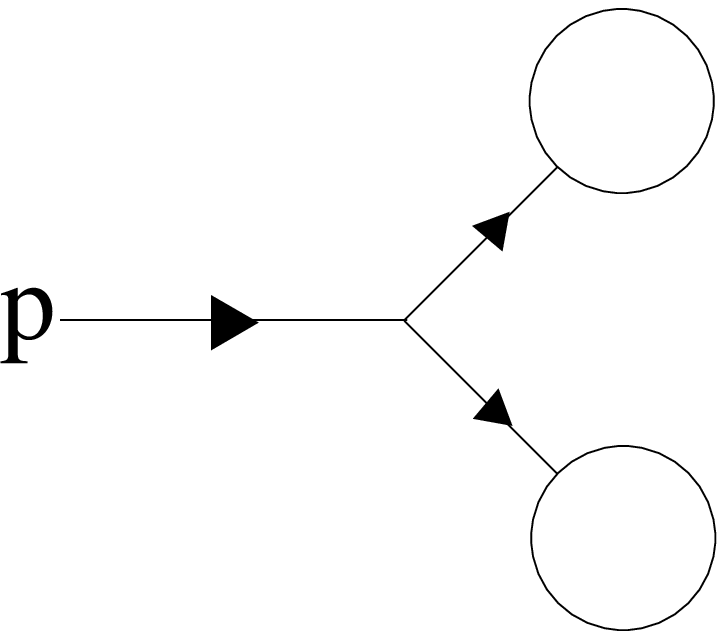}}
\end{array}
\virg
{\epsfxsize 2.2cm\epsffile{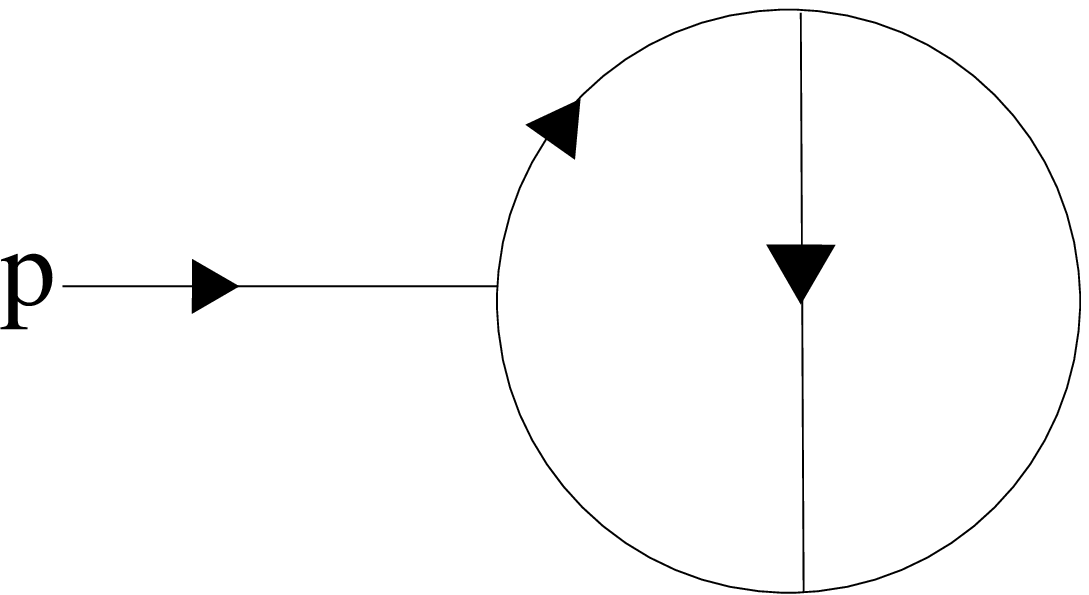}}
\end{array}
\virg
\begin{array}{r}
{\epsfxsize 3cm\epsffile{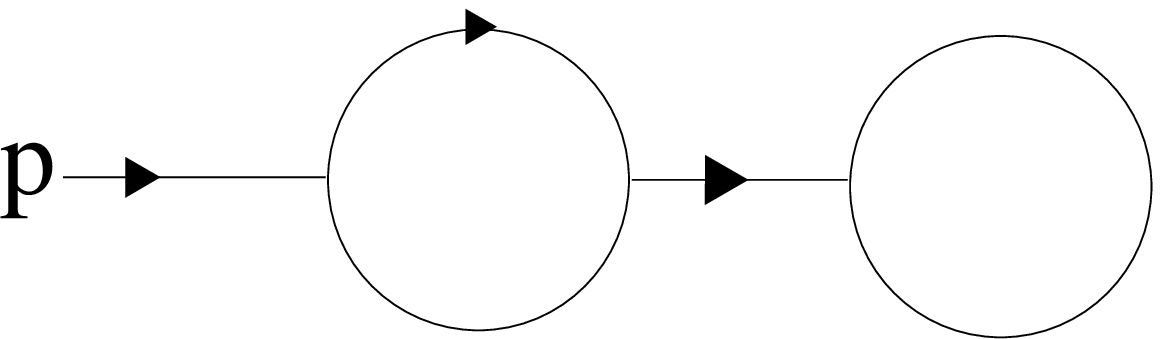}}
\end{array}.
\eeq

It just remains to specify the left and right children for each vertex. The only possibilities in accordance with rule 10) of def.\ref{defgraphs} are\footnote{ Note that the graphs are not
necessarily planar.}:
\bea
\begin{array}{r}
{\epsfxsize 2.2cm\epsffile{y123.eps}}
\end{array}
\virg
\begin{array}{r}
{\epsfxsize 2.5cm\epsffile{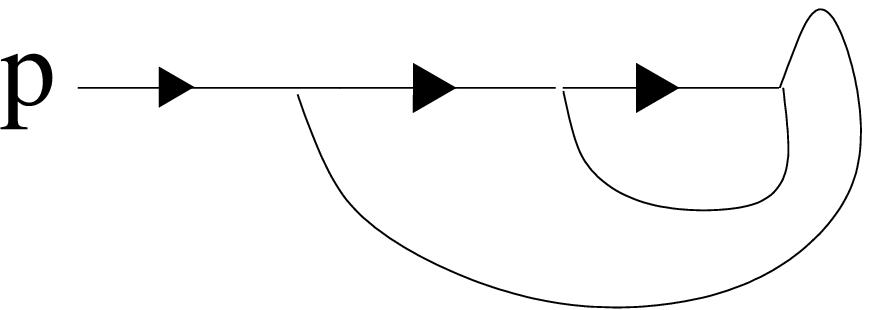}}
\end{array}
\virg
\begin{array}{r}
{\epsfxsize 3cm\epsffile{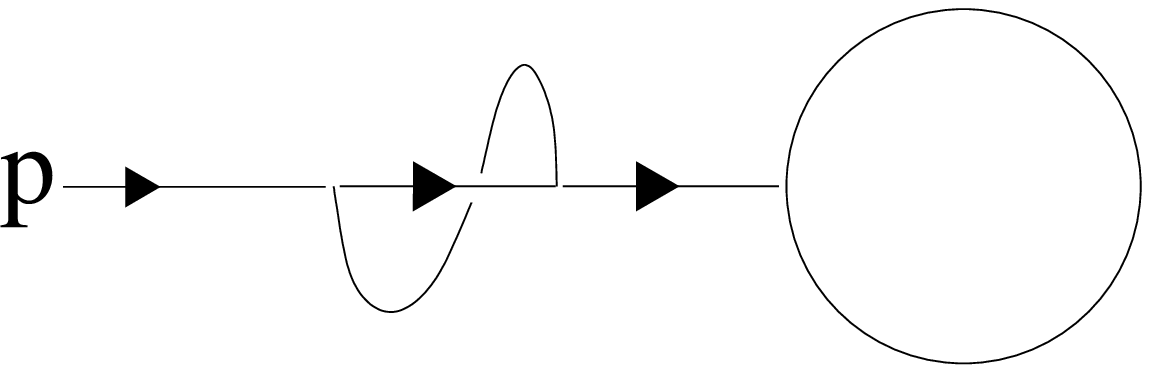}}
\end{array} \, ,\cr
\begin{array}{r}
{\epsfxsize 2.5cm\epsffile{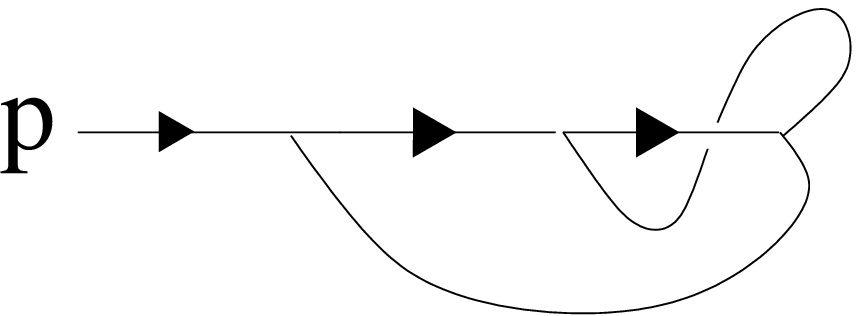}}
\end{array}
\virg
\begin{array}{r}
{\epsfxsize 3cm\epsffile{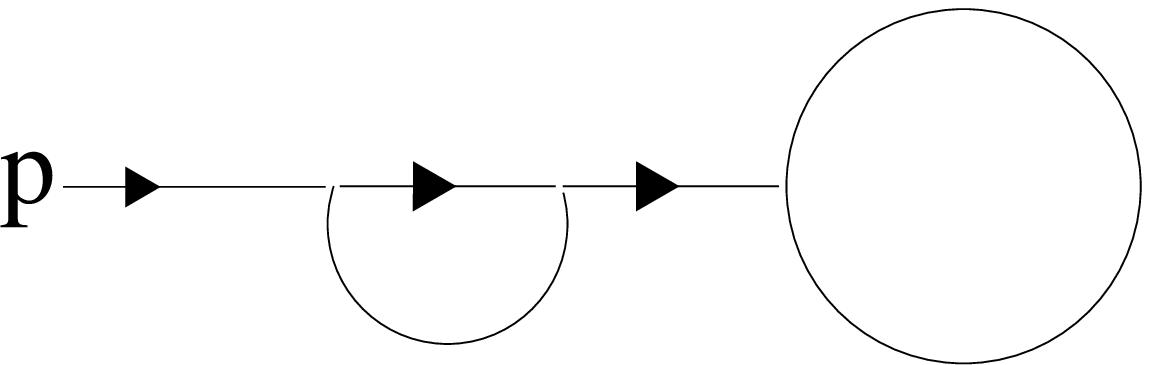}}
\end{array} .\cr
\eea

In order to simplify the drawing, we can draw a black dot to specify the right child. This way one gets only planar graphs:
\bea
\begin{array}{r}
{\epsfxsize 2.2cm\epsffile{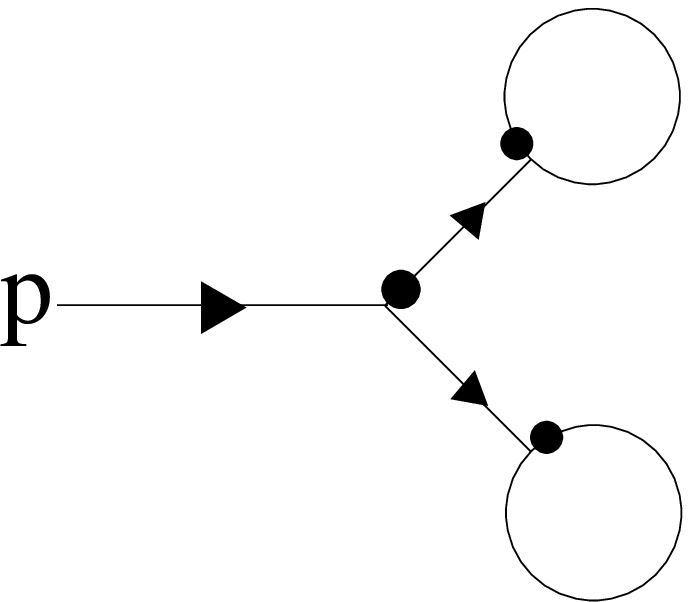}}
\end{array}
\virg
\begin{array}{r}
{\epsfxsize 2.5cm\epsffile{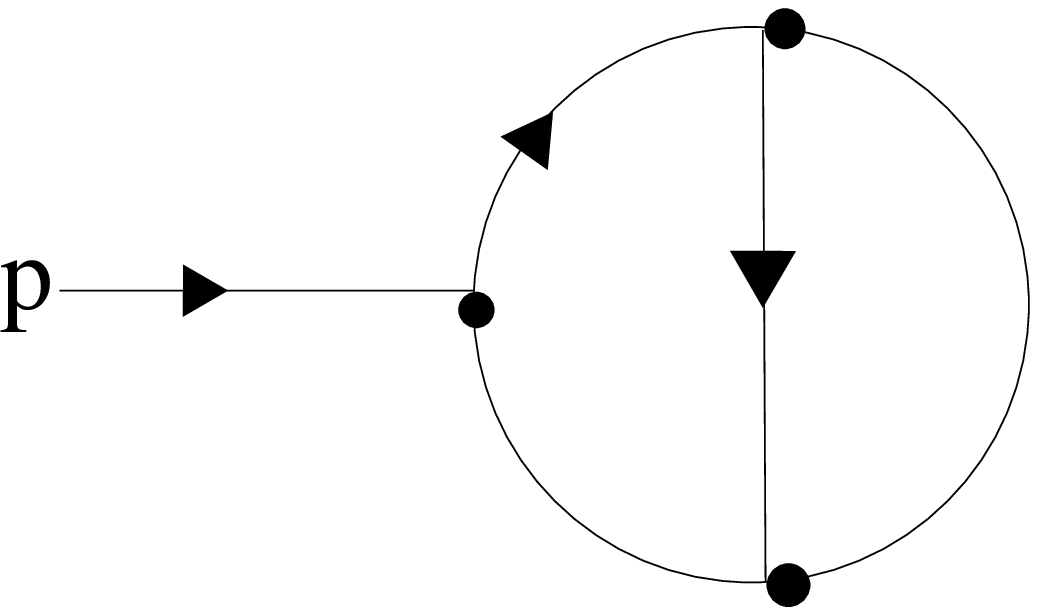}}
\end{array}
\virg
\begin{array}{r}
{\epsfxsize 3cm\epsffile{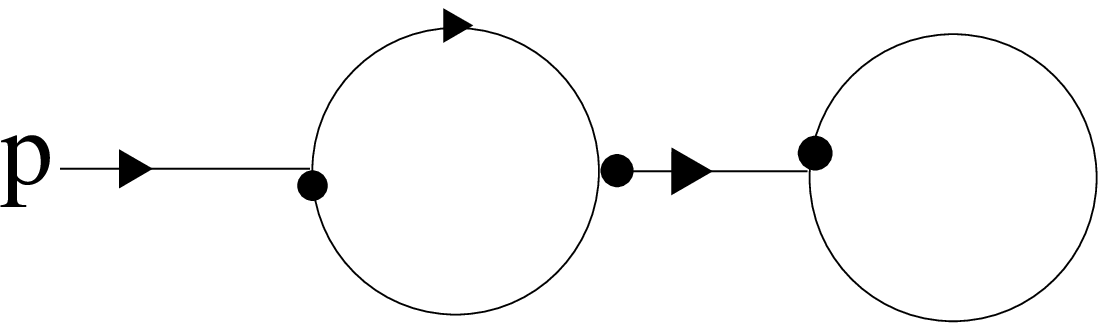}}
\end{array} \, ,\cr
\begin{array}{r}
{\epsfxsize 2.5cm\epsffile{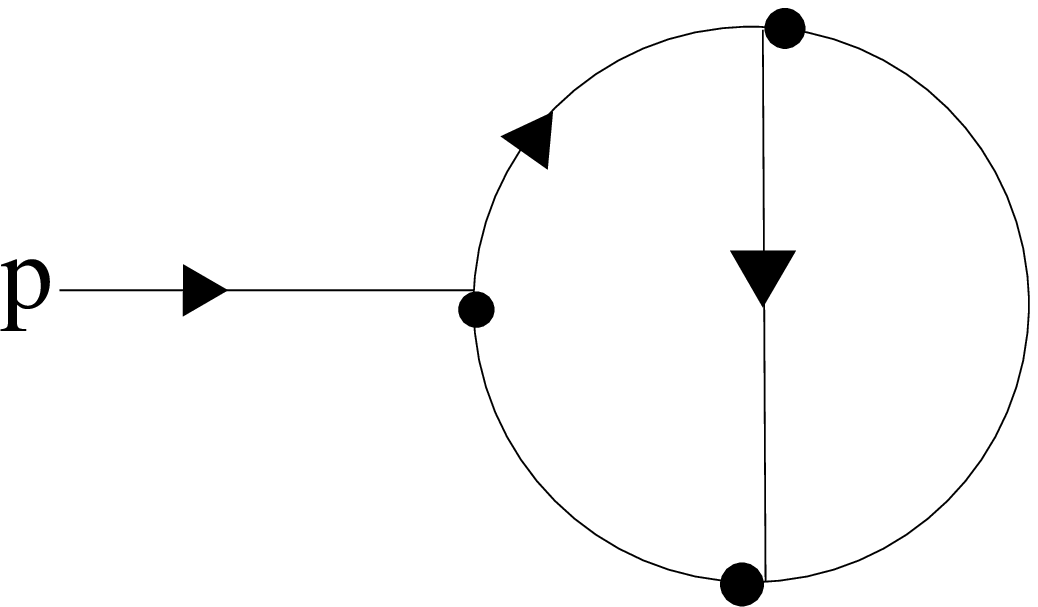}}
\end{array}
\virg
\begin{array}{r}
{\epsfxsize 3cm\epsffile{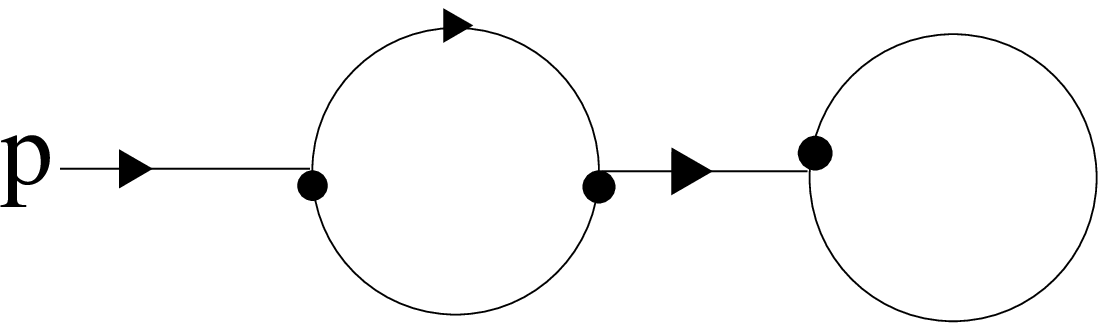}}
\end{array} \, .\cr
\eea
Remark that without the prescriptions 9) and 10), one would get 13 different graphs whereas we only have 5.

\subsubsection*{Weight of a graph}

Consider a graph $G\in {\cal G}_{k+1}^{(g)}(p,p_1,\dots,p_k)$.
Then, to each vertex $i=1,\dots,2g+k-1$ of $G$, we associate a label $q_i \in \spcurve$, and we associate $q_i$ to the beginning of the left child edge, and $\overline{q}_i$ to the right child edge.
Thus, each edge (arrowed or not), links two labels which are points on the spectral curve $\spcurve$.
\begin{itemize}
\item To an arrowed edge going from $q'$ towards $q$, we associate a factor $K(q',q)$.
\item To a non arrowed edge going between $q'$ and $q$ we associate a factor $B(q',q)$.
\item Following the arrows backwards (i.e.  from leaves to root), for each vertex $q$, we take the sum over all branchpoints $a_i$ of residues at $q\to a_i$.
\end{itemize}
After taking all the residues, we get the weight of the graph:
\beq
w(G)
\eeq
which is a multilinear form in $p,p_1,\dots,p_k$.

Similarly, we define weights of linear combinations of graphs by:
\beq
w(\alpha G_1 + \beta G_2) = \alpha w(G_1) + \beta w(G_2)
\eeq
and for a disconnected graph, i.e. a product of two graphs:
\beq
w( G_1  G_2) = w(G_1)  w(G_2)
.
\eeq

\bt\label{thdiagrepr}
We have:
\beq
\om_{k+1}^{(g)}(p,p_1,\dots,p_k) = \sum_{G\in {\cal G}_{k+1}^{(g)}(p,p_1,\dots,p_k)}\,\, w(G) = w\left(\sum_{G\in {\cal G}_{k+1}^{(g)}(p,p_1,\dots,p_k)}\,\, G\right) .
\eeq
\et
\proof{This is precisely what the recursion equations \ref{defWnginv} of def.\ref{defCorrnspinv} are doing.
Indeed, one can represent them diagrammatically by
\beq
\begin{array}{r}
{\epsfxsize 10cm\epsffile{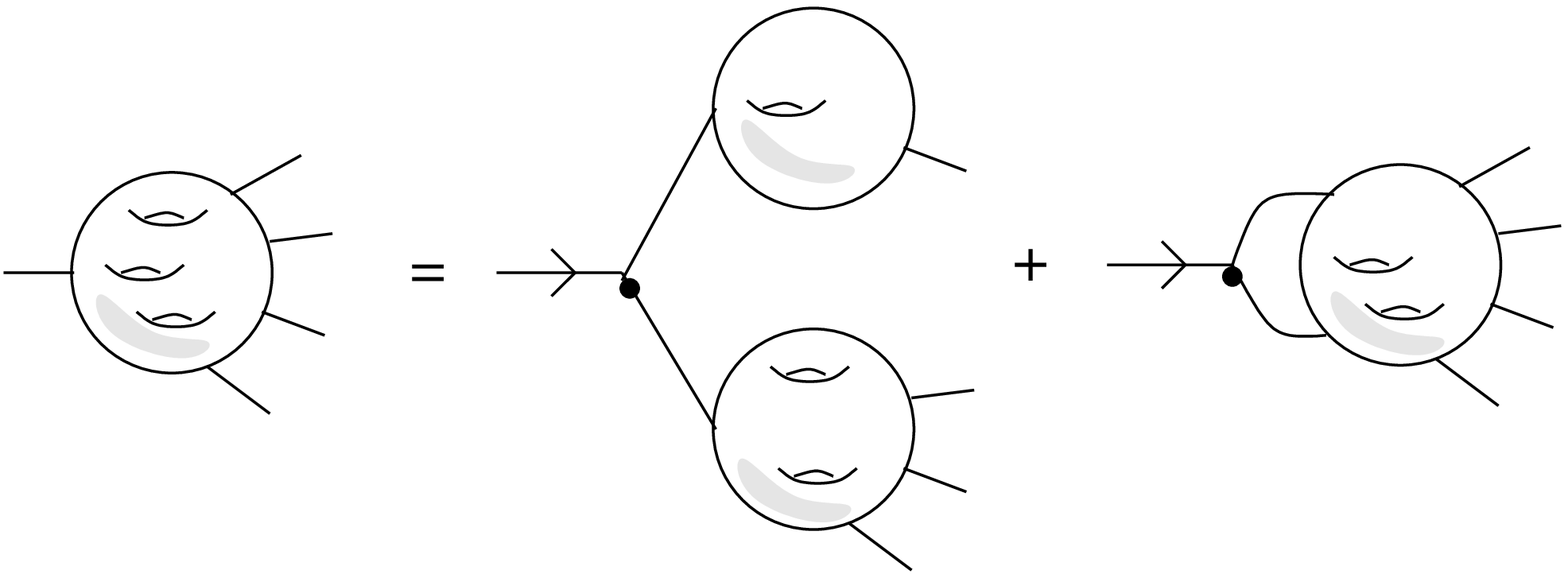}}
\end{array}.
\eeq
}

\bigskip

Such graphical notations are very convenient, and are a good support for intuition and even help proving some relationships.
It was immediately noticed after \cite{eynloop1mat} that those diagrams look very much like Feynman graphs, and there was a hope that they could be the Feynman's graphs for the Kodaira--Spencer quantum field theory.
But they ARE NOT Feynman graphs, because Feynman graphs can't have non-local restrictions like the fact that non oriented lines can join only a vertex and one of its descendent.

Those graphs are merely a notation for the recursive definition \ref{defWnginv}.

\bl\label{lemmasymfactor} {\bf Symmetry factor:}

The weight of two graphs differing by the exchange of the right and left children of a vertex are the same. Indeed, the distinction between right and left child is just a way of encoding symmetry factors.
\el

\proof{
This property follows directly from the fact that $K(z_0,z)=K(z_0,\bar{z})$.
}

\subsection{Examples.}

Let us present some examples of correlation functions and free energy for low orders.

\subsubsection{3-point function.}

\bea
\om_{3}^{(0)}(p,p_1,p_2)&=& \begin{array}{r}
{\epsfxsize 3.5cm\epsffile{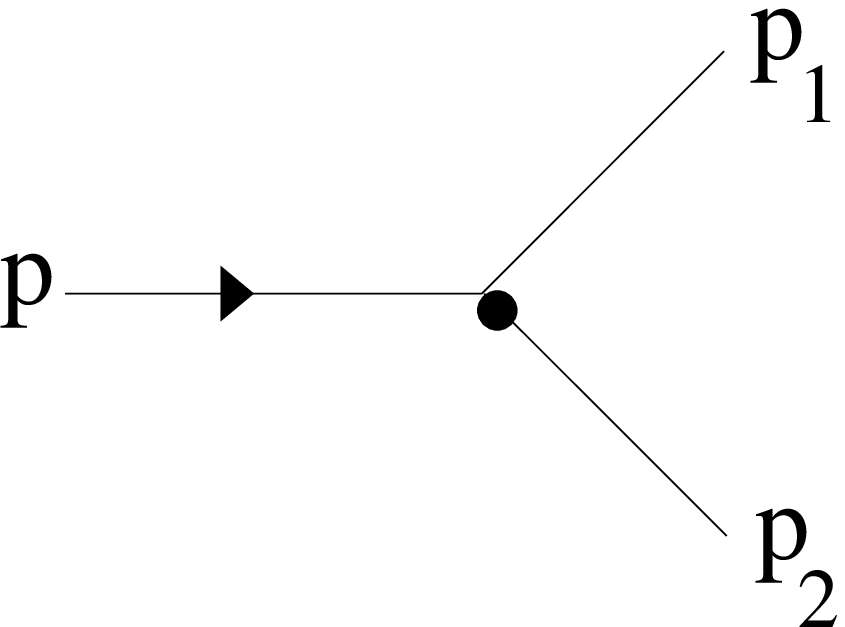}}
\end{array}
+
\begin{array}{r}
{\epsfxsize 3.5cm\epsffile{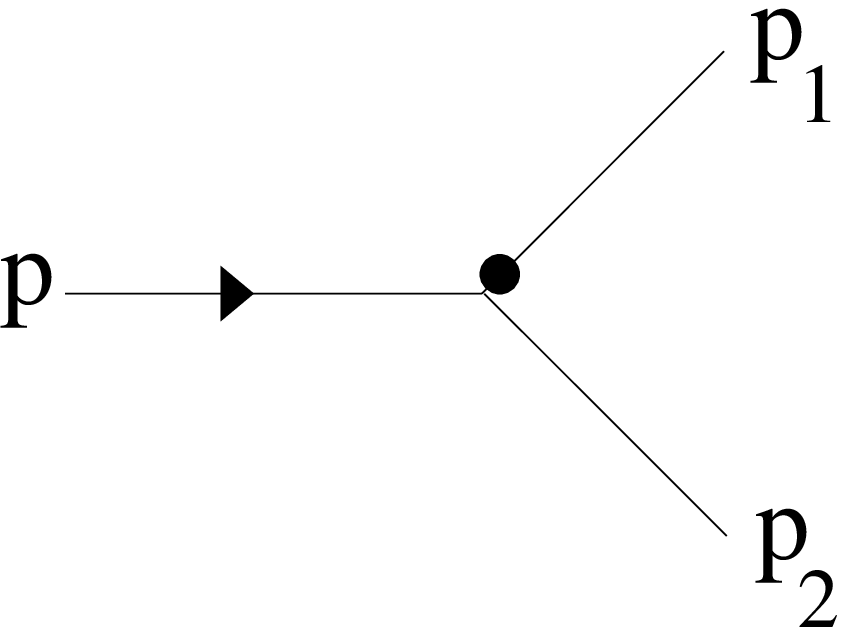}}
\end{array} \cr
&=& \Res_{q \to {\bf a}} K(p,q)\, \left[ B(q,p_1) B(\overline{q},p_2) + B(\overline{q},p_1) B(q,p_2) \right] \cr
&=& - 2\Res_{q \to {\bf a}} K(p,q)\, \left[ B(q,p_1)  B(q,p_2) \right] \cr
&=& \Res_{q \to {\bf a}} {B(q,p)\, B(q,p_1) \, B(q,p_2) \over dx(q)\,dy(q)} \cr
&=& \sum_i  {B(a_i,p)\, B(a_i,p_1) \, B(a_i,p_2) \over 2\, dy(a_i)\, dz_i(a_i)^2} \cr
\eea
where $z_i(z) =\sqrt{x(z)-z(a_i)}$ is a local coordinate near $a_i$.

\subsubsection{4-point function.}

\bea
\om_4^{(0)}(p,p_1,p_2,p_3)
&=&
\begin{array}{r}
{\epsfxsize 4cm\epsffile{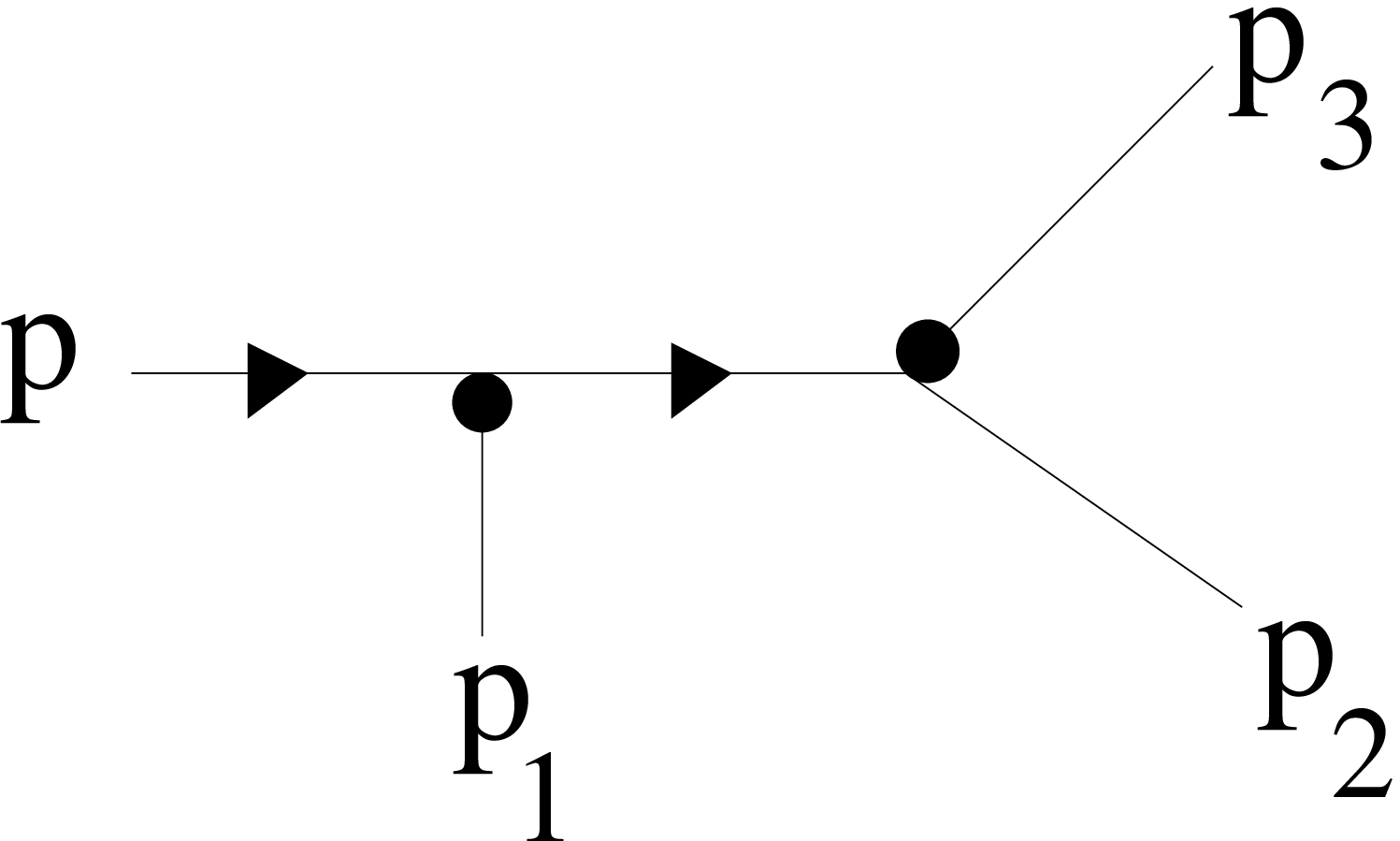}}
\end{array}
+ \;\;\;\; \hbox{5 permutations of (1,2,3)}\cr
&& + \begin{array}{r}
{\epsfxsize 4cm\epsffile{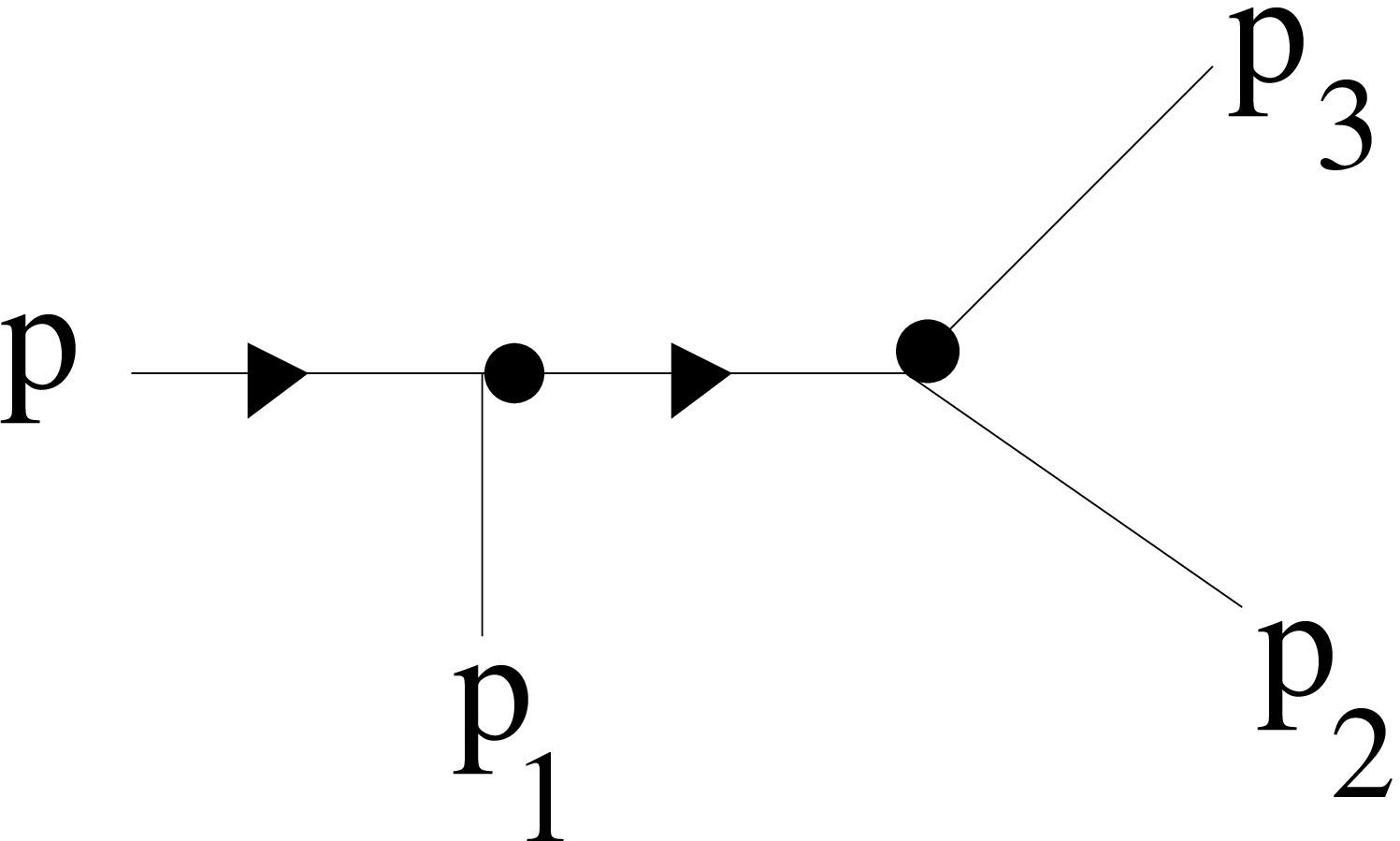}}
\end{array}
+ \;\;\;\; \hbox{5 permutations of (1,2,3)}\cr
&=& \Res_{q \to {\bf a}} \Res_{r \to {\bf a}} K(p,q)\,K(q,r)\,
\left[ B(\overline{q},p_1)B(r,p_2)B(\overline{r},p_3) \right. \cr
&& + B(\overline{q},p_1)B(\overline{r},p_2)B(r,p_3) + B(\overline{q},p_2)B(r,p_1)B(\overline{r},p_3) \cr
&& + B(\overline{q},p_2)B(\overline{r},p_1)B(r,p_3) + B(\overline{q},p_3)B(r,p_2)B(\overline{r},p_1) \cr
&& \left. + B(\overline{q},p_3)B(\overline{r},p_2)B(r,p_1)\right] \cr
&& + \Res_{q \to {\bf a}} \Res_{r \to {\bf a}} K(p,q)\,K(\overline{q},r)\,
\left[ B(q,p_1)B(r,p_2)B(\overline{r},p_3) \right. \cr
&& + B(q,p_1)B(\overline{r},p_2)B(r,p_3) + B(q,p_2)B(r,p_1)B(\overline{r},p_3)\cr
&& + B(q,p_2)B(\overline{r},p_1)B(r,p_3) + B(q,p_3)B(r,p_2)B(\overline{r},p_1) \cr
&& \left. + B(q,p_3)B(\overline{r},p_2)B(r,p_1)\right] .\cr
\eea

\subsubsection{1-point function to order 1.}

\bea
\om_{1}^{(1)}(p) &=& \begin{array}{r}
{\epsfxsize 4cm\epsffile{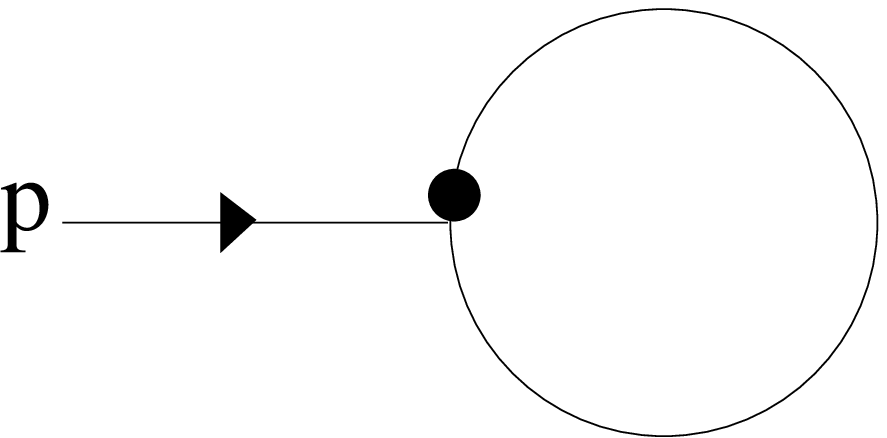}}
\end{array} \cr
&=& \Res_{q \to {\bf a}} K(p,q)\, B(q, \overline{q}) .\cr
\eea

\subsubsection{1-point function to order 2.}

\bea
\om_1^{(2)}(p)&=&
\begin{array}{r}
{\epsfxsize 4cm\epsffile{w121.eps}}
\end{array}
+
\begin{array}{r}
{\epsfxsize 4cm\epsffile{w123.eps}}
\end{array} \cr
&& +
\begin{array}{r}
{\epsfxsize 4cm\epsffile{w124.eps}}
\end{array}
+
\begin{array}{r}
{\epsfxsize 4cm\epsffile{w125.eps}}
\end{array} \cr
&& + \begin{array}{r}
{\epsfxsize 3.1cm\epsffile{w122.eps}}
\end{array}
\cr
&=&  \Res_{q \to {\bf a}} \Res_{r \to {\bf a}} \Res_{s \to {\bf a}} K(p,q)K(q,r)K(r,s) \,\, B(\overline{q},\overline{r}) B(s, \overline{s})\cr
&& +  \Res_{q \to {\bf a}} \Res_{r \to {\bf a}} \Res_{s \to {\bf a}} K(p,q)K(q,r)K(\overline{r},s) \,\, B(\overline{q},r) B(s, \overline{s})\cr
&& +  \Res_{q \to {\bf a}} \Res_{r \to {\bf a}} \Res_{s \to {\bf a}} K(p,q)K(q,r)K(r,s) \,\, B(\overline{q},\overline{s}) B(s, \overline{r})\cr
&& +  \Res_{q \to {\bf a}} \Res_{r \to {\bf a}} \Res_{s \to {\bf a}} K(p,q)K(q,r)K(r,s) \,\, B(\overline{q},s) B(\overline{s}, \overline{r})\cr
&& +  \Res_{q \to {\bf a}} \Res_{r \to {\bf a}} \Res_{s \to {\bf a}} K(p,q)K(\overline{q},r)K(q,s) \,\, B(\overline{r},r) B(\overline{s}, s)\cr
&=&
2 \begin{array}{r}
{\epsfxsize 4cm\epsffile{w121.eps}}
\end{array}
 +2
 \begin{array}{r}
{\epsfxsize 3cm\epsffile{w124.eps}}
\end{array}
+ \begin{array}{r}
{\epsfxsize 2.8cm\epsffile{w122.eps}}
\end{array}
\eea
where the last expression is obtained using lemma \ref{lemmasymfactor}.

\subsubsection{Free energy $F_2$.}

The second free energy reads
\bea
-2 F_2
&=& 2 \Res_{p \to {\bf a}} \Res_{q \to {\bf a}} \Res_{r \to {\bf a}} \Res_{s \to {\bf a}} \Phi(p)\, K(p,q)K(q,r)K(r,s) \,\, B(\overline{q},\overline{r}) B(s, \overline{s})\cr
&& +  2  \Res_{p \to {\bf a}} \Res_{q \to {\bf a}} \Res_{r \to {\bf a}} \Res_{s \to {\bf a}} \Phi(p)\, K(p,q)K(q,r)K(r,s) \,\, B(\overline{q},\overline{s}) B(s, \overline{r})\cr
&& + \Res_{p \to {\bf a}} \Res_{q \to {\bf a}} \Res_{r \to {\bf a}} \Res_{s \to {\bf a}} \Phi(p)\, K(p,q)K(\overline{q},r)K(q,s) \,\, B(\overline{r},r) B(\overline{s}, s) .\cr
\eea

\subsection{Teichmuller pants gluings}

Every Riemann surface of genus $g$ with $k$ boundaries can be decomposed into
$2g-2+k$ pants whose boundaries are $3g-3+k$ closed geodesics (in the Poincar\'e metric with constant negative curvature) \cite{Teichmuller}.
The number of ways (in the combinatorial sense) of gluing $2g-2+k$ pants by their boundaries is clearly the same as the number of diagrams of ${\cal G}_{k}^{(g)}$, and each diagram corresponds to one pants decomposition.

Indeed, consider the root boundary labeled by $p$, and attach a pair of pants to this boundary.
Draw an arrowed propagator from the boundary to the first pants.
Then, choose one of the other boundaries of the pair of pants (there are thus 2 choices, left or right), it must be glued to another pair of pants (possibly not distinct from the first one). If this pair of pants was never visited, draw an arrowed propagator, and if it was already visited, draw a non-arrowed propagator.
In the end, you get a diagram of ${\cal G}_k^{(g)}$.
This procedure is bijective (up to symmetry factors), and to a diagram of ${\cal G}_k^{(g)}$, one may associate a gluing of pants.

Example with $k=1$ and $g=2$:
$$
\om_1^{(2)}={\epsfxsize 3cm\epsffile{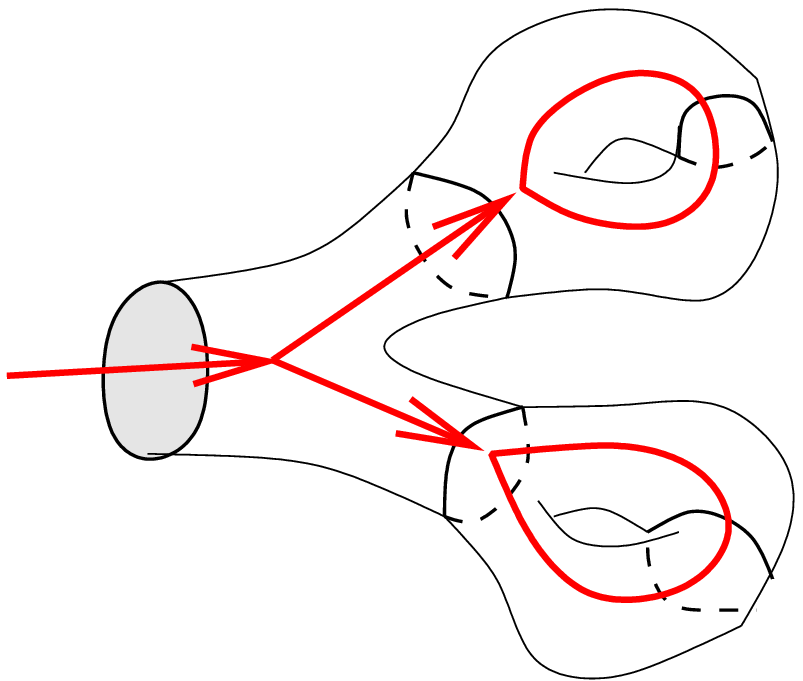}}+2 {\epsfxsize 4cm\epsffile{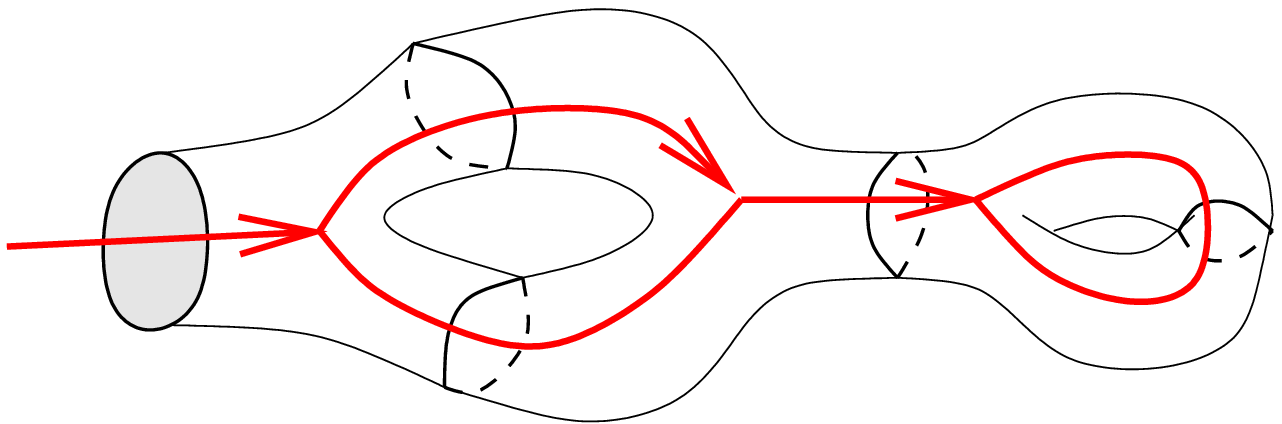}}+2 {\epsfxsize 4cm\epsffile{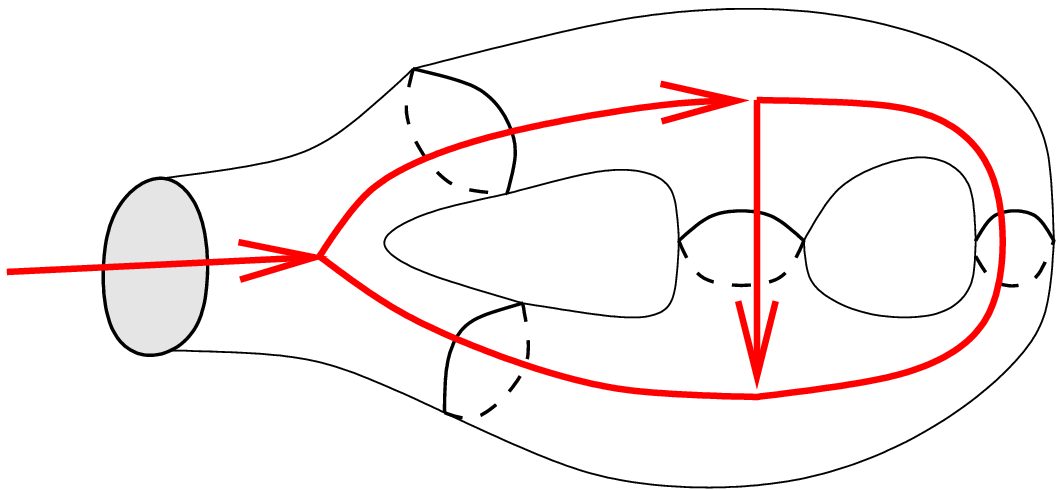}} .
$$

%%%%%%%%%%%%%%%%%%%%%%%%%%%%%%%%%%%%%%%%%%%%%%%%%%%%%%%%%%%%%%%%%%%%%%%%%%%%%%%%%%%%%%%%%%%%%%%%%%%%%%%%%%%%

\section{Main properties}

So, for every regular spectral curve ${\cal E}=(\spcurve,x,y)$ (and matrix $\kappa$ if $\spcurve$ has genus $\genus>0$) we have defined some meromorphic $n$-forms $\om_n^{(g)}$ and some complex numbers $F_g=\om_0^{(g)}$.
They have some remarkable properties (see \cite{EOinvariants}):
\begin{itemize}

\item $\om_n^{(g)}$ is symmetric in its $n$ variables (this is proved by recursion).

\item If $2g-2+n>0$, then $\om_n^{(g)}(z_1,\dots,z_n)$ is a meromorphic form (in $z_1$ for instance) with poles only at the branch-points, of degree at most $6g-6+2n+2$, and with vanishing residue.

\item If two spectral curves ${\cal E}=(\spcurve,x,y)$ and $\td{\cal E} =(\widetilde{\spcurve},\td{x},\td{y})$ are symplectically equivalent, they have the same $F_g$'s or $g>1$ (although they do not have the same $\om_n^{(g)}$'s in general)
\beq
dx\wedge dy = \pm\, d\td{x}\wedge d\td{y}
\qquad \to \quad
F_g({\cal E},\kappa)= F_g(\td{\cal E},\kappa) .
\eeq

\item if $\spcurve$ is of genus $\genus=0$, then $\tau=\exp{({\displaystyle \sum_{g=0}^\infty} N^{2-2g}F_g)}$ is  a formal tau function, it obeys Hirota's equation.
This theorem can be extended to $\genus>0$, with additional $\theta$-functions, see section \ref{secintegrability}.

\item Dilaton equation, for $2g-2+n>0$:
\beq
\sum_i \Res_{z_{n+1}\to a_i} \Phi(z_{n+1})\, \om_{n+1}^{(g)}(z_1,\dots,z_n,z_{n+1}) = (2-2g-n)\, \om_n^{(g)}(z_1,\dots,z_n) .
\eeq
This equation just reflects the homogeneity property, i.e. under a rescaling $y\to \l y$, we have $\om_n^{(g)}\to \l^{2-2g-n}\om_n^{(g)}$.

\item The derivatives of $\om_n^{(g)}$ with respect to many parameters on which the spectral curve may depend is computed below in section \ref{secvariat}.

\item The $\om_n^{(g)}$'s  have many other properties, for instance their modular behaviour satisfies the Holomorphic anomaly equations.

\end{itemize}

Let us study those properties in deeper details.

\subsection{Homogeneity}

If one changes the function $y(z)\to \l\, y(z)$, i.e. just a rescaling of the curve, then it is clear from \eq{defkernelK} that the kernel $K$ is changed to $K/\l$ and nothing else is changed.
Thus, $\om_n^{(g)}$ changes as:
\beq
\om_n^{(g)} \to \l^{2-2g-n}\,\, \om_n^{(g)},
\eeq
and in particular
\beq\label{eqFghomogeneity}
F_g(\spcurve,x,\l y) = \l^{2-2g}\, F_g(\spcurve,x,y).
\eeq
This implies that $F_g$ is a homogeneous function of the spectral curve, of degree $2-2g$.

\medskip
In particular, if one choses $\l=-1$ one gets (for $g\geq 2$):
\beq\label{Fg-yFgy}
F_g(\spcurve,x,-y,\kappa) = F_g(\spcurve,x,y,\kappa).
\eeq

\subsection{Symplectic invariance}\label{secsymplinv}

It is clear from the definitions, that $F_g$ and $\om_n^{(g)}$ depend on the spectral curve only through  the kernels $B$ and $K$, and the number and position of branchpoints.

\medskip

The Bergmann kernel $B$ depends only on the underlying complex structure of the Riemann surface $\spcurve$, thus it remains unchanged if we change the functions $x$ and $y$, as long as we don't change $\spcurve$.

\medskip

The kernel $K$, depends on the functions $x$ and $y$, only through the combination:
\beq
(y(z)-y(\bar{z}))\, dx(z).
\eeq
Therefore $K$ remains unchanged if we don't change this combination.

In particular, the kernel $K$, and therefore $F_g$ and $\om_n^{(g)}$ remain unchanged, if we change:

$\bullet$ $y\to y+R(x)$ where $R(x)$ is some rational function of $x$.

$\bullet$ $y\to \l y$, $x\to x/\l$ where $\l\in {\mathbb C}^*$.

$\bullet$ $x\to {ax+b\over cx+d}$, $y\to {(cx+d)^2\over ad-bc}\, y$.

%\medskip

%{\bf Remark:} {\it  if we change $y\to \l y$ and don't change $x$, $K$ is multiplied by $\l^{-1}$, and by an easy recursion, we find that $\om_n^{(g)}\to \l^{2-2g-n} \om_n^{(g)}$, and in particular $F_g\to \l^{2-2g}\, F_g$. This anticipates on the homogeneity property of section \ref{?}. Here it is useful because if we choose $\l=-1$ we see that $F_g$ is invariant under:

%$\bullet$ $x\to x, y\to -y$. }

\medskip

Those transformations, form a subgroup of the symplectomorphisms.
Indeed, in all those cases, the symplectic form $dx\wedge dy$ is unchanged.

\medskip
In order to have invariance under the full group of symplectomorphisms, we need to prove the invariance under the ${\pi\over 2}$ rotation in the $x,y$ plane, i.e. $x\to y$, $y\to -x$, which also conserves $dx\wedge dy$.
Using homogeneity \eq{Fg-yFgy}, we see that this is equivalent to consider the invariance under $x\to y$, $y\to x$.

This transformation however, does not conserve $K$, it does not conserve the number of branchpoints, and it does not conserve the $\om_n^{(g)}$'s with $n\geq 1$.
However, it was proved in \cite{symmetry} that it does conserve the $F_g$'s.
The proof of \cite{symmetry} is very technical. It is inspired from the loop equations for the 2-matrix model. It amounts to defining some mixed $n+m$-forms $\om_{n,m}^{(g)}$, where $x$ and $y$ play similar roles, and for which $\om_{n,0}^{(g)}$ coincides with the $\om_n^{(g)}$ for the spectral curve $(\spcurve,x,y)$, and $\om_{0,m}^{(g)}$ coincides with the $\om_m^{(g)}$ for the spectral curve $(\spcurve,y,x)$.
In particular $F_g=\om_{0,0}^{(g)}$ is both the $F_g=\om_0^{(g)}$ for the spectral curve $(\spcurve,x,y)$, and the $F_g=\om_0^{(g)}$ for the spectral curve $(\spcurve,y,x)$. The proof of \cite{symmetry} relies on the fact that $\om_{n+1,m}^{(g)}+\om_{n,m+1}^{(g)}$ is an exact form.

\medskip

That leads to:
\bt\label{thsymplinv}
Symplectic invariance

The $F_g$'s are invariant under the group of symplectomorphisms generated by:

$\bullet$ $y\to y+R(x)$ where $R(x)$ is some rational function of $x$.

$\bullet$ $y\to \l y$, $x\to x/\l$ where $\l\in {\mathbb C}^*$.

$\bullet$ $x\to {ax+b\over cx+d}$, $y\to {(cx+d)^2\over ad-bc}\, y$.

$\bullet$ $x\to y$, $y\to -x$.

\smallskip
In addition, the $F_g$'s are also invariant under:

$\bullet$ $x\to x$, $y\to -y$.

\et

This theorem is a powerful tool which allows to compare the $F_g$'s of models which look a priory very different.
We will see examples of applications in section \ref{seckontsevich}.

The $\om_n^{(g)}$'s with $n\geq 1$ are not conserved under symplectic transformation, instead they get shifted by exact forms.

\subsection{Derivatives}
\label{secvariat}

In this section, we study how the $F_g$'s and $\om_n^{(g)}$'s change under a change of spectral curve, and in particular under infinitesimal holomorphic changes.

\medskip

Consider an infinitesimal change $y\to y+\epsilon\delta y$ at fixed $x$, or in fact it is more appropriate to consider the variation of the differential form $ydx$:
\beq
ydx \to ydx + \epsilon \delta(ydx) +O(\epsilon^2) = ydx + \epsilon d\Omega +O(\epsilon^2)
\eeq
where $d\Omega$ is an analytical differential form on an open subset of $\spcurve$\footnote{Note that $ydx$ does not need to be a meromorphic form itself to be able to consider such deformations.}.
If instead of working at fixed $x$, we prefer to work with some local parameter $z$, we write:
\beq
\delta y(z) dx(z) -  \delta x(z) dy(z) =d\Omega(z).
\eeq
This shows that the set of holomorphic deformations of the spectral curve is equipped with a Poisson structure, but we shall not study it in details in this article, see \cite{Dubrovin1,Dubrovin2} for the Frobenius manifold structure.

\medskip

{\bf Classification of possible 1-forms $d\Omega$:}

The deformation $d\Omega$ is a 1-form. Here we shall consider only meromorphic deformations, and meromorphic 1-forms are classified as 1st kind (no pole), 2nd kind (multiple poles, without residues), and 3rd kind (only simple poles).

\begin{itemize}

\item First kind deformations are holomorphic forms on $\spcurve$, i.e. they are linear combinations of the $du_i$'s (see section \ref{defgenuscycle}):
\beq
du_i(z) = {1\over 2i\pi}\oint_{\ovl\bcycle_i} B(z,z')
\eeq
where $\ovl\bcycle_i = \bcycle_i - \sum_j \tau_{i,j}\acycle_j$.

\item 2nd kind deformations have double or multiple poles. They can be taken as linear combinations of Bergmann kernels or of their derivatives.
Choose a point $p\in \spcurve$. If $x$ is regular at $p$, choose the local parameter $\xi(z)=x(z)-x(p)$, and if $x$ has a pole of degree $d$ at $p$, choose $\xi(z)=x(z)^{-1/d}$, and define:
\beq\label{defxi}
B_k(z;p) = \Res_{z'\to p} B(z,z')\, \xi(z')^{-k}.
\eeq
All 2nd kind differentials are linear combinations of such $B_k(z;p)$.

\item 3rd kind differentials have only simple poles, and since the sum of residues must vanish, they must have at least 2 simple poles.
Choose two points $p_1$ and $p_2$ in the fundamental domain, and define:
\beq\label{defdS}
dS_{p_1,p_2}(z) = \int_{p_2}^{p_1} B(z,z').
\eeq
All 3rd kind differentials are linear combinations of such $dS$.

\end{itemize}

\bt
A general meromorphic differential form $d\Omega$ with poles $p_k$'s, can be written:
\beq\label{Omegalincomb123}
d\Omega(z)
= 2i\pi \sum_{i=1}^{\genus} \delta\epsilon_i\, du_i(z)
+ \sum_k \delta t_k \, dS_{p_k,o}(z)
 + \sum_k\sum_j \delta t_{k,j} \, B_j(z;p_k) .
\eeq
\et
It can be noticed that the coefficients $\delta\epsilon_i, \delta t_k, \delta t_{k,j}$ are the flat coordinates in the metrics of kernel $B$, of the corresponding Frobenius manifold structure.

\medskip
\proof{
Indeed, let $p_k$ be the poles of $d\Omega$, and write the negative part of the Laurent series of $d\Omega$ near its poles as:
\beq
d\Omega(z) \mathop{{\sim}}_{z\to p_k} \delta t_k {d\xi(z)\over \xi(z)} - \sum_{j\geq 1} j\, \delta t_{k,j} {d\xi(z)\over \xi(z)^{j+1}}.
\eeq
We see that
\beq
d\Omega(z) - \sum_k \delta t_k \, dS_{p_k,o}(z) - \sum_k\sum_j \delta t_{k,j} \, B_j(z;p_k)
\eeq
is a 1-form which has no poles, thus it is a holomorphic form, and it is a linear combination of the $du_i$'s.
}

If $\kappa=0$, i.e. if $B$ is the Bergmann kernel, it is normalized on the $\acycle$-cycles, and we have:
$\oint_{\acycle} dS = 0$, $\oint_{\acycle} B_\kappa = 0$. Thus, the $\delta \epsilon_i$ are easily computed as $\delta\epsilon_i = {1\over 2i\pi} \oint_{\acycle_i} d\Omega$.
However, if $\kappa\neq 0$, this is no longer true, we have  $\delta\epsilon = {1+\kappa\tau\over 2i\pi} \oint_{\acycle} d\Omega - {\kappa\over 2i\pi} \oint_{\bcycle} d\Omega$, and the variations $\delta t_{k,j}$ or $\delta t_k$ get mixed with the $\delta \epsilon_i$ through the variations of $\tau$.
The good way to undo this mixing, is by defining a covariant variation:

\bd
Covariant variation:
\beq
D_{d\Omega} \stackrel{{\rm def}}{=} \delta_{d\Omega} + \tr \left(\kappa\, \delta_{d\Omega}\tau\,\kappa\, {\partial\over \partial \kappa} \right)
\eeq
where $ \delta_{d\Omega}\tau$ is the variation of the Riemann matrix of periods $\tau$ under $ydx \to ydx + \epsilon d\Omega$.
\ed
Derivatives with respect to $\kappa$ are studied in details in section \ref{secderkappa} below.

\bigskip

The important point, is that $d\Omega$ can always be written as:
\beq
d\Omega(z) = \int_{\partial \Omega}\,\, B(z,z')\,\, \Lambda(z')
\eeq
where $\partial\Omega$ is some continuous path (a chain or a cycle, which is related to the Poincar\'e dual of $d\Omega$) on $\spcurve$, and $\Lambda(z')$ is an analytical function defined locally in a vicinity of  $\partial\Omega$.

\bigskip

The theorem is then:
\bt\label{thvariat} {\bf Variation of the spectral curve:}

Under an infinitesimal deformation $\delta y\, dx-\delta x\, dy = d\Omega(z)=\int_{\partial \Omega}\,\, B(z,z')\,\, \Lambda(z')$, the $\om_n^{(g)}$'s change by:
\beq
D_{d\Omega}\, \om_{n}^{(g)}(z_1,\dots,z_n) = \int_{\partial \Omega}\,\, \om_{n+1}^{(g)}(z_1,\dots,z_n,z')\,\, \Lambda(z') .
\eeq

\et

For example, in particular with $n=0$ we have:
\beq
D_{d\Omega} F_g = \int_{\partial \Omega}\,\, \om_{1}^{(g)}(z')\,\, \Lambda(z').
\eeq

\subsubsection{The loop operator}

This theorem can also be restated in terms of the "loop operator", which corresponds to $d\Omega(z) = B(z,z')$. We define:
\bd
The loop operator is:
\beq
D_{z'} \stackrel{{\rm def}}{=} D_{B(z,z')}.
\eeq
It satisfies:
\beq
D_{z'} \om_{n}^{(g)}(z_1,\dots,z_n) = \om_{n+1}^{(g)}(z_1,\dots,z_n,z') .
\eeq
The loop operator is a derivation, i.e. $D_{z'} (uv) = u D_{z'} v+v D_{z'}u$, and it is such that: $D_{z_1} D_{z_2} = D_{z_2} D_{z_1}$, and $D_{z_1} {\partial\over \partial z_2} = {\partial\over \partial z_2}D_{z_1}$.
\ed

In random matrix theory, the loop operator \cite{Ak96}, is most often written as a functional derivative with respect to the potential $V(x)$:
\beq
{\partial \over \partial V(x(z'))} \stackrel{{\rm def}}{=} D_{z'} .
\eeq

\subsubsection{Inverse of the loop operator}

The loop operator allows to find $\om_{n+1}^{(g)}$ in terms of $\om_n^{(g)}$, i.e. increases $n$ by $1$.
The inverse operator, which decreases $n$ by $1$ can also be written explicitly:

\bt
Let $\Phi$ be a primitive of $ydx$, i.e. a function defined on the fundamental domain such that $d\Phi=ydx$, then we have, if $2-2g-n<0$:
\beq
(2-2g-n)\,\om_n^{(g)}(z_1,\dots,z_n) = \sum_i \Res_{z\to a_i} \om_{n+1}^{(g)}(z_1,\dots, z_n,z)\,\, \Phi(z) .
\eeq
\et
This theorem is easily proved by recursion on $2g+n-2$.

This theorem is at the origin of definition \ref{defsympinvFg}, for $n=0$.

\subsection{Modular properties}

In section \ref{secBergmann} we have introduced a deformation of the Bergmann kernel with a symmetric matrix $\kappa$ of size $\genus\times \genus$. The reason to introduce this deformation, was that it encodes the modular dependence of the Bergmann kernel, i.e. how the Bergmann kernel changes under a change of choice of cycles $\acycle, \bcycle$.
Thus, studying the modular dependence of the $F_g$'s and $\om_n^{(g)}$'s amounts to studying their dependence on $\kappa$.

\smallskip

Also, in section \ref{secvariat}, we have seen that the covariant derivative involves the computation of derivatives with respect to $\kappa$.

\subsubsection{Dependence on $\kappa$} \label{secderkappa}

Since the kernels $B$ and $K$ depend linearly on $\kappa$, all the stable $\om_n^{(g)}$'s and $F_g$'s are polynomials in $\kappa$, of degree $3g-3+n$.

\medskip

Notice that $\partial B(z_1,z_2)/\partial \kappa_{i,j} = 2i\pi\,\, du_i(z_1)\,du_j(z_2)$ is factorized, i.e. a function of $z_1$ times a function of $z_2$. This simple observation, together with
\beq
du_i(z)  = {1\over 2i\pi}\,\oint_{\ovl\bcycle_i} B(z,z')
\virg \ovl\bcycle_i = \bcycle_i - \sum_j \tau_{i,j}\acycle_j
\eeq
leads, by an easy recursion, to the following theorem:
\bt
For  $2-2g-n \leq 0$:
\bea\label{domgdkappa}
2i\pi\,\partial \om_n^{(g)}(z_1,\dots,z_n) /\partial \kappa_{i,j}
&=& {1\over 2} \oint_{\ovl{\bcycle}_i} dz' \oint_{\ovl{\bcycle}_j} dz \,\,\, \Big[ \om_{n+2}^{(g-1)}(z_1,\dots,z_n,z,z')  \cr
&& + \sum_{h=0}^g \sum'_{I\subset J}  \om_{1+|I|}^{(h)}(z,I) \, \om_{1+n-|I|}^{(g-h)}(z',J/I)
\Big] \cr
\eea
where $J=\{z_1,\dots,z_n\}$, and $\sum_h\sum'_{I}$ means as usual that we exclude $(h,I)=(0,\emptyset),(g,J)$.
\et

This theorem can be applied recursively, to compute higher derivatives, and eventually recover a polynomial of $\kappa$ by its Taylor expansion at $\kappa=0$, of the form:
\beq\label{eqTaylorkappaFg}
F_g(\kappa) = \sum_{k=0}^{3g-3}\,\, {1\over k!}\,(\kappa)^k\,\, \partial^k F_g|_{\kappa=0} .
\eeq

According to theorem \ref{thvariat} of section \ref{secvariat}, the $\ovl\bcycle$-cycle integrals, computed at $\kappa=0$, are the derivatives with respect to coordinates $\epsilon_i$ of \eq{Omegalincomb123}:
\beq
\kappa=0 \quad \leftrightarrow \quad
{\partial \om_{n}^{(g)}(z_1,\dots,z_n) \over  \partial \epsilon_i} = \oint_{\ovl{\bcycle}_i} \om_{n+1}^{(g)}(z_1,\dots,z_n,z)
\eeq
and therefore we have:
\bea
\partial \om_n^{(g)}(z_1,\dots,z_n) /\partial \kappa_{i,j} |_{\kappa=0}
&=& {1\over 2}\,{\partial \over \partial \epsilon_i}\,{\partial \over \partial \epsilon_j} \,\,\, \om_{n}^{(g-1)}(z_1,\dots,z_n)  \cr
&& + {1\over 2}\,\sum_{h=0}^g \sum_{I\subset J}  {\partial \over \partial \epsilon_i}\,\om_{1+|I|}^{(h)}(I) \, {\partial \over \partial \epsilon_j}\,\om_{n-|I|}^{(g-h)}(J/I) . \cr
\eea

We can thus trade the $\kappa$ dependance into derivatives with respect to the coordinates $\epsilon_i$. For instance we have at $g=2$, and with $\genus=1$:
\beq
F_2(\kappa)
= \ovl{F_2}
+  {\kappa\over 2} \Big( \partial^2 \ovl{F}_1 + \partial \ovl{F}_1\,\partial \ovl{F}_1\Big)
 +  {\kappa^2 \over 8} \Big( \partial^4 \ovl{F}_0 + 4\,\,\partial^3 \ovl{F}_0\,\partial \ovl{F}_1\Big)
 +  {\kappa^3 \over 48} \Big( 10\,\, \partial^3 \ovl{F}_0\,\partial^3 \ovl{F}_0\Big)
\eeq
where $\ovl{F}_g = F_g(\kappa=0)$, and $\partial=\partial/\partial\epsilon$.

\medskip
This result is best interpreted graphically. For example with $g=2$ we have:
\bea
F_2(\kappa)
&=&
{\mbox{\epsfxsize=1.5truecm\epsfbox{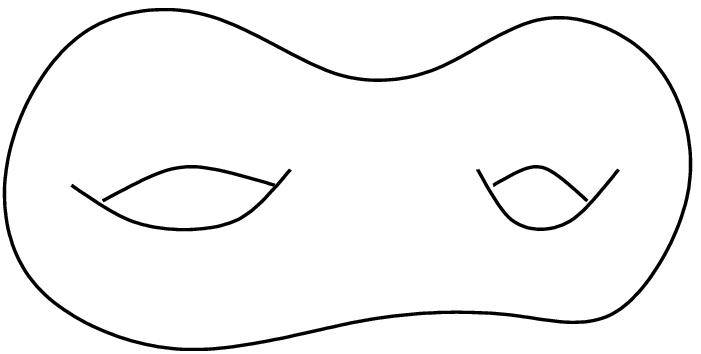}}}
+{\kappa\over 2}{\mbox{\epsfxsize=1truecm\epsfbox{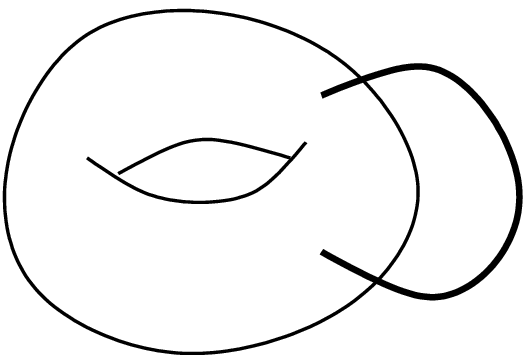}}}
+{\kappa\over 2}{\mbox{\epsfxsize=2truecm\epsfbox{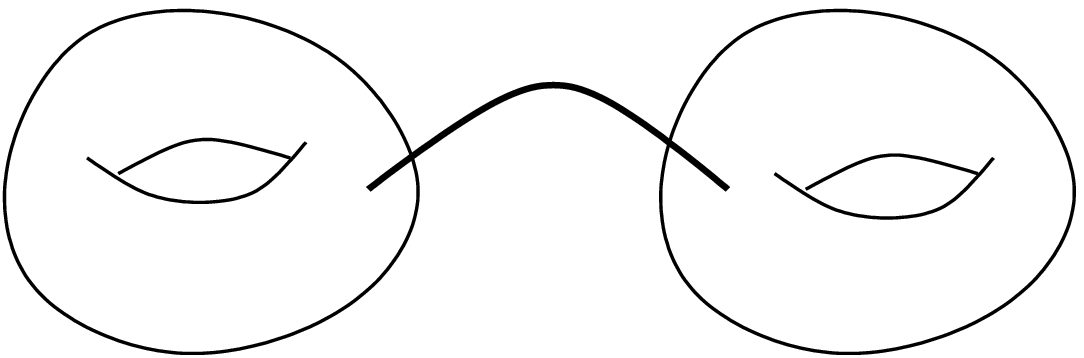}}}
+{\kappa^2\over 8}{\mbox{\epsfxsize=1.2truecm\epsfbox{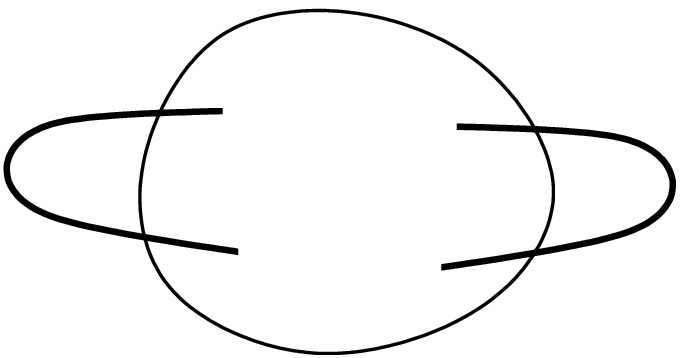}}}
+{\kappa^2\over 2}{\mbox{\epsfxsize=2.truecm\epsfbox{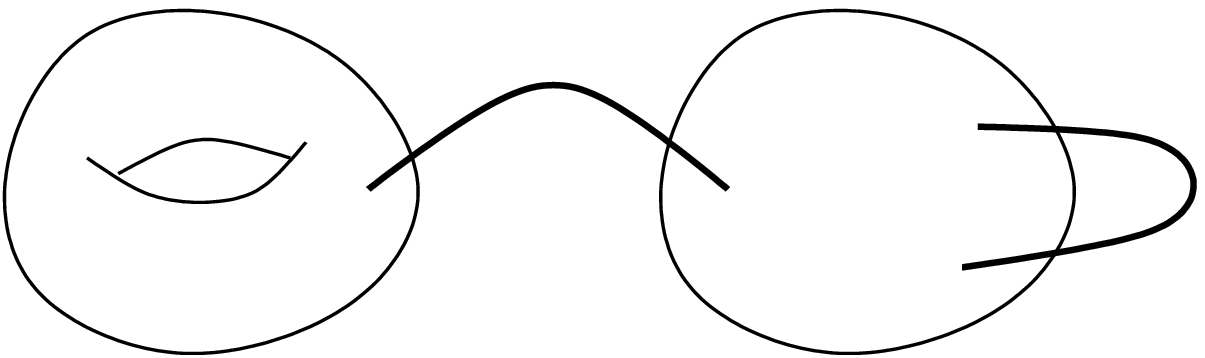}}} \cr
&& +{\kappa^3\over 8}{\mbox{\epsfxsize=2.2truecm\epsfbox{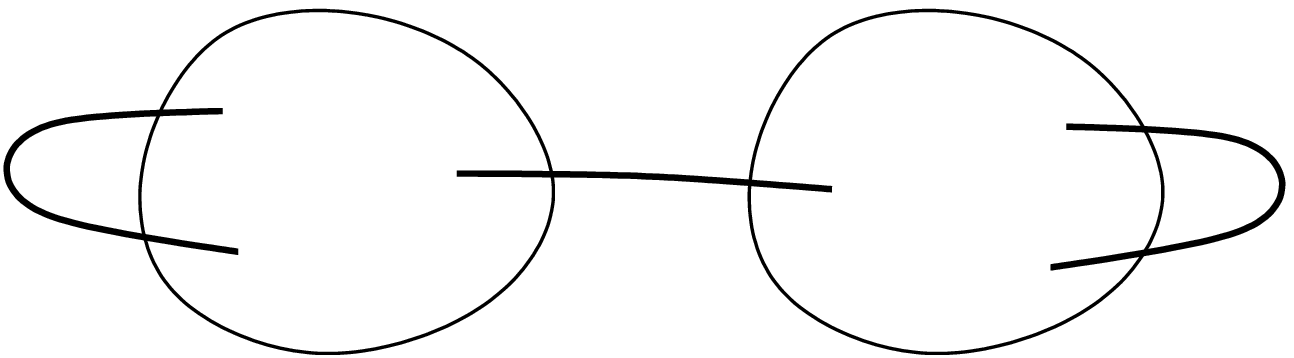}}}
+{\kappa^3\over 12}{\mbox{\epsfxsize=2.2truecm\epsfbox{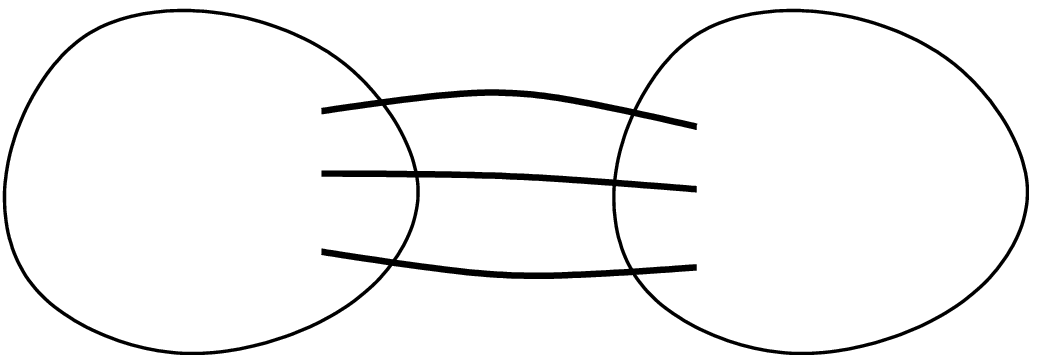}}}
\eea
where each line with endpoints $(i,j)$ is a factor $\kappa_{i,j}$, and each connected piece of Riemann surface of genus $h$, with $k$ punctures $i_1,\dots,i_k$ is a $\partial^k \ovl{F}_h/\partial \epsilon_{i_1} \dots\partial \epsilon_{i_k}$.
Each graph is a possible "stable" degeneracy of a genus $g$ Riemann surface (imagine each link   contracted to a point), stability means that each connected component of genus $h$ with $k$ marked points must have $2-2h-k<0$.
The prefactor is  $1/\# Aut$, i.e. the inverse of the number of automorphisms, for instance in the last graph ${\mbox{\epsfxsize=1.5truecm\epsfbox{S2k3v2.eps}}}$ we have a ${\mathbb Z}_2$ symmetry by exchanging the 2 spheres, and a $\sigma_3$ symmetry from permuting the 3 endpoints of the edges, i.e. $12=\#({\mathbb Z}_2\times \sigma_3)$ automorphisms.

\bigskip

More generally, by a careful analysis of the combinatorics of the $\partial_{\epsilon_i}$'s, one can see (this was done in \cite{EMO}, and coincides with the diagrammatics of \cite{ABK}) that the Taylor expansion \eq{eqTaylorkappaFg}, reconstructs the expansion of a formal Gaussian integral (i.e. order by order in powers of $N$):
\bea
\ee{{\displaystyle \sum_g} N^{2-2g} F_g(\epsilon,\kappa)}
&=& \int d\eta_1\dots \int\d\eta_{\genus} \,\, \ee{{\displaystyle \sum_g} N^{2-2g} \ovl{F}_g(\eta)\, }
\,\, \ee{-N^2{\displaystyle \sum_i} (\eta_i-\epsilon_i)\partial_{\epsilon_i} \ovl{F}_g} \cr
&& \,\, \ee{-{N^2\over 2} {\displaystyle \sum_{i,j}} (\eta_i-\epsilon_i)(\eta_j-\epsilon_j)\partial_{\epsilon_i}\partial_{\epsilon_j} \ovl{F}_g}
 \,\, \ee{-{N^2\over 4i\pi} {\displaystyle \sum_{i,j}} (\eta_i-\epsilon_i)(\eta_j-\epsilon_j)\,(\kappa^{-1})_{i,j}}
\eea
and the graphical representation above is just the Wick's expansion of the gaussian integral.

This diagrammatic expansion of modular transformations was first derived in \cite{ABK} in the context of topological strings.

\subsubsection{Holomorphic anomaly}\label{secholoano}

In particular, if we choose $\kappa$ to be the Zamolodchikov K\"ahler metric $\kappa = (\ovl\tau - \tau)^{-1}$, we have seen in section \ref{Schiffer}, that the Bergmann kernel becomes the Schiffer kernel and is modular invariant, which means that it is independent of the choice of cycles $\acycle, \bcycle$.
Since the only modular dependence of the $F_g$'s and $\om_n^{(g)}$'s is in the Bergmann kernel, we have:
\bt
If $\kappa$ is the Zamolodchikov K\"ahler metric $\kappa = (\ovl\tau - \tau)^{-1}$, then $F_g$ and $\om_n^{(g)}$'s are modular invariant.
\et

The price to pay to have modular invariant $F_g$'s, is that they are no longer analytical functions of $\tau$, i.e. analytical functions of the spectral curve, and in particular they are no longer analytical functions of the $\epsilon_i$'s.
However, since the only non-analytical dependence is polynomial in $\kappa$, and $\kappa^{-1}$ is linear in $\ovl\tau$ which is the only non-analytical term, and since we have relationships between derivatives with respect to $\kappa$ and derivatives with respect to $\epsilon$, by a simple chain rule, we obtain the following theorem \cite{EMO}:
\bt\label{thholaneq} The $\om_n^{(g)}$'s satisfy the Holomorphic anomaly equations
\bea
{\partial \om_n^{(g)}(J)\over \partial \ovl\epsilon_i}
&=& {-1\over (2i\pi)^3}\, \kappa {\partial^3 \ovl{F}_0\over \partial \ovl\epsilon^3}\,\kappa\,\, {1\over 2}\,\Big[
 {\partial^2 \om_{n}^{(g-1)}(J)\over \partial \epsilon^2}     + {\partial\tau \over \partial \epsilon} \,\kappa\, {\partial \om_{n}^{(g-1)}(J) \over \partial \epsilon} \cr
&& + \sum_{h=0}^g \sum_{I\subset J}  {\partial \om_{|I|}^{(h)}(I) \over \partial \epsilon} \, {\partial \om_{n-|I|}^{(g-h)}(J\backslash I) \over \partial \epsilon}
\Big].
\eea
In particular for $n=0$:
\beq
{\partial F_g\over \partial \ovl\epsilon_i}
= {-1\over (2i\pi)^3}\, \kappa {\partial^3 \ovl{F}_0\over \partial \ovl\epsilon^3}\,\kappa\,\, {1\over 2}\,\Big[
 {\partial^2 F_{g-1}\over \partial \epsilon^2}     + {\partial\tau \over \partial \epsilon} \,\kappa\, {\partial F_{g-1} \over \partial \epsilon}
 + \sum_{h=1}^{g-1}   {\partial F_h \over \partial \epsilon} \, {\partial F_{g-h} \over \partial \epsilon}
\Big].
\eeq

\et

This equation was first found by Bershadsky, Cecotti, Ooguri and Vafa (which we refer to as BCOV \cite{BCOV}) in the context of topological string theory. Here we see that the symplectic invariants $F_g$ always satisfy this equation, and it is tempting to believe that  the symplectic invariants $F_g$, should coincide with the string theory amplitudes, i.e. the Gromov-Witten invariants. This question is debated below in section \ref{sectopostring}.
Unfortunately, the holomorphic anomaly equations do not have a unique solution, and although this conjecture is almost surely correct, no proof exists at the present time, apart from a very limited number of cases\footnote{In \cite{Bmodel},
this conjecture is proposed as a new definition of the type IIB topological string
theory. The interested reader may find all the details of this conjecture as well as numerous checks in this paper.}.

\bigskip

Let us briefly sketch the idea of BCOV.
String theory partition functions represent "path integrals" over the set of all Riemann surfaces with some conformal invariant weight. In other words, they are integrals over moduli spaces of Riemann surfaces of given topology, and topological strings are integrals with a topological weight, they compute intersection numbers of bundles over moduli spaces (see \cite{MMtopo,Vonk} for introduction to topological strings).

Moduli spaces can be compactified by adding their "boundaries", which correspond to degenerate Riemann surfaces (for instance when a non contractible cycle gets pinched, or when marked points come together). The integrals have thus boundary terms, which can be represented by $\delta$-functions, and $\delta$-functions are not holomorphic.
%:
%\beq
%\delta^2(z,\overline{z}) = {1\over 2i\pi}\partial_{\overline{z}}\,{1\over z}
%\eeq
In other words, string theory partition functions contain non-holomorphic terms which count degenerate Riemann surfaces.

On the other hand, if one decides to integrate only on non-degenerate surfaces, one gets holomorphic patition functions, but not modular invariant, because the boundaries of the moduli spaces are associated to a choice of pinched cycles.
Modular invariant means independent of a choice of cycles.

To summarize, the holomorphic partition function is obtained after a choice of boundaries, i.e. a choice of a symplectic basis of non contractible cycles $\acycle_i\cap \bcycle_j=\delta_{i,j}$, and cannot be modular invariant.
The modular invariance is restored by adding the boundaries, but this breaks holomorphicity.

There is thus a relationship between holomorphicity and modular invariance.

\subsection{Background indepedence and non-perturbative modular invariance}\label{secbackground}

We have seen in the previous section, that the $F_g$'s are not modular invariant, unless we choose $\kappa = (\ovl\tau - \tau)^{-1}$, i.e. modular invariance can be restored by breaking holomorphicity.

In fact, there is another way of restoring modular invariance, without breaking holomorphicity. It exploits the fact that the modular transformations of $F_g$'s, i.e. \eq{domgdkappa}, is very similar to the modular transformation of theta-functions.
It was shown in \cite{EMnonperturbative}, that certain combinations of $\theta$-functions and $F_g$'s, are modular, and reconstruct a non-perturbative, modular partition function, which is also a Tau-function (see section \ref{secintegrability}), and which has a background independence property.

\medskip

Consider a characteristics $(\mu,\nu)$, and a spectral curve ${\cal E}=({\cal L},x,y)$, choose $\kappa=0$.
Following \cite{background,EMnonperturbative}, we introduce a {\it nonperturbative partition function} by summing over filling fractions,  defined by
\bea\label{exphatZTheta}
&& Z_{\cal E} (\mu,\nu;\epsilon) \cr
&=& \ee{{\displaystyle \sum_{g\geq 0}} N^{2-2g} F_g(\epsilon)}  \sum_{k} \sum_{l_i>0}\sum_{h_i>1-{l_i\over 2}} {N^{{\displaystyle \sum_i} (2-2h_i-l_i)}\over k! l_1!\,\dots\, l_k!}\,\,\, F_{h_1}^{(l_1)}\dots F_{h_k}^{(l_k)}  \,\, \Theta_{\mu,\nu}^{(\sum_i l_i)}(NF'_0,\tau) \cr
&=& \ee{ N^{2} F_0+F_1} \, \Theta_{\mu,\nu} \,
 \biggl\{ 1+ \sum_{j=1}^\infty N^{-j}\,Z_j(\mu,\nu;\epsilon)\biggr\} \cr
&=& \ee{N^2 F_0}\, \ee{F_1}\, \ee{(N^{-2} F_2 + N^{-4} F_3 +\dots)} \,\, \biggl\{ \Theta_{\mu,\nu} + {1\over N}\Bigl(\Theta'_{\mu,\nu} F_1' + {1\over 6} \Theta_{\mu,\nu}'''\,F_0'''\Bigr)  \cr
&& \quad +{1\over N^2}\Bigl({1\over 2}\Theta_{\mu,\nu}'' F_1'' + {1\over 2}\Theta_{\mu,\nu}'' F_1'^2 + {1\over 24}\Theta_{\mu,\nu}^{(4)} F_0''''+{1\over 6}\Theta_{\mu,\nu}^{(4)} F_0''' F_1' + {1\over 72}\Theta_{\nu,\mu}^{(6)} F_0'''^2 \Bigr) + \dots \biggr\}.  \cr
\eea
where $Z_j$ is the sum of all terms contributing to order $N^{-j}$.
In this partition function, the $F_g$'s are the symplectic invariants of the spectral curve ${\cal E}$, their derivatives are with respect to the background filling fraction $\epsilon$ and computed through theorem \ref{thvariat}, at:
\beq
\label{eps}
\epsilon = {1\over 2i\pi}\,\oint_{\acycle} y  d x.
\eeq
Notice that the $F_g({\cal E})$'s and their derivatives depend on the choice of a symplectic basis of $2\bar{g}$ one-cycles ${\cal A}_i,{\cal B}_j$ on ${\cal L}$. Finally, the theta function $\Theta_{\mu,\nu}$ of characteristics $(\mu,\nu)$ is defined by
\beq
\label{biget}
\Theta_{\mu,\nu}(u,\tau) = \sum_{n\in {\mathbb Z}^{\bar g}} \ee{(n+\mu-N \epsilon)u}\,\,\ee{i\pi  (n +\mu-N\epsilon)\tau (n+\mu-N\epsilon)}\,\,\ee{2 i\pi n \nu}
\eeq
and is evaluated at
\beq\label{defutau}
u=N F_0',
\qquad
F'_0  = \oint_{\bcycle} y(x) d x,
\qquad
\tau =  {1\over 2i\pi} F_0''.
\eeq
In (\ref{exphatZTheta}), the derivatives of the theta function (\ref{biget}) are w.r.t. $u$.
The derivatives of $\Theta$ and the derivatives of $F_g$, are written with tensorial notations.
For instance, ${1\over 6}\Theta_{\mu,\nu}^{(4)} F_0''' F_1' $ actually means:
\beq
{1\over 6}\Theta_{\mu,\nu}^{(4)} F_0''' F_1'
\equiv {1\over 2!\, 3!\, 1!}\,\, \sum_{i_1,i_2,i_3,i_4}\, {\partial^4 \Theta_{\mu,\nu}\over \partial u_{i_1}\partial u_{i_2}\partial u_{i_3}\partial u_{i_4}}\,\, {\partial^3 F_0\over \partial \epsilon_{i_1}\partial \epsilon_{i_2}\partial \epsilon_{i_3}}\,\, {\partial F_1\over \partial \epsilon_{i_4}}
\eeq
and the symmetry factor (here ${1\over 6}={2\over 2!\,3!\,1!}$) is the number of relabellings of the indices, giving the same pairings, and divided by the order of the group of relabellings, i.e. $k!\, l_1!\dots l_k!$, as usual in Feynmann graphs.

%Notice that $\Theta$ and its derivatives $\Theta^{(k)}$ are of order $\CO(1)$, in fact they are quasi-periodic functions of $N$, and are not growing for large $N$. This means that the $1/N$ expansion in (\ref{exphatZTheta}) is well defined.

%\smallskip

%\subsection{The Theta function}

The $\Theta$ function above is closely related to the standard theta function, which is defined by
\beq
\vartheta\bigl[^\mu_\nu\bigr](\xi|\tau)= \sum_{{\bf n}\in {\mathbb Z}^{\bar g}}
\exp\bigl[ i \pi (n+\mu) \tau (n+\mu) + 2i\pi  (n+\mu)(\xi+\nu)\bigr].
\eeq
It it easy to see that these two functions are related as follows
\beq
\label{Thetatheta}
\Theta_{\mu,\nu}(u,\tau) =\exp\biggl[ -N^2\Bigl(  \epsilon F_0'+{1\over 2} \epsilon^2 F_0'' \Bigr)\biggr]
\,\, \vartheta\bigl[^\mu_\nu\bigr](\xi|\tau)
\eeq
where
\beq
\label{xivalue}
\xi={N \over 2i\pi }  \oint_{\bcycle-\tau\acycle} y (x) d x =N \biggl( {F'_0 \over 2i\pi}  - \tau \epsilon \biggr).
\eeq

\subsubsection{Modularity}

\bt
All the terms $Z_j$ in \eq{exphatZTheta} are modular, i.e. they transform as the characteristics $(\mu,\nu)$.
More precisely, if we make a modular change of cycles $\acycle,\bcycle\to \td\acycle,\td\bcycle$, we have:
\beq
\ee{N^2 \tilde F_0 + \tilde F_1} \widetilde \Theta_{\tilde \mu,\tilde \nu} (\tilde u,\tilde \tau) =\zeta\bigl[^\mu_\nu\bigr]\, \, \ee{N^2 F_0 +F_1} \Theta_{ \mu, \nu} (u,\tau)
\eeq
and for all $j\geq 1$:
\beq
\td{Z}_j(\td\mu,\td\nu) = Z_j(\mu,\nu).
\eeq

\et
This theorem was proved in \cite{EMnonperturbative}, mostly using the diagrammatic representation of section \ref{sectiondiagrepresent}, and the diagrammatic representation of \cite{ABK}.

\medskip
For example the following quantities are modular:
\beq
Z_1 = {\Theta'_{\mu,\nu}\over \Theta_{\mu,\nu}} F_1' + {1\over 6} {\Theta_{\mu,\nu}'''\over \Theta_{\mu,\nu}}\,F_0'''\, ,
\eeq
\beq
Z_2 = F_2 + {1\over 2}{\Theta''_{\mu,\nu}\over \Theta_{\mu,\nu}} F_1'' + {1\over 2}{\Theta''_{\mu,\nu}\over \Theta_{\mu,\nu}} F_1'^2
+ {1\over 24}{\Theta^{(4)}_{\mu,\nu}\over \Theta_{\mu,\nu}} F_0''''
+ {1\over 6}{\Theta^{(4)}_{\mu,\nu}\over \Theta_{\mu,\nu}} F_0''' F'_1
+ {1\over 72}{\Theta^{(6)}_{\mu,\nu}\over \Theta_{\mu,\nu}} (F_0''')^2 .
\eeq

\subsubsection{Background independence}

\bt
The partition function  \eq{exphatZTheta} is independent of the background filling fraction $\epsilon$, i.e., for any two filling fractions $\epsilon_1$ and $\epsilon_2$:
\beq
Z_{\cal E}(\mu,\nu,\epsilon_1)=Z_{\cal E}(\mu,\nu,\epsilon_2) .
\eeq
\et
This theorem follows directly from the definition \eq{exphatZTheta}.
It has important consequences which we shall not study here \cite{EMnonperturbative}.

\subsection{Integrability}\label{secintegrability}

Out of the $F_g$'s, one can define a "formal tau-function". In this section, let us assume that $\spcurve = {\mathbb P}^1$, i.e. it has genus $\genus=0$. The higher genus case is discussed in section \ref{secintegrabilitygenus} below.

\medskip
\bd
The formal $\tau$-function is defined as a formal series in a variable $N$:
\beq
\ln\tau_N = \sum_{g=0}^\infty N^{2-2g}\, F_g .
\eeq
\ed

Now, we shall explain why it makes sense to call it a $\tau$-function.
$\tau$-functions are usually defined in the context of integrable systems, and they have several more or less equivalent definitions, see \cite{BBT}.

\smallskip
One possible definition of $\tau$-functions relies on Hirota equations \cite{Hirota, BBT}, and another one relies on a free fermion representations, i.e. determinantal formulae \cite{JM1,Kos, BBT}.

\subsubsection{Determinantal formulae}

In the following of this section, most of the functions have an obvious formal $N$ dependence. For the sake of brevity, we do
not write it explicitly as long as it is not needed.

Out of the $\om_n^{(g)}$'s, it is convenient to define the formal series:
\beq
\om_n(z_1,\dots,z_n) =
%-N\, y(z_1)dx(z_1)\delta_{n,1}
-\, {\delta_{n,2}\, dx(z_1)dx(z_2)\over (x(z_1)-x(z_2))^2} +  \sum_g N^{2-2g-n} \om_n^{(g)}(z_1,\dots,z_n)
\eeq
and also the "non-connected" correlators:
\beq
\ovl{\om}_n(J) = \sum_{k=1}^n \sum_{I_1\cup\dots\cup I_k=J}\,\, \prod_{i=1}^k \om_{|I_i|}(I_i) .
\eeq
For example:
\beq
\ovl{\om}_2(z_1,z_2) = \om_2(z_1,z_2) + \om_1(z_1)\om_1(z_2),
\eeq
\bea
\ovl{\om}_3(z_1,z_2,z_3)
&=& \om_3(z_1,z_2,z_3) + \om_1(z_1)\om_2(z_2,z_3)+ \om_1(z_2)\om_2(z_1,z_3) \cr
&& + \om_1(z_3)\om_2(z_1,z_2)+ \om_1(z_1)\om_1(z_2)\om_1(z_3).
\eea
In other words, the $\om_n$ are the cumulants of the $\ovl\om_n$'s.

\smallskip
The following proposition is proved in some cases (hyperelliptical spectral curves \cite{Eynbergere}), and in all matrix models, however, it is expected to hold for any spectral curve:

\bp\label{thdet}
There exists a (formal) kernel $H(z_1,z_2)$, such that:
\beq
\om_1(z) = \mathop{{\rm lim}}_{z'\to z}\, \left(H(z,z') - {\sqrt{dz\, dz'}\over z-z'}\right) ,
\eeq
\beq
\om_2(z_1,z_2) = -H(z_1,z_2)H(z_2,z_1) - {dx(z_1)\,dx(z_2)\over (x(z_1)-x(z_2))^2}
\eeq
and if $n\geq 3$:
\beq
\ovl\om_n(z_1,\dots,z_n) = "\det"_{i,j=1,\dots,n}(H(z_i,z_j))
\eeq
where the quotation mark $"\det"$ means the following:
write the determinant as a sum over permutations of products of $H$'s:
$\det(H(z_i,z_j))={\displaystyle \sum_\sigma} (-1)^\sigma {\displaystyle \prod_i} H(z_i,z_{\sigma(i)})$.
Then, every time a permutation has a fixed point $\sigma(i)=i$ we must replace $H(z_i,z_i)$ by $\om_1(z_i)$,
and every time a permutation has a length 2 cycle  $\sigma(i)=j, \sigma(j)=i$ we must replace the factor $H(z_i,z_j)H(z_j,z_i)$ by  $-\om_2(z_i,z_j)$.

This is equivalent to saying that for $n\geq 3$, the cumulants are given by:
\beq
\om_n(z_1,\dots,z_n) = \sum_{{\rm cyclic}\, \sigma} (-1)^\sigma \prod_{i=1}^n H(z_i,z_{\sigma(i)}) .
\eeq

\ep

{\bf Example:}
\beq
\om_3(z_1,z_2,z_3) = H(z_1,z_2)H(z_2,z_3)H(z_3,z_1)+H(z_1,z_3)H(z_3,z_2)H(z_2,z_1).
\eeq

The determinantal formulae for correlation functions were first found by Dyson and Mehta \cite{Dyson,Mehta2} in the context of random matrix theory, and have led to a huge number of applications.

\smallskip

Moreover the kernel $H$ can be written rather explicitly.
In all matrix cases, the kernel $H$ for the determinantal formulae above, coincides with the kernel $\hat H$ which we define below:

\bd\label{thexpkernel}
We define the formal kernel $\hat H$ as a formal spinor in $z_1$ and $z_2$, given by an exponential formula
\beq
\hat H(z_1,z_2) = {\sqrt{dx(z_1)\,dx(z_2)}\over x(z_1)-x(z_2)}\,\,\ee{{ {\displaystyle \sum_{n=1}^\infty}} {1\over n!}\,\, \int_{z_2}^{z_1} \dots \int_{z_2}^{z_1} \om_n } .
\eeq
\ed
This exponential formula for the kernel is to be understood order by order in powers of $N$, namely:
\bea
\hat H(z_1,z_2)
&=& {\ee{-N \int_{z_2}^{z_1} ydx}\over E(z_1,z_2)}\,\, \,\,  \Big[ 1 + N^{-1} \int_{z_2}^{z_1} \om_1^{(1)} \cr
&& + {N^{-1}\over 6}\int_{z_2}^{z_1}\int_{z_2}^{z_1}\int_{z_2}^{z_1} \om_3^{(0)}
\quad + O(N^{-2})
\Big],
\eea
where $E(z_1,z_2)$ is the prime form:
\beq
E(z_1,z_2) = {z_1-z_2\over \sqrt{dz_1\,\, dz_2}} .
\eeq
For example, one of the terms contributing to $\hat H$ to order $N^{-1}$  is:
\beq
\int_{z_2}^{z_1}\int_{z_2}^{z_1}\int_{z_2}^{z_1} \om_3^{(0)}
= \sum_i {(z_1-z_2)^3\over y'(a_i)\,x''(a_i)\,\, (a_i-z_1)^3\,(a_i-z_2)^3} .
\eeq

\medskip

In all matrix model cases, the kernel $H$ can be written as a Sato-formula, and coincides with $\hat H$, but this is not proved in general.

\subsubsection{Examples}

For example, if we consider the Airy curve $y=\sqrt{x}$, i.e. ${\cal E}=({\mathbb P}^1,z^2,z)$, we find that $\hat H$ is the Airy kernel:
\beq
\hat H(z_1,z_2)={Ai(z_1^2)Bi'(z_2^2)-Ai'(z_1^2)Bi(z_2^2)\over z_1^2-z_2^2}\, \sqrt{z_1dz_1\, z_2dz_2}.
\eeq
The corresponding Baker-Akhiezer function is the Airy function, and the correlators $\om_n$, are the correlators given by the determinantal Airy process.

 \subsubsection{Sato formula}

The theorem \ref{thvariat}, implies that under an infinitesimal change of spectral curve of the 3rd kind \eq{defdS}: $\delta y dx =  t \,\,dS_{z_1,z_2}$, we have:
\beq
{\partial \om^{(g)}_n(z'_1,\dots,z'_n) \over \partial t}= \int_{z_2}^{z_1}\,\om^{(g)}_{n+1}(z'_1,\dots,z'_n,z)
\eeq
and thus:
\beq
{\partial^n F_g  \over \partial t^n}= \int_{z_2}^{z_1}\dots \int_{z_2}^{z_1}\,\om^{(g)}_{n}(z'_1,\dots,z'_n) .
\eeq
The exponential formula of proposition \ref{thexpkernel}, is nothing but the Taylor expansion of $F_g(t)$ computed at $t=N^{-1}$ in terms of derivatives taken at $t=0$, i.e. $F_g(N^{-1}) = {\displaystyle \sum_k} {N^{-k}\over k!}\,\partial_t^k F_g(0)$. In other words:
\bt\label{thSato}
\beq
\hat H(z_1,z_2)  = {\sqrt{dx(z_1)\,dx(z_2)}\over x(z_1)-x(z_2)}\,\,{\tau_N\left(\spcurve, x, y +{1\over N}\, {dS_{z_1,z_2}\over dx}\right)\over \tau_N(\spcurve,x,y)} .
\eeq
\et
This theorem can be interpreted as Sato's formula \cite{sato} for integrable systems.
In the context of random matrix theory, it can be interpreted as Heine's formula \cite{Szego}.

\subsubsection{Baker-Akhiezer functions}

Let $\alpha_1,\dots, \alpha_m$ be the poles of the function $x(z)$, of respective degrees $d_1,\dots,d_m$.
Since $x$ is a meromorphic form of degree $d={\displaystyle \sum_i} d_i$, there are $d$ sheets, i.e. $d$ points on $\spcurve$, $z^1(x),\dots,z^d(x)$, such that $x(z^k)=x$.
The following matrix:
\beq
{\cal H}(x_1,x_2) = \left(\,\hat H(z^j(x_1),z^i(x_2))\, \right)_{i,j=1,\dots,d}
\eeq
is a square matrix of size $d\times d$.

The $\Psi$-function of the Lax system \cite{BBT}, is obtained by choosing $x_2=\infty$, i.e. $z^i(x_2)=\alpha_i$.
Since some poles $\alpha_i$ are multiple poles, in order to get an invertible matrix, we take linear combinations of rows, and we define:
\beq\label{defBA1}
i=1,\dots,m,\,\, j=1,\dots,d_i
\qquad \quad \psi_{i,j}(z) = \mathop{{\rm lim}}_{z_2\to \alpha_i} \left[
 \left(d\over d\xi_i(z_2)\right)^{j-1}\, { \ee{-N\int_o^{z_2} ydx}\,\,\hat H(z,z_2)\over \sqrt{d\xi_i(z_2)}} \right]
\eeq
where $o$ is an arbitrary basepoint, and $\xi_i(z_2)=x(z_2)^{-1/d_i}$ is the local parameter in the vicinity of $\alpha_i$.

Those functions are the Baker-Akhiezer functions.

We also have $d$ couples $I=(i,j)$ with $i=1,\dots,m$, $j=1,\dots, d_i$, and thus the following matrix is a square matrix:
\beq
\Psi(x) = \left(\psi_{I}(z^k(x))\right)_{I=1,\dots,d,\,\, k=1,\dots,d}.
\eeq
It is the $\Psi$-function of the corresponding Lax system \cite{BBT}.

\subsubsection{Hirota formula}

Notice that $\hat H(z_1,z_2)$ has a simple pole at $z_1=z_2$ and behaves like:
\beq
\hat H(z_1,z_2) \sim {\sqrt{dz_1\,\,dz_2}\over z_1-z_2}
\eeq
near $z_1=z_2$, and this holds for any ($\genus=0$) spectral curve ${\cal E}=(\spcurve,x,y)$.
In particular, we have, for any two such spectral curves ${\cal E}$ and $\td{\cal E}$:
\bt\label{thHirota}
\beq
\Res_{z\to z_2}\, \hat H(z_1,z;{\cal E})\hat H(z,z_2;\td{\cal E}) = \hat H(z_1,z_2;{\cal E}) .
\eeq
\et
If we consider that $\hat H$ is given by the Sato formula of theorem \ref{thSato}, this theorem is precisely the Hirota equation for the $\tau$-function $\tau_N$ \cite{BBT, KP}.

This theorem  justifies that we can call $\tau_N$ a Tau-function. By expanding locally the $dS_{z_1,z_2}$ in the vicinity of poles of $x$, we can see that it is the multicomponent Kadamtsev-Petviashvili (KP) tau-function. There is one set of component for each pole $\alpha_i$ of $x$.
In the case where ${\cal E}$ is an hyperelliptical curve, of type $y^2={\rm Pol}(x)$, the function $x$ has two poles, which are symmetric of oneanother, and everything can be written in terms of the expansion near only one pole. In that case $\tau_N$ reduces to the Kortweg-de-Vries (KdV) tau-function \cite{BBT}.

\subsubsection{Higher genus}
\label{secintegrabilitygenus}

So far, in this section, we were considering genus zero spectral curves, i.e. $z\in {\mathbb C}$, and $x(z)$ and $y(z)$ analytical functions of a complex variable.

\medskip
The integrability relied on the Sato formula, which gives the kernel $\hat H$ as the $\tau$-function of a shifted spectral curve, i.e. the exponential formula.

For higher genus $\genus\geq 1$, the problem is that the exponential formula does not define a well-defined spinor on $\spcurve$. Indeed, $\spcurve$ is not simply connected, and the abelian integrals $\int_{z_2}^{z_1} \dots \int_{z_2}^{z_1} \om_n$ are multivalued functions of $z_1$ and $z_2$ because there is not a unique integration path between $z_1$ and $z_2$.
The exponential formula has to be modified.
It was proposed to modify it with some theta functions (see section \ref{secbackground}).

\bd Given a characteristics $(\mu,\nu)$, the "tau-function" is defined by the non-perturbative partition function of section \ref{secbackground}:

\beq
\tau_N(\mu,\nu,{\cal E}) = Z_{\cal E} (\mu,\nu).
\eeq

\ed
Then define the spinor kernel $\hat H_{(\mu,\nu)}$ through the Sato formula:
\bd
\beq\label{thSatoenus}
\hat H_{(\mu,\nu)}(z_1,z_2)  = {\sqrt{dx(z_1)dx(z_2)}\over x(z_1)-x(z_2)}\,\,{\tau_N(\mu,\nu,\spcurve, x, y +{1\over N}\, {dS_{z_1,z_2}\over dx})\over \tau_N(\mu,\nu,\spcurve,x,y)}.
\eeq
%where $E(z_1,z_2)$ is the prime form.
\ed
With this definition, $\hat H_{(\mu,\nu)}$ is closely related to the Szeg\"o kernel \cite{Szego}.

\medskip

\bt
$\hat H_{(\mu,\nu)}(z_1,z_2)$ is well defined for $z_1,z_2\in \spcurve$.
\et

\proof{Integrals of $\om_n^{(g)}$'s are in principle defined only on the universal covering of $\spcurve$, and one needs to check that after going around an $\acycle$-cycle or $\bcycle$-cycle, $\hat H_{(\mu,\nu)}(z_1,z_2)$ takes the same value.

Notice that, if $z_1$ goes around an $\acycle$-cycle, then $dS_{z_1,z_2}$  is unchanged, and if $z_1$ goes around the cycle $\bcycle_i$, then $dS_{z_1,z_2}$ is shifted by a holomorphic differential:
\beq
dS_{z_1+\bcycle_i,z_2} = dS_{z_1,z_2} + 2i\pi\,du_i .
\eeq
However, it was proved in \cite{background} that the $\tau$ function above is background independent, which exactely means that, for any $\l$:
\beq
\tau_N(\mu,\nu,\spcurve,x,y+\l du_i/dx) = \tau_N(\mu,\nu,\spcurve,x,y)
\eeq
and therefore, we see that $\hat H_{(\mu,\nu)}(z_1,z_2)$ is unchanged if $z_1$ goes around a $\bcycle$-cycle.
}

Then, we see that \cite{EMnonperturbative}
\bt\label{thHirotagenus}
$\hat H_{(\mu,\nu)}(z_1,z_2)$ obeys the Hirota equation:
\beq
\Res_{z\to z_2}\, \hat H_{(\mu,\nu)}(z_1,z;{\cal E}) \hat H_{(\mu,\nu)}(z,z_2;\td{\cal E}) = \hat H_{(\mu,\nu)}(z_1,z_2;{\cal E}) .
\eeq
\et
If we consider that $\hat H_{(\mu,\nu)}$ is given by the Sato formula of theorem \ref{thSatoenus}, this theorem is precisely the Hirota equation for the Tau-function $\tau_N$.

\subsection{Virasoro constraints}\label{secvirasoro}

Its has been understood for a long time that the random matrix integrals are fundamentally linked to Virasoro
and ${\cal{W}}$-algebras through differentials equations on their moduli called Virasoro or ${\cal{W}}$-constraints.
The definition of the symplectic invariants and of the correlation functions themselves were inspired by these
constraints since they mimic the solution of the loop equations of random matrix models, the latter being considered
as equivalent to the Virasoro constraints.

In a series of papers \cite{AMM1,AMM2,AMM3}, Alexandrov, Mironov and Morozov go even further and propose to generalize the
notion of random matrix integrals by defining a general string partition function interpolating between different matrix models. This partition function is characterized as a "D-module" solution of some Virasoro constraints.
%Nevertheless, they could not give any explicit formula for this partition function.

It is natural to see the symplectic invariants and the $\tau$-function built from them as a good candidate for this
string partition function. It is thus interesting to clarify the arising of Virasoro constraints in the theory of
symplectic invariants by looking at the variations of the latter wrt the moduli of the spectral curve.

\subsubsection{Virasoro at the branch points}

One can slightly rewrite the recursive relations defining the correlation functions \eq{defWnginv} by moving all the terms to the same side
of the equation. One gets:
\bea
0
&=& \sum_i \Res_{z\to a_i}\, K(z_0,z)\,\Big[
\om_{n+2}^{(g-1)}(z,\bar{z},J)
+ \sum_{h=0}^g\sum_{I\subset J} \om_{1+|I|}^{(h)}(z,I) \om_{1+n-|I|}^{(g-h)}(\bar{z},J\backslash I) \cr
&& \qquad \qquad \qquad \qquad \qquad \qquad + ydx(z) \om_{n+1}^{(g)}(\bar{z},J)
+ ydx(\bar{z}) \om_{n+1}^{(g)}(z,J)\Big]
.\cr
\eea

By summing over the genus $g$ and interpreting the correlation function as the result of the loop insertion operator
on the symplectic invariants, this equation can be rephrased as a Virasoro constraint:
\bt\label{thglobconstrbranch}
For any point $z$ on the spectral curve, the partition function is a zero mode of the {\em global Virasoro
operator} $\widehat{\cal{V}}(z)$
\beq
\widehat{\cal{V}}(z) \tau_N = 0
\eeq
with
\beq\label{defglobbranch}
\widehat{\cal{V}}(z) = \sum_i \oint_{a_i} K(z,z') :{\cal{J}}(z'){\cal{J}}\left(\overline{z'}\right):
\eeq
and the {\em global current} is defined by:
\beq
{\cal{J}}(z)= N ydx(z) + {1 \over N} D_z
\eeq
for any point $z$ of the spectral curve.
\et
This means that the recursive definition of the correlation functions is nothing but a Virasoro constraint on
the $\tau$-function defined {\bf globally} on the spectral curve.

Let us now approach a particular branch point $a_i$ and blow up the spectral curve around this point (see section \ref{secsingFg}). A rational parametrization of the blown up curve can read
\beq
\left\{\begin{array}{l}
\tilde{x}(z) = z^2 \cr
\tilde{y}(z) = {\displaystyle \sum_{k=0}^\infty} T_k^{(i)} z^k\cr
\end{array}
\right.
\eeq
where the $T_k^{(i)}$'s are the coefficients of the Taylor expansion of $ydx$ around the branch points $a_i$
\beq
ydx(z) = \sum_{k=0}^\infty T_k^{(i)} \hat{\xi}_i^{k+1}(z) d\hat{\xi}_i(z)
\eeq
with the local coordinates
\beq
\hat{\xi}_i(z) = \sqrt{x(z) -x(a_i)}.
\eeq
From section \ref{secsingFg}, one knows that the projection of the correlation functions in the local patch around $a_i$ built from the local parameter $\hat{\xi}_i$ is given by the correlation functions of the blown up curve, i.e. the
correlation functions of the Kontsevich integral with times $T_k^{(i)}$, $k=0,\dots, \infty$.
Since the recursive definition of these correlation functions is equivalent to Kontsevich Virasoro constraints \footnote{See \cite{AMM2,Chekhov} for detail on these continuous Virasoro contraints.}, this means that the global Virasoro
operator projects to the continuous Virasoro operator in this local patch of coordinates around $a_i$:

\bt
For any branch point $a_i$, the partition function is a zero mode of a set
{\em local Kontsevich Virasoro operator} $\widehat{\cal{V}}_{i}(\hat{\xi}_i(z))$ for any point $z$ in a neighborhood
of $a_i$:
\beq\label{localbrancheq}
\forall i \, , \; \;  \widehat{\cal{V}}_{i}(\hat{\xi}_i(z)) \tau_N = 0
\eeq
where $\hat{\xi}_i(z) = \sqrt{x(z)-x(a_i)}$ and the Virasoro operator annihilates the Kontsevich $\tau$-function:
\beq
\widehat{\cal{V}}_{i}(\hat{\xi}) Z_K(T_k^{(i)}) = 0
\eeq
where
\beq
Z_K(T_k^{(i)}):= \int_{formal} e^{-N \Tr \left( {M^3 \over 3} - \Lambda^2 M\right)}
\eeq
with Kontsevich times
\beq
T_k^{(i)} := {1 \over N} \Tr \Lambda^{-k}.
\eeq

\et
Indeed, as it is exhibited in section \ref{seckontsevich}, the corresponding spectral curve has only
one branch point and all the moduli of the integral are summed up in the Taylor expansion of the differential form
$xdy$ at this branch point.

\subsubsection{Loop equations and Virasoro at the poles}

In the preceding section, we have translated the recursive definition of the correlation functions into a set of
Virasoro operators related to the moduli of the spectral curve at the branch points. One can proceed in a similar way
for the moduli at the poles of the one form $ydx$ by building a set of equations solved by the correlation functions
called "loop equations" since they mimic the loop equations of random matrix theory.

\bt\label{thglobalpole}
The correlation functions $\om_n$ are solutions of the loop equations:
\beq
\sum_{l=0}^{k}   \om_{l+1}(z,{\bf z_L}) \om_{k-l+1}(z,{\bf z_{K \backslash L}}) + {1 \over N^2} \om_{k+2}(z,z,{\bf z_K}) = P_{1,k}(z,{\bf z_K})dx(z)^2
\eeq
where the function
\beq
P_{1,k}(z,{\bf z_K}):= \sum_i \oint_{\alpha_i} {{\displaystyle \sum_{l=0}^{k}}   \om_{l+1}(z',{\bf z_L}) \om_{k-l+1}(z',{\bf z_{K \backslash L}}) + {1 \over N^2} \om_{k+2}(z',z',{\bf z_K})
\over (z_i(z)-z_i(z'))dx(z')}
\eeq
is a function of $z$ with poles only at the poles of $ydx$.
\et
In the matrix model case, these loop equations are often referred to as Virasoro constraints. They indeed encode a set of
Virasoro constraints build from the poles of $ydx$. Let us make this assertion clear in the general framework of the
symplectic invariants:

\bt\label{thglobconstrpole}
For any point $z \in \spcurve$, the $\tau$-function satisfies
\beq
{\cal{V}}(z) \tau_N = 0
\eeq
where one defines the {\em global Virasoro operator}
\beq\label{globalconstraint}
{\cal{V}}(z)= {1 \over N^2} :{\cal{J}}^2(z): +
\sum_i \oint_{\alpha_i} {  :{\cal{J}}^2(z'):
\over (\xi_i(z')-\xi_i(z))dx(z')}
\eeq
and $\xi_i(z)={1 \over \xi(z)}$ is a local parameter in the neighborhood of $\alpha_i$ (see \eq{defxi} for the definition of $\xi(z)$).
\et

The $\tau$-function can thus be seen as the zero mode of another Virasoro operator  globally defined on the spectral
curve. This new operator, equivalent to the loop equations, can be easily projected to a set of local Virasoro operators
in the vicinity of the poles of $ydx$ instead of the branch points for the first one. In order to follow this procedure
one has to restrict to $ydx$ which are holomorphic forms with poles $\alpha_i$ such that
$ydx(z) \sim_{z \to a_i} {\displaystyle \sum_k} k t_{i,k} \xi_i^k(z) d\xi_i(z)$.
\bt For any point $z$ in the neighborhood of a pole $\alpha_i$ of $ydx$:
\beq
{\cal{V}}_-^{(i)}(z) \tau_N =0
\eeq
where the {\em local Virasoro operator} is defined as the loop operator
\beq
{\cal{V}}_-^{(i)}(z) := \oint_{\alpha_i} {d\xi_i(z') \over \xi_i(z')-\xi_i(z)} :{J}^{(i)}(z')^2:
\eeq
with the local current
\beq
{J}^{(i)}(z):= \sum_{k\geq 0} \left[ {k t_{i,k}} \xi_i(z)^{k-1} d\xi_i(z) + {d\xi_i(z) \over \xi_i(z)^{k+1}} {\partial \over \partial t_{i,k}}\right].
\eeq
\et
Remark that these local Virasoro operators are indeed Laurent series in $\xi_i(z)$ with only negative powers whose
coefficients are differential operators satisfying Virasoro commutation relations:
\beq
{\cal{V}}_-^{(i)}(z) = \sum_{k>0} {\cal{V}}_k^{(i)} \xi_i(z)^{-k} (d\xi_i(z))^2
\eeq
and
\beq
\left[{\cal{V}}_j^{(i)},{\cal{V}}_l^{(k)}\right] = (j-l) {\cal{V}}_{j+l} \delta_{i,k}.
\eeq
These local operators around the poles have also a natural solution: the one hermitian matrix integral
\beq
Z_{1MM}:= \int_{formal} e^{-{N \over t} \Tr V(M)} dM
\eeq
with a polynomial potential
\beq
V(x):= {x^2\over 2} - {\displaystyle \sum_{k=3}^d} {t_k } x^k
\eeq
whose coefficients $t_k$ are identified with the moduli at the poles $t_{i,k}$ (see section \ref{sec1MM} for more details).

\subsubsection{Givental decomposition formulae}

Let us suppose in this short section that the spectral curve has genus 0, i.e. $\spcurve$=Riemann sphere.
In this case, the only moduli of the curve are:
\begin{itemize}
\item either the position of the poles and moduli $t_{i,k}$ at these poles;
\item either the position of the branch point and the moduli $T_k^{(i)}$.
\end{itemize}

Let us first focus on the branch points of the spectral curve. The dependence of $\tau_N$ on the moduli at the branch points
is constrained by the local Virasoro equations \eq{localbrancheq}. Thus, this $\tau$-function can be decomposed as a product of the
zero modes of the different local operators at the branch points, i.e. a product of Kontsevich integrals, up to a conjugation
operator mixing the moduli at the different branch points.

\bt\label{thdecompobranch}
$\tau_N$ can be decomposed into a product of Kontsevich integrals associated to the branch points $a_i$:
\beq\encadremath{
\tau_N({\bf T^{(1)}},{\bf T^{(2)}},\dots) = e^{\widehat{\cal{U}}} \prod_i {\cal{Z}}_K({\bf T^{(i)}}).}
\eeq
where the symbol ${\bf T^{(i)}}$ stands for the infinite family $\left\{T_k^{(i)}\right\}_{k=0}^\infty$,
with the intertwining operator $\widehat{\cal{U}}$ defined by
\beq
\widehat{\cal{U}}:=\sum_{i,j} \oint_{a_j} \oint_{a_i} \widehat{A}^{(i,j)}(z,z') \widehat{\Omega}_j(z') \widehat{\Omega}_i(z)
\eeq
where
\beq
\widehat{A}^{(i,j)}(z,z'):=
 B(z,z') - {d\hat{\xi}_i(z) d\hat{\xi}_j(z') \over (\hat{\xi}_i(z)-\hat{\xi}_j(z'))^2}
\eeq
and
\beq\label{defomeagi}
\widehat{\Omega}_i(z):= N \sum_k T_k^{(i)} \hat{\xi}_i^{k}(z) d\hat{\xi}_i(z) - {1 \over N} { d\hat{\xi}_i(z) \over k \hat{\xi}_i^{k}(z)} {\partial \over \partial T_{k}^{(i)}}.
\eeq

\et

One can proceed exactly in the same way by looking at the moduli at the poles: this time the decomposition is expressed
as a product of 1-hermitian matrix integrals.

\bt\label{thdecompopole}
$\tau_N$ can be decomposed into a product of one hermitian matrix integrals associated to the poles
$\alpha_i$ of the meromorphic form $ydx$
\beq\encadremath{
\tau_N({\bf t_1},{\bf t_2},\dots) = e^{\cal{U}} \prod_i {\cal{Z}}_{1MM}({\bf t_i}).}
\eeq
where ${\bf t_i}$ stands for the infinite set $\left\{t_{i,k}\right\}_{k=0}^\infty$,
with the intertwining operator ${\cal{U}}$ defined by
\beq
{\cal{U}}:=\sum_{i,j} \oint_{\alpha_j} \oint_{\alpha_i} A^{(i,j)}(z,z') \Omega_j(z') \Omega_i(z)
\eeq
where
\beq
A^{(i,j)}(z,z')
= B(z,z') - {d\xi_i(z) d\xi_j(z') \over (\xi_i(z)-\xi_j(z'))^2}
\eeq
and
\beq\label{defomegailoc}
\Omega_i(z):= N \sum_k t_{i,k} \xi_i^{k}(z) d\xi_i(z) - {1 \over N} { d\xi_i(z) \over k \xi_i^{k}(z)} {\partial \over \partial t_{i,k}}.
\eeq
\et

Remark that, in both cases, these decomposition formulae consist in writing a  KP tau function as a product of KdV tau-functions.
Indeed, these formula were already derived by Givental in the study of KP tau functions \cite{Giv1,Giv2}.

\subsubsection{Vertex operator and integrability}

In this paragraph, we do not consider the Baker-Akhiezer functions as defined in section \ref{secintegrability}. On the contrary, we define them as the images of the partition function under the action of some
global operator on the spectral curve.

Let us define the equivalent of the Baker-Akhiezer (BA) functions:
\bd
One defines the $x$-type global BA and dual BA functions as
\beq
{\bf \Psi}(z):=\exp\left(\Omega_x(z) \right) {\cal{Z}} \qquad \hbox{and} \qquad
{\bf \Psi}^*(z):=\exp\left(-\Omega_x(z) \right) {\cal{Z}}
\eeq
where
\beq
\Omega_x(q):= \int^q {\cal{J}}_x = N \int^q ydx(z) - {1 \over N} \int^q D_z .
\eeq
One also defines the x-type local BA functions as:
\beq
{\bf \Psi}_{i,0}(z):=\exp\left(\Omega_i(z) \right) {\cal{Z}} \qquad \hbox{and} \qquad
{\bf \Psi}_{i,0}^*(z):=\exp\left(-\Omega_i(z) \right) {\cal{Z}}
\eeq
where the operator $\Omega_i$ was defined in \eq{defomegailoc}.

We finally define the corresponding $y$-type BA functions:
\beq
\widetilde{\bf \Psi}(z):=\exp\left(\Omega_y(z) \right) {\cal{Z}} \qquad \hbox{and} \qquad
\widetilde{\bf \Psi}^*(z):=\exp\left(-\Omega_{y}(z) \right) {\cal{Z}}
\eeq
where
\beq
\Omega_y(q):= \int^q {\cal{J}}_y = N \int^q xdy(z) - {1 \over N} \int^q D_z .
\eeq

\ed

These functions correspond to deformations of the spectral curve and thus coincide with the Baker-Akhiezer functions of \eq{defBA1}.
\bl
The BA functions can be written in terms of the partition function as
\beq
{\bf \Psi}(z):=\exp\left(\Omega_x(z) \right) {\cal{Z}}({\bf t_i})= { {\cal{Z}}({\bf t_i}+ [z_i^{-1}]) \over {\cal{Z}}({\bf t_i})},
\eeq
\beq
{\bf \Psi}_{i,0}(z):=\exp\left(\Omega_i(z) \right) {\cal{Z}}(t_i)= {{\cal{Z}}(t_i+ [z_i^{-1}]) \over {\cal{Z}}(t_i)} ,
\eeq
\beq
\widetilde{{\bf \Psi}}(z):=\exp\left(\Omega_y(z) \right) {\cal{Z}}({\bf \tilde{t}_i})= { {\cal{Z}}({\bf \tilde{t}_i}+ [z_i^{-1}]) \over {\cal{Z}}({\bf \tilde{t}_i})}
\eeq
and
\beq
\widetilde{{\bf \Psi}}_{i,0}(z):=\exp\left(\Omega_{y,i}(z) \right) {\cal{Z}}(\tilde{t}_i)= {{\cal{Z}}(\tilde{t}_i+ [z_i^{-1}]) \over {\cal{Z}}(\tilde{t}_i)}
\eeq
where the $\tilde{t}_i$ are the coefficients of  the Taylor expansion of $xdy(z)$ as $z \to \alpha_i$ and
$\left[z^{-1}\right]$ is the usual Hirota symbol \cite{KP}.
\el

On the other hand, thanks to the pole structure of the BA functions, one gets
\bt The Baker-Akhizer functions satisfy the bilinear Hirota equation
\beq
\sum_i \oint_{\alpha_i} {\bf \Psi}(p|{\bf t}) e^{-N\int^p ydx+ N \int^p y'dx'}{\bf \Psi}^*(p|{\bf t'}) =
\sum_i \oint_{\alpha_i} {\bf \widetilde{\Psi}}(p|{\bf t}) e^{-N\int^p xdy+ N \int^p x'dy'} {\bf \widetilde{\Psi}}^*(p|{\bf t'})
\eeq
where $x'$ and $y'$ are functions on $\spcurve$ satisfying another algebraic equation
\beq
{\cal{E}}'(x'(z),y'(z) )=0
\eeq
compared to
\beq
{\cal{E}}(x(z),y(z) )=0.
\eeq
\et

\proof{The proof relies on the simple observation that
\beq
\Omega_x - N \int^z ydx = \Omega_y - N \int^z xdy = D_z .
\eeq
}

\bc
If $x$ has $q$ poles and $y$ has $p$ poles, the partition function ${\cal{Z}}({\bf t})$ is $\tau$-function of the multi-component $p+q$-KP hierarchy
since it satisfies the Hirota equations:
\bea
\sum_i \oint_{\alpha_i} {{\cal{Z}}(t_i+ [z_i^{-1}]) \over {\cal{Z}}(t_i)} {{\cal{Z}}(t_i'- [z_i^{-1}]) \over {\cal{Z}}(t_i')} e^{N {\displaystyle \sum_k} (t_{i,k}-t_{i,k}') z_i^{k-1}(p)} = \cr
 = \oint_{\alpha_i} {{\cal{Z}}(\tilde{t}_i+ [\tilde{z}_i^{-1}]) \over {\cal{Z}}(\tilde{t}_i)} {{\cal{Z}}(\tilde{t}_i'- [\tilde{z}_i^{-1}]) \over {\cal{Z}}(\tilde{t}_i')} e^{N {\displaystyle \sum_k} (\tilde{t}_{i,k}-\tilde{t}_{i,k}') \tilde{z}_i^{k-1}(p)} . \cr
\eea

\ec

\subsection{Singular limits}
\label{secsingFg}

The $F_g$'s and $\om_n^{(g)}$'s can be computed for any regular spectral curve, i.e. as long as the branchpoints are simple.
When the spectral curve is singular, the $F_g$'s are not defined.

Nevertheless, consider a 1-parameter family of spectral curves ${\cal E}(t)$, such that ${\cal E}(t_c)$
is singular, we prove below, that $F_g(t)$ diverges as $t\to t_c$, in the following form:
\beq
F_g(t) \sim (t-t_c)^{(2-2g)\mu}\, \td{F}_g .
\eeq
The goal of this section is to prove this divergent behavior, and compute the exponent $\mu$ and the prefactor $\td{F}_g$. These aymptotics are very important in many applications in mathematics and physics, for instance Witten's conjecture relates the asymptotics of large discrete surfaces, to integrals over moduli spaces of continuous Riemann surfaces.
Asymptotic formulae play also a key role in the universal limits of random matrix eigenvalues statistics, or in the study of universality in the statistics of non-intersecting Brownian motions (see section \ref{secbrownian}).

In the context of matrix models quantum gravity, the prefactor $\td{F}_g$ is called the ``double scaling limit'' of $F_g$, and the exponent $2-2\mu$ is called
\beq
2-2\mu = \gamma_{{\rm string}} = {\rm "string \, susceptibility\, exponent"}.
\eeq
It is such that $F_0''$ formally diverges with the exponent $-\gamma_{{\rm string}}$:
\beq
{d^2 F_0 \over dt^2} \sim (t-t_c)^{-\gamma_{{\rm string}}}.
\eeq

\subsubsection{Blow up of a spectral curve}

Consider a one parameter family of spectral curves ${\cal E}(t)=(\spcurve(t),x(z,t), y(z,t))$, such that ${\cal E}(t)$ is regular in an interval $]t_c,t_0]$.
For the moment we do not assume that ${\cal E}(t_c)$ is singular, i.e. it may be either regular or singular.
In a small vicinity of $t_c$, we can, to leading orders, parameterize ${\cal E}(t)$ in terms of $\spcurve=\spcurve(t_c)$.

Moreover, let $a$ be a branchpoint, and let us study the correlators $\om_n^{(g)}(z_1,\dots,z_n)$ in the vicinity of $a$. We choose a rescaled local coordinate $\zeta$ in the vicinity of $a$. Let us write:
\beq
z = a + (t-t_c)^\nu \zeta + o((t-t_c)^\nu).
\eeq
We want to compute the asymptotic behavior of $\om_n^{(g)}(a+(t-t_c)^\nu \zeta_1,\dots,a+(t-t_c)^\nu \zeta_n;t)$ in the limit $t\to t_c$.

\medskip

First, let us study the behavior of $x$ and $y$, by Taylor expansion. Let $q$ be the first non-trivial power in the Taylor expansion of $x$, i.e.:
\beq
x(a+(t-t_c)^\nu\zeta;t) = x(a;t_c) + (t-t_c)^{q\nu} \td{x}(\zeta) + o((t-t_c)^{q\nu})
\eeq
and similarly, there is an exponent $p$ such that:
\beq
y(a+(t-t_c)^\nu\zeta;t) = y(a;t_c) + (t-t_c)^{p\nu} \td{y}(\zeta) + o((t-t_c)^{p\nu}).
\eeq
This means that at $t=t_c$ the curve ${\cal E}(t_c)$ behaves like $y\sim y(a) + (x-x(a))^{p/q}$. It is regular if ${p\over q}={1\over 2}$ and singular otherwise.

The rescaled curve $\td{{\cal E}}=(\td\spcurve,\td{x},\td{y})$ is called the blow up of the spectral curve near the branchpoint $a$, in the limit $t\to t_c$.

\medskip

The choice of the exponent $\nu$, must be such that $\td{{\cal E}}$ is a regular spectral curve.
We cannot give a general formula for $\nu$, since it depends on the explicit choice of a 1-parameter family of spectral curves ${\cal E}(t)$, and how it is parametrized.
Also, here we consider only algebraic singularities of type $y\sim x^{p/q}$, but the method could certainly be extended to other types of singularities.

\bigskip

{\bf Examples:}

\medskip

$\bullet$ The following spectral curve arises in the enumeration of quadrangulated surfaces (see section \ref{secquadr}):
\beq
\spcurve={\mathbb P}^1
\virg
x(z) = \gamma\left(z+{1\over z}\right)
\virg
y(z) = {-1\over \gamma z} + {t_4 \gamma^3\over z^3}
\eeq
where
\beq
\gamma^2 = {1-\sqrt{1-12t_4}\over 6 t_4} .
\eeq
The branchpoints $x'(a)=0$ are $a=\pm 1$.
Consider the branchpoint $a=1$, and introduce an auxillary scaling variable $t$:
\beq
z = 1 + t \zeta ,
\eeq
such that $x$ and $y$ are independent of $t$, and let us study the vicinity of $t\to t_c=0$.

We whish to study the behaviour of $\om_n^{(g)}(1+t\zeta_1,\dots,1+t\zeta_n)$ in the vicinity of $t\to 0$, i.e. the behaviour of $\om_n^{(g)}$ in the vicinity of the branchpoint $a=1$.

In the limit $t\to 0$, we Taylor expand $x$ and $y$:
\beq
x(z) = 2\gamma + \gamma t^2 \zeta^2 + O(t^3) ,
\eeq
\beq
y(z) = {-1\over \gamma} + {t_4\gamma^3} + t \zeta ({1\over \gamma} - 3 t_4 \gamma^3) + O(t^2).
\eeq
Notice that we have $q=2$ and $p=1$, which means that our curve is not singular at $t=0$ (which was expected since it is actually independent of $t$).
\smallskip

The blow up is:
\beq
\td{x}(\zeta) = \gamma\, \zeta^2
\virg
\td{y}(\zeta) =  \left({1\over \gamma} - 3 t_4 \gamma^3\right)\, \zeta .
\eeq
Up to a rescaling, it is the Airy spectral curve (see section \ref{secdsl} and example \eq{Airy}).
%Therefore:
%\beq
%\om_n^{(g)}(1+t\zeta_1,\dots,1+t\zeta_n) \sim (t^3\,\,(1-3 t_4 \gamma^4))^{2-2g-n}\,\, \td\om_n^{(g)}(\zeta_1,\dots,\zeta_n)_{\rm Airy}
%\eeq
%where ${\td\om_n^{(g)}}{}_{\rm Airy}$ are the $ \om_n^{(g)}$'s for the Airy curve $\td{y}=\sqrt{\td{x}}$ studied in section \ref{}.

\medskip

$\bullet$ In the previous example, at $t_4={1\over 12}$, we have $(1 - 3 t_4 \gamma^4)=0$, and thus, one needs to go further in the Taylor expansion.
Let us now choose $t=1-12t_4$, $t_c=0$, and $a=1$.
We have in the limit $t\to t_c$:
\beq
x\left(1+ {1\over \sqrt{2}}\,t^{1\over 4}\zeta\right) = 2\sqrt{2} + {\sqrt{t}\over \sqrt{2}}\,\, (\zeta^2-2) + o(\sqrt{t}) ,
\eeq
\beq
y\left(1+{1\over \sqrt{2}}\,t^{1\over 4}\zeta\right) = -{\sqrt{2}\over 3} +  {t^{1\over 2}\over 2\sqrt{2}}\, (\zeta^2-2) -{t^{3\over 4}\over 12}\,(7\zeta^3-12\zeta)  + o(t^{3\over 4}),
\eeq
and, in fact, what we really need is the asymptotic behavior of $y(z)-y(\bar{z})$, i.e.:
\beq
y\left(1+{1\over \sqrt{2}}\,(t_c-t)^{1\over 4}\zeta\right)-y\left({1 \over (1+{1\over \sqrt{2}}\,(t_c-t)^{1\over 4}\zeta)}\right)
= -{2\,t^{3\over 4}\over 3}\,(\zeta^3-3\zeta)  + o(t^{3\over 4}) .
\eeq

The blow up of the spectral curve in this limit is thus:
\beq
\td{x}(\zeta) = {1\over \sqrt{2}}\,(\zeta^2-2)
\virg
\td{y}(\zeta) =  -{1\over 3} \,(\zeta^3-3\zeta) .
\eeq
Notice that it is proportional to the "pure gravity" spectral curve $(p,q)=(3,2)$, see section \ref{secdsl}, and see the first example in section \ref{secexspcurves}. It is the spectral curve which arises everytime we have a $y\sim x^{3/2}$ cusp singularity.

\subsubsection{Asymptotics}

In order to study the asymptotics of the $\om_n^{(g)}$'s, we need to study the asymptotics of the kernels $B$ and $K$, in the limit $t\to t_c$.

\medskip
We have:
\beq
B(z_0,z) \sim
\begin{tabular}{|l|l|l|}
\hline
 & $z$ near $a$ & $z$ far from $a$ \cr
\hline
 $z_0$ near $a$ & $\td{B}(\zeta_0,\zeta)$ & $O((t-t_c)^\nu)$ \cr
 $z_0$ far from $a$ & $O((t-t_c)^\nu)$ & $O(1)$ \cr
\hline
\end{tabular}
\times (1+O((t-t_c)^\nu)),
\eeq
where $\td{B}(\zeta_0,\zeta)$ is the Bergmann kernel of the blown up spectral curve $(\td\spcurve,\td{x},\td{y})$:
\beq
\td{B}(\zeta_0,\zeta) = {d\zeta_0 \,\, d\zeta\over (\zeta-\zeta_0)^2}.
\eeq

Similarly, the kernel $K$ behaves like:
\beq
K(z_0,z) \sim
\begin{tabular}{|l|l|l|}
\hline
 & $z$ near $a$ & $z$ far from $a$ \cr
\hline
$z_0$ near $a$ & $ (t-t_c)^{-(p+q)\nu} \td{K}(\zeta_0,\zeta)$ & $ O(1)$ \cr
$z_0$ far from $a$ & $ O((t-t_c)^{-(p+q-1)\nu})$ & $O(1)$ \cr
\hline
\end{tabular}
\times(1+O((t-t_c)^\nu)),
\eeq
where $\td{K}(\zeta_0,\zeta)$ is the recursion kernel of the blown up spectral curve $(\td\spcurve,\td{x},\td{y})$:
\beq
\td{K}(\zeta_0,\zeta) = {1\over 2}\, \left( {1\over \zeta_0-\zeta} - {1\over \zeta_0+\zeta}\right)\,\, {1\over 2\td{y}(\zeta)\, \td{x}'(\zeta)} .
\eeq

%( dz_0/(z_0-z)-1/(z_0-1/z)  ) 1/(2y(z) x'(1/z))

Therefore, we see that the leading contribution to $\om_{n}^{(g)}(1+\delta \zeta_0,\dots,1+\delta \zeta_n)$ is given by the terms where all residues are taken near $a$, and the leading contribution can be computed only in terms of $\td{B}$ and $\td{K}$. By an easy recursion, on gets:

\bt\label{thsinglimomng}

Singular limit of $\om_n^{(g)}$ with $2-2g-n<0$ and $n\geq 1$.

If $a$ is a branchpoint, the asymptotics of $\om_n^{(g)}$ in the vicinity of $t\to t_c$ and $z_i\to a$, are given by:
\beq
\om_n^{(g)}(a+(t-t_c)^\nu\zeta_1,\dots,a+(t-t_c)^\nu\zeta_n)
\sim
(t-t_c)^{(2-2g-n)(p+q)\nu}\,\,\, \td\om_n^{(g)}(\zeta_1,\dots,\zeta_n)
\eeq
where $\td\om_n^{(g)}$ are the correlators of the blown up spectral curve $\td{\cal E}=(\td\spcurve,\td{x},\td{y})$.
\et
In this theorem, the exponent $\nu$ is unspecified, it depends on our choice of 1-parameter family of curves, i.e. it depends on each example. We recall that it must be chosen such that the blown up curve $\td{\cal E}$ is regular. We see examples in section \ref{secdsl}.

This theorem implies in particular, that all correlation functions in the vicinity of a regular branchpoint, are, to leading order, the same as the Airy process correlation functions, we recover the universals Airy law near regular branchpoints. This is related to the universal Tracy-Widom law \cite{TWlaw}.

\medskip
One may extend this theorem to $\om_0^{(g)}=F_g$'s:
\bt\label{thsinglimFg}

Singular limit of $F_g$.

If at $t=t_c$, the branchpoint $a$ becomes singular, and no other branchpoint becomes singular, then the asymptotics of $F_g, g\geq 2$ in the vicinity of $t\to t_c$ are given by:
\beq
F_g \sim (t-t_c)^{(2-2g)(p+q)\nu}\,\,\, \widetilde{F}_g \quad +o((t-t_c)^{(2-2g)(p+q)\nu})
\eeq
where $\widetilde{F}_g$ are the symplectic invariants of the blown up spectral curve $\td{\cal E}=(\td\spcurve,\td{x},\td{y})$.
\et

In fact this theorem holds also in the case where the branchpoint $a$ is not singular, but it becomes useless. Indeed, if $a$ is not singular, the Blown up spectral curve is the Airy curve, and the $\td{F}_g$'s of the Airy spectral curve vanish for all $g\geq 1$, and therefore all what the theorem says in this case, is that $F_g$ does not diverge as $(t-t_c)^{(2-2g)(p+q)\nu}$. And this is obvious since $F_g$ is not divergent at all.

\medskip
Also, the condition that $a$ is the only singular branchpoint, is not so necessary.
In fact, if several branchpoints become singular simultaneously at $t_c$, with the same exponents $\nu$ and $(p,q)$, then the asymptotics of $F_g$ is the sum of contributions of most singular branchpoints.
The most basic example is a symmetric spectral curve $y(x)=y(-x)$, for which branchpoints come by pairs. The leading order of $F_g$ then gets a factor $2$:
$F_g \sim 2\,\, (t-t_c)^{(2-2g)(p+q)\nu}\,\,\, \widetilde{F}_g \quad +o((t-t_c)^{(2-2g)(p+q)\nu})$.

%%%%%%%%%%%%%%%%%%%%%%%%%%%%%%%%%%%%%%%%%%%%%%%%%%%%%%%%%%%%%%%%%%%%%%%%%%%%%%%%%%%%%%%%%

\section{Application to matrix models}\label{secMM}

The recursion relations defining the symplectic invariants and their correlation functions were originally found
in the study of the one-Hermitian random matrix model \cite{eynloop1mat,ec1loopF} where they appeared as the solution to the so-called
loop equations.
It is thus interesting to remind to which extent the symplectic invariants give a solution to the computation
of the free energies and correlation functions' topological expansions in different matrix models. It is also interesting
to emphasize the special properties of the spectral curves obtained from matrix models to remind that they represent only
a particular subcase in the whole framework for symplectic invariants.

\subsection{1-matrix model}\label{sec1MM}

The formal 1-matrix integral is defined as a formal power series in a variable $t$.

Consider a polynomial $V(x)$ (called "potential") of degree $d+1>2$ as well as its $d$ stationary points $\xi_i$:
\beq
\forall i= 1, \dots,d \; , \; \; V'(\xi_i)=0
\eeq
and the non-quadratic part of its Taylor expansion around these points
\beq
\delta V_i(x) = V(x) - V(\xi_i) - {V''(\xi_i) \over 2} (x-\xi_i)^2.
\eeq
Let us also consider a $d$-partition of N, i.e. a set of $d$ integers $n_i$ satisfying
\beq
\sum_{i=1}^d n_i = N.
\eeq

\bd Formal 1-hermitian matrix integral.

The formal 1-matrix integral is defined as a formal power series in a variable $t$:
\bea\label{defZformal1MM}
Z_{\rm 1MM} &=& \ee{-{N\over t}{\displaystyle \sum_i} n_iV(\xi_i)}\,\,{\displaystyle \sum_{k=0}^\infty} {(-1)^k N^k\over t^k\,\, k!}\,\, \int_{H_{n_1}}\dots \int_{H_{n_d}}
\prod_{i=1}^d dM_i \,\, \left(\sum_i \Tr \delta V_i(M_i)\right)^k  \cr
&&  \ee{-{N\over 2t} {\displaystyle \sum_i} V''(x_i) \Tr  \left(M_i-\xi_i {\bf 1}_{n_i}\right)^2} \,\,\,
\prod_{i>j} \det\left(M_i \otimes {\bf 1}_{n_j} - {\bf 1}_{n_i} \otimes M_j \right)^2\cr
&=& 1+ \sum_{k=1}^\infty t^k\, A_k \cr
\eea
where each $A_k$ is a (well defined) polynomial moment of a Gaussian integral.

It  is denoted by:
\beq
Z_{\rm 1MM}=
\int_{\rm formal} \ee{-{N\over t}\Tr V(M)}\, dM .
\eeq
\ed
This last notation comes from the exchange of the Taylor expansion of $\exp{-{N\over t} {\displaystyle \sum_i} \Tr \delta V_i(M_i)}$ and the integral. Once
this commutation is performed, the integral obtained corresponds to the formal expansion of the integral $\int \ee{-{N\over t}\Tr V(M)}\, dM$
around a saddle point $\tilde{M}$ solution of $V'(\td{M})=0$, which we choose as:
\beq
\tilde{M}=\hbox{diag} \left(\overbrace{\xi_1,\dots,\xi_1}^{n_1},\dots,\overbrace{\xi_i,\dots,\xi_i}^{n_i},\dots,\overbrace{\xi_{d},\dots,\xi_{d}}^{n_{d}}\right).
\eeq
The integers $n_i$ thus correspond to choosing the number of eigenvalues of the saddle matrix located at a particular solution of the saddle-point equation $V'(\xi_i)=0$.

\smallskip

However, in general, the Taylor expansion and the integral do not commute, and the formal matrix integral is different from the usual convergent matrix integral:
\beq
\int_{\rm formal} \ee{-{N\over t}\Tr V(M)}\, dM
\neq \int \ee{-{N\over t}\Tr V(M)}\, dM.
\eeq
In fact, typically, convergent matrix integrals are obtained for $V>0$, whereas formal matrix integrals have combinatorical interpretations for $V<0$.

\br The definition of the matrix integral does not depend only on the potential (i.e. the coefficients of this polynomial)
but also on the filling fractions
\beq
\epsilon_i := {t n_i \over N} \virg i=1,\dots ,d
\virg
\sum_{i=1}^d \epsilon_i = t.
\eeq
\er

\bigskip

The formal logarithm of $Z_{{\rm 1 MM}}$ is also a formal power series in $t$ of the form:
\beq
\ln{Z_{\rm 1MM}} = \sum_{k=0}^\infty t^k \,\widetilde{A}_k
\eeq
and it can be seen from general properties of polynomial moments of Gaussian matrix integrals (observation first made by 't Hooft \cite{thooft}), that each coefficient $N^{-2} \widetilde{A}_k$ is a {\bf polynomial} in $1/N^2$:
\beq
\widetilde{A}_k = N^2\,\sum_{g=0}^{g_{\rm max}(k)}\, \widetilde{A}_k^{(g)}\, N^{-2g} .
\eeq
Therefore, we may collect together the coefficients of given powers of $N$, and define a formal power series:
\beq
F_g = t^{2-2g}\,\sum_{k=1}^\infty \td{A}_k^{(g)}\,\, t^k .
\eeq
We have:
\bt
\beq
\ln{Z_{\rm 1MM}} = \sum_{g=0}^\infty (N/t)^{2-2g} \, F_g .
\eeq
\et
This theorem is an equality between formal power series of $t$. This means the coefficients in the small $t$ power series expansions of both sides are the same. And for a given power of $t$, the sum in the RHS is in fact a finite sum.

\medskip

\subsubsection{Loop equations}

We may also define the following formal correlation functions:
\beq
W_n(x_1,\dots,x_n) = \left< \Tr{1\over x_1-M} \dots \Tr{1\over x_n-M}\right>_c
\eeq
where the subscript $c$ means the cumulant and the notation $\Tr{1 \over x-M}$ stands for the formal series:
\beq\label{defformalTr1/x-M}
\Tr {1 \over x-M}
\equiv
\sum_{i=1}^d \,\sum_{k=0}^{\infty} \,\, \Tr{(M_i-\xi_i\,{\mathbf 1}_{n_i})^k \over (x-\xi_i)^{k+1}}
\eeq
to be inserted in the integrand of \eq{defZformal1MM}.
Again, $W_n(x_1,\dots,x_n)$ is defined as a formal power series in $t$, whose coefficients are polynomial moments of Gaussian integrals.
Moreover, one may notice that the coefficient of $t^k$ is a polynomial in $1/N$, and is a rational fraction of $x_1,\dots,x_n$ with poles at the $\xi_i's$.

\medskip

\br
Each coefficient of $W_n(x_1,\dots,x_n)$ is a rational fraction of the $x_j$'s with poles at the $\xi_i$'s, and from \eq{defformalTr1/x-M}, one sees that simple poles can appear only when $k=0$ in \eq{defformalTr1/x-M}, i.e. terms independent of $M_i$. This implies that the cumulants $W_n(x_1,\dots,x_n)$ can have no simple poles when $n>1$, and for $W_1$, the only residue is:
\beq\label{ResW1xiini1MM}
\Res_{x\to \xi_i} W_1(x) dx = n_i
\eeq
and when $n>1$:
\beq\label{ResWnxiini1MM}
\Res_{x_1\to \xi_i} W_n(x_1,x_2,\dots x_n) dx_1 = 0.
\eeq
Again those two equalities are equalities between the coefficients of formal series of $t$.
\er

\medskip

We may collect together coefficients with the same power of $N$. That allows to write:
\beq\label{WnWngtoexp1MM}
W_n(x_1,\dots,x_n) = \sum_{g\geq 0} \left({N \over t}\right)^{2-2g-n} \,\, W_n^{(g)}(x_1,\dots,x_n)
\eeq
where each $W_n^{(g)}$ is a formal power series in $t$, whose coefficients are rational fractions of $x_1,\dots,x_n$ with poles at the $\xi_i's$.
\Eq{WnWngtoexp1MM} is an equality of formal power series of $t$.

For further convenience we also define in a similar manner:
\beq
P_n(x_1;x_2,\dots,x_n) = \left< \Tr{V'(x_1)-V'(M)\over x_1-M}\,\, \Tr{1\over x_2-M} \dots \Tr{1\over x_n-M}\right>_c
\eeq
which is a polynomial in the variable $x_1$.
Again, we may collect together coefficients with the same power of $N^{-1}$, and define $P_n^{(g)}$ such that:
\beq
P_n(x_1;x_2,\dots,x_n) = \sum_{g \geq 0} \left({N\over t}\right)^{2-2g-n} \,\, P_n^{(g)}(x_1;x_2,\dots,x_n).
\eeq

Then, the Schwinger-Dyson equations imply:
\bt
We have the loop equations, $\forall n,g$:
\bea\label{loopeq1MM1cut}
&& W^{(g-1)}_{n+2}(x,x,J) + \sum_{h=0}^g\sum_{I\subset J} W^{(h)}_{1+|I|}(x,I) W^{(g-h)}_{1+n-|I|}(x,J\backslash I) \cr
&& + \sum_{j=1}^n {\partial \over \partial x_j}\,{W^{(g)}_{n}(x,J\backslash \{x_j\})-W^{(g)}_{n}(J)\over x-x_j} \cr
&=& V'(x) W^{(g)}_{n+1}(x,J) - P^{(g)}_{n+1}(x;J)
\eea
where $J=\{x_1,\dots,x_n\}$.

\et

\proof{This theorem can be proved by integrating by parts the gaussian integrals for each power of $t$, see \cite{eynloop1mat}. It is called Schwinger-Dyson equations, or loop equations, or sometimes Ward identities...
cf \cite{Migdalloop, Kazakovloop, Virasoro, ZJDFG}.}

Remark also that this theorem corresponds to the global Virasoro constraints of theorem \ref{thglobalpole}.

Loop equations were initially used to find the topoloical expansion of the 1-matrix model in the special case of 1-cut \cite{ACKM} and also to the first orders 2-cuts \cite{akeman}.

\subsubsection{Spectral curve}

For $n=0$ and $g=0$ the loop equation \eq{loopeq1MM1cut} reduces to an algebraic equation for $W_1^{(0)}(x)$ sometimes known as the master loop equation:
\beq\label{eqspcurve1MMW10}
(W_1^{(0)}(x))^2 = V'(x)\,W_1^{(0)}(x) - P_1^{(0)}(x)
\eeq
where $P_1^{(0)}(x)$ is a polynomial of $x$ of degree at most $d-1$.
The one point function is thus given by
\beq
W_1^{(0)}(x) = {V'(x)\over 2} - \sqrt{{V'(x)^2\over 4}- P_1^{(0)}(x)}.
\eeq
We define:
\beq
y = {V'(x)\over 2} - W_1^{(0)}(x) = \sqrt{{V'(x)^2\over 4}- P_1^{(0)}(x)},
\eeq
and the master loop equation implies that the function $y$ is solution of
the algebraic equation (hyperelliptical)
$H_{1MM}(x,y)=0$ where
\beq
H_{1MM}(x,y):=y^2 - \left({V'(x)\over 2 }\right)^2 + P_1^{(0)}(x)
\eeq
which is called the spectral curve associated to the one hermitian matrix model.

\medskip

From \eq{ResW1xiini1MM}, we see that, if one chooses the $\acycle_i$-cycle to be a circle, independent of $t$, around $\xi_i$, then we have (order by order in $t$):
\beq
{1 \over 2 i \pi} \oint_{{\cal{A}}_i} W_1^{(0)}(x) dx = {t\, n_i\over N} =  \epsilon_i \virg i = 1, \dots, d .
\eeq
This last equation gives $d$ constraints, i.e. the same number as the coefficients of the polynomial $P_1^{(0)}$ (which is of degree $d-1$),
therefore it determines $P_1^{(0)}(x)$.
Just by looking at the first terms in the small $t$ expansion, one has:
\beq\label{eqP10formal1MMasympt}
P_1^{(0)}(x) = \sum_{i=1}^d \epsilon_i\,\,{V'(x)\over x-\xi_i} + O(t^2).
\eeq
The data $n_i/N$ are thus equivalent to the data of $P_1^{(0)}$.
Notice that $\epsilon_i=t n_i/N$ is of order $O(t)$ in the small $t$ expansion.

\medskip

In case some $\epsilon_i$'s are vanishing, we define $\genus+1=$ the number of non-vanishing $\epsilon_i$'s, and we assume that $\epsilon_1,\dots,\epsilon_{\genus+1}$ are non-vanishing and $\epsilon_{\genus+2},\dots,\epsilon_{d}$ are zero.

Order by order in $t$ we have:
\beq\label{y1MMfirstordert}
y = \prod_{i=\genus+2}^d (x-\xi_i-A_i(t)) \, \sqrt{\prod_{i=1}^{\genus+1} ((x-\xi_i-B_i(t))^2-4 C_i(t))}
\eeq
where $A_i(t), B_i(t), C_i(t)$ are formal power series of $t$.
To the first orders:
\beq
A_i = {4\over V''(\xi_i)}\,\,\sum_{j=1}^{\genus+1}\, {\epsilon_j\over \xi_i-\xi_j} + O(t^2),
\eeq
\beq
B_i = {1\over 2 V''(\xi_i)}\,\sum_{j\neq i} {\epsilon_j\over \xi_i-\xi_j}\quad - {\epsilon_i\over 4}\,\,{V'''(\xi_i)\over V''^2(\xi_i)} + O(t^2),
\eeq
\beq
C_i = {\epsilon_i\over V''(\xi_i)} + O(t^2).
\eeq

%
%and
%\beq
%{1 \over 2 i \pi} \oint_{{\cal{A}}_i} W_n(x_1,x_2,\dots,x_n) dx_1 = 0 \virg i = 1, \dots, d .
%\eeq

\bigskip

Let us study the specificities of the 1-matrix model spectral curve $(\spcurve_{1MM},x,y)$.

\medskip

\noindent {\bf Genus of $\spcurve_{1MM}$}

\noindent The Riemann surface $\spcurve_{1MM}$ has genus $\genus$ lower than $d -1$:
\beq
\genus \leq d-1 .
\eeq

\medskip

\noindent {\bf Sheeted structure}

\medskip

\noindent The polynomial $H_{1MM}(x,y)$ has degree $2$ in $y$. This means that the embedding of $\spcurve_{1MM}$
is composed by $2$ copies of the Riemann sphere, called sheets, glued by $\genus+1$ cuts so that the resulting Riemann surface $\spcurve_{1MM}$ has genus $\genus$.
Each copy of the Riemann sphere corresponds to one particular branch of the solutions of the equation
$H_{1MM}(x,y)=0$.
Since there are only two sheets in involution, this spectral curve is said to be hyperelliptic. It also means that the application
$z \to \overline{z}$ is globally defined since it is the map which exchanges both sheets:
\beq
y(\overline{z}) = - y(z) .
\eeq

\medskip

\noindent {\bf Pole structure}

\medskip

\noindent The function $x(z)$ on the Riemann surface $\spcurve_{1MM}$ has two simple poles (call them $\alpha_+$ and $\alpha_-$), one in each sheet.
Near $\alpha_\pm$, $y(z)$ behaves like:
\beq
y(z) \mathop{\sim}_{z\to \alpha_\pm} \pm \, {1\over 2}V'(x(z))  \mp {t\over x(z)} + O(1/x(z)^2).
\eeq

\subsubsection{The 2-point function}

For $n=1$ and $g=0$, the loop equation \eq{loopeq1MM1cut} reads:
\beq
{\partial \over \partial x_1}\,{W^{(0)}_{1}(x)-W^{(0)}_{1}(x_1)\over x-x_1} = (V'(x)-2W_1^{(0)}(x)) W^{(0)}_{2}(x,x_1) - P^{(0)}_{2}(x;x_1)
\eeq
i.e.:
\bea\label{eqW201MM}
W_2^{(0)}(x,x_1)
&=& {{\partial \over \partial x_1}\,{W^{(0)}_{1}(x)-W^{(0)}_{1}(x_1)\over x-x_1} + P^{(0)}_{2}(x;x_1)\over 2 y(x)} \cr
&=& -{1\over 2(x-x_1)^2} +  {{1\over 2}\,{\partial \over \partial x_1}\,{V'(x)-V'(x_1)+2y(x_1)\over x-x_1} + P^{(0)}_{2}(x;x_1)\over 2 y(x)}. \cr
\eea
This equation shows that  $W_2^{(0)}(x,x_1)$ is a meromorphic function of $z$ and $z_1$ on the spectral curve $\spcurve_{1MM}$. It is a multivalued function of $x,x_1$, but it is a monovalued function in the variables $z,z_1$. Therefore let us write:
\beq
W_2^{(0)}(x(z),x(z_1)) \, dx(z) dx(z_1) = \bar\om_2^{(0)}(z,z_1)
\eeq
where $\bar\om_2^{(0)}(z,z_1)$ is a meromorphic form of $z$ and $z_1$.
It is clear from the first line of \eq{eqW201MM}, that $\bar\om_2^{(0)}(z,z_1)$ has no pole at $z=z_1$, but it has a pole at $z=\bar{z}_1$.
Moreover, since $dx(z)/y(z)$ has no pole at branchpoints, we see that $\bar\om_2^{(0)}(z,z_1)$ has no pole when $z$ approaches a branchpoint.
By looking at the behavior at large $x$, we see, from  $  \deg P_2^{(0)} \leq d-2$, that $\bar\om_2^{(0)}(z,z_1)$ has no pole at the two infinities $\alpha_\pm$.

From \eq{eqW201MM}, it may seem that $\bar\om_2^{(0)}(z,z_1)$ could have simple poles at the zeroes of $y(z)$, but the residues are computed by \eq{ResWnxiini1MM} and they vanish.
Thus, we see that the only possible pole of $\bar\om_2^{(0)}(z,z_1)$ can be at $z=\bar{z}_1$.

Then, notice that the second line of \eq{eqW201MM} is the sum of a term which is even under $z\to \bar{z}$, and a term which is odd under $z\to \bar{z}$.
Since the sum of those two terms must have no pole at $z=z_1$, we see that the pole at $z=\bar{z}_1$ must be twice the pole of the even part.
Therefore we find that $\bar\om_2^{(0)}(z,z_1)$ has a double pole at $z=\bar{z}_1$, with no residue, and no other pole.
Moreover, \eq{ResWnxiini1MM} implies that on every $\acycle$-cycle we have:
\beq
\oint_{z\in {\cal A}_i}\, \bar\om_2^{(0)}(z,z_1) = 0 .
\eeq
The only meromorphic differential having all those properties is the Bergmann kernel:
\beq
\bar\om_2^{(0)}(z,z_1) = -B(z,\bar{z}_1) = B(z,z_1)  - {dx(z)\, dx(z_1)\over (x(z)-x(z_1))^2} .
\eeq
(we choose $\kappa=0$).

\subsubsection{Higher correlators}

Similarly to what we just did with $W_2^{(0)}$, we are going to compute every $W_n^{(g)}$ and relate it to the symplectic invariants of the curve $y(x)$.

\smallskip
First, notice that the loop equations \eq{loopeq1MM1cut} imply recursively, that each $W_n^{(g)}$ is in fact a meromorphic function on the spectral curve, and thus we prefer to rewrite:
\bd
\bea
\om_n^{(g)}(z_1,\dots,z_n)
&=& W_n^{(g)}(x(z_1),\dots,x(z_n))\,dx(z_1)\dots dx(z_n) \cr
&&  + \delta_{n,2}\delta_{g,0}\,{dx(z_1)dx(z_2)\over (x(z_1)-x(z_2))^2}.
%- \delta_{n,1}\delta_{g,0}W_1^{(0)}(x(z_1))\,dx(z_1) .
\eea
\ed
Indeed, correlation functions $W_n^{(g)}$ are multivalued functions of the complex variable $x$ whereas  the
$\om_n^{(g)}$ are monovalued $n$-forms on the spectral curve $\spcurve$.
Somehow, these forms are built to choose
one particular branch of solution of the master loop equation \eq{eqspcurve1MMW10}.

\medskip

{\bf Structure of the form $\omega_n^{(g)}$:}

One clearly sees from loop equation \eq{loopeq1MM1cut}, that $\om_{n+1}^{(g)}(z,z_1,\dots,z_n)$ can possibly have poles only at branchpoints, or at coinciding points $x(z)=x(z_j)$, i.e. at $z=z_j$ or $z=\bar{z}_j$, or also at the zeroes of $y(z)$.

From the degrees $\deg P_n^{(g)}\leq d-2$, one can see that there is no poles at the infinities $\alpha_\pm$.
Also, there is manifestly no pole at $z=z_j$, and one may notice that $\om_{n+1}^{(g)}$ is an odd function when $z\to \bar{z}$, thus it can also have no pole when $z=\bar{z}_j$.

The zeroes of $y(z)$ are either the branchpoints, or double points, which are, order by order in $t$, of the form (see \eq{y1MMfirstordert}):
\beq
\xi_i + A_i(t) + O(t^2) \virg A_i(t)=O(t).
\eeq
Let us assume by recursion on $2g+n$, that each $\om_n^{(g)}$ has no poles at those double points.
This is true for $\om_1^{(0)}$ and $\om_2^{(0)}$. Assume that it is true for every $\om_{n'}^{(g')}$ with $2g'+n'\leq 2g+n$, let us prove it for $2g+n+1$.
From the loop equation \eq{loopeq1MM1cut}, one sees that $\om_{n+1}^{(g)}(z,z_1,\dots,z_n)$ could have at most a simple pole at such double points. But because of \eq{ResWnxiini1MM} and \eq{ResW1xiini1MM}, the residue must vanish, and thus there is no pole.

\bigskip
Therefore we obtain the following structure for the $\om_n^{(g)}$'s:

\bl\label{lemmapolesbponly1MM}
$\om_n^{(g)}$ can have poles only at branchpoints when $2g+n\geq 3$.

Moreover, we see from \eq{ResWnxiini1MM} and \eq{ResW1xiini1MM}, that if $2g+n\geq 3$ we have:
\beq\label{ointacycleomng1MM}
\oint_{x_1\in\acycle_i} \om_n^{(g)}(x_1,\dots,x_n) = 0 .
\eeq
\el

\subsubsection{Symplectic invariants}

Let:
\beq
dS_{z_1,z_2}(z) = \int_{z_2}^{z_1} B(z,z')
\eeq
be the 3rd kind differential in $z$, having a simple pole at $z=z_1$ with residue $+1$, and a simple pole of residue $-1$ at $z=z_2$, and no other poles, and normalized on $\acycle$-cycles:
\beq\label{dSacyclevanish}
\oint_{\acycle_i} dS_{z_1,z_2} = 0 .
\eeq
The fact that it has only simple poles with residue $+1$ at $z_1$ allows to write the Cauchy formula on the spectral curve ($o$ being an arbitrary base point on $\spcurve_{1MM}$):
\beq\label{Cauchyformula1MM}
\om_{n+1}^{(g)}(z,z_1,\dots,z_n) = \Res_{z' \to z} dS_{z',o}(z)\, \om_{n+1}^{(g)}(z',z_1,\dots,z_n).
\eeq
On the other hand, the differential form $\om_{n+1}^{(g)}(z,z_1,\dots,z_n)$ has poles only at the branch points $z \to a_i$, and thus the Riemann bilinear identity tells us that:
\bea\label{Cauchyformula1MMRiemanbil}
&& \Res_{z' \to z} dS_{z',o}(z)\, \om_{n+1}^{(g)}(z',z_1,\dots,z_n) + \sum_i \Res_{z' \to a_i} dS_{z',o}(z)\, \om_{n+1}^{(g)}(z',z_1,\dots,z_n) \cr
&=& \sum_i \oint_{z'\in\acycle_i} B(z,z') \oint_{z'\in \bcycle_i} \om_{n+1}^{(g)}(z',z_1,\dots,z_n) \cr
&& - \sum_i \oint_{z'\in\bcycle_i} B(z,z') \oint_{z'\in \acycle_i} \om_{n+1}^{(g)}(z',z_1,\dots,z_n). \cr
\eea
Due to \eq{dSacyclevanish} and \eq{ointacycleomng1MM}, the right hand side vanishes and thus:
\beq
\Res_{z' \to z} dS_{z',o}(z)\, \om_{n+1}^{(g)}(z',z_1,\dots,z_n)
= -  \sum_i \Res_{z' \to a_i} dS_{z',o}(z)\, \om_{n+1}^{(g)}(z',z_1,\dots,z_n).
\eeq
Finally, one can plug in the loop equation \eq{loopeq1MM1cut} and remind that the polynomial $P_1^{(g)}$ has no pole at the branch points, and thus we find:
\bea
&& \om_{n+1}^{(g)}(z,z_1,\dots,z_n)  \cr
&=&
\sum_i \Res_{z'\to a_i}\, K(z,z')\,\Big[
\om_{n+2}^{(g-1)}(z',\bar{z'},z1, \dots , z_n) \cr
&&  \qquad \qquad + \sum_{h=0}^g\sum'_{I\subset \left\{z_1,\dots,z_n\right\}} \om_{1+|I|}^{(h)}(z',I) \om_{1+n-|I|}^{(g-h)}(\bar{z'}, \left\{z_1,\dots,z_n\right\} \backslash I) \Big] . \cr
\eea
where $K(z,z')$ is the kernel:
\beq
K(z,z') = {dS_{\bar z',z'}(z)\over 2(y(z')-y(\bar z'))\,dx(z')}.
\eeq
In other words:
\bt
$\om_n^{(g)}(z_1,\dots,z_n)$ and $F^{(g)}$ are the correlators and symplectic invariants of the spectral curve  ${\cal E}_{\rm 1MM}$ of equation $H_{\rm 1MM}(x,y)=0$.
\et

To get this theorem, we also recover the $F_g$'s from theorem \ref{thvariat}, or just by homogeneity, see \cite{ec1loopF} for details.
For the 1-matrix model, $F_0$ has been known from the origin of random matrices, $F_1$ was  first found in \cite{ACKM} for the 1-cut case, and in \cite{Chekh} for the multicut case. The other $F_g$'s were first found in \cite{ec1loopF}.

%
%\br

%{\bf Enumeration of maps}

%When the spectral curve has genus 0, these matrix integrals have a simple combinatorics interpretation. One can see it for
%example the case where $\epsilon_1=N$ and al other filling fractions vanish.
%It can then be easily seen by computing Gaussian matrix integrals with Wick's theorem, and this is the seminal work of Brezin-Itzykson-Parisi-Zuber \cite{BIPZ}, that the $F_g$'s are generating functions (see chapter \ref{secmaps} for more details) counting closed connected discrete surfaces of genus $g$, with $v$ vertices:
%\beq
%F_g = \sum_v t^v \, \sum_{\Sigma}\, {t_3^{n_3(\Sigma)}\dots t_d^{n_d(\Sigma)}\over \#{\rm Aut}(\Sigma)}
%\eeq
%where $\Sigma$ belongs to the set of discrete surfaces of genus $g$, with $v$ vertices, made of $n_3$ triangles, $n_4$ quadrangles, $n_5$ pentagons, ...

%If the surface has higher genus, there also exists such an interpretation but slightly more complicated. Indeed, if the
%genus is higher than 0, one do not only count maps made of polygons glued by their edges but also singular
%discrete surfaces where one allows polygons to be glued by their centers. This new type of gluing mimics singular surfaces
%obtained by pinching a non-trivial cycle on it. This more general link with combinatorics is out of the scope of this review
%and the interested reader can find a good discussion in \cite{Eynformal}.

%\er

%

\subsection{2-matrix model}
\label{sec2MMformal}

The method of loop equations can also be used to solve the formal 2-matrix model.
One of the main applications and reasons for introducing the 2-matrix model, was the problem of counting Ising model configurations on random discrete surfaces, or in other words bi-colored maps (see section \ref{secmaps}), it was first introduced and solved by V.Kazakov \cite{KazakovIsing}.
It corresponds to a formal 2-matrix integral.
It can be rephrased in terms of symplectic invariants too.

\medskip

For this purpose, one generalizes the notion of formal matrix integral to integrals over two normal matrices.
\bd
Let $N$ be an integer and $V_1$ and $V_2$ two polynomial potentials:
\beq
V_1(x) = - \sum_{k=2}^{d_1+1} {t_k\over k}\, x^k
\virg
V_2(y) = - \sum_{k=2}^{d_2+1} {\td{t}_k\over k}\, y^k.
\eeq
Let $d=d_1 d_2$, and let $\vec{n}$ be a  $d$-partition of $N$:
\beq
\vec{n}:= \{n_1, n_2, \dots, n_{d}\} \qquad \qquad \hbox{such that} \qquad \qquad \sum_{i=1}^{d} n_i = N.
\eeq

Let $\{(\xi_i,\eta_i)\}_{i=1}^{d_1 d_2}$ be the $d=d_1 d_2$ solutions of the system of equations
\beq
\left\{
\begin{array}{l}
V'_1(\xi_i)=\eta_i\cr
V'_2(\eta_i)=\xi_i
\end{array}\right. .
\eeq

One defines the non-quadratic part of the Taylor expansions of the potentials around these saddle points
\beq
\delta V_{1,i}(x) = V_1(x) - V_1(\xi_i) - {V_1''(\xi_i)\over 2}(x-\xi_i)^2
\eeq
and
\beq
\delta V_{2,i}(y) = V_2(y) - V_2(\eta_i) - {V_2''(\eta_i)\over 2}(y-\eta_i)^2.
\eeq

For all $l$, one defines the polynomial in $t$:
\beq\label{Akl}
\begin{array}{l}
{\displaystyle \sum_{k= l/2}^{ld}} A_{k,l} t^k = \cr
= {(-1)^l N^l\over l!\, t^l}\,\int dM_1 \dots dM_d d\td{M}_1 \dots d\td{M}_d (\sum_i \Tr \delta V_{1,i}(M_i)+ \delta V_{2,i}(\td{M}_i))^l \cr
 \,\, {\displaystyle \prod_{i=1}^{d}} \ee{-{N \over t}\left(\Tr {V_1''(\xi_i)\over 2}(M_i-\xi_i\,{\bf 1}_{n_i})^2+{V_2''(\eta_i)\over 2}(\td{M}_i-\eta_i\,{\bf 1}_{n_i})^2 - (M_i-\xi_i\,{\bf 1}_{n_i})(\td{M}_i-\eta_i\,{\bf 1}_{n_i}) \right)} \cr
 \,\,\,{\displaystyle \prod_{i>j}} \det(M_i\otimes {\bf 1}_{n_j} - {\bf 1}_{n_i} \otimes M_j)
\,\, {\displaystyle \prod_{i>j}} \det(\td{M}_i\otimes {\bf 1}_{n_j} - {\bf 1}_{n_i} \otimes \td{M}_j) \cr
\end{array}
\eeq
as a gaussian integral over hermitian matrices $M_i$ and $\widetilde{M}_i$ of size $n_i \times n_i$.

The formal 2-matrix model partition function is then defined as a formal power series in $t$ (cf \cite{eynform,Eynhab}):
\bea\label{defZform2MM}
Z_{\rm 2MM} &:=& \sum_{k=0}^\infty t^k \big( \sum_{j=0}^{2k} A_{k,j} \big).
%\cr
%&=&\sum_{l=0}^\infty {(-1)^l N^l\over l!\, t^l}\,\int {\displaystyle \prod_{\alpha=1}^d} dM_\alpha d\td{M}_\alpha  \left(\sum_i \Tr \delta V_{1,i}(M_i)+ \delta V_{2,i}(\td{M}_i)\right)^l \cr
%&& \,\, {\displaystyle \prod_{i=1}^{d}} \ee{-{N \over t}\left(\Tr {V_1''(\xi_i)\over 2}(M_i-\xi_i\,{\bf 1}_{n_i})^2+{V_2''(\eta_i)\over 2}(\td{M}_i-\eta_i\,{\bf 1}_{n_i})^2 - (M_i-\xi_i\,{\bf 1}_{n_i})(\td{M}_i-\eta_i\,{\bf 1}_{n_i})\right)} \cr
%&& \,\,\,{\displaystyle \prod_{i>j}} \det(M_i\otimes {\bf 1}_{n_j} - {\bf 1}_{n_i} \otimes M_j)
%\,\,\prod_{i>j} \det(\td{M}_i\otimes {\bf 1}_{n_j} - {\bf 1}_{n_i} \otimes \td{M}_j). \cr
\eea

As in the 1-matrix model, one uses the notation
\beq
Z_{\rm 2MM}=\int_{\rm formal} \ee{-{N\over t}\Tr V_1(M_1)+V_2(M_2)-M_1 M_2}\, dM_1\, dM_2.
%= \ee{\sum_g (N/t)^{2-2g}\, F_g}.
\eeq
\ed

One is also interested in the formal logarithm of the partition function: the free energy
\beq
F_{2MM}:= \ln Z_{2MM}
\eeq
which has a topological expansion (due again to 't Hooft's observation \cite{thooft})
\beq
F_{2MM}= \sum_{g \geq 0} \left({N \over t}\right)^{2-2g} F_g,
\eeq
where each $F_g$ is a formal power series of $t$.

\subsubsection{Loop equations and spectral curve}

As in the 1-matrix model (cf \eq{defformalTr1/x-M}), one also defines correlation functions by
\beq
W_{k,l}(x_1,\dots,x_k,y_1,\dots,y_l):= \left<\prod_{i=1}^k \Tr{1 \over x_i -M_1} \prod_{i=1}^l \Tr{1 \over y_i -M_2} \right>_c
\eeq
denoted as the non-mixed correlation functions\footnote{There exists more general correlation functions mixing the two types
of matrices $M_1$ and $M_2$ inside the same trace, for example $<\Tr (M_1^k M_2^l)>$, but their study is too far from the main topic of this review to be
treated here. It is studied in \cite{EOBethe,EOallmixed}.}.
These correlation functions also admit a topological expansion
\beq
W_{k,l}(x_1,\dots,x_k,y_1,\dots,y_l)= \sum_{g\geq 0} \left({N\over t}\right)^{2-2g-k} W_{k,l}^{(g)}(x_1,\dots,x_k,y_1,\dots,y_l).
\eeq
One also needs the polynomials in $x_1$ and $y$
\beq
P_n(x_1,y;x_2,\dots,x_n) = \left<\Tr\left({V_1'(x_1)- V_1'(M_1) \over x_1 - M_1}{V_2'(y)- V_2'(M_2) \over y - M_2}\right)
\prod_{i=2}^n \Tr {1 \over x_i - M_1} \right>_c
\eeq
as well as
\beq
U_n(x_1,y;x_2,\dots,x_n) = \left<\Tr\left({1 \over x_1 - M_1}{V_2'(y)- V_2'(M_2) \over y - M_2}\right)
\prod_{i=2}^n \Tr {1 \over x_i - M_1} \right>_c
\eeq
which are polynomials in $y$ only.

\subsubsection{Loop equations}

Loop equations proceed from integration by parts in the formal matrix integral (i.e. integration by parts in each gaussian integral for each power of $t$), or by writing invariance under changes of variables, i.e. they are Schwinger-Dyson equations.

The loop equations for the 2-matrix model were first studied by M. Staudacher \cite{staudacher}, and then written in a more concise form in \cite{eynm2m,eynmultimat}.
The loop equations for the 2-matrix model are (where $J=\{x_1,\dots,x_n\}$):
\bea\label{loop2MMv1}
&& {N\over t}(y-V'_1(x))\,U_{n+1}(x,y;J)  + U_{n+2}(x,y;x,J) \cr
&& +  \sum_{I\subset J} W_{1+|I|,0}(x,I)\,U_{1+n-|I|}(x,y;J/I) \cr
&& + \sum_{j=1}^n {\partial \over \partial x_j}\, {U_{n}(x,y;J/\{x_j\})-U_{n}(x_j,y;J/\{x_j\})\over x-x_j} \cr
&=& - {N\over t}P_{n+1}(x,y;J) +{N^2\over t^2} .
\eea
And then, identifying the coefficients of polynomials of $1/N^2$ for each power of $t$, we have:
\beq
(y-V'_1(x)+ W_{1,0}^{(0)}(x))\,U^{(0)}_1(x,y) = (V'_2(y)-x)\,W_{1,0}^{(0)}(x) - P_1^{(0)}(x,y) + 1
\eeq
and for $(g,n)\neq (0,0)$:
\bea\label{loop2MM}
&& (y-V'_1(x)+W_{1,0}^{(0)}(x))\,U_{n+1}^{(g)}(x,y;J) + W_{n+1,0}^{(g)}(x,J)\,U_{1}^{(0)}(x,y)  \cr
&& + U_{n+2}^{(g-1)}(x,y;x,J) + \sum_{h=0}^g \sum'_{I\subset J} W_{1+|I|,0}^{(h)}(x,I)\,U_{1+n-|I|}^{(g-h)}(x,y;J/I) \cr
&& + \sum_{j=1}^n {\partial \over \partial x_j}\, {U_{n}^{(g)}(x,y;J/\{x_j\})-U_{n}^{(g)}(x_j,y;J/\{x_j\})\over x-x_j} \cr
&=& - P_{n+1}^{(g)}(x,y;J) .
\eea
where $\sum_h\sum_I'$ means that we exclude the terms $(h,I)=(0,\emptyset)$ and $(g,J)$.

\subsubsection{Spectral curve}

Consider the first loop equation:
\beq
(y-V'_1(x)+ W_{1,0}^{(0)}(x))\,U^{(0)}_1(x,y) = (V'_2(y)-x)\,W_{1,0}^{(0)}(x) - P_1^{(0)}(x,y) + 1 .
\eeq
It is valid for any $x$ and $y$, and in particular we may choose:
\beq
y=y(x) = V'_1(x)-W_{1,0}^{(0)}(x) .
\eeq
Since $U^{(0)}_1(x,y)$ is a polynomial in $y$, it cannot have a pole at this value of $y=y(x)$, and thus, for this value of $y=V_1'(x) -  W_{1,0}^{(0)}(x)$, we have the algebraic equation:
\beq
H_{2MM}(x,y(x))=  (V'_2(y(x))-x)\,(V'_1(x)-y(x)) - P_1^{(0)}(x,y(x)) + 1 =0
\eeq
known as the 2-matrix model spectral curve.

\medskip

Let us study the specificities of the 2-matrix model spectral curve $(\spcurve_{2MM},x,y)$ (see \cite{KazMar}).

\medskip

\noindent {\bf Genus of $\spcurve_{2MM}$}

\noindent The Riemann surface $\spcurve_{2MM}$ has genus $\genus$ lower than $d_1 d_2 -1$:
\beq
\genus \leq d_1 d_2 -1 .
\eeq

\medskip

\noindent {\bf Sheeted structure}

\medskip

\noindent The polynomial $H_{2MM}(x,y)$ has degree $d_2+1$ (resp. $d_1+1$)  in $y$ (resp. $x$). This means that the embedding of $\spcurve_{2MM}$ as a branch-covering of the $\mathbb P^1$ $x$-plane (resp. $y$-plane)
is composed by $d_2+1$ (resp. $d_1+1$) copies of the Riemann sphere $\mathbb P^1$, called $x$-sheets (resp. $y$-sheets), glued by cuts, so that the resulting Riemann surface $\spcurve_{2MM}$
has genus $\genus$. Each copy of the Riemann sphere corresponds to one particular branch of the solutions of the equation
$H_{2MM}(x,y)$ in $y$ (resp. $x$).

\smallskip

Since there are $d_2+1$ $x$-sheets, this means that there are $d_2+1$ points in $\spcurve_{2MM}$ corresponding to the same value of $x$. We write:
\beq\label{defzisheets2MM}
x(z^i) = x(z) \,\, , \quad i=0,\dots,d_2
\eeq
We will take the convention that $z^0=z$.

\medskip

Branchpoints are zeroes of $dx$, and they are also places where 2 sheets merge $z\to z^i$ for some $i$. By convention, we call this other point $z^i$, and thus $\bar{z}$ is one of the $z^i$'s.
In general, $\overline{z}$ is not globally defined but only defined locally around the branch points.

\medskip

\noindent {\bf Pole structure of the functions $x(z)$ and $y(z)$}

\medskip

\noindent The function $x(z)$ (resp. $y(z)$) on the Riemann surface $\spcurve_{2MM}$ has two poles: one of degree $1$ (resp. degree $d_1$)
at a pole called $\infty_x$ and one of degree $d_2$ (resp. degree $1$) at a pole called $\infty_y$. It means that $d_2$ $x$-sheets merge at $\infty_y$,
and only one $x$-sheet contains $\infty_x$ alone.

Near $\infty_x$, a local parameter is $1/x$, and we have:
\beq
y(z) \mathop{{\sim}}_{z\to\infty_x} \,\, V'_1(x(z)) - {t\over x(z)} + O(1/x(z)^2) .
\eeq
And near $\infty_y$, a local parameter is $1/y$, and we have:
\beq
x(z) \mathop{{\sim}}_{z\to\infty_y} \,\, V'_2(y(z)) - {t\over y(z)} + O(1/y(z)^2) .
\eeq

\medskip

According to section \ref{secvirasoro}, the fact that we have two poles, means that the tau-function built from the symplectic invariants
of this curve is the tau-function of the $1+1$-KP hierarchy\footnote{Actually, one should take
$d_1$ and $d_2$ arbitrary large to obtain the 1+1-KP tau-function. The times of the hierarchy being given by the coefficients of the Laurent expansions of $ydx$ and $xdy$ around $\infty_x$ and $\infty_y$ respectively.}.

\subsubsection{Preliminaries to the solution of loop equations}

As in the preceding section, we promote the correlation functions to differential forms on the spectral curve to make them monovalued:
\bd
\bea
\om_n^{(g)}(z_1,\dots,z_n)
&=& W_{n,0}^{(g)}(x(z_1),\dots,x(z_n))\,dx(z_1)\dots dx(z_n) \cr
&&  + \delta_{n,2}\delta_{g,0}\,{dx(z_1)dx(z_2)\over (x(z_1)-x(z_2))^2}.
%- \delta_{n,1}\delta_{g,0}W_1^{(0)}(x(z_1))\,dx(z_1) .
\eea
\ed

Exactly like in the 1-matrix formal model, the very definition of the model as a formal series in $t$, implies that each $\om_n^{(g)}$ with $2g+n\geq 3$ is a meromorphic $n$-form with poles only at the branchpoints (i.e. the zeroes of $dx$), and with vanishing $\acycle$-cycle integrals.
The only exception is $\om_2^{(0)}(z_1,z_2)$ which can have a pole only at $z_1=z_2$, and which is found to be the Bergmann kernel (see \cite{DKK}):
\beq
\om_2^{(0)}(z_1,z_2) = B(z_1,z_2).
\eeq

Before proving that the solution of loop equations for the $\om_n^{(g)}$'s are the symplectic invariant's correlators, we need a small lemma.
Consider the "full spectral curve":
\beq
E(x,y) = (V'_1(x)-y)(V'_2(y)-x) - {t\over N} P_1(x,y) + 1
\eeq
where $P_1(x,y) = \sum_g (N/t)^{1-2g} P_1^{(g)}(x,y)$.
We also consider its descendants:
\beq
E_{n+1}(x,y;z_1,\dots,z_n) = \delta_{n,0}(V'_1(x)-y)(V'_2(y)-x) - {t\over N} P_{n+1}(x,y;z_1,\dots,z_n) + \delta_{n,0}.
\eeq

We have:
\bl\label{completecurve}
\bea\label{expE}
&& E_{n+1}(x(z),y;z_1,\dots,z_n)  \cr
&=& -\td{t}_{d_2+1}\,"\Big<\prod_{i=0}^{d_2} \left(y-V'_1(x(z^i))+{t \over N} \Tr{1\over x(z^i)-M_1}\right)\,\cr
&& \prod_{j=1}^n \Tr{1\over x(z_j)-M_1} \Big>_{c,\{x_1,\dots,x_n\}}",\cr
\eea
and
\bea\label{expU}
&& {t\over N}U_{n+1}(x(z),y) - \delta_{n,0} (V'_2(y)-x(z)) \cr
&=& -\td{t}_{d_2+1}\,"\Big<\prod_{i=1}^{d_2} \left(y-V'_1(x(z^i))+ {t \over N} \Tr{1\over x(z^i)-M_1}\right)\, \cr
&& \prod_{j=1}^n \Tr{1\over x(z_j)-M_1}  \Big>_{c,\{x_1,\dots,x_n\}}" .\cr
\eea
where $\td{t}_{d_2+1}$ is the leading coefficient of $V'_2(y)$, $z^i$ are the preimages of $x(z)$ (see \eq{defzisheets2MM}),
the subscript ${}_{c,\{x_1,\dots,x_n\}}$ means that we take the connected part with respect to the $\Tr{1\over x(z_j)-M_1}$ terms, but not the $\Tr{1\over x(z^i)-M_1}$ terms,
and the inverted comas $"\left<.\right>"$ mean that, every time one encounters a two-point function
in the cumulant expansion, one replaces it by\footnote{Remark that this notation reminds the notation $"det"$ in th.\ref{thdet}.}
\beq
W_{2,0}(x,x'):= \left< \Tr{1 \over x - M_1} \Tr{1 \over x' - M_1}\right> +{1\over (x-x')^2}.
\eeq
\el
For example formula \eq{expE}, for $n=1$, reads to the first subleading order in $t/N$:
\bea
P_1^{(1)}(x,y;z_1)
&=& \td{t}_{d_2+1}\sum_{j=0}^{d_2} W_{2,0}^{(1)}(z^j,z_1)\, \prod_{i\neq j, i=0}^{d_2} (y-y(z^j)) \cr
&&+ \td{t}_{d_2+1}\sum_{j\neq k=0}^{d_2} W_{2,0}^{(0)}(z^j,z_1)\,W_{1,0}^{(1)}(z^k)\, \prod_{i\neq j,k, i=0}^{d_2} (y-y(z^j)) \cr
&& + \td{t}_{d_2+1}\sum_{j\neq k=0}^{d_2} W_{3,0}^{(0)}(z^j,z^k,z_1)\, \prod_{i\neq j,k, i=0}^{d_2} (y-y(z^j)) \cr
&& + \td{t}_{d_2+1}\sum_{j\neq k\neq l=0}^{d_2} W_{2,0}^{(0)}(z^j,z_1)\,W_{2,0}^{(0)}(z^k,z^l)\, \prod_{i\neq j,k,l, i=0}^{d_2} (y-y(z^j)) . \cr
\eea
Formula \eq{expU} would be almost the same, but with the indices $i,j,k,l\geq 1$ instead of $\geq 0$.

This lemma was proved in \cite{CEO} and relies on the fact that the loop equation \eq{loop2MM} has a unique solution admitting a topological expansion. The way to prove this lemma mostly follows from Lagrange interpolation formula for polynomials, as well as Cauchy residue formula on $\spcurve_{2MM}$.

\medskip

Since $E_{n+1}(x,y;z_1,\dots,z_n)$ is a polynomial of $y$ of degree $\geq d_2+1$, by expanding $E_{n+1}(x,y;z_1,\dots,z_n)$ in powers of $y$, this lemma gives $d_2+1$ equations.
In particular, the term in $y^{d_2}$ gives (if $2g+n\geq 2$):
\beq\label{eq2MMsumi1}
\sum_{i=0}^{d_2} \om_{n+1,0}^{(g)}(z^i,z_1,\dots,z_n) = 0.
\eeq
And the term in $y^{d_2-1}$ gives a bilinear equation in the correlation functions:
\bea\label{eq2MMsumi2}
&& \sum_{i\neq j}^{d_2}
\om_{n+2,0}^{(g-1)}(z^i,z^j,J) + \sum_{h=0}^g\sum'_{I\subset J}\om_{1+|I|,0}^{(h)}(z^i,I)  \om_{1+n-|I|,0}^{(g-h)}(z^j,J/I)  \cr
&=& \sum_{i\neq j} y(z^i) \om_{n+1,0}^{(g)}(z^j,J) + y(z^j) \om_{n+1,0}^{(g)}(z^i,J)
-  f_n^{(g)}(x(z),J)\, dx(z)^2
\eea
where $f_n^{(g)}(x(z),J)$ is a rational function of $x(z)$ with no pole when $z$ approaches a branchpoint.

\subsubsection{Solution of loop equations and symplectic invariants}

Let us write Cauchy formula, exactly like for the 1-matrix model \eq{Cauchyformula1MM}:
\bea\label{Cauchyformula2MM}
\om_{n+1}^{(g)}(z_0,J)
&=& -\Res_{z\to z_0} dS_{z,0}(z_0)\,\,  \om_{n+1}^{(g)}(z,J) \cr
&=& \sum_i \Res_{z\to a_i} dS_{z,0}(z_0)\,\,  \om_{n+1}^{(g)}(z,J) \cr
\eea
where we have moved the integration contours using Riemann bilinear identity like for the 1-matrix model \eq{Cauchyformula1MMRiemanbil}.

Now, notice that near a branchpoint $a_i$, $\om_{n+1,0}^{(g)}(z,J)$ and $\om_{n+1,0}^{(g)}(\bar{z},J)$ have a pole at $z=a_i$ and all the $\om_{n+1,0}^{(g)}(z^i,J)$ such that $z^i\neq z,\bar{z}$ have no pole at $z\to a_i$.
According to \eq{eq2MMsumi1}, we have:
\beq
\om_{n+1,0}^{(g)}(z,J)+\om_{n+1,0}^{(g)}(\bar{z},J) = {\rm regular}
\eeq
and from \eq{eq2MMsumi2}, we have near $a_i$:
\bea\label{eq2MMsumi3}
 (y(z)-y(\bar{z})) \om_{n+1,0}^{(g)}(z,J)
&=& \om_{n+2,0}^{(g-1)}(z,\bar{z},J) + \sum_{h=0}^g\sum'_{I\subset J}\om_{1+|I|,0}^{(h)}(z,I)  \om_{1+n-|I|,0}^{(g-h)}(\bar{z},J/I)  \cr
&& + {\rm regular} .
\eea
Inserting this last equation into the Cauchy formula \eq{Cauchyformula2MM}, we find:
\beq
\om_{1+n}^{(g)}(z_0,J) = \sum_i \Res_{z \to a_i} K(z_0,z)\left[\om_{n+2,0}^{(g-1)}(z,\bar{z},J) + \sum_{h=0}^g\sum'_{I\subset J}\om_{1+|I|,0}^{(h)}(z,I)  \om_{1+n-|I|,0}^{(g-h)}(\bar{z},J/I)\right]
\eeq
where
\beq
K(z_0,z) = {dS_{\bar z,z}(z_0)\over 2(y(z)-y(\bar z))\,dx(z)}.
\eeq

This gives the theorem:
\bt
$\om_n^{(g)}(z_1,\dots,z_n)$ and $F_g$ are the correlators and symplectic invariants of the spectral curve  $({\spcurve}_{\rm 2MM},x,y)$.
\et

Here we have only briefly sketched the proof of \cite{CEO}, and we refer the reader to details there, in particular for finding the $F_g$'s.

\smallskip
$F_0$ was found for example in \cite{Kri, MarcoF}, $F_1$ was found in \cite{eynm2m,eynm2mg1, EKK}, and the other $F_g$'s in \cite{CEO}.

\medskip

\br
The correlation functions $\om_n^{(g)}$ which we consider here, are expectation values of traces of only the matrix $M_1$ (in terms of combinatorics of maps of section \ref{secmaps}, they are generating functions for bicolored maps, whose boundaries are of color $1$ only).
There exists a generalization of symplectic invariants for all other possible expectation values, i.e. all possible boundary conditions, but this is largely outside of the scope of this review.
We refer the reader to \cite{EOallmixed} for further details.

\medskip

An obvious remark, is that color $1$ and $2$, i.e. functions $x$ and $y$ play similar roles.
We can obtain generating functions for bicolored maps, whose boundaries are of color $2$ only, by just exchanging the roles of $x$ and $y$, i.e. by computing residues at the zeroes of $dy$.

\medskip

In particular we may compute generating functions for bicolored maps with no boundaries (i.e. the $F_g$'s), with either $x$ or $y$. In other words the $F_g$'s are unchanged if we exchange $x$ and $y$.
This is a special case of the symplectic invariance property $F_g(\spcurve,x,y) = F_g(\spcurve,y,x)$.

In fact, the general proof of symplectic invariance consists in defining some mixed generating functions for bicolored maps, whose boundaries are bicolored, it was done in \cite{symmetry}.

\er

\subsection{Chain of matrices in an external field}\label{secchain}

Another matrix model which can be solved with the same technics is the chain of matrices matrix model.
\smallskip

Consider the model of an arbitrary long open chain of matrices in an external field, which includes the one and two matrix models as particular cases.

Consider $m$ potentials $V_k(x) = -{\displaystyle \sum_{j=2}^{d_k+1}} {t_{k,j}\over j}\, x^j\ ,\,\,\,  k=1,\dots,m$. The formal chain of matrices matrix integral is:
\beq
Z_{\rm chain}=\int_{\rm formal} \ee{-{N\over t}\Tr \left( {\displaystyle \sum_{k=1}^m} V_k(M_k) - {\displaystyle \sum_{k=1}^{m}} c_{k,k+1}\,M_k M_{k+1} \right)}\, dM_1\dots\, dM_m
= \ee{{\displaystyle \sum_g} (N/t)^{2-2g}\, F_g}
\eeq
where the integral is a formal integral in the sense of the preceding sections and $M_{m+1}$ is a constant given diagonal matrix
 $M_{m+1}=\L$ with $s$ distinct eigenvalues $\l_i$ with multiplicities $l_i$:
\beq
M_{m+1}=\L= \hbox{diag}\left(\overbrace{\l_1,\dots,\l_1}^{l_1},\dots,\overbrace{\l_i,\dots,\l_i}^{l_i},\dots,\overbrace{\l_s,\dots,\l_s}^{l_s}\right)
\eeq
with ${\displaystyle \sum_i} l_i = N$.

Note also that we may choose $c_{m,m+1}=1$ since it can be reabsorbed as a rescaling of $\L$.

Once again, in the definition of the formal integral, one has to choose around which saddle point one expands.
Saddle points are solutions of:
\beq
\forall k=1,\dots,m,\qquad V'_k(\xi_k) = c_{k-1,k} \xi_{k-1}+c_{k,k+1}\xi_{k+1}
\virg
\exists j,\, \xi_{m+1} = \l_j .
\eeq
This system is an algebraic equation with $D= s d_1 d_2 \dots d_m $ solutions.

Therefore the choice of a saddle point is
encoded in the choice of a set of filling fractions
\beq
\epsilon_i = T {n_i \over N}
\eeq
for $i=1,\dots, D$ with $D= d_1 d_2 \dots d_m s$ and $n_i$ arbitrary integers satisfying
\beq
\sum_i n_i = N.
\eeq

\subsubsection{Definition of the correlation functions}

The loop equations of the chain of matrices were derived in \cite{eynmultimat, Eynpratts}, and they 
require the definition of several quantities, as follows:

For convenience, we introduce in the sense of \eq{defformalTr1/x-M}:
\beq
G_i(x_i):={1 \over x_i - M_i} .
\eeq

We also consider the minimal polynomial of $\L$, such that $S(\L)=0$, i.e.:
\beq
S(z)= {\displaystyle \prod_{i=1}^s} (z-\l_i)
\eeq
and we introduce the following polynomial in $z$
\beq
Q(z)= {1 \over c_{n,n+1}}\, {S(z) - S(\L) \over z-\L}.
\eeq
We also define the polynomials $f_{i,j}(x_i,\dots,x_j)$ by $f_{i,j}=0$ if $j<i-1$, $f_{i,i-1}=1$, and
\beq
f_{i,j}(x_i,\dots,x_j) = \det\pmatrix{ V'_i(x_i) & -c_{i,i+1} x_{i+1} & & 0 \cr
-c_{i,i+1} x_i & V'_{i+1}(x_{i+1}) & \ddots & \cr
& \ddots & \ddots & -c_{j-1,j} x_j \cr
0 & & -c_{j-1,j} x_{j-1} & V'_j(x_j)}
\eeq
if $j\geq i$.
They satisfy the recursion
\beq
c_{i-1,i} f_{i,j}(x_i,\dots,x_j) = V_i'(x_i) f_{i+1,j}(x_{i+1},\dots,x_j) - c_{i,i+1} \, x_i \, x_{i+1} \, f_{i+2}(x_{i+2},\dots,x_j).
\eeq

\bigskip

Let us then define the correlation functions and auxiliary functions:
\beq
W_0(x) = \left< \Tr G_1(x) \right>.
\eeq
For $i=2,\dots , m$, we define:
\beq
W_i(x_1,x_i,\dots,x_m,z)=  {\displaystyle \hbox{Pol}_{x_i,\dots,x_m}} f_{i,m}(x_i,\dots,x_m) \left< \Tr \left( G_1(x_1) G_i(x_i) \dots G_m(x_m)Q(z)\right)\right>,
\eeq
which is a polynomial in variables $x_i,\dots,x_m,z$, but not in $x_1$.
And for $i=1$, we define:
\beq
W_1(x_1,x_2,\dots,x_m,z)=  \hbox{Pol}_{x_1,\dots,x_m} f_{1,m}(x_1,\dots,x_m) \left< \Tr \left( G_1(x_1) G_2(x_2) \dots G_m(x_m)Q(z)\right)\right>.
\eeq
which is a polynomial in all variables.

We also define:
\bea
&& W_{i;1}(x_1,x_i,\dots,x_m,z;x_1') \cr
&=& {\displaystyle \hbox{Pol}_{x_i,\dots,x_m}} f_{i,m}(x_i,\dots,x_m) \left< \Tr \left(G_1(x_1')\right) \Tr \left( G_1(x_1) G_i(x_i) \dots G_m(x_m)Q(z)\right)\right>_c . \cr
\eea
All these functions admit a topological expansion, for example:
\beq
W_0 = \sum_g (N/t)^{1-2g} W_0^{(g)} \virg
W_1 = \sum_g (N/t)^{1-2g} W_1^{(g)} \virg
W_i = \sum_g (N/t)^{1-2g} W_i^{(g)}
\eeq
and
\beq
W_{i;1} = \sum_g (N/t)^{-2g} W_{i;1}^{(g)}.
\eeq

\subsubsection{Loop equations and spectral curve}

In this model, the master loop equation reads \cite{eynmultimat, Eynpratts}:
\bea\label{masterloopchain1}
&& W_{2;1}(x_1,\dots ,x_{m+1};x_1) \cr
&& + (c_{1,2} x_2 - V_1'(x_1) + {t\over N}W_0(x_1))
\Big( {t\over N}W_2(x_1, \dots ,x_{m+1}) - S(x_{m+1})\Big)  \cr
&&= -{t\over N} W_1(x_1, \dots, x_{m+1}) + \left( V_1'(x_1) - c_{1,2} x_2 \right) S(x_{m+1}) + \cr
&&  \qquad \qquad + {t\over N} \sum_{i=2}^m
\left(V_i'(x_i)-c_{i-1,i} x_{i-1} - c_{i,i+1} x_{i+1}\right) W_{i+1}(x_1,x_i,\dots,x_{m+1}). \cr
\eea
This equation is valid for any set of variables $x_1,x_2,\dots,x_{m+1}$, however, it can be simplified by choosing special values for those variables, in particular values for which
the last terms in the RHS vanishes.
For this purpose, one defines some $\hat x_i(x_1,x_2)$ as functions of the two first variables $x_1$ and $x_2$, as follows:
\beq
\hat x_1(x_1,x_2)=x_1
\virg
\hat x_2(x_1,x_2)=x_2,
\eeq
and for $i=2,\dots,m$:
\beq\label{recxichainmat}
c_{i,i+1} \hat{x}_{i+1}(x_1,x_2) = V_i'(\hat{x}_i(x_1,x_2)) - c_{i-1,i} \hat{x}_{i-1}(x_1,x_2) .
\eeq
Choosing $x_i=\hat x_i(x_1,x_2)$, reduces the master loop equation to an equation in $x_1$ and $x_2$:
\beq
 \widehat{W}_{2;1}(x_1,x_2;x_1) + {t\over N}\left(c_{1,2} x_2 - Y(x_1)\right) \widehat{U}(x_1,x_2) = \,\widehat{E}(x_1,x_2)
\eeq
where
\beq
Y(x) = V_1'(x) - {t\over N}W_0(x) \virg \widehat{U}(x_1,x_2) = W_2(x_1,x_2,\hat{x}_3,\dots,\hat{x}_{m+1}) - {N\over t}S(\hat{x}_{m+1}),
\eeq
\beq
\widehat{W}_{2;1}(x_1,x_2;x_1) = W_{2;1}(x_1,x_2,\hat{x}_3,\dots,\hat{x}_{m+1};x_1)
\eeq
and
\beq
\widehat{E}(x_1,x_2) = - {t\over N}\widehat{W}_1(x_1,x_2) + \left(V_1'(x_1)-c_{1,2} x_2\right) \widehat{S}(x_1,x_2)
\eeq
with
\beq
\widehat{S}(x_1,x_2) = S(\hat{x}_{m+1})  \virg \widehat{W}_1(x_1,x_2) = W_1(x_1,x_2,\hat{x}_3,\dots,\hat{x}_{m+1}) .
\eeq
Notice that $\widehat{W}_1(x_1,x_2)$, and thus $\widehat{E}(x_1,x_2)$ is a polynomial in both $x_1$ and $x_2$.

Finally, the leading order in the topological expansion gives
\beq\label{masterloopleading}
\widehat{E}^{(0)}(x_1,x_2) = \left(c_{1,2} x_2 - Y^{(0)}(x_1)\right) \widehat{U}^{(0)}(x_1,x_2).
\eeq

\medskip
We may notice that this equation is more or less the same as in the 2-matrix model, and it is solved in the same way.

Again, this equation is valid for any $x_1$ and $x_2$, and if we choose $x_2$ such that $c_{1,2} x_2 = Y^{(0)}(x_1)$, we get:
\beq\label{spcurvechainmat}
H_{chain}(x_1,x_2) := \widehat{E}^{(0)}(x_1,x_2) = 0.
\eeq
This algebraic equation is the spectral curve of our model.

\bigskip

{\bf Study of the spectral curve}

The algebraic plane curve $H_{chain}(x_1,x_2)=0$, can be parameterized by a variable $z$ living on a compact Riemann surface $\spcurve_{chain}$ of some genus $\genus$, and two meromorphic functions $x_1(z)$ and $x_2(z)$ on it. Let us study it in greater details.

\medskip

\noindent {\bf Genus of $\spcurve_{chain}$}

\noindent The Riemann surface $\spcurve_{chain}$ has genus $\genus$ lower than $D -s$:
\beq
\genus \leq D -s ,
\eeq
where $D=s\,d_1\dots d_m$.

\medskip

\noindent {\bf Sheeted structure}

\medskip

\noindent The polynomial $H_{2MM}(x_1,x_2)$ has degree $1+{D\over d_1} $ (resp. $d_1+ {D\over d_1 d_2}$)  in $x_2$ (resp. $x_1$).
This means that the embedding of $\spcurve_{chain}$ is composed by $1+{D\over d_1}$ (resp. $d_1+{D\over d_1d_2}$)
copies of the Riemann sphere, called $x_1$-sheets (resp. $x_2$-sheets), glued by cuts so that the resulting Riemann surface $\spcurve_{chain}$
has genus $\genus$. Each copy of the Riemann sphere corresponds to one particular branch of the solutions of the equation
$H_{chain}(x_1,x_2)=0$ in $x_2$ (resp. $x_1$).

\medskip

\noindent {\bf Pole structure}

\medskip

\noindent

In the preceding cases (1 and 2-matrix models), one was interested in the pole structure of only two functions $x$ and $y$ on the spectral curve.
In the case of the chain of matrices, the problem is slightly richer since one can consider, not only the meromorphic
functions $x_1$ and $x_2$, but also all the $x_i(p):=\hat{x}_i(x_1(p),x_2(p))$ as meromorphic functions on $\spcurve_{chain}$. Their
negative divisors are given by
\beq
\left[x_k(p)\right]_- = - r_k \infty - s_k \sum_{i=1}^s \hat{\l}_i
\eeq
where $\infty$ is the only point of $\spcurve_{chain}$ where $x_1$ has a simple pole, the $\hat{\l}_i$ are the preimages
of $\l_i$ under the map $x_{m+1}(p)$:
\beq
x_{m+1}(\hat{\l}_i) = \l_i
\eeq
and the degrees $r_k$ and $s_k$ are integers given by
\beq
r_1:=1 \, , \; r_k:= d_1 d_2 \dots d_{k-1} \, , \; s_{m+1}:=0 \, , \; s_m:=1 \quad \hbox{and} \quad s_k:= d_{k+1} d_{k+2} \dots d_m\, s .
\eeq

Note that the presence of an external matrix creates as many poles as the number of distinct eigenvalues of this external
matrix $M_{m+1}=\L$\footnote{The cases of matrix models without external field correspond to a totally degenerate external matrix
$\L= c\, {\rm Id}$ with only 1-eigenvalue. There are thus two poles as in the 1 or 2 matrix models studied earlier.}.

\medskip
\br
This matrix model has also a combinatorics interpretation in terms of counting colored surfaces. This interpretation is
discussed in chapter \ref{secmaps}.
\er

\subsubsection{Solution of the loop equations}

The loop equations have been solved in \cite{Eynpratts} by the same method as the 2-matrix model. It proceeds in three steps.
One first shows that the loop equations \eq{masterloopchain1} have a unique solution admitting a topological expansion. One then propose
an Ansatz of solution and prove that it is indeed right. This gives
\bt\label{completecurvechain}
\beq
E(x(z),y) = -\td{t}_{d_2+1}\,"\left<\prod_{i=0}^{d_2} (y-V'_1(x(z^i))+{t \over N} \Tr{1\over x(z^i)-M_1}) \right>".
\eeq
\et
One finally develops this expression as a polynomial in $y$ to get the bilinear relation
\beq
\om_1^{(g)}(z_0) = \sum_i \Res_{z \to a_i} K(z_0,z)\left[\om_2^{(g+1)}(z,\overline{z}) +
\sum_{h=0}^g \om_1^{(h)}(z) \om_1^{(g-h)}(\overline{z})\right]
\eeq
where, as in the preceding section,
\bea
\om_n^{(g)}(z_1,\dots,z_n)
&=& W_{n}^{(g)}(x(z_1),\dots,x(z_n))\,dx(z_1)\dots dx(z_n) \cr
&&  + \delta_{n,2}\delta_{g,0}\,{dx(z_1)dx(z_2)\over (x(z_1)-x(z_2))^2} .
%- \delta_{n,1}\delta_{g,0}W_1^{(0)}(x(z_1))\,dx(z_1) .
\eea
This allows to obtain the theorem:
\bt\label{thFgchainmat}
$\om_n^{(g)}(z_1,\dots,z_n)$ and $F_g$ are the correlators and symplectic invariants of the spectral curve  $({\spcurve}_{chain},c_{1,2} x_1, x_2)$.
\et

Since $c_{1,2} x_1+c_{2,3} x_3 = V'_2(x_2)$, we may use the symplectic invariance theorem \ref{thsymplinv}, and homogeneity theorem \eq{eqFghomogeneity}, which allows also to write:
\bea
F_g
&=&  F_g({\spcurve}_{chain},c_{1,2} x_1,x_2)  \cr
&=&  F_g({\spcurve}_{chain},-c_{2,3} x_3,x_2)  \cr
&=&  F_g({\spcurve}_{chain},c_{2,3} x_3,x_2)  \cr
&=&  F_g({\spcurve}_{chain}, x_2, c_{2,3} x_3)  \cr
&=&  F_g({\spcurve}_{chain}, c_{2,3} x_2,  x_3) .  \cr
\eea
And by an easy recursion, for any $k=1,\dots,m$:
\beq\label{Fgchainmatanyk}
F_g
= \,F_g({\spcurve}_{chain},c_{k,k+1}\, x_k,x_{k+1})
= \,F_g({\spcurve}_{chain}, x_k, c_{k,k+1}\,x_{k+1}) .
\eeq
In other words, the $F_g$'s can be computed by choosing the spectral curve of any two consecutive $x_k$'s. It means it doesnot depends on $k$, it doesnot depend on where we are in the chain.

\subsubsection{Matrix quantum mechanics}

Matrix quantum mechanics is the limit of an infinitely long chain of matrices $m\to \infty$.
This model is very useful in string theory \cite{matrixQM}.

In this limit, the index $k$ of the matrix $M_k$, becomes a continuous time variable $t$.
The coefficients are scaled in a way such that the coupling term $\Tr (M_k-M_{k+1})^2$ becomes a kinetic energy $\Tr (dM/dt)^2$.

More explicitly, consider the chain of $m$ matrices
\beq
Z = \int dM_1\dots dM_m\,\, \ee{-N\Tr \left[ {\displaystyle \sum_{k=1}^m} \eta V_k(M_k) + {\mu\over 2\eta} {\displaystyle \sum_{k=1}^{m-1}} (M_k-M_{k+1})^2 \right]}
\eeq
and take the $\eta\to 0$ limit, and $T=m\eta$ of order 1. The index $k$ becomes a time $t=k\eta$, and in the $\eta\to 0$ limit we have:
\beq
Z = \int D[M(t)]\,\,\, \ee{-N\Tr \int_0^T [V(M(t),t)+ {\mu\over 2}\, (dM/dt)^2 ]\,dt}
\eeq
The spectral curve is characterized as before:

find a compact Riemann surface $\spcurve$, and some time-dependent function $x(z,t)$ analytical in the variable $z$ on some domain of $\spcurve$, which satisfy \eq{recxichainmat} which become Newton's equations of motion:
\beq\label{eqmotiontimeM}
\mu\, \ddot{x}(z,t)  = -V'(x(z,t),t)
\eeq
and such that the initial and final impulsions
\beq
p(x(z,0),0)  = \mu\dot{x}(z,0)
\virg
p(x(z,T),T) = \mu\dot{x}(z,T)
\eeq
are analytical outside some cuts.

Therefore, theorem \ref{thFgchainmat} gives $\forall t\in [0,T]$:
\beq
\ln{Z} = \sum_g N^{2-2g}\,\, F_g(\spcurve,x(z,t),\mu\dot{x}(z,t)) .
\eeq
In other words, the spectral curve is here the realtionship between impulsion and position for the classical equation of motion. Although the spectral curve depends on time $t$, the $F_g$'s are independent of $t$.

In the case where the potential $V(x,t)=V(x)$ does not depend on time $t$, the equations of motion \eq{eqmotiontimeM} can be integrated and give the energy conservation:
\beq
E(z) = {\mu\over 2}\,\dot{x}^2(z,t) + V(x(z,t))
\eeq
i.e.
\beq
\mu\dot{x}(z,t) = \sqrt{2\mu(E(z)-V(x(z,t))}
\eeq
and we have:
\beq
\ln{Z} = \sum_g N^{2-2g}\,\, F_g(\spcurve,x(z,t),\sqrt{2\mu(E(z)-V(x(z,t))}) .
\eeq

\subsection{1-Matrix model in an external field}
\label{sec1Mext}

As a special example of the chain of matrices above, let us consider the special case $m=1$, i.e. 1-matrix model with an external field.

\medskip

The formal 1-matrix model in an external field $\hat\L$ is defined as \cite{PZinnmatext}:
\beq
Z_{M\,{\rm .ext}}(\hat\L) = \int_{\rm formal}\, \ee{-{N\over t}\,\Tr(V(M)-\hat\L M)}\, dM
\virg
\hat\L = \hbox{diag} \left(\overbrace{\hat\L_1,\dots,\hat\L_1}^{m_1}, \dots, \overbrace{\hat\L_s,\dots,\hat\L_s}^{m_s}\right)
\eeq
where {\it formal}, as usual means that we Taylor expand near a critical value and then exchange the order of Taylor expansion and gaussian integral. A critical point is a matrix $M_0$ solution of $V'(M_0)=\hat\L$.
Let us assume that $\hat\L$ has $s$ distinct eigenvalues $\hat\L_i$ of multiplicities $m_i$.
Its minimal polynomial is:
\beq
S(y) = \prod_{i=1}^s (y-\hat\L_i).
\eeq
For each $\hat\L_i$, let $\xi_{i,j}, j=1,\dots, \deg V'$ be the $d=\deg V'$ solutions of $V'(\xi_{i,j})=\hat\L_i$.
A critical point $M_0$ is characterized by $s$ partitions of the $m_i$'s into at most $d=\deg V'$ parts
\beq
m_i = \sum_{j=1}^d n_{i,j} .
\eeq
It is of the form:
\beq
M_0 =\hbox{diag} \left(\overbrace{\xi_{i,j},\dots,\xi_{i,j}}^{n_{i,j}}\right).
\eeq
The parameters:
\beq
\epsilon_{i,j} = {t\,n_{i,j}\over N}
\eeq
are called the filling fractions, and they parametrize which formal integral we are considering.

\medskip
The 1-matrix model in an external field is of course a special case of the chain of matrices described in the previous section  \ref{secchain}, and therefore one finds that it has a topological expansion given by the symplectic invariants of a spectral curve:
\beq
\ln{(Z_{M\,{\rm .ext}}(\hat\L))} = \sum_{g=0}^\infty (N/T)^{2-2g}\, F_g({\cal E}_{M\,{\rm .ext}})
\eeq
and
\beq
\left<\Tr {dx(z_1)\over x(z_1)-M}\dots \Tr {dx(z_n)\over x(z_n)-M}\right>_c =
\sum_g (N/t)^{2-2g-n}\,\,\om_{n}^{(g)}(z_1,\dots,z_n).
\eeq

However, let us make the results of section \ref{secchain} a little bit more explicit in that case.

\subsubsection{Spectral curve}

The spectral curve ${\cal E}_{M\,{\rm .ext}}$, obeys the equation:
\beq\label{spcurve1MMext}
0 = {\cal E}_{M\,{\rm .ext}}(x,y)
= (V'(x)-y) - {t\over N}\sum_{j=1}^s {P_j(x)\over y-\hat\L_j}
\eeq
where $P_j(x)$ is a polynomial of degree at most $d-1$.
Such a spectral curve is typically of genus $\genus\leq sd-s$.

All the coefficients of all the $P_j$'s are fixed by the filling fractions requirement that (order by order in $t$):
\beq
{1\over 2i\pi}\,\oint_{\acycle_I} ydx = \epsilon_{I}
\eeq
where $I=(i,j)$ runs through the values $i=1,\dots, s$, $j=1,\dots,d$, and $\acycle_I$ is a small circle around $\xi_{i,j}$. We remind that the $\epsilon_{i,j}$ are not independent, only $sd-s$ of them are independent:
\beq
\sum_{j=1}^d \epsilon_{i,j} = {t\,m_i\over N}.
\eeq

\medskip

The two functions $x(z)$ and $y(z)$ are characterized by the fact that $x(z)$ has a simple pole at $\infty$, simple poles at some $\l_i$ such that $y(\l_i)=\hat\L_i$, and $y(z)$ has a pole of degree $d$ at $\infty$.
And we have:
\beq
V'(x(z))-y(z) \mathop{\sim}_{z\to\infty}\,\, {1\over x(z)} + O(1/x(z)^2)
\eeq
and near $\l_i$, the residues of $x$ are such that:
\beq
\Res_{z\to \l_i}\, xdy = -{t\,m_i\over N}
\eeq
and the cycle integrals:
\beq
{1\over 2i\pi}\oint_{\acycle_{i,j}}\, ydx = \epsilon_{i,j} = {t\, n_{i,j}\over N} .
\eeq

\subsubsection{Rational case}

It is interesting to study the case of a rational spectral curve.
The two rational functions $x(z)$ and $y(z)$ are of the form:
\beq
{\cal E}_{M\,{\rm ext}} = \left\{
\begin{array}{l}
x(z) = z - {t\over N}\Tr {1\over Q'(\L)(z-\L)} \cr
y(z) = Q(z)
\end{array}
\right.
\eeq
where $\L$ is a diagonal matrix determined by:
\beq
Q(\L) = \hat\L
\eeq
and $Q$ is a polynomial of degree $d=\deg V'$, determined by:
\beq
V'(x(z)) = Q(z) + {1\over z}+O(z^{-2}).
\eeq

\subsection{Convergent matrix integrals}

So far, we have been discussing formal matrix integrals, which consist in exchanging integration and the small $t$ Taylor series of $\ee{-{N\over t} \Tr V(M)}$. Formal matrix integrals always have a "topological expansion" of the type:
\beq
\ln{Z} = \sum_g (N/t)^{2-2g}\, F_g(t).
\eeq

Now, let us consider a "convergent" matrix integral:
\beq\label{defconvintmat}
Z = \int_{H_N(\gamma)}\, dM\,\, \ee{-{N\over t}\,\Tr V(M)}
\eeq
where the integration domain $H_N(\gamma)$ is the set of normal matrices with eigenvalues on a path $\gamma$:
\beq
H_N(\gamma) = \{ M\,\, / \, M=U\L U^\dagger, U\in U(N)\, , \L={\rm diag}(\L_1,\dots,\L_N)\, , \L_i\in \gamma\}
\eeq
equipped with the complex $U(N)$ invariant measure
\beq
dM = \prod_{i>j} (\L_i-\L_j)^2\,\,\,dU\,\, \prod_i d\L_i
\eeq
where $dU$ is the Haar measure on $U(N)$, and $d\L_i$ is the curviline measure along $\gamma$.

\smallskip
$\bullet$ For example $H_N({\mathbb R}) = H_N$ is the set of hermitian matrices.

$\bullet$ For example $H_N(S_1) = U(N)$ is the set of unitary matrices ($S_1$ is the unit circle).

In eigenvalues, the convergent matrix integral is:
\beq\label{Zgamma}
Z(\gamma) = {1\over N!}\,\int_{\gamma^N}\,\, dx_1\dots\,dx_N\,\, \prod_{i>j}(x_i-x_j)^2\,\, \prod_{i=1}^N \ee{-{N\over t} V(x_i)} .
\eeq

\medskip

Imagine that $V$ is a polynomial of degree $d+1$ (the present discussion can be extended easily to a 2-matrix model, or a chain of matrices, and to all cases where $V'$ is a rational fraction \cite{Marcopath}).
There are $d$ homologically independent paths on which the integral $\int \ee{-V(x)}\, dx$ is convergent, let us call $\gamma_1,\dots,\gamma_d$, a basis of such paths (a choice of basis is not unique). See the figure for the example of a quartic potential ($d=3$):
$$\epsfxsize=10cm\epsfbox{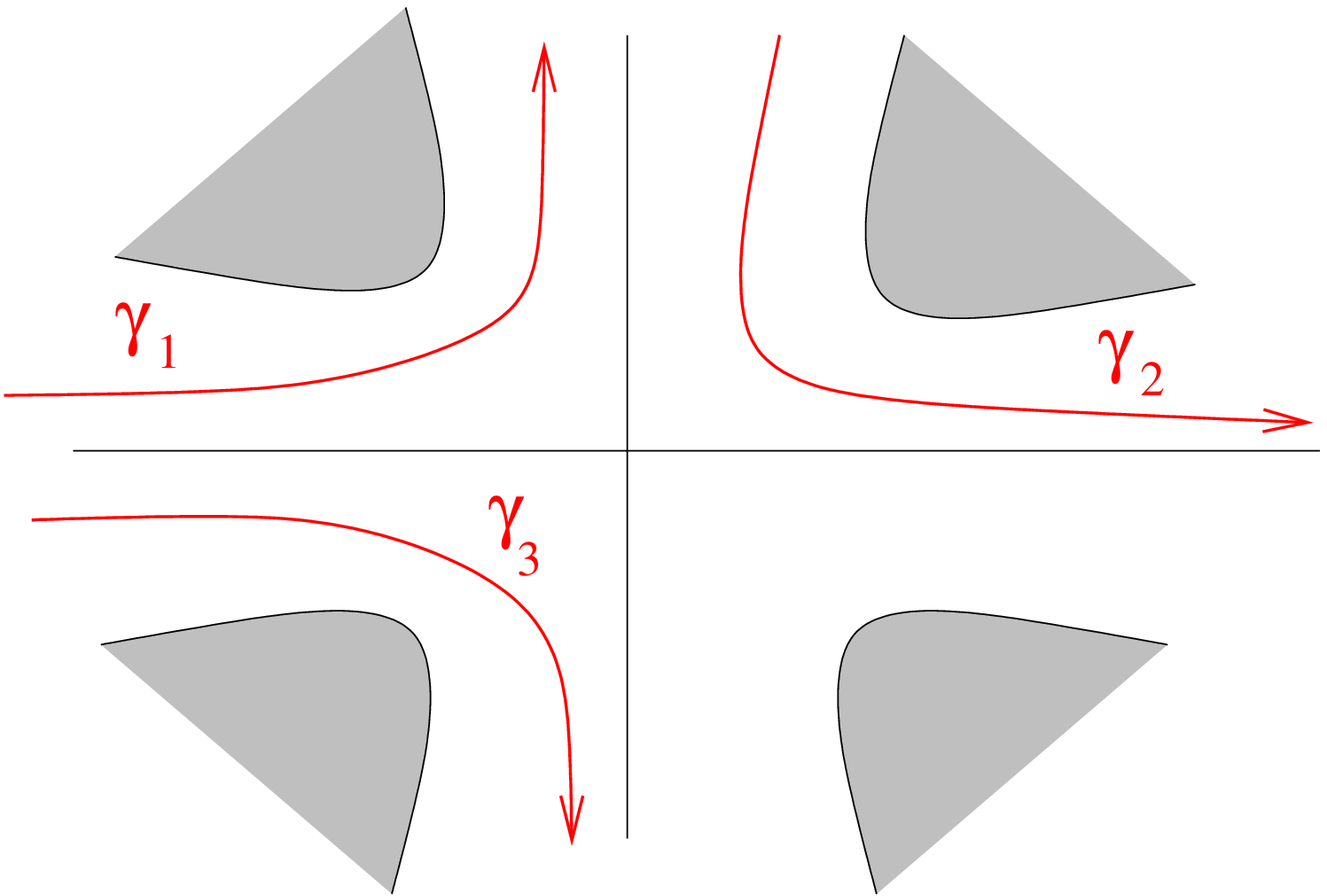}$$

The path $\gamma$ in the matrix integral \eq{defconvintmat} is a linear combination of such paths:
\beq
\gamma=\sum_{i=1}^d c_i \gamma_i
\eeq
and the convergent matrix integral \eq{Zgamma} can be written:
\beq
Z(\gamma) = \sum_{n_1+\dots+n_d=N}\,\, {c_1^{n_1}\dots c_d^{n_d}\over n_1!\dots n_d!}\,\int_{\gamma_1^{n_1}\times\dots\times \gamma_d^{n_d}}\,\, dx_1\dots\,dx_N\,\, \prod_{i>j}(x_i-x_j)^2\,\, \prod_{i=1}^N \ee{-{N\over t} V(x_i)} .
\eeq
This leads us to define the convergent matrix integral with {\bf fixed filling fractions} $n_i$, as:
\beq
\hat{Z}_{n_1,\dots,n_d} \stackrel{{\rm def}}{=} {1\over n_1!\dots n_d!}\,\int_{\gamma_1^{n_1}\times\dots\times \gamma_d^{n_d}}\,\, dx_1\dots\,dx_N\,\, \prod_{i>j}(x_i-x_j)^2\,\, \prod_{i=1}^N \ee{-{N\over t} V(x_i)},
\eeq
and thus we have:
\beq\label{ZammasumZhatni}
Z(\gamma) = \sum_{n_1+\dots+n_d=N}\,\, c_1^{n_1}\dots c_d^{n_d}\,\, \hat{Z}_{n_1,\dots,n_d}.
\eeq
This holds for any choice of basis $\gamma_1,\dots,\gamma_d$.

\medskip

There is a conjecture\footnote{This conjecture is proved in some cases, and in particular proved for the 1-matrix model with arbitrary $\gamma$ and arbitrary polynomial $V$ (the proof follows from M. Bertola's work \cite{bertolaboutroux}), but, at the time this article is being written, it is not proved in more general cases, for instance not proved for the general 2-matrix model.} that there exists a "good" basis of paths $\gamma_1,\dots,\gamma_d$ \cite{Deift}, such that $\hat{Z}_{n_1,\dots,n_d}$ is a formal matrix integral (and thus it has a topological expansion in powers of $t/N$)!

\smallskip

The "good" paths $\gamma_1,\dots,\gamma_d$ can be seen as "steepest descent" paths, they should be such that the effective potential:
\beq
V_{\rm eff}(x) = V(x) - {t\over N}  \left<\ln{(\det (x-M))}\right>
\eeq
is such that along each $\gamma_i$, the real part of the large $N$ leading order of $V_{\rm eff}$ is decreasing, then constant and finally increasing, and at the same time, the imaginary part of $V_{\rm eff}$ is constant, then increasing, then constant:
$$\epsfxsize=10cm\epsfbox{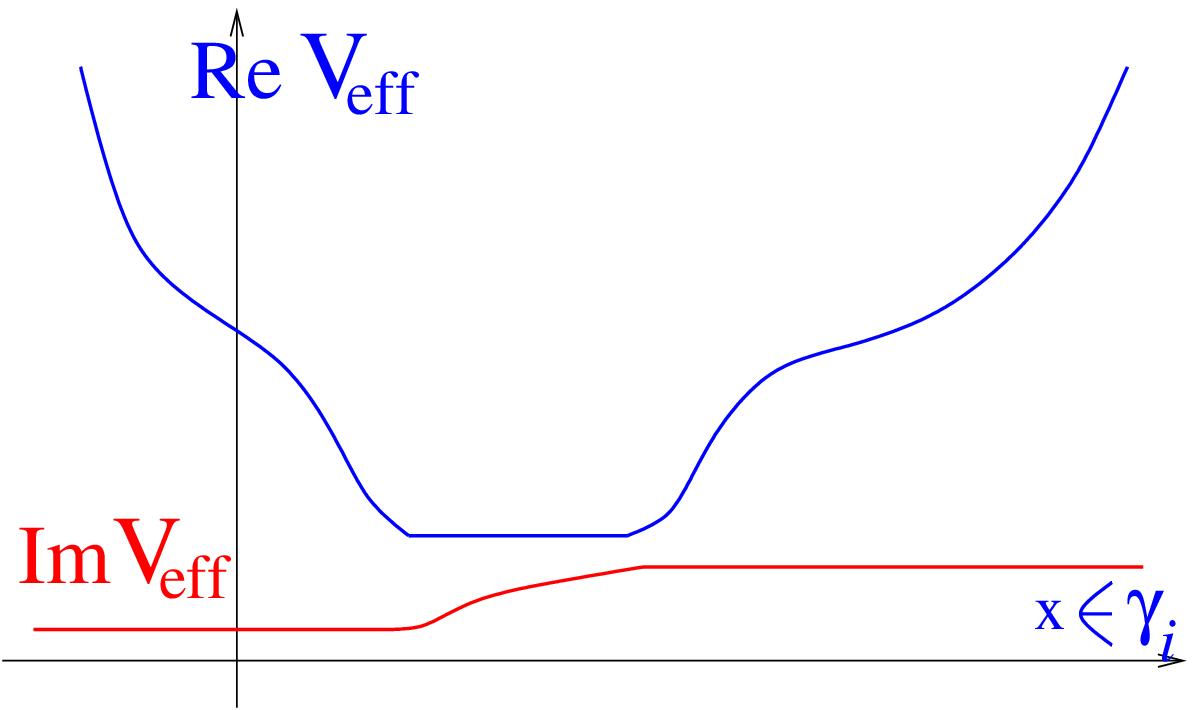}$$

If such paths exist, then we can write:
\beq\label{Zhatnitopexp}
\hat{Z}_{n_1,\dots,n_d} = \ee{{\displaystyle \sum_g} (N/t)^{2-2g}\, F_g(\epsilon_i)}
\eeq
where $\epsilon_i$ are the filling fractions:
\beq
\epsilon_i = {t\,\, n_i\over N}
\eeq
and the coefficients $F_g$ in the expansion, are the symplectic invariants $F_g$ of the corresponding formal matrix integral.
Since the $F_g$'s are analytical functions of the filling fractions, we have, for any "background" filling fraction $\eta=(\eta_1,\dots,\eta_{d-1})$
\beq
F_g(\epsilon_i) = \sum_{k=0}^\infty {(\epsilon-\eta)^k\over k!}\,\, \partial_\eta^k\,\, F_g(\eta)
\eeq
where we assume tensorial notations and sums over indices. For simplicity, the unfamiliar reader may assume $d=2$, i.e. $\eta$ and $\epsilon$ are scalar.

\Eq{Zhatnitopexp} thus becomes:
\bea
\hat{Z}_{n_1,\dots,n_d}
&=& \ee{{\displaystyle \sum_g} (N/t)^{2-2g}\, F_g(\eta)}\,\,
\ee{(n-N\eta/t)NF'_0/t}\ee{{1\over 2}(n-N\eta/t)^2\,F''_0}\,\, \cr
&& \ee{{\displaystyle \sum_g} {\displaystyle \sum_{k\geq 2-2g}} {(N/t)^{2-2g-k}\over k!}\, (n-N\eta/t)^k\, F_g^{(k)}(\eta)} \cr
&=& \ee{{\displaystyle \sum_g} (N/t)^{2-2g}\, F_g(\eta)}\,\,
\ee{(n-N\eta/t)NF'_0/t}\ee{{1\over 2}(n-N\eta/t)^2\,F''_0}\,\, \cr
&& \sum_{l} {1\over l!}\,\sum_{g_1,\dots,g_l}\sum'_{k_1,\dots,k_l}
 {(N/t)^{\sum_i (2-2g_i-k_i)}\over k_1!\dots k_l!}\, \, \prod_i F_{g_i}^{(k_i)}(\eta)\,\,\,\, (n-N\eta/t)^{\sum k_i} \cr
\eea
where we have separated the terms with a positive power of $N$ from those with a negative power of $N$ ($\sum'$ means that we consider only terms with $k_i>0$ and $2g_i+k_i-2>0$).
Then, writing:
\beq
c_i =\ee{2i\pi \nu_i},
\eeq
 we perform the sum over filling fractions in \eq{ZammasumZhatni}, and we get \cite{BDE, background}:
\bea
Z(\gamma)
&\sim& \ee{{\displaystyle \sum_g} (N/t)^{2-2g}\, F_g(\eta)}\,\,
\sum_{n\in {\mathbb Z}^{d-1}}\,\,  \ee{2i\pi n\nu}\,\ee{(n-N\eta/t)NF'_0/t}\ee{{1\over 2}(n-N\eta/t)^2\,F''_0}\,\, \cr
&& \sum_{l} {1\over l!}\,\sum_{g_1,\dots,g_l}\sum'_{k_1,\dots,k_l}
 {(N/t)^{{\displaystyle \sum_i} (2-2g_i-k_i)}\over k_1!\dots k_l!}\, \, \prod_i F_{g_i}^{(k_i)}(\eta)\,\,\,\, (n-N\eta/t)^{\sum k_i} . \cr
\eea
That is, we find the non-perturbative partition function of section \ref{secbackground} (see \cite{background, EMnonperturbative}):
\bea\label{ZgammalargeN}
Z(\gamma)
&\sim& \ee{{\displaystyle \sum_g} (N/t)^{2-2g}\, F_g(\eta)} \cr
&& \sum_{l} {1\over l!}\,\sum_{g_1,\dots,g_l}\sum'_{k_1,\dots,k_l}
 {(N/t)^{{\displaystyle \sum_i} (2-2g_i-k_i)}\over k_1!\dots k_l!}\, \, \prod_i F_{g_i}^{(k_i)}(\eta)\,\,\,\,
 \Theta^{(\sum k_i)}_{(0,\nu)}(NF'_0/t,F''_0) \cr
\eea
where
\beq
\Theta_{(\mu,\nu)}(u,F''_0) = \sum_{n\in {\mathbb Z}^{d-1}}\,\, \ee{2i\pi n\nu}\,\ee{2i\pi (n-N\eta/t+\mu)u}\, \ee{{1\over 2}\,(n-N\eta/t)^2\, F''_0}.
\eeq
This formula is expected to give the large $N$ expansion of convergent matrix models.
It was proved in several cases, but a general proof is still missing.

To the first orders, \eq{ZgammalargeN} reads:
\beq
Z(\gamma)
\sim \ee{{N^2\over t^2}F_0}\,\ee{F_1}\,\,  \Big[\Theta+{t\over N} \Big(
\Theta'\,F'_1 + {\Theta'''\,F_0'''\over 6}\Big) + \dots \Big].
\eeq

We have to make several remarks:

$\bullet$ background independence:

Formula \eq{ZgammalargeN} is independent of the background $\eta$. This was discussed in section \ref{secbackground}, and it is related to the fact that the non-perturbative partition function is modular.

This also implies that \eq{ZgammalargeN} cannot be a good large $N$ asymptotic expansion for all values of $\eta$. The conjecture is that one should choose $\eta$ as a real minimum of $\Re F_0$.
Notice that if $\eta$ is real, then
\beq
 \Re F_0'' = 2\pi\, \Im\tau >0
\eeq
where $\tau$ is the Riemann matrix of periods of the spectral curve, and thus $\Re F_0$ is a convex function of $\eta$, within each cell of the moduli space of spectral curves corresponding to the potential $V$.
Unfortunately, this moduli space is not very well known, and it is not known how to find such $\eta$ in general.

If such $\eta$ can be found we have ${1\over 2i\pi}\oint_{\acycle} ydx=\eta\in{\mathbb R}$, and $F'_0=\oint_{\bcycle} ydx$, and thus, for any contour ${\cal C}$ we have:
\beq
\Re\oint_{\cal C} ydx = 0.
\eeq
A spectral curve with that property is called a "Boutroux" curve
(See \cite{bertolaboutroux}).

\bigskip

$\bullet$ Characteristics $(\mu,\nu)$:

We see here, that the characteristics $(\mu,\nu)$ of the $\Theta$-function, is related to the path $\gamma$ chosen at the beginning to define the convergent integral.
It is not fully understood how to associate a path to a characteristic and vice-versa.

\section{Non-intersecting Brownian motions}\label{secbrownian}

\subsection{Dyson motions and integrability: introduction}

Let us consider $N$ Brownian motions on the real line whose positions at time $t$ are denoted by $x_i(t)$ for $i=1, \dots, N$. Let us constrain them not to intersect and fix their starting and ending points: the particle $i$
goes from $a_i$ at $t=0$ to $b_i$ at $t=1$:
\beq
x_i(0) = a_i \virg x_i(1) = b_i
\eeq
with
\beq
a_1 \leq a_2 \leq \dots \leq a_N \qquad \hbox{and} \qquad
b_1 \leq b_2 \leq \dots \leq b_N .
\eeq
Once these parameters are fixed, one is interested in the statistic of these Brownian movers at a given time $t \in (0,1)$. They are given by the correlation functions:
\beq
R_k(\l_1,\l_2, \dots,\l_k|t) = {1 \over N^k} \left<\displaystyle{\prod_{i=1}^k} \Tr \delta(\l_i-M)\right>_t
\eeq
where we denote $M={\rm diag}(x_1,\dots,x_N)$.

These correlation functions can be written under a determinantal form \cite{Dysondet,Mehta}
\beq
R_k(x_1,\dots,x_k|t) = \det \left[H_{N,t}(x_i,x_j)\right]_{i,j=1}^k
\eeq
for some kernel $H_{N,t}(x_i,x_j)$ depending on time $t$.
In particular, the density of Brownian movers at time $t$ is given by
\beq
R_1(x) = H_{N,t}(x,x).
\eeq

Let us now consider a particular case where some of the starting and ending points merge in groups:
\beq
(a_1,\dots,a_N) = (\overbrace{\alpha_1,\dots, \alpha_1}^{n_1},\overbrace{\alpha_2,\dots, \alpha_2}^{n_2},\dots,\overbrace{\alpha_2,\dots, \alpha_2}^{n_2},\dots,\overbrace{\alpha_p,\dots, \alpha_p}^{n_p})
\eeq
and
\beq
(b_1,\dots,b_N) = (\overbrace{\beta_1,\dots, \beta_1}^{\tilde{n}_1},\overbrace{\beta_2,\dots, \beta_2}^{\tilde{n}_2},\dots,\overbrace{\beta_2,\dots, \beta_2}^{\tilde{n}_2},\dots,\overbrace{\beta_q,\dots, \beta_q}^{\tilde{n}_q})
\eeq
with
\beq
\sum_{i=1}^p n_i = \sum_{i=1}^q \tilde{n}_i = N.
\eeq
It means that one considers $p$ groups of $n_i$ particles starting from distinct points $\alpha_i$ at $t=0$, merging for
intermediate times and splitting into $q$ groups of $\tilde{n}_i$ movers reaching $\beta_i$ at $t=1$. Remark \cite{KP} that
the kernel reduces to the kernel of the $p+q$ mutli-component KP integrable hierarchy:
\beq
H_{N,t}(x,x') = H_{N,t}^{(p,q)}(x,x')
\eeq
since it satisfies the corresponding Hirota equation.

\begin{figure}
  \includegraphics[width=10cm,angle=-90]{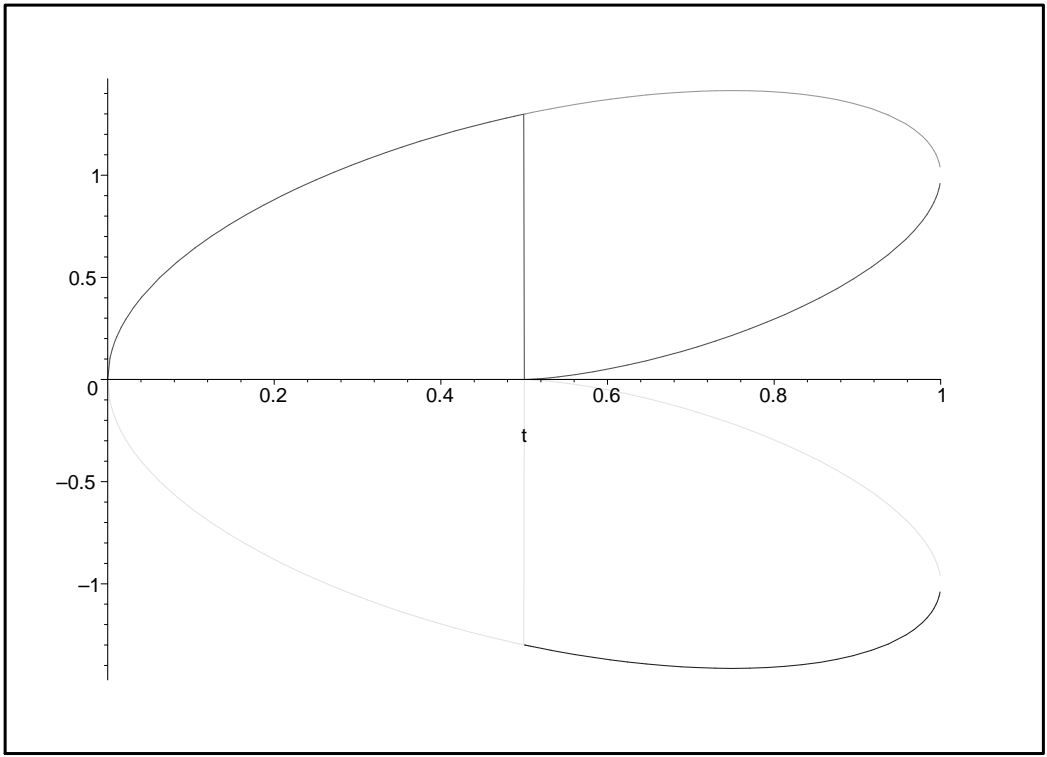}\\
  \caption{Example of one starting point at 0 and two ending points at +1 and -1 half of the movers going to +1 and the
  other half to -1( the particles go from the left to the right). In the large $N$ limit, the Brownian movers
  fill the sector of the space time delimited by the figure. One sees that before $t={1 \over 2}$, they are distributed along
  a unique segment which splits into two disjoint segments for $t>{1 \over 2}$. }\label{brown1}
\end{figure}

In the following one studies the behavior of this phenomenon as the number of particles goes to infinity while the ratios
\beq
\epsilon_i = {n_i \over N} \qquad \hbox{and} \qquad \tilde{\epsilon}_i = {\tilde{n}_i \over N}
\eeq
are kept fixed and finite.
In this case, the Brownian movers form clouds which fill a connected region of the complex plane describing the space time.
For example, for 1 starting point and two ending points one typically gets a configuration of the type depicted in figure
\ref{brown1}: all movers leave the origin and begin to flee from one another. It creates a larger and larger segment of the
space filled by the Brownian movers. As the time grows, because the movers want to reach different points, they split into two groups
heading towards these two end-points.

As often in the study of such integrable system, the kernel exhibits universal behaviors : in any point of the space time, one can rescale the kernel so that one obtains a universal kernel independent of the position of the considered
point; typically, one recovers the Sine, Airy and Pearcy kernels.
The purpose of this part is to emphasize the role played by the spectral curve and algebraic geometry in the study of these universality properties.

\subsection{Particles starting from one point: a matrix model representation}

Let us consider the particular case of one starting point $p=1$. This means that all the Brownian movers start from the same point which we may assume to be located at the origin 0 of the real axis. It was proved \cite{Dyson} that this case is
equivalent to a matrix model. More precisely, the statistic of the Brownian movers at a given time $t$ is the
same as the statistic of the rescaled eigenvalues of an hermitian random matrix $M$ of size $N \times N$ submitted to an external field $A(t)$. This
matrix model is given by the partition function
\beq
Z\left(A(t)\right) = \int dM e^{-N \Tr \left(M^2 + A(t) M \right)}
\eeq
and
\beq
A(t) = \hbox{diag} \left( \overbrace{A_1(t),\dots,A_1(t)}^{\tilde{n}_1}, \overbrace{A_2(t),\dots,A_2(t)}^{\tilde{n}_2} , \dots , \overbrace{A_q(t),\dots,A_q(t)}^{\tilde{n}_q} \right)
\eeq
with the time dependent elements
\beq
A_i(t) =  \sqrt{2 t \over t ( t-1)} \beta_i.
\eeq
The limit of a large number of particles corresponds to the large matrix limit.

This model can also be expressed in terms of the eigenvalues $(x_1,\dots,x_N)$ of the random matrix $M$ using the sell known HCIZ integral formual \cite{HC,IZ}. They are
submitted to a probability measure
\beq
d\mu(x_1,\dots, x_N) = \prod_{i=1}^N dx_i \Delta(x)^2 e^{-N \sum_i \left({x_i^2 \over 2} - x_i a_i\right)}.
\eeq
After the rescaling
\beq
x_i \to x_i \sqrt{t(1-t)},
\eeq
they have the same statistic as $N$ Brownian movers starting from the origin.

The correlation functions
\beq
R_k(x_1,x_2, \dots,x_k) = {1 \over N^k} \left<\displaystyle{\prod_{i=1}^k} \Tr \delta(x_i-M)\right>,
\eeq
have also a determinantal expression
\beq
R_k(x_1,\dots,x_k) = \det \left[H_{N,t}(x_i,x_j)\right]_{i,j=1}^k,
\eeq
 in terms of a kernel $H_{N,t}$. As $N$ goes to infinity, these eigenvalues
merge into dense intervals of the real axis and the density
$R_1(x)$
is supported by a finite number of segments $[z_{2i-1},z_{2i}]$. A classical result of the study of random matrices states that
these segments correspond to the discontinuity of the resolvent $W(x)  = \left< {\displaystyle \sum_i} {1 \over x-x_i}\right>$:
\beq\label{defdiscont}
W_+(x) - W_-(x) = R_1(x)
\eeq
for $x \in [z_{2i-1},z_{2i}]$.

\subsubsection{Gaussian matrix model in an external field}
Let us now quickly remind this matrix model's loop equations and large $N$ solution. This model is a special case of the matrix model in an external field studied in section \ref{sec1Mext}.

The resolvent $W(x)$ is thus the solution of the algebraic equation \eq{spcurve1MMext} with $V'(x)=x$, i.e.:
\beq\label{brownianspcurve}
E(x,Y(x))= Y(x)-x + \sum_{i=1}^q {\epsilon_i \over Y(x)-a_i(t)} = 0
\eeq
where
\beq
Y(x) = W(x) - x.
\eeq
This equation can be seen as the embedding of a Riemann surface $\spcurve$ into $\mathbb{CP}^1 \times \mathbb{CP}^1$.

First of all, one can see that this spectral curve has always genus 0: it admits a simple rational parametrization
\beq
\left\{ \begin{array}{l}
x(z) = z + {\displaystyle \sum_{i=1}^q} {n_i \over N (z-a_i(t))} \cr
y(z) = z \cr
\end{array}
\right. .
\eeq

It is composed of $q+1$ sheets, one of which contains the pole $z= \infty$. This sheet is called the physical sheet.

For a fixed number $q$ of distinct eigenvalues of the external matrix, it may have up to $q$ cuts linking the physical
sheet to the $q$ others. These cuts correspond to the discontinuity of the resolvent giving rise to the density of
eigenvalues \eq{defdiscont}. The cuts $[z_{2i-1},z_{2i}]$ are thus the support of the eigenvalues.
Each $z_i$ is solution of $x'(z_i)=0$. 
To be precise, using the notations
of section \ref{secsymplectic}, the density of eigenvalues on the cut $i$ is given by
\beq
\rho(x(p)) = y(p^{(0)}) - y(p^{(i)}) \virg \hbox{for} \qquad x(p) \in [z_{2i-1},z_{2i}]
\eeq
where one labels the physical sheet by $0$ and the label $i$ corresponds to the sheet linked to the physical one by the
cut $[z_{2i-1},z_{2i}]$.

Let us now study the evolution of the structure of this spectral curve as the time evolves from $0$ to $1$: one is particularly
interested in the time evolution of the position of the branch points $z_i(t)$.

For a given time $t$, these branch points are the simple real roots of the equation $\left. \partial_y E(x,y) \right|_{y= y(x)} = 0$:
\beq
x'(z_j) = 1 - \sum_{i=1}^k {n_i \over N (z_j-a_i(t))^2} = 0.
\eeq

\begin{figure}
\hspace{3cm}  \includegraphics[width=8cm]{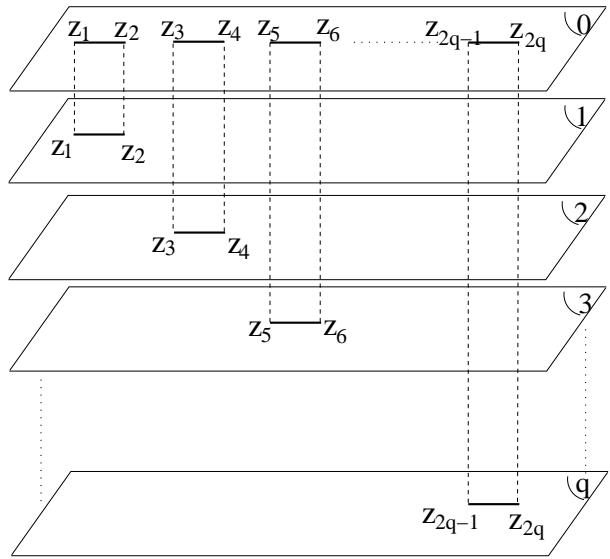}\\
  \caption{For large times, the spectral curve is composed of $q+1$ sheets linked by $q$ real cuts $[z_{2i-1},z_{2i}]$.
  These cuts can be seen as the section of the support of the Brownian movers at fixed time. In the example depicted in
  figure \ref{brown1}, it corresponds to times greater than ${1 \over 2}$.}\label{largespec}
\end{figure}

Let us first consider large times close to $1$. In this case, the eigenvalues $a_i(t)$ become large and are far apart from one
another. Thus, the equation has $2k$ distinct real roots in $z$: the spectral curve has $k$ distinct cuts $[z_{2i-1},z_{2i}]$ (see
figure \ref{largespec}).

\begin{figure}
\hspace{3cm}  \includegraphics[width=8cm]{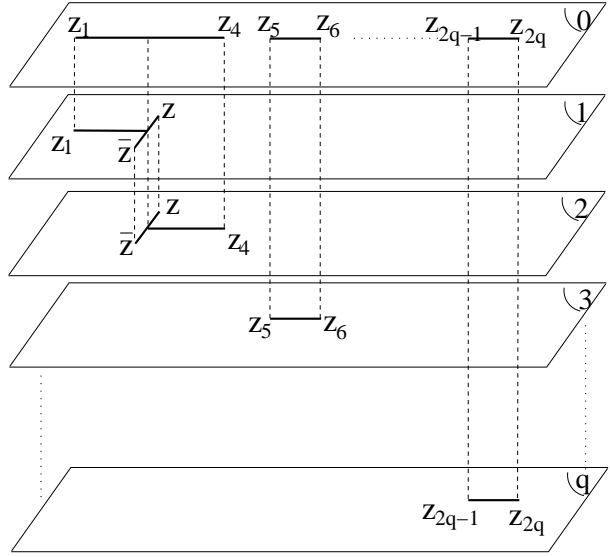}\\
  \caption{For intermediate times some of the real cuts have merged and created imaginary cuts linking two non-physical sheets.
  In this example, $z_2$ and $z_3$ have collapsed and given rise to the imaginary cut $[z,\overline{z}]$.}\label{interspec}
\end{figure}

Then, as the time decreases, the branch points come closer from one another and merge for some critical time before
becoming complex conjugated with an increasing imaginary part. It means that two real cuts merge into one and
an imaginary cut linking two non physical sheets appear (see figure \ref{interspec}).

Finally, as the time becomes small and approaches zero, all the branch point are coupled complex conjugated numbers except
two of them: the ones with the smallest and the largest real parts: there is only one real cut left.

This time evolution of the spectral curve has a simple interpretation in terms of statistic of the eigenvalues. Indeed, thanks
to \eq{defdiscont}, the real cuts are the support of the random matrix eigenvalues (whereas imaginary cuts just follow from the interaction
between the different groups of eigenvalues). In terms of non intersecting Brownian motions, these real cuts
are the segments filled by the Brownian movers at a given time $t$: it is the constant time section of the region of space-time
filled by the Brownian movers.

The time evolution of the spectral curve can thus be interpreted as follows. For times close to one, the movers form
 $q$ groups lying on segments centered around the $q$ end points and whose extremities are the branch points of the
spectral curve. As the time decreases, the branch points come closer to each other and finally some of them merge, i.e.
two of the disjoints segments supporting the Brownian movers merge into one and the sector filled by the Brownian
movers exhibits a cusp. Then the different segments keep on merging
as time decreases until they give a simply connected support for $t \to 0$. This follows the intuition that all the particles
leave 0 in one group which step by step splits into smaller groups to end up with $q$ groups reaching the end points at $t=1$
(see figure \ref{brown1} for $q=2$).

\br
For the following, it is interesting to note that the critical times when two disjoint segments merge correspond to
singular spectral curves in the sense of definition \ref{defregular}. Indeed, at this time, the spectral curve has a double branch
point at the location where the two simple branch points merge.
\er

\subsubsection{Replica formula and spectral curve}

Let us now follow another approach, exact for finite $N$, exhibiting the role played by the spectral curve directly in the formulation
of the kernel $H_N$. For this purpose, we sketch the derivation of a double integral representation of the kernel using the
replica method developed in this context by Br\'ezin and Hikami \cite{Brezin2,Brezin3,Brezin4,Brezin5}.

Let us first consider the "Fourier" transforms of the correlation functions
\beq
U_l(t_1,t_2, \dots,t_l) = \left<\displaystyle{\prod_{i=1}^l} \Tr e^{iNt_iM}\right>
\eeq
and, in particular, one gets the Fourier transform of the two points correlation function:
%\beq
%U_2(t_1,t_2) = {1 \over Z(A) N^2 } \sum_{\alpha_1,\alpha_2=1}^N \int \left(\prod_{j=1}^N dx_j\right) {\Delta(x) \over %\Delta(a)}
%e^{-N {\displaystyle \sum_{j=1}^N} \left( {x_j^2 \over 2} - x_j a_j\right) + i N (t_1 x_{\alpha_1} + t_2 %x_{\alpha_2})},
%\eeq
%i.e.
\beq
U_2(t_1,t_2) = {1 \over Z(A) N^2} \sum_{\alpha_1,\alpha_2=1}^N \int \left(\prod_{j=1}^N dx_j\right) {\Delta(x) \over \Delta(a)}
e^{-N {\displaystyle \sum_{j=1}^N} \left[ {x_j^2 \over 2} - x_j \left(a_j+ i t_1 \delta_{j,\alpha_1}+ i t_2 \delta_{j,\alpha_2}\right)\right]}.
\eeq
One can now integrate the variables $x_j$ by noting that
\beq
\int \left(\prod_{j=1}^N dx_j\right) \Delta(x) e^{-N {\displaystyle \sum_{j=1}^N} \left[ {x_j^2 \over 2} + x_j b_j \right]} = \Delta(b) e^{{N \over 2} {\displaystyle \sum_{j=1}^N} b_j^2}
\eeq
and using the expansion $\Delta(x) = \prod_{i \neq j} (x_i-x_j)$:
\bea
U_2(t_1,t_2) &=& \sum_{\alpha_1,\alpha_2=1}^N  e^{N \left(it_1 a_{\alpha_1} + it_2 a_{\alpha_2} - {t_1^2 + t_2^2 \over 2} - t_1 t_2 \delta_{\alpha_1,\alpha_2}\right)} \times \cr
&& \times { \displaystyle{\prod_{1\leq l < m \leq N}} (a_l-a_m+ i t_1(\delta_{l,\alpha_1}-\delta_{m,\alpha_1})
+ i t_2(\delta_{l,\alpha_2}-\delta_{m,\alpha_2})) \over \displaystyle{\prod_{1\leq l < m \leq N}} (a_l-a_m)}
 . \cr
\eea
One can see that this can be written as a double contour integral
\bea
U_2(t_1,t_2)&=& {e^{-N {t_1^2 + t_2^2 \over 2}} \over t_1 t_2} \oint \oint {du dv \over (2i \pi )^2} e^{Ni(t_1u+t_2v)}
{(u-v+it_1-i t_2)(u-v) \over (u-v+it_1)(u-v-it_2)} \times \cr
&& \;\;\;\;\;\;\;\;\;\;\;\;\;\;\;\;\;\;\;\; \times \prod_k \left(1+{it_1 \over u-a_k}\right)\left(1+{it_2 \over v-a_k}\right)\cr
\eea
or
\bea
U_2(t_1,t_2)&=& {e^{-N {t_1^2 + t_2^2 \over 2}} \over t_1 t_2} \oint \oint {du dv \over (2i \pi )^2} e^{Ni(t_1u+t_2v)}
\left(1-{ t_1 t_2 \over (u-v+it_1)(u-v-it_2)}\right) \times \cr
&& \;\;\;\;\;\;\;\;\;\;\;\;\;\;\;\;\;\;\;\; \times  \prod_k \left(1+{it_1 \over u-a_k}\right)\left(1+{it_2 \over v-a_k}\right)\cr
\eea
where the integration contours encircle all the eigenvalues $a_k$ and the pole $v = u -i t_1$\footnote{see for instance \cite{Brezin2} for more details around eq.(2-20) and eq.(4-40).}.

We can now go back to the correlation function
\beq
R_2(\lambda,\mu) = \int_{-\infty}^\infty \int_{-\infty}^\infty {dt_1 dt_2 \over 4 \pi^2} e^{-iN(t_1\lambda+t_2\mu)} U(t_1,t_2) .
\eeq
By first integrating on $t_1$ and $t_2$ with the shifts $t_1 \to t_1 - iu$ and $t_2 \to t_2 - iu$, we can show that:
\beq
R_2(\lambda,\mu) = K_N(\lambda,\lambda) K_N(\mu,\mu) - K_N(\mu,\lambda) K_N(\lambda,\mu)
\eeq
where the kernel is defined by
\beq
K_N(\lambda,\mu) = \int {dt \over 2 \pi} \oint {dv \over 2i \pi} \displaystyle \prod_{k=1}^N \left({it-a_k \over v-a_k}\right) {1 \over v-it}
e^{-N\left({v^2+t^2 \over 2}+it\lambda-v\mu\right)}
\eeq
where the integration contour for $v$ goes around all the points $a_k$ and the and the integration for $t$ is parallel to the
real axis and avoids the $v$ contour. Moreover, it is straightforwardly proven that  any $k$-point function can be written as the Fredholm determinant:
\beq
R_k(x_1,\dots,x_k) = \det \left[K_N(x_i,x_j)\right]_{i,j=1}^k.
\eeq

By Wick rotating the integration variable $t \to i t$, one gets
\beq
K_N(\lambda,\mu) = \int {dt \over 2 i \pi} \oint {dv \over 2i \pi} \displaystyle \prod_{k=1}^N \left({t-a_k \over v-a_k}\right) {1 \over v-t}
e^{-N\left({v^2-t^2 \over 2}+t\lambda-v\mu\right)}
\eeq
where the integration contour for $t$ is now parallel to the imaginary axis.
One can then rewrite it under a more factorized form:
\beq
K_N(\lambda,\mu) = \int {dt \over 2 i \pi} \oint {dv \over 2i \pi} e^{-N(S(\mu,v)-S(\lambda,t))} {1 \over v-t}
\eeq
where
\beq
S(x,y) = {y^2 \over 2} - x y + \sum_{i=1}^q \epsilon_i \ln (y-a_i)
\eeq
with the "filling fractions" given by
\beq
\epsilon_i:={n_i \over N}.
\eeq

How to compute such an integral ? Let us use the saddle point method for both integrals. The saddle points of the first
exponential are given by the $y$ solutions of
\beq
\partial_y S(x,y) = y -x + \sum_{i=1}^k {\epsilon_i \over y-a_i}=0
\eeq
which is nothing but the equation of the spectral curve \eq{brownianspcurve} !

In this setup, the spectral curve can thus be seen as the location of the saddle points of the action $S(x,y)$
in this formulation of the kernel. Remark also that this formulation of the kernel is very similar to
the formulation of th.\ref{thexpkernel} in terms of symplectic invariants.

\subsubsection{From the replica formula to the symplectic invariants formalism}

Let us now start from the expression of th.\ref{thexpkernel} for the kernel. For this purpose, one has to compute the one
form $ydx$. Using the global parameterization
\beq
\left\{\begin{array}{l}
x(z) = z + \sum_i {\epsilon_i \over z-a_i(t)} \cr
y(z) = z \cr
\end{array} \right. ,
\eeq
one gets
\beq
ydx(z)=  z dz + \sum_i {\epsilon_i dz \over z-a_i(t)}.
\eeq
Integrating by parts, one can see that
\beq
\int_{z_2}^{z_1} ydx = - \int_{z_2}^{z_1} x(z) dy(z) + \left[x(z) y(z) \right]_{z_2}^{z_1}.
\eeq
Moreover, the function $y(z)$ can also be considered as a global coordinate on the spectral curve since it
coincides with the $z$ coordinate. The previous equation can hence be written
\beq
\int_{y_2}^{y_1} y dx(y) = - \int_{y_2}^{y_1} x(y) dy + \left[x(y) y \right]_{y_2}^{y_1}.
\eeq

On the other hand, the spectral curve has a particular form: it has only one $x$-sheet and takes the form
\beq
H(x,y) = x - x(y) = 0
\eeq
with the function
\beq
x(y) = y + \sum_i {\epsilon_i \over y-a_i(t)} .
\eeq
Thus the action, i.e. the integral of the spectral curve wrt $y$, reads
\beq
S(x,y) = x y - \int x(y) dy
\eeq
and further
\bea
S(x_1,y_1) - S(x_2,y_2) &=& x_1 y_1 - x_2 y_2 - \int_{y_1}^{y_2} x(y) dy \cr
&=& \int_{y_2}^{y_1} y dx(y) + y_1 \left[x_1 - x(y_1)\right] + y_2 \left[x_2 - x(y_2)\right] . \cr
\eea
It is now possible to compare the kernel built from the symplectic invariants and the spectral curve
\beq
H_N(x_1,x_2) = {1 \over x_1-x_2} e^{N \int_{y(x_2)}^{y(x_1)} y dx(y)} \left[ 1 + O\left({1 \over N}\right) \right]
\eeq
and the kernel following the replica formula
\beq
\widetilde{H}_N(x_1,x_2) = \oint \oint dy_1 dy_2 {1 \over y_1 - y_2} e^{N \left[S(x_1,y_1)-S(x_2,y_2)\right]}.
\eeq
Indeed, the later reads
\beq
\widetilde{H}_N(x_1,x_2) = \oint \oint dy_1 dy_2 H_N(x(y_1),x(y_2)) e^{N\left\{y_1 \left[x_1 - x(y_1)\right] + y_2 \left[x_2 - x(y_2)\right]\right\}}.
\eeq
The saddle point equation directly then states the equality of both kernels in the large $N$ limit if the integration contours are good steepest descent contours.

\br
Note that this particular representation of the kernel under the form of a double integral on the complex plane
(or more precisely the Riemann sphere) is a direct consequence of the specific parameterization of the spectral curve $x= x(y)$.
\er

\subsubsection{Critical behaviors and singular spectral curves}

Since both kernels coincide, one can use the singular limits derived in the symplectic invariant setup to obtain some
critical universal behaviors of the exclusion process.

\medskip

\noindent $\bullet$ {\bf Universality on the Edge: the Airy kernel}

\medskip

Let us study the statistic of Brownian movers around the edge of their limiting support. It means that one is
interested in a point $(x,t)$ of the space time such that $(x,y(x))$ is close to a branch point $(x_c,y_c)= (x(z_c),y(z_c))$ of the spectral curve at time $t_c$. Following the study of section \ref{secsingFg}, let us rescale the global coordinate $z = z_c + {1 \over N^{1 \over 3}} Z$ and one gets the blown up spectral curve
\beq
\left\{\begin{array}{l}
x(z) = x(z_c) + {Z \over N^{1 \over 3}} x'(z_c) + {Z^2 \over 2 N^{2 \over 3}} x''(z_c) + O\left({1 \over N}\right) \cr
y(z)  =  y(z_c) + {Z \over N^{1 \over 3}} y'(z_c) + O\left({1 \over N^{2 \over 3}}\right) \cr
\end{array}
\right. .
\eeq
Since the point $z_c$ is a branch point, one has $x'(z_c)=0$, and the blown up curve reduces to
\beq
\left\{\begin{array}{l}
x_{airy}(Z) =  {1 \over N^{2 \over 3}} {x''(z_c)\over 2} Z^2  \cr
y_{airy}(Z)  =  {1 \over N^{1 \over 3}} y'(z_c) Z \cr
\end{array}
\right.
\eeq
to leading order as $N \to \infty$. Remark that the scaling is such that $ydx(Z)  = O\left({1 \over N}\right)$.
This curve is the Airy curve described in the second example of section \ref{secexspcurves}. Theorem \ref{thsinglimomng} states that, in terms of the rescaled variable
$Z$, i.e. the distance from the considered critical point,  the kernel and the correlation functions reduce to the one of
the Airy curve, independently of the position of the critical point.

Note that this Airy curve has also the form $x = x(y)$ where the function $x(y) = {x''(z_c) \over 2 y'(z_c)^2} y^2$. The kernel
can thus also be written under a double integral form
\beq
H_{Airy}(x_1,x_2) = \oint \oint {dy_1 dy_2 \over y_1 -y_2} e^{S_{Airy}(x_1,y_1)-S_{Airy}(x_1,y_1)}
\eeq
with $S_{Airy}(x,y) = x y - {x''(z_c) \over y'(z_c)^2} {y^3 \over 6}$, i.e. the Airy kernel.

\medskip

\noindent $\bullet$ {\bf Universality at the cusp: the Pearcy kernel}

\medskip

Let us finally consider a point where two groups of Brownian movers merge. This is obtained when two edges merge or, in the spectral
curve formalism, when two branch points merge as time decreases. At this critical time $t_c$, the corresponding spectral curve
is singular since the merging of two simple branch points gives rise to a double branch point. Let us be more specific:
one considers a cusp at the position $(x_c,t_c)$ in the space time. Let us blow up the space time around this point using the
rescaling
\beq
\left\{
\begin{array}{l}
t:= t_c + {\alpha_t T \over N^{1 \over 2}} \cr
z:= z_c + {\alpha_y Z \over N^{1 \over 4}} \cr
\end{array}
\right.
\eeq
where $z$ is the global parameter of the spectral curve. The rational parameterization of the spectral curve thus reads
\beq
\left\{\begin{array}{l}
x(z,t) = x(z_c,t_c) + {\alpha_y Z \over N^{1 \over 4}} \partial_z x(z_c,t_c) + {\alpha_y^2 Z^2 \over 2 N^{1 \over 2}} \left(\partial_z\right)^2 x(z_c,t_c)
+ {\alpha_y^3 Z^3 \over 6 N^{3 \over 4}} \left(\partial_z\right)^3 x(z_c,t_c) \cr
 \qquad \qquad + {\alpha_y \alpha_t Z T \over N^{3 \over 4}} \partial_z \partial_t x(z_c,t_c)
 + O\left({1 \over N}\right) \cr
y(z)  =  y(z_c) + {\alpha_y Z \over N^{1 \over 4}} y'(z_c) + O\left({1 \over N^{2 \over 3}}\right) \cr
\end{array}
\right. .
\eeq
Since the critical point is a double branch point, the blown up curve reduces to
\beq
\left\{\begin{array}{l}
x(Z,T) =
{1 \over N^{3 \over 4}} \left[{\alpha_y^3 Z^3 \over 6} \left(\partial_z\right)^3 x(z_c,t_c) + \alpha_y \alpha_t Z T  \partial_z \partial_t x(z_c,t_c)\right]
\cr
y(Z)  ={\alpha_y Z \over N^{1 \over 4}} \partial_z y(z_c)\cr
\end{array}
\right.
\eeq
which could be called the Pearcy curve. Indeed, since this curve also has the form $x = x(y)$, the associated kernel
has the double contour integral representation
\beq
H_{Pearcy}(x_1,x_2) = \oint \oint {dy_1 dy_2 \over y_1 -y_2} e^{S_{Pearcy}(x_1,y_1)-S_{Pearcy}(x_1,y_1)}
\eeq
with $S_{Pearcy}(x,y) = x y - \left({ \alpha_y^2 \left(\partial_z\right)^3 x(z_c,t_c) \over \left(\partial_z y(z_c)\right)^3}\right){y^4 \over 24}
- \left({\alpha_t \partial_z \partial_t x(z_c,t_c) \over 2 \partial_z y(z_c)}\right) y^2 T$. This is noting but the Pearcy kernel
once the right integration contour is found and the rescaling coefficients are fixed by:
\beq
{ \alpha_y^2 \left(\partial_z\right)^3 x(z_c,t_c) \over \left(\partial_z y(z_c)\right)^3} = 6
\eeq
and
\beq
{\alpha_t \partial_z \partial_t x(z_c,t_c) \over 2 \partial_z y(z_c)} = 1.
\eeq

%-------------------------- debut surfaces discretisees ---------------------------------------------------------

\section{Enumeration of discrete surfaces or maps}
\label{secmaps}

The symplectic invariants provide a solution to Tutte's equations for counting discrete surfaces (also called maps), for arbitrary topologies. Indeed, it was found by Brezin-Itzykson-Parisi-Zuber \cite{BIPZ}, and further developed by
\cite{ambjornrmt,David,Kazakov}, that generating functions for discrete surfaces can be written as formal matrix models.
 Let us review how to enumerate various ensembles of discrete surfaces.

\subsection{Introduction}

\bd
Let ${\mathbb M}_n^{(g)}$ be the set of connected orientable discrete surfaces of genus $g$ obtained by gluing together polygonal faces, namely $n_3$ triangles, $n_4$ quadrangles, ... $n_k$ $k$-angles, as well as $n$ marked polygonal faces of perimeters $l_1,\dots,l_n$, each of the marked faces having one marked edge on its boundary.
Let us call $v$ the number of vertices of a discrete surface, and let ${\mathbb M}_n^{(g)}(v)$ be the set of discrete surfaces in ${\mathbb M}_n^{(g)}$, with $v$ vertices.

We require that unmarked faces have perimeter $\geq 3$, whereas marked faces are only required to have perimeter $l_i\geq 1$.
\ed

Marked faces are also called "boundaries".

Notice that nothing in our definition prevents from gluing a side of a polygon, to another side of the same polygon.

\bt
${\mathbb M}_n^{(g)}(v)$ is a finite set.
\et

\proof{
Let $e$ be the number of edges of a discrete surface in ${\mathbb M}_n^{(g)}(v)$.
The total number of half edges is:
\beq
2e = \sum_{j\geq 3} jn_j + \sum_{i=1}^n l_i .
\eeq
The Euler characteristics is:
\beq
\chi=2-2g = v-e+n+\sum_{j\geq 3} n_j = v+n-{1\over 2} \sum_{j\geq 3} (j-2)n_j -{1\over 2}\sum_{i=1}^n l_i .
\eeq
This implies:
\beq\label{eqchidiscrsurf}
 {1\over 2} \sum_{j\geq 3} (j-2)n_j + {1\over 2}\sum_{i=1}^n l_i = 2g-2+n+v
\eeq
and therefore the $n_j$'s are bounded, and the $l_i$'s are bounded. There is then a finite number of possible discrete surfaces having a finite number of faces, edges and vertices.
}

In order to enumerate discrete surfaces, we define the generating functions:
\bd
The generating function is the formal power series in $t$:
\bea
W_n^{(g)}(x_1,\dots,x_n;t_3,\dots,t_d;t)
&=& {t\over x_1}\delta_{n,1}\delta_{g,0}  \cr
&&+ \sum_{v=1}^\infty t^v
\sum_{S\in {\mathbb M}_n^{(g)}(v)}\, {1\over \#{\rm Aut}(S)}\,\,{t_3^{n_3(S)}\dots t_d^{n_d(S)}\over x_1^{l_1(S)}\dots x_n^{l_n(S)}}\,\,\prod_{i=1}^n {1\over x_i} .\cr
\eea
Most often, we will write only the dependance in the $x_i$'s explicitly, and write:
\beq
W_n^{(g)}(x_1,\dots,x_n;t_3,\dots,t_d;t) = W_n^{(g)}(x_1,\dots,x_n) .
\eeq

\medskip

The generating functions counting surfaces with marked faces of given perimeters $l_1,\dots,l_n$ are by definition:
\beq
T^{(g)}_{l_1,\dots,l_n} = (-1)^n \Res_{x_1\to\infty}\dots \Res_{x_n\to\infty}\,\, x_1^{l_1}\dots x_n^{l_n}\,\, W_n^{(g)}(x_1,\dots,x_n)\,dx_1\dots dx_n .
\eeq
\ed

Notice that rooted discrete surfaces, i.e. surfaces with only 1 marked edge, have no non-trivial automorphisms, and thus $\#{\rm Aut}=1$ when $n=1$.

\subsection{Tutte's recursion equations}

Tutte's equations are recursions on the number of edges \cite{tutte,tutte2}.
If one erases the  marked edge on the 1st marked face whose perimeter is $l_1+1$, several mutually exclusive possibilities may occur:
\begin{itemize}
\item the marked edge separates the marked face with some unmarked face (let us say a $j$-gon with $j\geq 3$), and removing that edge is equivalent to removing a $j$-gon (with weight $t_j$). We thus get a discrete surface of genus $g$ with the same number of boundaries, and the length of the first boundary is now $l_1+j-1$.

\item the marked edge separates two distinct marked faces (face $1$ and face $m$ with $2\leq m\leq n$, ), thus the marked edge of the first boundary is one of the $l_m$ edges of the $m^{\rm th}$ boundary.
We thus get a discrete surface of genus $g$ with $n-1$ boundaries. The other $n-2$ boundaries remain unchanged,  and there is now one boundary of length $l_1+l_m-1$.

\item the same marked face lies on both sides of the marked edge, therefore by removing it, we disconnect the boundary.
Two cases can occur: either the discrete surface itself gets disconnected into two discrete surfaces of genus $h$ and $g-h$, one having $|J|+1$ boundaries of lengths $j,J$, where $J$ is a subset of $K=\{l_2,\dots,l_n\}$, and the other discrete surface having $k-|J|$ boundaries of lengths $l_1-1-j,K/J$, or the discrete surface remains connected because there was a handle connecting the two sides, and thus by removing the marked edge, we get a discrete surface of genus $g-1$, with $n+1$ boundaries of lengths $j,l_1-j-1,K$.

\end{itemize}

This procedure is (up to the symmetry factors) bijective, and all those possibilities correspond to the following recursive equation:
\bea\label{loopeqex1mmTTbis}
&& \sum_{j=0}^{l_1-1} \Big[ \sum_{h=0}^g \sum_{J\subset K} \TT^{(h)}_{j,J} \TT^{(g-h)}_{l_1-1-j,K/J}
+
\TT^{(g-1)}_{j,l_1-1-j,K} \Big]
 + \sum_{m=2}^{n} l_m \, \TT^{(g)}_{l_m+l_1-1,K/\{ l_m\} }  \cr
&=& \TT^{(g)}_{l_1+1,K} - \sum_{j=3}^d t_{j} \TT^{(g)}_{l_1+j-1,K} .
\eea
This equation is illustrated as follows (where the 1st marked face is the "exterior face"):
$$
{\mbox{\epsfxsize=12.truecm\epsfbox{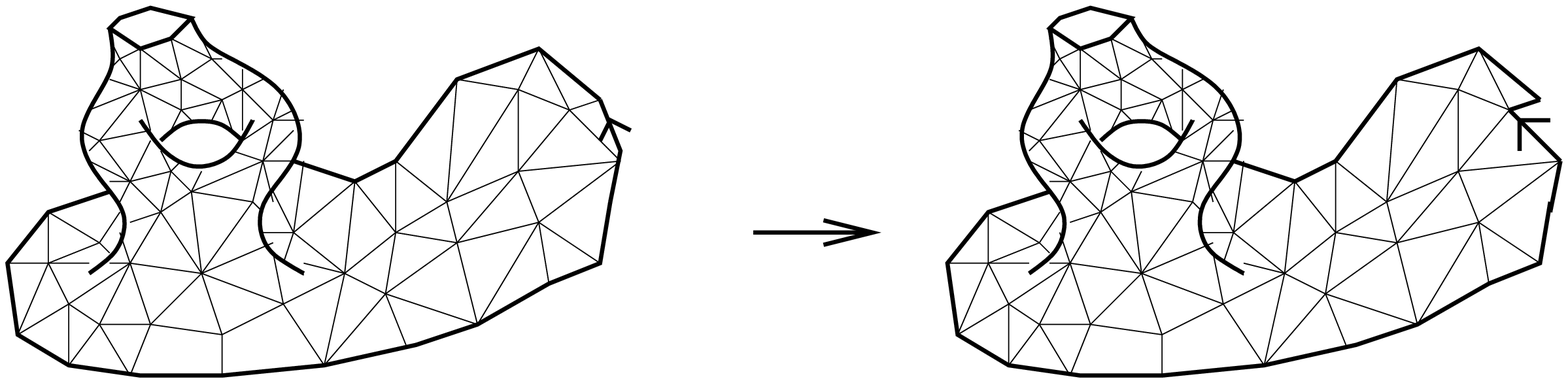}}}
$$
$$
{\mbox{\epsfxsize=12.truecm\epsfbox{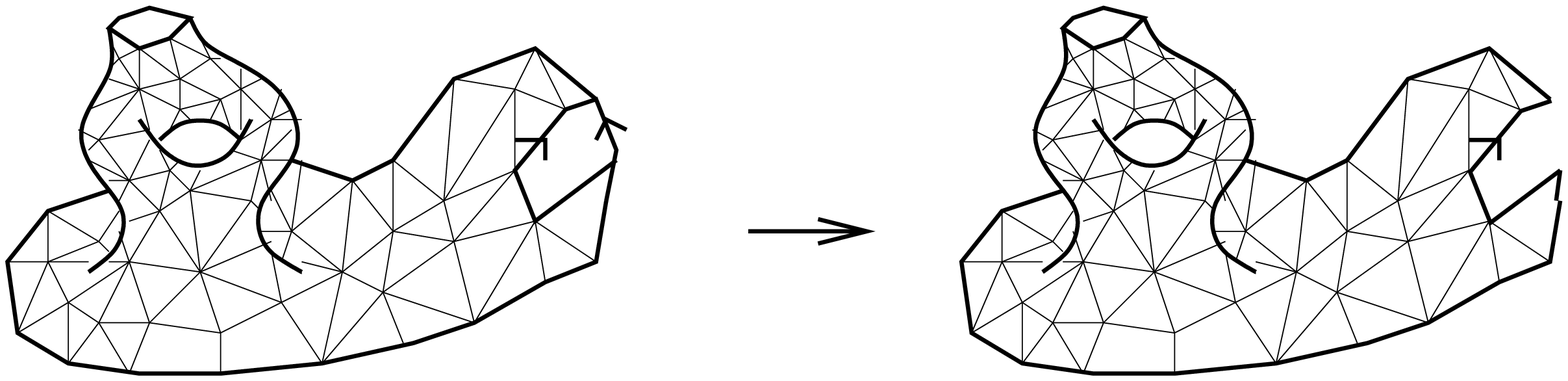}}}
$$
$$
{\mbox{\epsfxsize=12.truecm\epsfbox{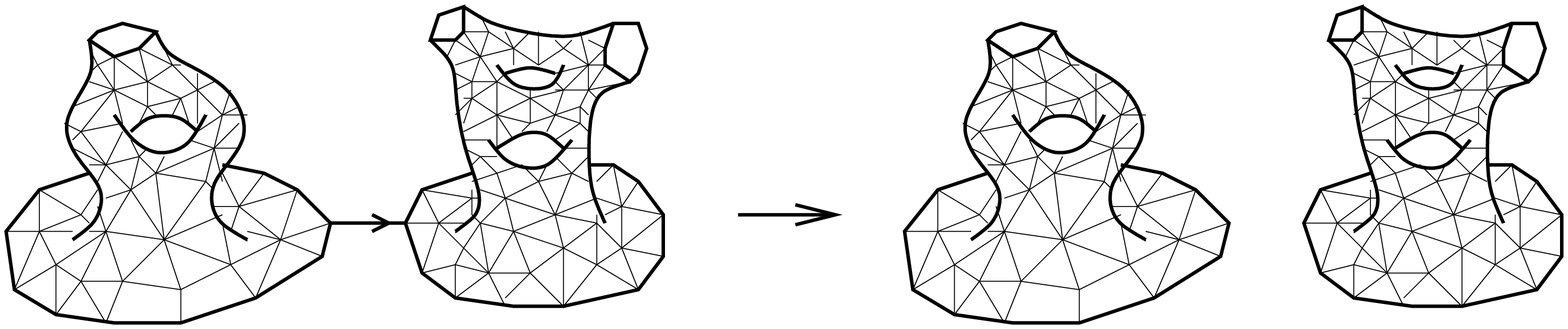}}}
$$
$$
{\mbox{\epsfxsize=12.truecm\epsfbox{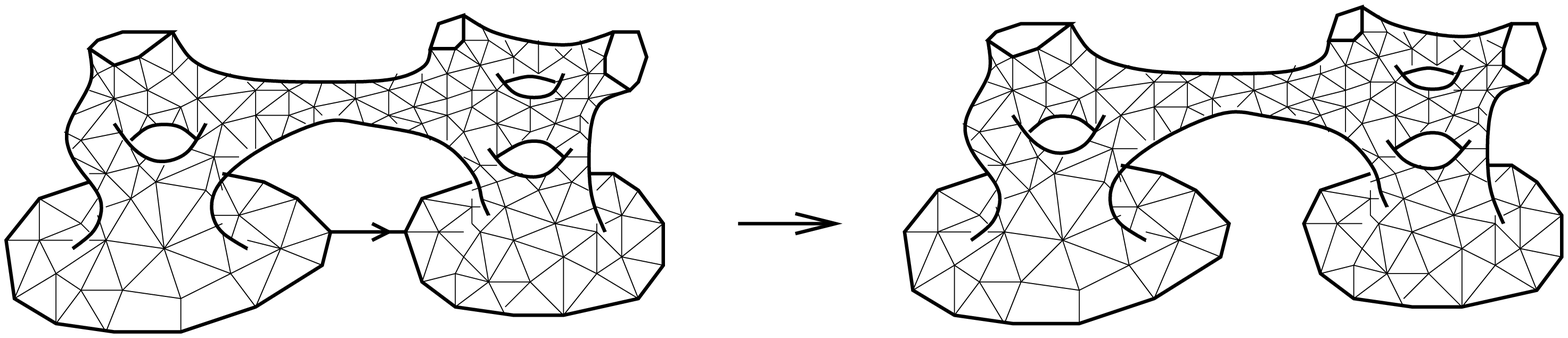}}}
$$

\subsection{Loop equations}

Rewritten in terms of the $W_n^{(g)}$'s, Tutte's equations \eq{loopeqex1mmTTbis} read:

\bt
Loop equations.
For any $n$ and $g$, and $L=\{x_2,\dots,x_n\}$, we have:
\bea\label{loopeqex1W}
&& \sum_{h=0}^g \sum_{J\subset L} W^{(h)}_{1+|J|}(x_1,J) W^{(g-h)}_{n-|J|}(x_1,L/J)
+
W^{(g-1)}_{n+1}(x_1,x_1,L) \cr
&&  + \sum_{j=2}^{n} {\partial \over \partial x_j}\, { W^{(g)}_{n-1}(x_1,L/\{ j\}) -W^{(g)}_{n-1}(L) \over x_1-x_j }  \cr
&=& V'(x_1) W^{(g)}_n(x_1,L) - P^{(g)}_n(x_1,L)
\eea
where
\beq
V'(x) = x-\sum_{j\geq 3} t_j x^{j-1}
\eeq
where $P^{(g)}_n(x_1,L)$ is a polynomial in $x_1$, of degree $d-3$ (except $P_1^{(0)}$ which is of degree $d-2$):
\beq
P^{(g)}_n(x_1,x_2,\dots,x_n)
=   - \sum_{j=2}^{d-1} t_{j+1} \,\,\sum_{i=0}^{j-1}\, x_1^i\,\,  \sum_{l_2,\dots,l_n=1}^\infty
 {\TT^{(g)}_{j-1-i,l_2,\dots,l_n}\over x_2^{l_2+1}\dots x_n^{l_n+1}}  \,\, +t\, \delta_{g,0}\delta_{n,1} .
\eeq

\et

\proof{Indeed, if we expand both sides of \eq{loopeqex1W} in powers of $x_1\to\infty$, and identify the coefficients on both side, we find that
the negative powers of the $x_i$'s give precisely the loop equations \eq{loopeqex1mmTTbis},
whereas the coefficients of positive powers of $x_1$ cancel due to the definition of $P_n^{(g)}$,
which is exactely the positive part of $V'(x_1) W^{(g)}_n$:
\beq
P_n^{(g)}(x_1,x_2,\dots,x_n) = \mathop{{\rm Pol}}_{x_1\to\infty}\,\,\left( V'(x_1)\, W_n^{(g)}(x_1,x_2,\dots,x_n) \right)
\eeq
where Pol means that we keep only the polynomial part, i.e. the positive part of the Laurent series at $x_1\to\infty$.
}

\medskip

We see that Tutte's equations \eq{loopeqex1W} are identical to the matrix model loop equations \eq{loopeq1MM1cut}.
The solution is thus the same, and it is expressed in terms of symplectic invariants.
One only has to find the corresponding spectral curve.

\subsubsection{Spectral curve and disc amplitude}

The spectral curve is given by the function $W_1^{(0)}(x)$.

\medskip
With $n=1$ and $g=0$, the loop equation \eq{loopeqex1W} reads:
\beq
(W_1^{(0)}(x))^2 = V'(x)\,W_1^{(0)}(x) - P_1^{(0)}(x)
\eeq
which implies:
\beq
W_1^{(0)}(x) = {1\over 2}\,\left( V'(x) - \sqrt{V'(x)^2-4P_1^{(0)}(x)}\,\,\right)
\eeq
where $P_1^{(0)}(x)$ is a polynomial of degree $\deg V'-1$ in $x$, namely:
\beq
P^{(0)}_1(x)
= t - \sum_{j=2}^{d-1} t_{j+1} \,\,\sum_{i=0}^{j-1}\, x^i\,\, \TT^{(0)}_{j-1-i}.
\eeq
Notice that \eq{eqchidiscrsurf} implies that discrete surfaces with $g=0$ and $n=1$ must have $v\geq 2$, and thus:
\beq
\TT^{(0)}_j = t\delta_{j,0} + O(t^2).
\eeq
Therefore $P_1^{(0)}(x)$ is a formal series in $t$ such that:
\beq\label{eqsmalltP10discrsurf}
P_1^{(0)}(x) = t \,{V'(x)\over x}+O(t^2) .
\eeq

A general spectral curve with filling fractions $\epsilon_i = {-1\over 2i\pi}\oint_{\acycle_i} W_1^{(0)}(x) dx$, would correspond to (see \eq{eqP10formal1MMasympt}):
\beq
P_1^{(0)}(x) = \sum_{i=1}^{d-1} \epsilon_i \,\, \,{V'(x)\over x-X_i}+O(t^2)
\eeq
where $X_i$, $i=1,\dots,d-1$ are the zeroes of $V'(x)$.

In other words, the spectral curve counting discrete surfaces, has only one non-vanishing filling fraction, it is a 1-cut spectral curve, or equivalently, it is a genus $\genus=0$ spectral curve.

More precisely, \eq{eqsmalltP10discrsurf} implies that the zeroes of $V'(x)^2-4P_1^{(0)}(x)$ have the following small $t$ behaviour:\\
$\bullet$ two zeroes $a,b$ are of the form:
\beq
a\sim 2\sqrt{t} + O(t)
\virg
b\sim -2\sqrt{t} + O(t);
\eeq
$\bullet$ there are $d-2$ double zeroes of the form $X_i \pm O(t)$, where $V'(X_i)=0$, and $X_i\neq 0$.

%Then, notice that $W_1^{(0)}(x)$ is a formal series in $t$, whose coefficients are polynomials in $1/x$, thus:
%\beq
%{1\over 2i\pi\, t}\,\oint_{{\cal C}} W_1^{(0)}(x) dx =
%\left\{\begin{array}{l}
%1\quad {\rm if}\,\,{\cal C}\,{\rm surround}\, 0\cr
%0\quad {\rm otherwise}
%\end{array}\right.
%\eeq
%This implies that all the pairs of zeroes not centered around zero, are double zeroes.
%Only the pair near $0$ can be a pair of simple zeroes.
Therefore there exists $a,b$ and a polynomial $M(x)$ such that:
\beq
V'(x)^2-4P_1^{(0)}(x) = (x-a)(x-b)\,M(x)^2
\eeq
and thus:
\beq\label{eqW10Msqrt}
W_1^{(0)}(x) = {1\over 2}\,\left( V'(x)-M(x)\sqrt{(x-a)(x-b)}\,\right)
\eeq
where $a=2\sqrt{t}+O(t)$, $b=-2\sqrt{t}+O(t)$, $M(x)={V'(x)\over x}+O(t)$.

\subsubsection{Rational parametrization}\label{sec1MMrational}

Since the spectral curve has only one cut $[a,b]$, it has genus $\genus=0$, and thus it is a rational spectral curve, and it can be parameterized by rational functions of a complex variable.
Here, this can be done very explicitly.

\medskip

We parameterize $x$ as:
\beq
x(z) = {a+b\over 2} + {a-b\over 4}\,\left(z+{1\over z}\right)=\alpha+\gamma\left(z+{1\over z}\right)
\virg
\alpha={a+b\over 2}\, , \,\,\, \gamma={a-b\over 4}.
\eeq
This parametrization is convenient because we have:
\beq
\sqrt{(x-a)(x-b)}=\gamma\left(z-{1\over z}\right)
\eeq
and therefore, from \eq{eqW10Msqrt}, we see that $W_1^{(0)}(x(z))$ is a rational fraction of $z$.

Since $x(z)=x(1/z)$, we can find some complex numbers $u_0,u_1,\dots,u_{d-1}$ such that:
\beq
V'(x(z)) = \sum_{k=0}^{d-1} u_k (z^k+z^{-k})
\eeq
and similarly:
\beq
M(x(z))\,\sqrt{(x(z)-a)(x(z)-b)} = \sum_{k=1}^{d-1} \td{u}_k (z^k-z^{-k}).
\eeq
Thus we have:
\beq
W_1^{(0)}(x(z)) = u_0+ {1\over 2}\sum_{k\geq 1} (u_k-\td{u}_k)\,z^k + {1\over 2}\sum_{k\geq 1} (u_k+\td{u}_k)\,z^{-k}.
\eeq
Since, by definition, $W_1^{(0)}(x(z))$ contains only negative powers of $x$, $W_1^{(0)}(x)\sim {t\over x}+O(1/x^2)$, it must contain only negative powers of $z$, and therefore we must have $u_0=0$ and $\td{u}_k=u_k$,  i.e. $W_1^{(0)}(x(z))$ is a polynomial in $1/z$:
\beq
W_1^{(0)}(x(z)) = \sum_{k=1}^{d-1} u_k\,z^{-k}.
\eeq
Since $W_1^{(0)}(x)\sim {t\over x}$ at large $x$, the coefficient of $1/z$ must be $u_1=t/\gamma$.
Therefore, $a$ and $b$ are determined by the two equations:
\beq
u_0=0 \virg u_1 = {t\over \gamma}.
\eeq

All this can be summarized by the theorem:
\bt\label{thspcurvediscrsurf}
Let $V'(x) = x-{\displaystyle \sum_{k=2}^{d-1}} t_{k+1} x^k$ where $t_k$ is the Boltzmann weight for $k$-gons.
For an arbitrary $\alpha$ and $\gamma$, we write:
\beq
V'(\alpha+\gamma(z+1/z)) = \sum_{k=0}^{d-1} u_k(z^k+z^{-k}).
\eeq
The coefficients $u_k$ are thus polynomials of $\alpha$ and $\gamma$.
We determine $\alpha$ and $\gamma$ by:
\beq
u_0=0
\virg
u_1={t\over \gamma}
\eeq
and by the conditions that $\alpha=O(t)$ and $\gamma^2 = t+O(t^2)$ at small $t$.

Then the spectral curve ${\cal E}=({\mathbb P}^1,x,y)$ is:
\beq
x(z)=\alpha+\gamma(z+1/z)
\virg
y(z)=-W_1^{(0)}(x(z)) = -\sum_k u_k z^{-k}.
\eeq

\et

\subsubsection{Generating function of the cylinder, annulus}

For $n=2$ and $g=0$, Tutte's equations give:
\beq
2W_1^{(0)}(x_1)W_2^{(0)}(x_1,x_2)+{\partial \over\partial x_2}\,{W_1^{(0)}(x_1)-W_1^{(0)}(x_2)\over x_1-x_2} = V'(x_1)W_2^{(0)}(x_1,x_2) - P_2^{(0)}(x_1,x_2)
\eeq
and one can prove that:
\beq
W_2^{(0)}(x(z_1),x(z_2)) = {1\over (z_1-z_2)^2\,x'(z_1)x'(z_2)}- {1\over (x(z_1)-x(z_2))^2}
\eeq
i.e.
\beq
W_2^{(0)}(x(z_1),x(z_2))dx(z_1)dx(z_2) + {dx(z_1)dx(z_2)\over (x(z_1)-x(z_2))^2} = B(z_1,z_2) = {dz_1 dz_2\over (z_1-z_2)^2}
\eeq
which is the Bergmann kernel on ${\mathbb P}^1$, i.e. the Bergmann kernel of the spectral curve  ${\cal E}=({\mathbb P}^1,x,y)$.

\subsubsection{Generating functions of discrete surfaces of higher topologies}

For arbitrary $n$ and $g$, the generating functions $W_n^{(g)}$ counting discrete surfaces of genus $g$ with $n$ boundaries, are obtained from the $\om_n^{(g)}$'s of the spectral curve ${\cal E}=({\mathbb P}^1,x,y)$, by:
\bea
\om_n^{(g)}(z_1,\dots,z_n)
&=& W_n^{(g)}(x(z_1),\dots,x(z_n))\,\, dx(z_1)\dots dx(z_n)  \cr
&& \cr
&&  +\delta_{n,2}\delta_{g,0}\,\,{dx(z_1)dx(z_2)\over (x(z_1)-x(z_2))^2}.
\eea

In particular, with $n=0$, the generating function for counting surfaces of genus $g$ and no boundary, is given by the symplectic invariants of ${\cal E}$:
\bea
 \sum_{v=1}^\infty t^v
\sum_{S\in {\mathbb M}_0^{(g)}(v)}\, \,\,{t_3^{n_3(S)}\dots t_d^{n_d(S)}\over \#{\rm Aut}(S)}
=F_g({\cal E}).
\eea

\subsection{Example quadrangulations}\label{secquadr}

If we count only quadrangulations, we choose $t_4\neq 0$, and all other $t_k=0$, i.e.:
\beq
V'(x) = x-t_4 x^3.
\eeq
The spectral curve is found from theorem \eq{thspcurvediscrsurf}.
We write:
\beq
x(z)=\alpha+\gamma\left(z+{1\over z}\right)
\eeq
and
\bea
&& V'(x(z)) \cr
&=& x(z) - t_4 x^3(z) \cr
&=& \alpha+\gamma(z+z^{-1}) -t_4\Big( \alpha^3+3\alpha^2\gamma(z+z^{-1}) \cr
&& +3\alpha\gamma^2(z^2+2+z^{-2})  + \gamma^3(z^3+3z+3z^{-1}+z^{-3}) \, \Big)\,. \cr
\eea
In other words:
\beq
2 u_0 = \alpha-t_4(\alpha^3+6\alpha\gamma^2)
\virg
u_1 = \gamma-3t_4(\alpha^2\gamma+\gamma^3)
\eeq
\beq
u_2 = -3t_4\alpha\gamma^2
\virg
u_3= -t_4\gamma^3.
\eeq
The condition $u_0=0$ implies:
\beq
0=\alpha(1-t_4(\alpha^2+6\gamma^2)) .
\eeq
Since we must choose a solution where $\alpha=O(t)$ and $\gamma^2=O(t)$ at small $t$, we must choose $\alpha=0$.
Then the condition $u_1=t/\gamma$ gives:
\beq
t= \gamma^2-3t_4\gamma^4
\eeq
i.e.
\beq
\gamma^2 = {1-\sqrt{1-12 t t_4}\over 6 t_4} .
\eeq
The spectral curve is then:
\beq
x(z) = \gamma\left(z+{1\over z}\right)
\virg
y(z) = -{t\over \gamma z} + {t_4\gamma^3\over z^3}.
\eeq

\subsubsection{Rooted planar quadrangulations}

The number of planar ($g=0$) quadrangulations with only 1 boundary of length $2l$ is $\TT^{(0)}_{2l} = {\displaystyle \Res_\infty} x(z)^{2l} \,\, y(z) dx(z)$ and thus:
\beq
\TT^{(0)}_{2l} = \gamma^{2l}\,\,{(2l)!\over l!\,(l+2)!}\,\,\, (2(l+1)t-l\gamma^2).
\eeq
In particular, if we require all faces, including the marked face of the boundary, to be quadrangles, we choose $2l=4$, and we find the generating function counting planar quadrangulations with one marked edge (rooted quadrangulations)\footnote{Notice that discrete surfaces with $n=1$ marked face and marked edge can have no non-trivial symmetry conserving the marked edge, and thus $\#{\rm Aut}=1$.}:
\bea
\TT^{(0)}_{4}
&=& \sum_v t^v\,\, \sum_{S\in {\mathbb M}_1^{(0)}(v), l(S)=4}\, t_4^{n_4(S)} \cr
&=& \gamma^{4}\,\, (3t-\gamma^2) \cr
&=& t^3\, \sum_n {2\,\,\, 3^n\, (2n)!\over n!\, (n+2)!}\,\, (t t_4)^{n-1}.
\eea
Thus we recover the famous result of Tutte \cite{tutte2} that the number of rooted planar quadrangulations with $n=n_4+1$ faces is:
\beq
{{2\,\,\, 3^n\, (2n)!\over n!\, (n+2)!}}.
\eeq

\subsubsection{Quadrangulations of the annulus}

The generating function counting quadrangulations of the annulus $g=0, n=2$ is given by the Bergmann kernel:
\bea
W_2^{(0)}(x_1,x_2)
&=& \sum_v t^v\,\, \sum_{S\in {\mathbb M}_2^{(0)}(v)}\, {t_4^{n_4(S)}\over x_1^{l_1(s)}\,x_2^{l_2(s)}\,\,\,\, \#{\rm Aut}(S)} \cr
&=& {1\over (z_1-z_2)^2\,\,x'(z_1)x'(z_2)} -{1\over (x_1-x_2)^2}
\eea
and thus, if we fix the perimeter lengths of the 2 boundaries as $2l_1$ and $2l_2$, we have:
\bea
\TT^{(0)}_{2l_1,2l_2}
&=& \sum_v t^v\,\, \sum_{S\in {\mathbb M}_2^{(0)}(v), l_1(S)=2l_1,l_2(S)=2l_2}\, {t_4^{n_4(S)}\over \#{\rm Aut}(S)} \cr
&=& \Res_{z_1\to\infty}\Res_{z_2\to\infty}\,\, {x(z_1)^{2l_1}\,\,x(z_2)^{2l_2}\over (z_1-z_2)^2}\,\, dz_1 dz_2 \cr
&=& \gamma^{2(l_1+l_2)}\,\,\sum_{j=1}^{{\rm min}(l_1,l_2)}\,2j\,\,{(2l_1)!\,(2l_2)!\over (l_1+j)!(l_1-j)!(l_2+j)!(l_2-j)!} .\cr
\eea
If we require the marked boundary faces to be quadrangles, we choose $2l_1=2l_2=4$ and thus:
\bea
\TT^{(0)}_{4,4}
&=& \sum_v t^v\,\, \sum_{S\in {\mathbb M}_2^{(0)}(v), l_1(S)=4,l_2(S)=4}\, {t_4^{n_4(S)}\over \#{\rm Aut}(S)} \cr
&=& 36\,\gamma^{8}\cr
&=& 36\, t^4\, \, \sum_{n\geq 2} {(2n+2)!\over n!\, (n+2)!}\,\, (3 t t_4)^{n-2}
\eea
i.e. the number of annulus quadrangulations with $n=n_4+2$ faces, where all faces including the 2 marked faces are quadrangles, is $4\,\times\, 3^n\, {(2n+2)!\over n!\, (n+2)!}$.

\subsubsection{Quadrangulations on a pair of pants}

The generating function for quadrangulations on the pair of pants $g=0,n=3$, is given by $\om_3^{(0)}$:
\beq
\om_3^{(0)}(z_1,z_2,z_3) =
 {1\over 2\gamma y'(1) }\,\,\Big({dz_1\over (z_1-1)^2}\,{dz_2\over (z_2-1)^2}\,{dz_3\over (z_3-1)^2}-{dz_1\over (z_1+1)^2}\,{dz_2\over (z_2+1)^2}\,{dz_3\over (z_3+1)^2}\Big)
\eeq
and for instance we find:
\beq
T^{(0)}_{4,4,4} = (12)^3 \,\,{\gamma^{12}\over 2t}\,\, \left(1+{1\over \sqrt{1-12  t t_4}}\right)
= t^5\,\sum_n 2^6\, 3^n\, {(2n+1)!\over (n+2)!\,(n-1)!}\,\,(t t_4)^{n-3}
\eeq
i.e. the number of quadrangulations on a pair of pants, where all $n$ faces, including the 3 marked faces, are quadrangles, is $2^6\, 3^n\, {(2n+1)!\over (n+2)!\,(n-1)!}$.

\subsubsection{Quadrangulations on a genus 1 disc}

The generating function for quadrangulations on the genus 1 disc $g=1,n=1$, is given by $\om_1^{(1)}$:
\beq
\om_{1}^{(1)}(z) = {-z+8z^3-z^5 + \gamma^2 (z-5z^3+z^5)\over 3(\gamma^2-2)^2(z^2-1)^4}
\eeq
and:
\beq
\TT_4^{(1)} = {\gamma^6\over (\gamma^2-2)^2} = {1\over 6t_4}\left( {1\over 1-12t_4} - {1\over \sqrt{1-12t_4}}\right)
\eeq
\beq
\TT_4^{(1)}
= 2\, \left( 1  + \pmatrix{-1/2\cr n} (-1)^{n-1} \right) (12 t_4)^{n-1}
= 2\, \left( 1  - {(2n-1)!!\over n!\, 2^n}  \right) (12 t_4)^{n-1}
\eeq
i.e. the number of rooted quandrangulations of genus 1 with $n$ faces is:
\beq
{1\over 6}\, \left( 1  - {(2n-1)!! \over n!\, 2^{n}}  \right) (12)^{n}
={3^{n}\over 6}\, \left( 2^{2n}  - {(2n)! \over n!\, n!}  \right).
\eeq

\subsection{Colored surfaces}

Exactly like the 1-matrix integral is related to the enumeration of discrete surfaces, the 2-matrix integral and the chain of matrices are also related to enumeration of discrete surfaces carrying colors (1 color per matrix).

\subsubsection{The Ising model on a discrete surface}

The Ising model is a problem of enumeration of bicolored discrete surfaces, and it is related to the 2-matrix model. It was introduced by Kazakov \cite{KazakovIsing}.

\medskip

Consider ${\mathbb M}_n^{(g)}=$ the set of connected orientable discrete surfaces of genus $g$ obtained by gluing together polygonal faces of two possible colors (or spin) $\pm$, namely $n_{3+}$ triangles of color $+$, $n_{4+}$ quadrangles of color $+$, ... $n_{k+}$ $k$-angles of color $+$,
 and also $n_{3-}$ triangles of color $-$, $n_{4-}$ quadrangles of color $-$, ... $n_{k-}$ $k$-angles of color $-$, as well as $n$ marked polygonal faces of color $+$ of perimeters $l_1,\dots,l_n$, each of the marked faces having one marked edge on its boundary.
Let us call $v$ the number of vertices of a discrete surface, and let ${\mathbb M}_n^{(g)}(v)$ be the set of discrete surfaces in ${\mathbb M}_n^{(g)}$, with $v$ vertices.
We call $n_{++}$ the number of edges with $+$ on both sides, $n_{--}$ the number of edges with $-$ on both sides, and  $n_{+-}$ the number of edges separating faces of different colors.

We require that unmarked faces have perimeter $\geq 3$, whereas marked faces are only required to have perimeter $l_i\geq 1$. Moreover, we recall that marked faces are required to have color $+$.

\bigskip
We define their generating functions as follows:

the generating function is the formal power series in $t$:
\bea
&& W_n^{(g)}(x_1,\dots,x_n;t_2,t_3,\dots,t_d;\td{t}_2,\td{t}_3,\dots,\td{t}_{\td{d}};t) \cr
&=& {t\over x_1}\delta_{n,1}\delta_{g,0}  \cr
&&+ \sum_{v=1}^\infty t^v
\sum_{S\in {\mathbb M}_n^{(g)}(v)}\, {1\over \#{\rm Aut}(S)} \cr
&& \qquad \,\,{t_3^{n_{3+}(S)}\dots t_d^{n_{d+}(S)}\,\,\td{t}_3^{n_{3-}(S)}\dots {\td{t}}_d^{n_{\td{d}-}(S)}\over x_1^{1+l_1(S)}\dots x_n^{1+l_n(S)}}\,\,{t_2^{n_{--}(S)}\,\td{t}_2^{n_{++}(S)}\over (t_2 \td{t}_2-1)^{n_{++}(S)+n_{++}(S)+n_{+-}(S)}} .\cr
\eea

\bigskip
For instance if we want to consider the Ising model on a random triangulation, we choose $t_k=0$ for $k\geq 4$, and $\td{t}_k=0$ for $k\geq 4$.

\bigskip
One may write Tutte-like recursion equations on the number of edges, which are identical to the loop equations of the 2-matrix model \cite{staudacher,eynm2m}, and for which we may use the results of section \ref{sec2MMformal}.
Like in the 1-matrix model, the spectral curve is found by the requirement that $W_1^{(0}(x)$ be a 1-cut solution of some algebraic equation, and have a good small $t$ expansion.
The recipe for finding the correct spectral curve is summarized in the following theorem:

\bt
We define the potentials:
\beq
V'_1(x) = t_2 x - \sum_{j=2}^{d-1}\, t_{j+1} x^j
\virg
V'_2(y) = \td{t}_2 y - \sum_{j=2}^{\td{d}-1}\, \td{t}_{j+1} y^j.
\eeq
For arbitrary coefficients $\gamma,\alpha_i,\beta_i$, we define the two rational functions:
\beq
x(z) = \gamma z + \sum_{i=0}^{\td{d}-1} \alpha_i z^{-i}
\virg
y(z) = \gamma z^{-1} + \sum_{i=0}^{d-1} \beta_i z^{i}.
\eeq
The coefficients $\gamma,\alpha_i,\beta_i$ are uniquely determined by the conditions:
\beq
V'_1(x(z))-y(z)\sim {t\over \gamma z} + O(z^{-2})
\virg
V'_2(y(z))-x(z)\sim {tz\over \gamma} + O(z^{2})
\eeq
and such that $\gamma^2$ as well as $\alpha_i$,$\beta_i$ are power series in $t$ which behave like $O(t)$ for $t\to 0$.

\medskip
Then, the spectral curve is:
\beq
{\cal E}_{\rm Ising} = ({\mathbb CP}^1,x,y),
\eeq
the function $W_1^{(0)}(x)$ is given by:
\beq
W_1^{(0)}(x(z)) = V'_1(x(z))-y(z),
\eeq
the function $W_2^{(0)}(x_1,x_2)$ is given by:
\beq
W_2^{(0)}(x(z_1),x(z_2))\, x'(z_1) x'(z_2) = {1\over (z_1-z_2)^2} - {x'(z_1)x'(z_2)\over (x(z_1)-x(z_2))^2}
\eeq
and all the stable $W_n^{(g)}$'s are given by the $\om_n^{(g)}$'s of the spectral curve ${\cal E}_{\rm Ising}$ by:
\beq
W_n^{(g)}(x(z_1),\dots,x(z_n))\,dx(z_1)\dots dx(z_n) = \om_n^{(g)}(z_1,\dots,z_n).
\eeq
In particular, the generating function for counting bicolored maps with no boundaries are the symplectic invariants $F_g({\cal E}_{\rm Ising})$.
\et

\medskip

{\bf Example: Ising model on quadrangulations.}

\medskip

We have
\beq
V'_1(x) = t_2 x - t_4 x^3
\virg
V'_2(y) = \td{t}_2 y - \td{t}_4 y^3.
\eeq
We find:
\beq
x(z) = \gamma z + \alpha_1 z^{-1} + \alpha_3 z^{-3}
\virg
y(z) = \gamma z^{-1} + \beta_1 z + \beta_3 z^{3}
\eeq
with the equation:
\beq
\beta_3 = -t_4 \gamma^3
\virg
\beta_1 = t_2 \gamma - 3 t_4 \gamma^2 \alpha_1
\virg
\alpha_3 = -\td{t}_4 \gamma^3,
\eeq
\beq
\alpha_1 = \td{t}_2 \gamma - 3 \td{t}_4 \gamma^2 \beta_1
\virg
\alpha_1 \beta_1 + 3 \alpha_3 \beta_3 = \gamma^2+t.
\eeq

That gives an algebraic equation for $\gamma^2$:
\beq
3 t_4 \td{t}_4 \gamma^4+{(t_2-3\td{t}_2t_4\gamma^2)(\td{t}_2-3t_2\td{t}_4\gamma^2)\over (1-9\td{t}_4t_4\gamma^4)^2} = 1+{t\over \gamma^2}
\eeq
and we choose the unique solution such that:
\beq
\gamma^2 \sim {t\over t_2\td{t}_2-1} + O(t^2).
\eeq

\subsubsection{The chain of matrices discrete surfaces}

A chain of matrices (see section \ref{secchain}), with $m$ matrices $M_1,\dots, M_m$ can also be interpreted as a generating function for enumerating discrete surfaces with $m$ possible colors.
The "colors" are labeled $1,\dots,m$.

\medskip

We are going to consider discrete surfaces, whose unmarked faces can have any color, and are at least triangles, and marked faces have color $1$.

\medskip

Consider ${\mathbb M}_n^{(g)}$ be the set of connected orientable discrete surfaces of genus $g$ obtained by gluing together polygonal faces of $m$ possible colors $k=1,\dots,m$ (let $n_{j,k}$ be the number of faces of size $j$ and color $k$, we assume $j\geq 3$),
and $n$ marked faces of color $1$, with a marked edge, and of size $l_i\geq 1$, $i=1,\dots,n$.

Let $n_{<i,j>}$ be the number of edges such that the two sides are faces of color $i$ and $j$.

Let $C$ be the following Toeplitz matrix of color couplings:
\beq
C^{-1} =
\left(\begin{array}{llllr}
t_{2,1} & -1         &  0   & \dots & 0 \cr
-1         & t_{2,2} & -1  &            &     \cr
   0           & \ddots & \ddots & \ddots & 0 \cr
& & -1 & t_{2,m-1} & -1 \cr
0 & & 0& -1 & t_{2,m}
\end{array}\right).
\eeq

\bigskip
We define their generating functions as follows.
The generating function is the formal power series in $t$:
\bea
&& W_n^{(g)}(x_1,\dots,x_n;t_{i,j};t) \cr
&=& {t\over x_1}\delta_{n,1}\delta_{g,0}  \cr
&&+ \sum_{v=1}^\infty t^v
\sum_{S\in {\mathbb M}_n^{(g)}(v)}\,
 \,\,{\prod_{i\geq 3}\prod_{j=1}^m t_{i,j}^{n_{i,j}(S)} \over x_1^{1+l_1(S)}\dots x_n^{1+l_n(S)}}\,\,\,\,{\prod_{i,j=1}^m\,\, C_{i,j}^{n_{<i,j>}(S)}\over \#{\rm Aut}(S)} . \cr
\eea

\bigskip
Again, one may write Tutte-like recursion equations on the number of edges, which are identical to the loop equations of the chain of matrices model \cite{eynmultimat}, and we may apply the results of section \ref{secchain}.
Like in the 1-matrix model, the spectral curve is found by the requirement that $W_1^{(0}(x)$ be a 1-cut solution of some algebraic equation, and have a good small $t$ expansion.
The recipe for finding the correct spectral curve is summarized in the following theorem:

\bt
We define the potentials $V_1,\dots, V_m$ by:
\beq
V'_k(x) = t_{2,k} x - \sum_{j=2}^{d_k-1}\, t_{j+1,k} x^j
,\qquad \quad k=1,\dots,m.
\eeq
For arbitrary coefficients $\alpha_{i,k}$, we define the rational functions:
\beq
x_k(z) = \sum_{j=-s_k}^{r_k} \alpha_{j,k} z^j
\eeq
with:
\beq
r_1=1,\qquad r_{k+1} = r_k .(d_k-1),
\eeq
\beq
s_m=1,\qquad s_{k-1} = s_k .(d_k-1).
\eeq
The coefficients $\gamma,\alpha_{i,k}$ are uniquely determined by the conditions:
\beq
\alpha_{1,1}=\alpha_{-1,m}=\gamma,
\eeq
\beq\label{eqV'xxchmatsurf}
\forall\, k=2,\dots,m-1\, , \qquad \quad
V'_k(x_k(z)) = x_{k-1}(z)+x_{k+1}(z),
\eeq
\beq
\forall\, k=1,\dots,m-1\, , \qquad \quad
\sum_{j} j\, \alpha_{j,k+1}\alpha_{-j,k} = t
\eeq
and such that $\gamma^2$ as well as $\alpha_{i,k}$ are power series in $t$ which behave like $O(t)$ for small $t$.

\medskip
Then, the spectral curve is:
\beq
{\cal E}_{\rm ch.mat} = ({\mathbb CP}^1,x_1,x_2),
\eeq
the function $W_1^{(0)}(x)$ is given by:
\beq
W_1^{(0)}(x_1(z)) = V'_1(x_1(z))-x_2(z),
\eeq
 the function $W_2^{(0)}(x_1(z_1),x_1(z_2))$ is given by:
\beq
W_2^{(0)}(x_1(z_1),x_1(z_2))\, x'_1(z_1) x'_2(z_2) = {1\over (z_1-z_2)^2} - {x'_1(z_1)x'_1(z_2)\over (x_1(z_1)-x_1(z_2))^2}
\eeq
and all the stable $W_n^{(g)}$'s are given by the $\om_n^{(g)}$'s of the spectral curve ${\cal E}_{\rm ch.mat}$ by:
\beq
W_n^{(g)}(x_1(z_1),\dots,x_1(z_n))\,dx_1(z_1)\dots dx_1(z_n) = \om_n^{(g)}(z_1,\dots,z_n),
\eeq
i.e. $F_g$ is independent of $k$.

\et

\br
Because of \eq{eqV'xxchmatsurf}, we see that the spectral curves
$({\mathbb CP}^1,x_k,x_{k+1})$ are all symplectically equivalent for any $k=1,\dots,m-1$, and thus, we have:
\beq
\forall\, k=1,\dots,m-1\, , \qquad \quad
F_g = F_g(({\mathbb CP}^1,x_k,x_{k+1}))  = F_g(({\mathbb CP}^1,x_{k+1},x_k)).
\eeq
\er

%-------------------------- fin surfaces discretisees ---------------------------------------------------------

\section{Double scaling limits and large maps}\label{secdsl}

\subsection{Minimal models and continuous surfaces}

In the preceding section, we explained how the symplectic invariants can be used to count discrete surfaces.
The theorem \ref{thsinglimFg}, allows to find various limits of symplectic invariants, and here, it can be used to find the asymptotics of generating functions of large discrete surfaces \cite{KazakovRMTcrit}.

The conjecture \cite{Witten} was that large discrete surfaces tend towards continuous surfaces weighted with the Liouville theory action, possibly coupled to some conformal matter fields.

The idea is to count discrete surfaces made of large numbers of polygons, and send the size of polygons to 0, so that the total area remains finite \cite{Brezindsl,Douglas,Migdal}.

In section \ref{secmaps} we found the generating function for counting maps of genus $g$ as the symplectic invariants $F_g({\cal E})$ of some rational spectral curve ${\cal E}$.
In this enumeration of maps, the expectation value of the
number of $k$-gons in the considered maps is given by
\beq
\left<n_k\right>= t_k {\partial \, \ln{F_g} \over \partial t_k}
\eeq
and the expectation value of the number of vertices is:
\beq
\left<v\right>= t {\partial \, \ln{F_g} \over \partial t}.
\eeq
Large discrete surfaces are obtained when those numbers diverge, i.e. when the parameters $t_k$ (or $t$) approach a singularity, for which the spectral curve is singular (in the sense of \eq{defregular}).
%More precisely, one should approach this singularity proportionally to a power of the typical size of the polygons $\epsilon$:
%\beq
%t = t_c + \l \epsilon^{2 \over p}
%\eeq
%for some critical exponent $p$.

Let us now study the blow up of the matrix models' spectral curves around these singularities.

\subsection{Minimal model $(p,q)$ and KP hierarchy}

For discrete surfaces (formal 1-matrix model), or bicolored discrete surfaces with an Ising model (2-matrix model), or for the formal chain of matrices, the spectral curve depends on the parameter $t$ in an algebraic way. We choose potentials $V_k$ as well as a critical $t=t_c$, such that at $t=t_c$ the spectral curve has a cusp singularity of the type $y(z)\sim (x(z)-x(a))^{p/q}$ (see \cite{DKK} for a list of critical potentials of minimal degrees, for the 2-matrix model).

\medskip

We expand ${\cal E}$ in the vicinity of the critical branchpoint $z \to a$ and $t\to t_c$:
\beq
z=a+(t-t_c)^\nu \, \zeta
\eeq
\beq\label{eqblowupPQdsl}
{\cal E} \sim \left\{
\begin{array}{l}
x(z) = x(a) + (t-t_c)^{q\nu}\,Q(\zeta) + o((t-t_c)^{q\nu}) \cr
y(z) = y(a) + (t-t_c)^{p\nu}\,\,P(\zeta) + o((t-t_c)^{p\nu}) \cr
%\nu={1\over p+q-1}
\end{array}
\right.
\eeq
where $Q$ and $P$ are polynomials of the complex variable $\zeta$, of respective degrees $q$ and $p$, and where the exponent $\nu$ is a scaling exponent such that the blown up spectral curve is regular.

We define the double scaling limit \cite{KazakovRMTcrit} spectral curve as the blow-up of the singularity:
\beq\label{spcurvepq}
{\cal E}_{(p,q)} = \left\{
\begin{array}{l}
x(\zeta) = Q(\zeta) \cr
y(\zeta) = P(\zeta)
\end{array}
\right. .
\eeq

In order to find the exponent $\nu$, one may notice that for the chain of matrices (and thus also 1-matrix and 2-matrix model), the derivative with respect to $t$ of the form $y dx$ is:
\beq
{d\over d t} (y dx(z)) = {dz\over z} \sim  (t-t_c)^\nu \, {1\over a} \, d\zeta .
\eeq
Comparing this with \eq{eqblowupPQdsl}, it is easy to see that this implies:
\beq
\nu  = {1\over p+q-1}
\eeq
and:
\beq
p P(\zeta) Q'(\zeta) - q Q(\zeta) P'(\zeta) = {1\over a} .
\eeq

Theorem \ref{thsinglimFg} of section \ref{secsingFg}, imply that the symplectic invariants defined in eq.\ref{defFginv} give the double scaling limit:
\beq
F_{g}({\cal E}) \sim (t-t_c)^{(2-2g){p+q\over p+q-1}}\, F_{g}({\cal E}_{(p,q)}) \,\,\, (1+ o(1)).
\eeq

\smallskip
The spectral curve \eq{spcurvepq}, is the spectral curve of the $(p,q)$ minimal model in conformal field theory \cite{BookPDF}, coupled to gravity \cite{KPZ}. It corresponds to a finite dimensional irreducible representation of the group of conformal transformations, it has a central charge:
\beq
c=1-6\,{(p-q)^2\over pq}.
\eeq
The exponent $\nu={1\over p+q-1}$ or more precisely the exponent "$\gamma-$string":
\beq
\gamma=2-2(p+q)\nu
\eeq
is given by the famous KPZ formula \cite{BookPDF, KPZ}.
Notice that the symplectic invariance of the $F_g$'s under $x\leftrightarrow y$, is related to the $(p,q)\leftrightarrow (q,p)$ duality \cite{KharMar}.

\medskip

The corresponding tau function is:
\beq
\tau(N) = \exp{\left({\displaystyle \sum_{g=0}^\infty} N^{2-2g}\, F_g\left({\cal{E}}_{(p,q)}\right)\right)} .
\eeq
It is a tau function of the $(p,q)$ reduction of the integrable hierarchy of Kadamtsev-Petviashvili (KP) \cite{BBT}.
The function
\beq
F(t) = \sum_{g=0}^\infty t^{(2-2g){p+q\over p+q-1}} F_g\left({\cal{E}}_{(p,q)}\right) = \ln{(\tau(t^{p+q-1\over p+q}))}
\eeq
and its second derivative
\beq
u(t)=F''(t)
\eeq
can be found as follows:
find two differential operators  $\hat{P}=d^p + p u d^{p-2} + \dots$ and $\hat{Q}=d^q+u d^{q-2} + \dots$ (where $d=d/dt$) satisfying the string equation:
\beq
[\hat{P},\hat{Q}]=1.
\eeq
This equation implies the differential equation of KP for the function $u(t)$.

For example, for pure gravity $(p,q)=(3,2)$ (with central charge $c=0$), we have:
\beq
\hat Q=d^2-2u
\virg
\hat P=d^3-3 ud - {3\over 2} u'
\eeq
and $[\hat P,\hat Q]=1$ imply the Painlev\'e equation for $u=F''$:
\beq
  3u^2 -{1\over 2} u'' =  t .
\eeq

\subsection{Minimal model $(p,2)$ and KdV}

As a special case, we consider the 1-matrix model.
In that case, the spectral curve is always hyperelliptical $y^2={\rm Pol}(x)$, and thus the only cusp singularities must be half integers $y\sim x^{p/2}$, i.e. $q=2$ and $p=2k+1$.

\smallskip

The operators $\hat{Q}$ and $\hat{P}$ are such that:
\beq
\hat{Q}=d^2-2 u(t),
\eeq
\beq
\hat{P} = d^{2k+1} - (2k+1) u(t) d^{2k-1} + \sum_{j=0}^{2k-2} v_{j}(t) d^j .
\eeq
The string equation $[\hat{P},\hat{Q}]=1$ implies a differential equation for $u(t)$:
\beq
R_{k+1}(u) = t
\eeq
where $R_k$ is the $k^{\rm th}$ Gelfand-Dikii differential polynomial (see \cite{BookPDF, ZJDFG, BBT}).
They obey the recursion:
\beq
R_0=2 \virg R_{j+1}' = -2 u \, R_j' - u'\,R_j + {1\over 4}\,R_j'''.
\eeq
The first few of them are:
\bea
R_0&=&2\cr
R_1&=&-2u\cr
R_2&=&3u^3-{1\over 2}u''\cr
R_3&=&-5u^3+{5\over 2}u u'' - {1\over 4} u'^2 - {1\over 8}u'''\cr
\vdots\cr
\eea

We have seen in the previous section, that the spectral curve \eq{spcurvepq} of the $(2k+1,2)$ scaling limit of the 1-matrix model, is:
\beq\label{specKdV}
{\cal E}_{(2k+1,2)} = \left\{
\begin{array}{l}
\td{x}(z) = z^2- 2u_0 \cr
\td{y}(z) = {\displaystyle \sum_{j=0}^k} \ovl{t}_k\, u_0^{k-j}\, z^{2j+1}
\end{array}
\right. .
\eeq
It is the classical limit of $(\hat P,\hat Q)$, where $d=d/dt$ becomes a complex number $z$:
\bea
\hat Q = d^2-2u \qquad &\to& \qquad  x=z^2-2u_0  \cr
\hat P = d^{2k+1}-(2k+1)ud^{2k-1} + \dots \qquad &\to& \qquad  y=z^{2k+1}-(2k+1)u_0 z^{2k-1} +\dots
 \cr
\eea

\smallskip
The symplectic invariants defined in eq.\ref{defFginv} give the double scaling limit:
\beq
F_{g}({\cal E}_{\rm 1MM}) \sim (t-t_c)^{(2-2g)(2k+3)\over(2k+2)}\, F_{g}({\cal E}_{(2k+1,2)}) + o((t-t_c)^{(2-2g)(2k+3)\over(2k+2)})
\eeq
and, if $z_1,\dots, z_n$ all lie in the vicinity of $1+ O((t-t_c)^{1\over 2k+2})$, we have:
\beq
\om_n^{(g)}({\cal E}_{\rm 1MM})(z_1,\dots,z_n) \sim (t-t_c)^{(2-2g-n)(2k+3)\over(2k+2)}\, \om_n^{(g)}({\cal E}_{(p,2)})(\zeta_1,\dots,\zeta_n) + o((t-t_c)^{(2-2g-n)(2k+3)\over(2k+2)}) .
\eeq
where $z_i=1+(t-t_c)^{1\over 2k+2} \zeta_i$

The formal function:
\beq
F(\xi) = \sum_g \xi^{(2-2g)(2k+3)\over(2k+2)}\, F_{g}({\cal E}_{(2k+1,2)})
\eeq
is such that its second derivative
\beq
u(\xi)=F''(\xi)
\eeq
satisfies the $k+1^{\rm st}$ Gelfand-Dikii equation:
\beq
R_{k+1}(u)=\xi.
\eeq

%%%%%%%%%%%%%%%%%%%%%%%%%%%%%%%%%%%%%%%%%%%%%%%%%%%%%%%%%%%%%%%%%%%%%%%%%%%%%%%%%%%%%%%%%%%%%%%%%%%%%%%%%%%%

\section{Partitions and Plancherel measure}

In many different problems of mathematics or physics, one needs to count partitions with the Plancherel Weight.

$$
{\mbox{\epsfxsize=8.truecm\epsfbox{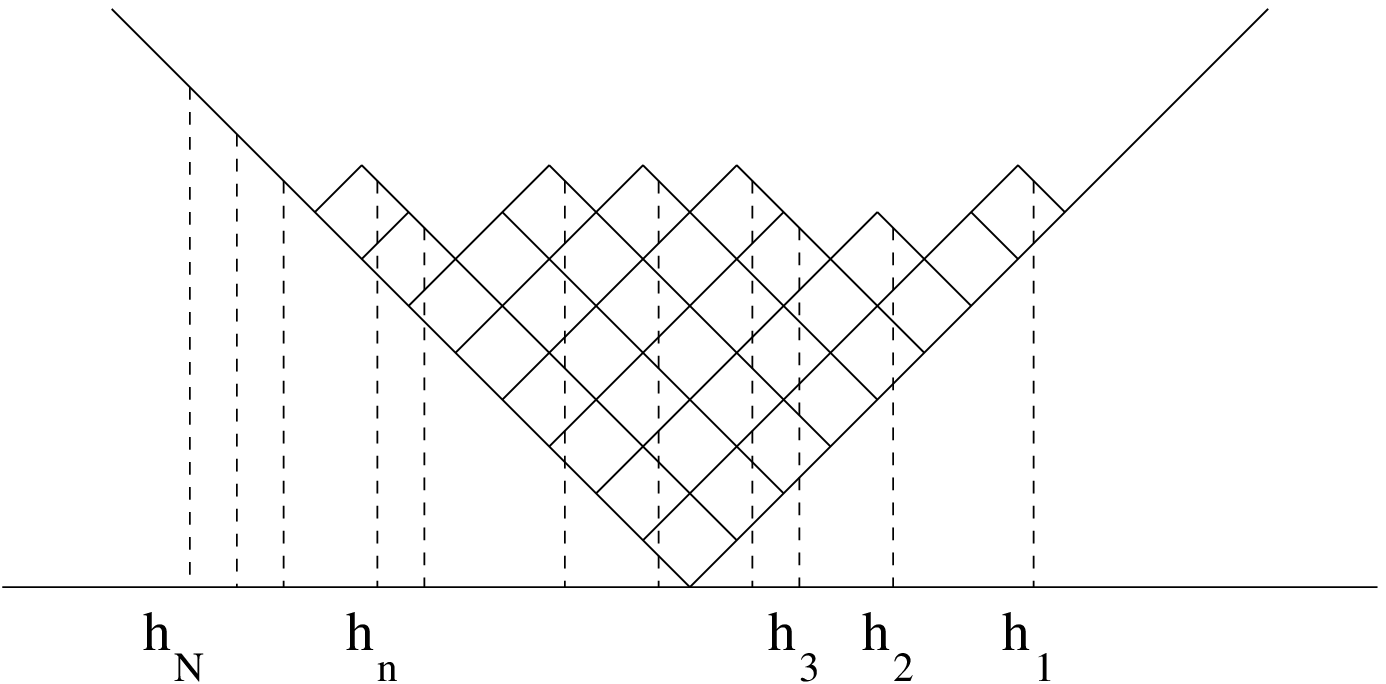}}}
$$

Given a partition $\l_1\geq\l_2\dots \geq\l_N\geq 0$, we define:

$\bullet$ its weight $|\l|=\l_1+\l_2+\dots+\l_N$.

$\bullet$ its length $n(\l)=\#\{ i,\, \l_i\neq 0\}$, we have $n(\l)\leq N$.

$\bullet$ its Plancherel weight:
\beq
{\cal P}(\l) = \left({\rm dim}(\l)\over |\l|! \right)^2 = {\prod_{i>j} (h_i-h_j)^2\over \prod_i (h_i !)^2}
\eeq
where the $h_i$'s correspond to the down-right edges of the partition rotated by $\pi/4$:
\beq
h_i = \l_i-i+N
\virg
h_1>h_2>\dots>h_N\geq 0 .
\eeq

$\bullet$ The Casimirs:
\beq
C_k(\l) = \sum_{i=1}^N \left(h_i-{N-1\over 2}\right)^k - \sum_{i=1}^N \left({1\over 2}-i\right)^k + (1-2^{-k}) \, \zeta(-k) .
\eeq
For example
\beq
C_1(\l) = |\l|-{1\over 24}
\virg
C_2(\l) = \sum_{i=1}^N \l_i(\l_i-2i+1) .
\eeq

\subsection{The partition function}

The partition function one would like to compute is:
\beq
Z_N(Q) = \sum_{n(\l)\leq N}\, \left({\rm dim}(\l)\over |\l|! \right)^2\,\, Q^{2|\l|}
\eeq
and particularly its large $Q$ expansion:
\beq
\ln{Z_N(Q)} = \sum_g Q^{2-2g}\, F_g .
\eeq
Indeed, since $Q$ is coupled to the weight $|\l|$ of partitions, the large $Q$ limit corresponds to the limit of large partitions.

More generally, one is also interested in expectation values of moments of Casimirs, and one whishes to compute the partition function:
\beq\label{ZPlancherel}
Z_N(Q,t_k) = \sum_{n(\l)\leq N}\, \left({\rm dim}(\l)\over |\l|! \right)^2\,\, Q^{2|\l|}\,\,\, \ee{{\displaystyle \sum_{k=1}^d} {t_k\,\, Q^{1-k} \over k} C_k(\l)}
\eeq
where we have taken into account the scaling of the Casimirs in the large $Q$ limit.
Again, we are interested in the large size expansion:
\beq
\ln{Z_N(Q,t_k)} \sim \sum_g Q^{2-2g}\, F_g(t_k,N) .
\eeq

\bigskip

In the case where $t_k=0$ for all $k$, the answer has been known for a long time, and we have the following identity:
\beq
\sum_{\l}\, \left({\rm dim}(\l)\over |\l|! \right)^2\,\, Q^{2|\l|} = \ee{Q^2}
\eeq
and therefore:
\beq
F_g =\delta_{g,0} .
\eeq
When some of the $t_k$'s are turned on, the answer can be written in terms of a matrix model and symplectic invariants of an appropriate spectral curve.

\bigskip

Some applications of this model include the statistics of the longest increasing subsequence of a random sequence, which is equivalent to the statistical physics of growing 2D crystals, indeed the Plancherel measure ${\cal P}(\l)$ is precisely the probability to obtain shape $\l$ by dropping square boxes from the sky at random times and random positions, see \cite{Okounkov2,PS}.

Another application concerns algebraic geometry and topological string theory. The partition function $Z_N(Q,t_k)$ is also the generating function for counting ramified branched coverings of ${\mathbb P}^1$, i.e. the Hurwitz numbers, see \cite{MMtor,Nekrasov,NO,Okounkov,Okounkov2}.

\medskip

It has been observed in many works \cite{BaikRains,Johansson,PS}, that locally, in the large $Q$ limit, the Plancherel statistics of partitions, shows many similarities with universal statistical laws observed in various limits of matrix models.
In fact, it was found in \cite{partitions}, that the similarity is much stronger, and in fact $Z_N(Q,t_k)$ {\bf is a matrix model} for all $Q$ (not only large $Q$).
In particular this shows that the $F_g$'s are again symplectic invariants.

\subsection{Matrix model for counting partitions}

Consider the contour ${\cal C}$ which surrounds all positive integer points.
$$
{\mbox{\epsfxsize=8.truecm\epsfbox{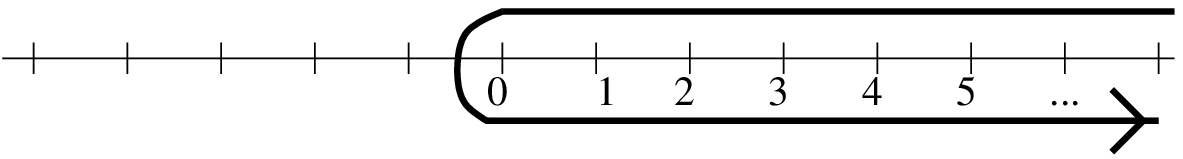}}}
$$
The following matrix integral\footnote{$H_N({\cal C})$ stands for the set of normal matrices whose eigenvalues lie on a contour ${\cal C}$.}:
\beq
Z = \int_{H_N({\cal C})} \, dM\,\, \ee{-Q \Tr V(M)}
\eeq
can be written in eigenvalues:
\beq
Z = {1\over N!}\,\int_{{\cal C}^N} dx_1\dots dx_N\,\, \prod_{i>j} (x_i-x_j)^2\,\,\, \prod_i \ee{-Q V(x_i)}
\eeq
where we choose:
\beq\label{defVPlancherel}
\ee{-Q V(x)} = {\ee{i\pi Q x}\over \sin{(\pi Q x)}}\,\,\, {Q^{2Q x}\over \Gamma(Qx+1)^2}\,\, \ee{- Q {\displaystyle \sum_k} {t_k\over k}\, (x-(N-{1\over 2})/Q)^k} .
\eeq
The integration over ${\cal C}$ picks up residues at the poles of $\ee{-QV(x)}$, i.e. at the poles of $1\over \sin{\pi  Q x}$, and thus forces $Q x_i\in {\mathbb N}$, and the factor $\prod_{i>j} (x_i-x_j)^2$ forces $x_i\neq x_j$, thus we have $Q x_i=h_i\in {\mathbb N}$, and since they are dum variables, we can always reorder them so that $h_1>h_2>\dots >h_N\geq 0$.
Therefore we have:
\bea
Z
&=& \int_{H_N({\cal C})} \, dM\,\, \ee{-Q \Tr V(M)}  \cr
&=& \sum_{h_1>\dots> h_N\geq 0}\,\, {{\displaystyle \prod_{i>j}}(h_i-h_j)^2\over {\displaystyle \prod_i} (h_i!)^2}\,\, Q^{2{\displaystyle \sum_i} h_i}\,\, \ee{{1\over Q}\, {\displaystyle \sum_k \sum_i} {t_k\over k} Q^{-k}\,(h_i-N+{1\over 2})^k} \cr
&\propto& Z_N(Q,t_k),
\eea
i.e. we recover the Plancherel generating function $Z_N(Q,t_k)$ for partitions \eq{ZPlancherel}.
Therefore, $Z_N(Q,t_k)$ can be written as a normal matrix integral with eigenvalues supported on ${\cal C}$.
Since the loop equations are independent of the integration contour (provided there is no boundary term  when integrating by parts), we have the same loop equations as for any other matrix model, and thus the same solution in terms of symplectic invariants.

\medskip

The spectral curve, i.e. the equilibrium density of eigenvalues of the matrix, is computed like for any other matrix model, like in section \ref{secMM}.
Here, the potential $V(M)$ may look quite complicated, and the corresponding spectral curve also looks at first sight quite complicated.
However, due to the special properties of the $\Gamma$ functions, it simplifies considerably, and reduces to a rather simple expression, which coincides with a "naive" large $Q$ limit.
The naive large $Q$ limit has been computed in many works \cite{MarNekr,Okounkov,Okounkov2,VershikKerov}, and has been known for some time.
It has the property that it is nearly independent of $N$.
Let us just state the final result, and refer the reader to \cite{partitions} for details.

The spectral curve is obtained from the following recipe:

\bt
Define:
\beq
U(x) = \sum_{k=1}^{d-1} t_{k+1} x^k .
\eeq
Then define the coefficients $u_0,u_1,\dots, u_{d-1}$ by the equation:
\beq\label{spcurvePlancherel1}
U( \ee{-u_0}(z+1/z-\alpha)) = \sum_{k=0}^{d-1} u_k (z^k+z^{-k}) .
\eeq

Then the spectral curve is:
\beq
\encadremath{
{\cal E}_{\rm Plancherel}=
\left\{
\begin{array}{l}
x(z) = {N-{1\over 2}\over Q} +  \ee{-u_0}( z+1/z-u_1) \cr
y(z) = \ln{z} + {1\over 2} {\displaystyle \sum_{k=1}^{d-1}}\, u_k (z^k-z^{-k})
\end{array}
\right.
}\eeq
and the large $Q$ expansion of $Z_N(Q,t_k)$ is given by:
\beq\label{PlanchZNFg}
\ln{Z_N(Q,t_k)} \sim \sum_{g=0}^\infty Q^{2-2g}\, F_g({\cal E}_{\rm Plancherel})
\eeq
where $F_g$ are the symplectic invariants defined in section \ref{secsymplectic}.

\et

\br
Since $x\to x-{N-{1\over 2}\over Q}$ is a symplectic transformation, $F_g$ is independent of $N$, and therefore \eq{PlanchZNFg} seems to be independent of $N$. In fact, the $N$ dependence is smaller than any power of $Q$, it is exponentially small.
This phenomenon is known as the arctic circle property.
All partitions with size $n(\l)>Q$ seem to be "frozen" and contribute only exponentially to the partition function.
\er

\subsubsection{Example no Casimir}

When all $t_k=0$ we have $U=0$, i.e. $u_0=0$ and $u_1=0$.
The spectral curve is thus:
\beq
\left\{
\begin{array}{l}
x(z) =  {N-{1\over 2}\over Q} + z+1/z \cr
y(z) = \ln{z} \qquad = {\rm Arcosh}((x-{N-{1\over 2}\over Q})/2)
\end{array}
\right. .
\eeq
For that spectral curve one finds:
\beq
F_0=1, F_1=F_2=F_3=\dots=0
\eeq
which is in agreement with
\beq
Z(Q)=\ee{Q^2} .
\eeq

\subsubsection{Example: Plancherel measure with the 2nd Casimir}

If we choose $t_2\neq 0$ and all other $t_k=0$, we have from \eq{spcurvePlancherel1}:
\beq
t_2 \ee{-u_0}\,\left(z+{1\over z}-u_1\right) = 2 u_0 + u_1\left(z+{1\over z}\right)
\eeq
and thus:
\beq
t_2^2 = -2 u_0 \ee{2u_0}
\virg
u_1 = \sqrt{- 2 u_0}.
\eeq
That gives:
\beq
2 u_0 = L(-t_2^2)
\eeq
where $L(x)$ is the Lambert function, solution of $L\ee{L}=x$.

The spectral curve is thus:
\beq
\left\{
\begin{array}{l}
x(z) =  {N-{1\over 2}\over Q} + \ee{-u_0}(z+1/z-u_1) \cr
y(z) = \ln{z} + {1\over 2}u_1(z-{1\over z})
\end{array}
\right. .
\eeq
For that spectral curve one finds:
\beq
F_0= {\ee{-2u_0}\over 2}\, (1+u_0)(2-u_0),
\eeq
\beq
F_1= {1\over 24} \ln{(\ee{-2u_0}(1+2u_0))},
\eeq
\beq
F_2={\ee{2u_0}\over 180}\, {u_0^3(1-12 u_0)\over (1+2u_0)^5},
\eeq
and so on...

\subsection{$q$-deformed partitions}\label{secqdeformpart}

The $q$-deformed Plancherel weight is obtained by replacing integer numbers $h_i\in \mathbb N$ by the $q-$numbers $[h_i]={q^{h_i\over 2}-q^{-h_i\over 2}\over q^{1\over 2}-q^{-{1\over 2}}}$, and thus:
\beq
{\cal P}_q(\l) = \left({\rm dim}_q(\l)\over [|\l|]! \right)^2
= \left(\prod_{i>j} [h_i-h_j]\over \prod_{i=1}^N\prod_{j=1}^{h_i} [j]\right)^2
= {\prod_{i>j} (q^{h_i-h_j\over 2} - q^{h_j-h_i\over 2})^2\over \prod_{i=1}^N \prod_{j=1}^{h_i} (q^{j\over 2}-q^{-j\over 2})^2}
\eeq
which can also be written:
\beq
{\cal P}_q(\l)
= \prod_{i>j} (q^{h_i}-q^{h_j})^2\,\,
\prod_{i=1}^N  q^{(1-N)h_i}\,\, q^{h_i(h_i+1)\over 2}\,\, \left(g(q^{-h_i})\over g(1)\right)^2
\eeq
where $g(x)$ is the $q$-product:
\beq
g(x) = \prod_{n=1}^\infty \left(1-{1\over x}\, q^n\right).
\eeq

Again, our goal, for various applications in physics and mathematics, is to compute the following sum:
\beq\label{defZqPl}
Z_N(q;t_k) = \sum_{n(\l)\leq N}\,{\cal P}_q(\l) \,\,\, \ee{{1\over \ln{q}}\,{\displaystyle \sum_k }{t_k\over k} \, (\ln{q})^k\, C_k(\l)}
\eeq
and one would like to compute it in the $q\to 0$ limit, i.e.
\beq
\ln{Z_N} \sim \sum_{g=0}^\infty (\ln{q})^{2-2g}\, F_g .
\eeq
Again, we will find that the $F_g$'s are the symplectic invariants of some spectral curve.

\medskip

The main application concerns algebraic geometry and topological strings. $Z_N(q;t_k)$ is the partition function for the Gromov-Witten invariants of some family of Calabi-Yau 3-fold, see \cite{MMtor}.

\subsubsection{Matrix model}

Again, the idea is to represent the sum \eq{defZqPl} as a matrix integral.

Consider the contour ${\cal C}$ which surrounds all points of the form $1,q,q^2,q^3,\dots$, i.e. ${\cal C}$ is a circle of radius $1<r<|q^{-1}|$:
$$
{\mbox{\epsfxsize=5.truecm\epsfbox{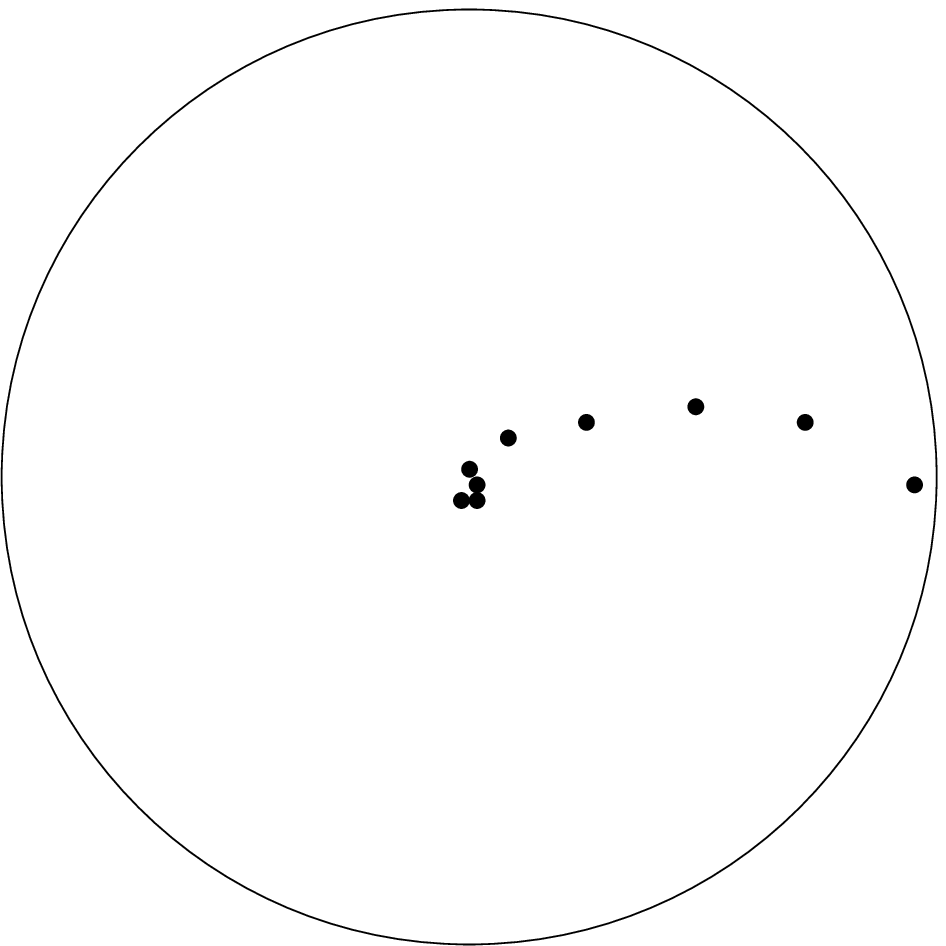}}}
$$
The following matrix integral:
\beq
Z = \int_{H_N({\cal C})} \, dM\,\, \ee{{1\over \ln{q}} \Tr V(M)}
\eeq
can be written in eigenvalues:
\beq
Z = {1\over N!}\,\int_{{\cal C}^N} dx_1\dots dx_N\,\, \prod_{i>j} (x_i-x_j)^2\,\,\, \prod_i \ee{{1\over \ln{q}} V(x_i)}
\eeq
where we choose:
\beq\label{defVPlancherelq}
\ee{{1\over \ln{q}} V(x)}
= \ee{{1\over \ln{q}} \sum_k {t_k\over k}\,(\ln{x}-(N-{1\over 2})\ln{q})^k}\,\, \ee{(1-N)\ln{x}}\, {\sqrt{x}\, g(1/x)^2\over g(1)^2}
\,\, \ee{{(\ln{x})^2\over 2\ln{q}}}\,\, f(x)
\eeq
with
\beq
f(x) = -\, {\ee{i\pi {\ln{x}\over \ln{q}}}\, g(1)^2\,\, \ee{{(\ln{x})^2\over 2\ln{q}} }\over \sqrt{x} (1-x)\,g(x)\,g(1/x)} .
\eeq

The integration over ${\cal C}$ picks up residues at the poles of $\ee{{1\over \ln{q}}V(x)}$, i.e. at the poles of $f(x)$, and thus forces ${\ln{x_i}\over \ln{q}}\in {\mathbb N}$, and the factor ${\displaystyle \prod_{i>j}} (x_i-x_j)^2$ forces $x_i\neq x_j$, thus we have $x_i=q^{h_i}$ where $h_i\in {\mathbb N}$, and since they are dum variables, we can always reorder them so that $h_1>h_2>\dots >h_N\geq 0$.
The residues are such that we have:
\beq
 Z_N(q;t_k)
\propto \int_{H_N({\cal C})} \, dM\,\, \ee{{1\over \ln{q}} \Tr V(M)}.
\eeq
Therefore, as in the preceding case, $Z_N(q,t_k)$ can be written as a normal matrix integral with eigenvalues supported on ${\cal C}$.
Once again, the small $q$ expansion of the free energy is thus given by the corresponding symplectic invariants.

\subsubsection{Spectral curve}
It remains to compute the spectral curve, i.e. the equilibrium density of eigenvalues for the matrix potential:
\bea
V(x)
&=& \sum_k {t_k\over k}\,(\ln{x}-(N-{1\over 2})\ln{q})^k + (\ln{x})^2 + i\pi \ln{x} \cr
&& +(1-N)\ln{q}\,\ln{x} + \ln{q}\,\ln{g(1/x)\over g(x)} - \ln{q}\ln{(1-x)} .
\eea

Again, the spectral curve, i.e. the equilibrium density of eigenvalues of the matrix, is computed like for any other matrix model, like in section \ref{secMM}.
Here, the potential $V(M)$ may look quite complicated, and the corresponding spectral curve also looks at first sight quite complicated.
However, due to the special properties of the $q$-product $g(x)$, it simplifies considerably, and reduces to a rather simple expression, which coincides with a "naive" large $\ln{q}$ limit.
The naive large $\ln{q}$ limit has been computed in many works \cite{MMtor}, and has been known for some time.
It has the property that it is nearly independent of $N$.
Let us just state the final result, and refer the reader to \cite{partitions} for details.

%

%The spectral curve is obtained as usual, by looking for a function $W(x)$ analytical in the complex plane with some cuts, and such that the discontinuity along the cuts is:
%\beq\label{eqspcurveW10pmPl}
%W(x+i0)+W(x-i0)=V'(x)\qquad \quad , \,\, x\in\, {\rm cuts}\, .
%\eeq
%This equation has many solutions, and the one which leads to the correct spectral curve is the one such that the cuts are compatible with the integration path (see section \ref{}). Here, the correct solution is the one which gives a power series in $\ln{q}$.

%We have:
%\bea
%x\,V'(x)
%&=& \sum_k t_{k+1}\,(\ln{x}-(N-{1\over 2})\ln{q})^k + 2 \ln{x} + i\pi  \cr
%&& +(1-N)\ln{q} - \ln{q}\, \left({xg'(x)\over g(x)} + {{1\over x} g'(1/x)\over g(1/x)})\right) + \ln{q}\,{x\over 1-x}
%\eea

%and we use the Stirling asymptotic for the $q$-product $g(x)$:
%\bea
%\ln{q}\,{xg'(x)\over g(x)}
%&=& -Li_1(x^{-1}) + {1\over 2 \ln{q}} Li_0(x^{-1}) - \sum_{m=2}^\infty Li_{1-2m}(x^{-1})\, \, {B_{2m}\over (2m)!}\, (\ln{q})^{-2m}  \cr
%\eea
%where $Li_m$ is the polylogarithm, and $B_m$ is the $m^{\rm th}$ Bernouilli number:
%\beq
%Li_1(x) = -\ln{(1-x)}
%\virg
%Li_0(x) = {x\over 1-x}
%\eeq
In principle,
%the solution of \eq{eqspcurveW10pmPl}
the spectral curve could be found for all $t_k$'s, but for simplicity, we compute it explicitly for $t_3=t_4=\dots=0$, i.e. for only $t_1$ and $t_2$. We write:
\beq
t_1=t
\virg
t_2 = p-1
\virg
t_3=t_4=\dots=0 .
\eeq

%In that case, the correct solution of \eq{eqspcurveW10pmPl}, is then found to be $W(x(z))=\om(z)$, where:
%\bea
%x(z) &=& (z_0-z)(z_0-1/z )/(z_0+1)^2 \cr
%x(z) \om(z)
%&=&  -2\, (\ln{(1+1/z)}-\ln{(1+1/z_0)}) \cr
%&& + p\,(\ln{(1-1/z z_0)}-\ln{(1-1/z_0^2)})  \cr
%&& +  \sum_{m=1}^\infty  Li_{1-2m}(x(z)) \,  {B_{2m}\over 2m!} \,\, (\ln{q})^{-2m} + {x(z)\over 2\ln{q}\,(x(z)-1)}  \cr
%\eea
%where
%\beq
%\ee{-t} = {1\over z_0^2}\,\,(1-{1\over z_0^2})^{p(p-2)}
%\eeq
%Indeed, $\om(z)$ is analytical outside the unit disc $|z|>1$, i.e. $W(x)$ is analytical outside $[a,1]$ where $a=(z_0-1)^2/(z_0+1)^2$.

%\medskip
In that case the spectral curve is:
%then $y(z) ={1\over 2}(\om(1/z)-\om(z))$, i.e.
\beq\label{spcurvePlq}
{\cal E}=\left\{
\begin{array}{l}
x(z)  =  {(1-{z\over z_0})(1-{1\over z z_0})\over (1+{1\over z_0})^2} \cr
\cr
y(z) = {1\over x(z)}\,\left(- \ln{z}  + {p\over 2}\,\ln{\left(1-z/z_0\over 1-1/z z_0\right)}  \right)
\end{array}\right.
\quad ,\,\, \ee{-t} = {1\over z_0^2}\,\,\left(1-{1\over z_0^2}\right)^{p(p-2)}
\eeq
and thus we have:
\beq
Z = \sum_{n(\l)\leq N}\,\, \left({\dim_q(\l)\over [|\l|]!}\right)^2\,\,\ee{-t|\l|}\,\, q^{{p-1\over 2}\,C_2(\l)}
\sim \ee{\sum_g (\ln{q})^{2-2g}\,\, F_g({\cal E})} .
\eeq
The large $\ln{q}$ expansion of $\ln{Z}$ is given by the symplectic invariants of curve ${\cal E}$, and this expansion is independent of $N$, provided that $N>\ovl{n}$, where:
\beq
\ovl{n} = {1\over 2}+\ln{q}\,\left(p\ln{(1-1/z_0)}+(p-2)\ln{(1+1/z_0)}\right) .
\eeq

In fact, this means that the $N$ dependence is in non-perturbative terms, smaller than any power of $\ln{q}$. This is again the arctic circle phenomenon, the partitions of size $>\ovl{n}$ have an exponentially small probability, the size of the system seems to be frozen to $\ovl{n}$.

\subsubsection{Mirror curve}

In topological strings (see \cite{topovertex,Hori,MMbook,Bmodel,Vonk} and section \ref{sectopostring}), it is known that Gromov-Witten invariants of some Toric-Calabi-Yau manifolds, can be written as sums over partitions, this is called the topological vertex method.

In particular, for the Calabi-Yau manifold $X_p={\cal O}(-p)\oplus{\cal O}(p-2)\to{\mathbb P}^1$, which is a rank 2 bundle over ${\mathbb P}^1$, the Gromov Witten invariants ${\cal N}_{g,d}(X_p)$ which count the number of Riemann surfaces of genus $g$ and degree $d$ which can be embedded in $X_p$ and going through given points, are given by:
\beq
\sum_{g} (\ln{q})^{2-2g}\,\sum_d \ee{-td}\, {\cal N}_{g,d}(X_p)
= \ln {Z}
\eeq
where
\beq
Z=\sum_{\l}\,\, \left({\dim_q(\l)\over [|\l|]!}\right)^2\,\,\ee{-t|\l|}\,\, q^{{p-1\over 2}\,C_2(\l)}.
\eeq
Therefore, we have found in the previous section that the Gromov-Witten invariants of $X_p$ are given by the symplectic invariants of ${\cal E}$:
\beq
\sum_d \ee{-td}\,{\cal N}_{g,d}(X_p) = F_g({\cal E}).
\eeq
This result is interesting by itself, since it already gives a practical way of computing the Gromov-Witten invariants of $X_p$.

\medskip
But we can go further.
Notice that $x(z)=u(z)$ is a rational function of $z$, and $y(z)={1\over u(z)}\,\ln{v(z)}$ where $v(z)$ is also a rational function of $z$, and notice that:
\beq
dx\wedge dy = d\ln{u}\wedge d\ln{v}
\eeq
thus, ${\cal E}$ is symplectically equivalent to the following spectral curve:
\beq\label{spcurvePlqmirror}
\td{\cal E}=\left\{
\begin{array}{l}
\td{x}(z)  =  \ln{\left((1-{z\over z_0})(1-{1\over z z_0})\right)} \cr
\cr
\td{y}(z) = \ln{\left({1\over z}\,\left({1-z/z_0\over 1-1/z z_0}\right)^{p\over 2}  \right)}
\end{array}\right.
\quad ,\,\, \ee{-t} = {1\over z_0^2}\,\,(1-{1\over z_0^2})^{p(p-2)}
\eeq
and thus we have:
\beq
F_g({\cal E})=F_g(\td{\cal E}).
\eeq

Notice that $u=\ee{\td x}$ and $v=\ee{\td y}$ are both rational fractions of $z$, and thus, by eliminating $z$, there exists an algebraic relationship between them, i.e. there exists a polynomial $H$ such that
\beq
H(u,v)=0.
\eeq
%\beq
%v+{1\over v}
%={z_0\over u^{p/2}}\,\, \left[ (1-b)\,a^p + (1-a)\, b^p\right]
%\eeq

This curve is known in the context of topological strings \cite{Hori}:
\beq
H(\ee{\td x},\ee{\td y})=\om_+\,\om_-
\eeq
is a Calabi-Yau 3-fold $\td{X}_p$, which is mirror of $X_p$ under mirror symmetry , and $H(\ee{x},\ee{y})=0$ is the singular locus of $\td{X}_p$.
Therefore, we have obtained that, in agreement with the "remodelling of the B-model proposal" (see \cite{Bmodel} and section \ref{sectopostring}), we have:

{\bf the Gromov-Witten invariants of $X_p$, are the symplectic invariants of the singular curve of its mirror $\td{X}_p$.}

%%%%%%%%%%%%%%%%%%%%%%%%%%%%%%%%%%%%%%%%%%%%%%%%%%%%%%%%%%%%%%%%%%%%%%%%%%%%%%%%%%%%%%%%%%%%%%%%%%%%%%%%%%

\section{Intersection numbers and volumes of moduli spaces}

\subsection{Kontsevich integral and intersection numbers}\label{seckontsevich}

\subsubsection{Matrix integral}

Let $\L$ be a diagonal matrix $\L={\rm diag}(\l_1,\dots,\l_N)$.

Kontsevich's integral \cite{kontsevich} is the following formal matrix integral (defined as a formal power series in large $\L$)
\beq
Z_{\rm Kontsevich}(\L) = \int\, dM\,\, \ee{-N\Tr {M^3\over 3}+\L M^2}
\eeq
and we consider its topological expansion:
\beq
\ln{Z_{\rm Kontsevich}(\L)} = \sum_{g=0}^\infty N^{2-2g}\, F_g .
\eeq
Notice that, upon shifting $M\to M-\L$ we have
\beq
Z_{\rm Kontsevich}(\L) = \ee{{N\over 3}\Tr \L^3}\,\,
\int\, dM\,\, \ee{-N\Tr \left[ {M^3\over 3} -\L^2 M\right] }
\eeq
and thus, this integral is a special case of 1-matrix integral with an external field (see section \ref{sec1Mext}), which implies that the coefficients $F_g$ are the symplectic invariants of the corresponding spectral curve.

\subsubsection{Kontsevich's Spectral curve}

We have seen in section \ref{sec1Mext}, that the topological expansion of a matrix integral with external field, of the type:
\beq
Z =  \int\, dM\,\, \ee{-N\Tr V(M)-\L^2 M}
\eeq
is given by the symplectic invariants of its spectral curve:
\beq
\ln{Z} = \sum_{g} N^{2-2g} F_g({\cal E})
\eeq
where the spectral curve ${\cal E}$ is characterized by the algebraic equation:
\beq\label{eqspcurveKontsSinv}
V'(x)-y = {1\over N}\,\sum_i {P_i(x)\over y-\l_i^2}
\eeq
where $P_i(x)$ is a polynomial of $x$ of degree at most $\deg V''$, which behaves at large $x$ like $V'(x)/x$.
Here we have $V(x)={x^3\over 3}$, hence $V'(x)=x^2$, and thus $P_i(x)=x+P_i(0)$.

The fact that $Z$ is to be understood as a formal power series at large $\L$, means that we look for a rational spectral curve, and this determines all the polynomials $P_i(x)$.

Here we find that the rational spectral curve of the form \eq{eqspcurveKontsSinv} is:
\beq
{\cal E}_{K} = \left\{
\begin{array}{l}
x(z) = z - {1\over N}\Tr {1\over 2\hat\L(\hat\L-z)} \cr
y(z) = z^2 + t_1
\end{array}
\right.
\eeq
where $t_1={1\over N}\Tr {1\over \hat\L}$ and
\beq
\L^2=\hat\L^2+ t_1.
\eeq

From now on, for simplicity, we shall assume that $t_1=0$:
\beq
t_1 = 0 =  {1\over N}\Tr {1\over \hat\L}
\eeq
and therefore we have
\beq
\hat\L=\L
\eeq
and the spectral curve is:
\beq
{\cal E}_{K} = \left\{
\begin{array}{l}
x(z) = z - {1\over N}\Tr {1\over 2\L(\L-z)} \cr
y(z) = z^2
\end{array}
\right. .
\eeq

\subsubsection{Symplectic invariants}

To compute the $F_g$'s of the spectral curve ${\cal E}_K$, we need to consider the branchpoints, i.e. the zeroes of $x'(z)$, and they are quite complicated.

However, we may use symplectic invariance, and compute the $F_g$'s after exchanging the roles of $x$ and $y$, and thus, consider the spectral curve:
\beq
\td{\cal E}_{K} = \left\{
\begin{array}{l}
x(z) = z^2  \cr
y(z) = z - {1\over N}\Tr {1\over 2\L(\L-z)}
\end{array}
\right. .
\eeq
This spectral curve has now only one branchpoint solution of $x'(z)=0$, which is located at $z=0$.
Since the $F_g$'s are obtained by computing residues near $z=0$, we may Taylor expand $y(z)$ near $z=0$, and we have:
\beq\label{Kontsspcurvetd}
\td{\cal E}_{K} = \left\{
\begin{array}{l}
x(z) = z^2  \cr
y(z) = z - {1\over 2} \sum_{k=0}^\infty t_{k+2}\,z^k\,
\end{array}
\right.
\eeq

Now, it is rather easy to compute the first few symplectic invariants:
\beq
\om_1^{(1)}(z) =  - {dz\over 8 (2-t_3)} \,\left({1\over z^4}+{t_5\over (2-t_3)z^2}\right),
\eeq
\beq
\om_3^{(0)}(z_1,z_2,z_3)
= - {1 \over 2-t_3}\, {dz_1\,dz_2\,dz_3\over z_1^2 z_2^2 z_3^2},
\eeq
\bea
 {\om_2^{(1)}(z_1,z_2) }
&=& {dz_1 \, dz_2 \over 8 (2-t_3)^4 z_1^6 z_2^6} \Big[ (2-t_3)^2 ( 5 z_1^4 + 5 z_2^4+3z_1^2 z_2^2) \cr
&& \qquad + 6 t_5^2 z_1^4 z_2^4 + (2-t_3) \big(6 t_5 z_1^4 z_2^2 +6 t_5 z_1^2 z_2^4 + 5 t_7 z_1^4 z_2^4\big) \Big] , \cr
\eea
\bea
{\om_1^{(2)}(z) } &=& - { dz \over 128 (2-t_3)^7 z^{10}} \Big[ 252\, t_5^4 z^8 + 12\, t_5^2 z^6 (2-t_3) (50\, t_7 z^2 + 21\, t_5) \cr
&& \quad + z^4 (2-t_3)^2 ( 252\, t_5^2 + 348\, t_5 t_7 z^2 + 145\, t_7^2 z^4 + 308\, t_5 t_9 z^4) \cr
&& \qquad + z^2 (2-t_3) (203\, t_5 +145\, z^2 t_7 + 105\, z^4 t_9 +105\, z^6 t_{11}) \cr
&& \qquad \quad + 105\, (2 -t_3)^4 \Big] .\cr
\eea
And so on...

For example, the first and second order free energies are:
\beq
F_{\rm Kontsevitch}^{(1)} = - {1 \over 24} \ln\left(1-{t_3\over 2} \right)
\eeq
and
\beq\label{F2Konts}
F_{\rm Kontsevitch}^{(2)} = {1 \over 1920}\, {252\, t_5^3 + 435\, t_5 t_7 (2-t_3) + 175\, t_9 (2-t_3)^2 \over (2 -t_3)^5 }.
\eeq

\br
The fact that the symplectic invariants depend only on odd $t_k$'s can be understood in terms of symplectic invariance: indeed, adding to $y(z)$ any rational function of $x(z)$ (i.e. any rational function of $z^2$) is a symplectic transformation. It does not change the $F_g$'s, and therefore the $F_g$'s depend only on the odd part of $y(z)$, i.e. only on the odd $t_k$'s.
\er

\subsubsection{Intersection numbers}

The Kontsevich integral is important because it computes intersection numbers of Chern classes of line bundles over the moduli spaces of Riemann surfaces \cite{Witten, kontsevich}.

\smallskip
Let ${\cal M}_{g,n}$ be the moduli space of Riemann surfaces $\Sigma$ of genus $g$, with $n$ marked points $p_1,\dots,p_n$.
This moduli space is a complex manifold of dimension:
\beq
\dim {\cal M}_{g,n} = d_{g,n} = 3g-3+n .
\eeq
This moduli space can be compactified into a compact space $\ovl{\cal M}_{g,n}$ by adding all stable nodal surfaces (stable means that each component has a Euler characteristics $<0$).

The cotangent bundle ${\cal L}_i$ is the bundle over ${\cal M}_{g,n}$, whose fiber is the cotangent space of $\Sigma$ at the point $p_i$. Let
\beq
\psi_i=c_1({\cal L}_i)
\eeq
be its first Chern class. ${\cal L}_i$ and $\psi_i$ can be extended to $\ovl{\cal M}_{g,n}$.

Chern classes $\psi_i$ provide useful information on the topology of $\ovl{\cal M}_{g,n}$.
Indeed if one computes the integral of $\psi_i$ over a cycle in $\ovl{\cal M}_{g,n}$, this integral tells how many times the cotangent space rotates.
The intersection numbers are defined as:
\beq
<\tau_{d_1}\dots \tau_{d_n}> = \int_{\ovl{\cal M}_{g,n}} \, \psi_1^{d_1}\psi_2^{d_2}\dots \psi_n^{d_n}
\eeq
and are non-zero only if ${\displaystyle \sum_i} d_i=d_{g,n}=3g-3+n$.

\medskip

The Kontsevich integral is a generating function for those numbers:
\bea\label{Kontssumnintnb}
&& \ln{Z_{\rm Kontsevich}(\L) } \cr
%&=& \sum_{n,g}\,\, \sum_{G\in {\cal G}_{g,n}}\,{N^{2-2g-n}\over \#{\rm Aut}(G)}\,\, \prod_{e={\rm edges}}\,\, {1\over \l_{e\,{\rm left}}+\l_{e\,{\rm right}}} \cr
%&=& \sum_{n,g}\,\, \sum_{i_1,\dots, i_n=1}^N\,\, {N^{2-2g-n}\over \rho_{g,n}}\,\, \hat{U}_{g,n}(\l_{i_1},\dots,\l_{i_n}) \cr
&=& \sum_g (N/2)^{2-2g}\,\sum_{n}\,\, \sum_{d_1+\dots+d_n=d_{g,n}}\,\,
\,\, \prod_i (2d_i-1)!!\, {t_{2d_i+1}\over 2}\,\, <\tau_{d_1}\dots \tau_{d_n}> \cr
\eea
where the sum is restricted to $d_i>0$ because we have assumed $t_1=0$.

For example we have:
\bea
F_2
&=& \sum_k (t_3/2)^k \left({105\,t_9\over 2} <\tau_1^k\tau_4>  +{45\, t_5 t_7\over 4}<\tau_1^k\tau_2\tau_3> + {27\, t_5^3\over 8}\,<\tau_1^k \tau_2^3>\right). \cr
\eea
%and if we compare with \eq{F2Konts}, we find:
%\beq
%<\tau_1^k\tau_4> =  7 (k+1)(k+2) / 2^9
%\eeq

\subsubsection{Correlators and unmarked faces}

The symplectic invariants of the spectral curve \eq{Kontsspcurvetd} are the $F_g$'s. They generate intersection numbers of $\psi$-classes, i.e. Chern classes of cotangent line bundles over marked points, i.e. they generate the intersection numbers of the type:
\beq
<\psi_1^{d_1}\dots \psi_n^{d_n}>.
\eeq

The correlators $\om_n^{(g)}$s of that spectral curve, also have some interpretation in terms of integrals of some classes over moduli-spaces.

Notice that Kontsevich integral \eq{Kontssumnintnb} contains a summation over $n$, i.e. over the number of marked points.
One may whish to distinguish some of those marked points, and fix marked faces around them, and perform the sum over the other marked points, in some sense forget the other marked points.

The forgetful map, is the map from $\ovl{\cal M}_{g,n+m}$ to $\ovl{\cal M}_{g,n}$, which consists in forgetting $m$ marked points. Under this map, $\psi$ classes project to the Mumford $\kappa$-classes:
\beq
\int_{\ovl{\cal M}_{g,n+m}}\, \psi_1^{d_1}\dots \psi_n^{d_n}\,\, \prod_{k=1}^m\, \psi_{n+k}^{a_k+1}
=
\int_{\ovl{\cal M}_{g,n}}\, \psi_1^{d_1}\dots \psi_n^{d_n}\,\, \sum_{\sigma\in \Sigma_m}\, \prod_{c={\rm cycles\,of\,}\sigma}\,\, \kappa_{(\sum_{i\in c\,}\,a_i)} .
\eeq
For examples with $m=1$ and $m=2$:
\beq
\int_{\ovl{\cal M}_{g,n+1}}\, \psi_1^{d_1}\dots \psi_n^{d_n}\,\,  \psi_{n+1}^{a+1}
=
\int_{\ovl{\cal M}_{g,n}}\, \psi_1^{d_1}\dots \psi_n^{d_n}\,\,  \kappa_{a},
\eeq
\beq
\int_{\ovl{\cal M}_{g,n+2}}\, \psi_1^{d_1}\dots \psi_n^{d_n}\,\,  \psi_{n+1}^{a_1+1}\,\psi_{n+2}^{a_2+1}
=
\int_{\ovl{\cal M}_{g,n}}\, \psi_1^{d_1}\dots \psi_n^{d_n}\,\,  (\kappa_{a_1+a_2}+\kappa_{a_1}\kappa_{a_2}) .
\eeq

One finds \cite{volumes}, that the correlators $\om_n^{(g)}(z_1,\dots,z_n)$, are the generating functions for $\kappa$ classes, coupled to $n$ $\psi$-classes:
\bea
\om^{(g)}_{n}(z_1,\dots,z_n)
&=& 2^{- d_{g,n}}(t_3-2)^{2-2g-n}\!\!\!\!  \sum_{d_0+d_1+\dots+d_n=d_{g,n}}\,
 \sum_{k=1}^{d_0} {1\over k!}\,\sum_{b_1+\dots+b_k =d_0, b_i>0}  \cr
 && \qquad \qquad \prod_{i=1}^n {2d_i+1!\over d_i!}\, {dz_i\over z_i^{2d_i+2}}\,\, \prod_{l=1}^k \td{t}_{b_l} <\prod_{l=1}^k \kappa_{b_l} \prod_{i=1}^n  \psi_i^{d_i}>_{g,n}  \cr
\eea
where the coefficients $\td{t}_b$ are the Schur transform of the $t_k$'s:
\beq\label{formulafortdtb}
\td{t}_b  = \sum_{l=1}^b\,  {(-1)^l\over l}  \,\sum_{a_1+\dots+a_{l}=b, a_i>0}\,\,  \prod_{j} {2 a_{j}+1!\over a_{j} !}\,\,{t_{2a_{j}+3}\over t_3-2} .
\eeq
Their generating function is obtained by:
\beq\label{ttdKonts}
f(z) = \sum_{a=1}^\infty {2 a+1!\over a !}\,\,{t_{2a+3}\over 2-t_3} \,z^a
\virg
 -\ln{(1-f(z))}  = \sum_{b=1}^\infty \td{t}_b\, z^b .
\eeq

{\bf Example:}
\beq
\td{t}_1 = -6\, {t_5\over t_3-2}
\virg
\td{t}_2 = -60\, {t_7\over t_3-2} + 18 {t_5^2\over (t_3-2)^2}
\,\, , \, \dots .
\eeq

$\om_{n}^{(g)}$  is the Laplace transform of:
\bea\label{Vngkappa}
&& 2^{d_{g,n}}\,(t_3-2)^{2g-2+n}\,\, V_{g,n}(L_1,\dots,L_n)  \cr
&=& \!\!\!\! \!\!\!\!   \sum_{d_0+d_1+\dots+d_n=d_{g,n}}\, \prod_{j=1}^n {L_j^{2 d_j}\over d_j !}\,\, \sum_k {1\over k!}  \, \sum_{b_1+b_2+\dots+b_k=d_0, b_i>0}\, \, \prod_{i=1}^k \td{t}_{b_i}  \,\,<\prod_{i=1}^k \kappa_{b_i} \prod_j \psi_F^{d_j}>  .\cr
\eea
$V_{g,n}$ can be interpreted as the generating function for counting intersection numbers of $\kappa$-classes, on the moduli-space of Riemann surfaces with $n$ boundaries (discs removed from the surface), of perimeters $L_1,\dots,L_n$.

\subsection{Application: Weil-Petersson volumes}\label{secWP}

Consider the stable moduli space ${\cal M}_{g,n}$ of Riemann surfaces of genus $g$ with $n$ boundaries (stability means $2-2g-n<0$).
Every surface in ${\cal M}_{g,n}$ has a negative Euler-characteristics, and thus a negative average curvature. It can be equipped with a unique constant negative curvature metric, called Poincar\'e metric, such that the boundaries are geodesics.
Let $L_1,\dots,L_n$ be the geodesic lengths of the boundaries.

Every Riemann surface in ${\cal M}_{g,n}$ can be decomposed into $2g-2+n$ pairs of pants, whose boundaries are geodesics (such a decomposition is not unique).
Conversely, $2g-2+n$ pairs of pants with fixed given boundary perimeters can be glued together to form a Riemann surface of ${\cal M}_{g,n}$ provided that the lengths of boundaries which are glued together match.
The gluing is not unique, because 2 circles of the same perimeter can be glued in many ways, twisted by an arbitrary angle.
$$
{\epsfxsize 10cm\epsffile{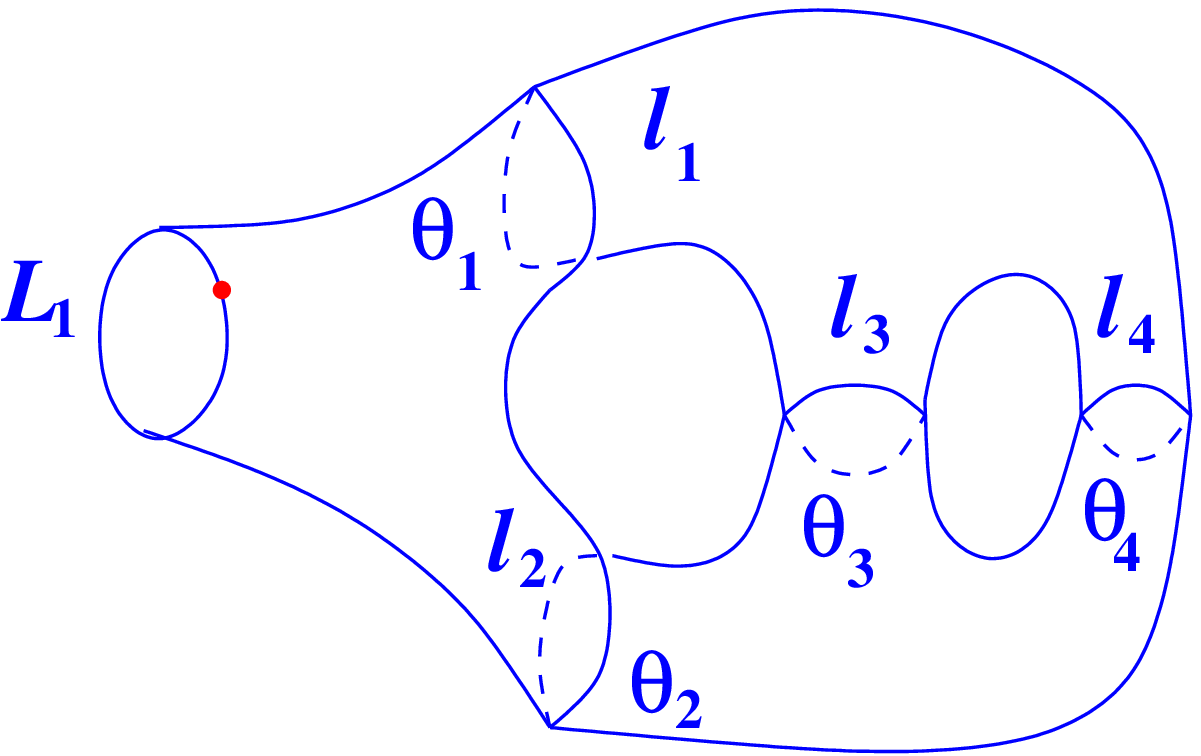}}
$$

The $3g-3+n$ geodesic lengths of the glued boundaries $l_1,\dots,l_{3g-3+n}$, together with the $3g-3+n$ twisting angles $\theta_1,\dots,\theta_{3g-3+n}$, provide a local set of coordinates parameterizing ${\cal M}_{g,n}$.
It turns out that, although we don't have uniqueness of the decomposition, the corresponding symplectic form, called Weil-Petersson symplectic metric is well defined on ${\cal M}_{g,n}$:
\beq
\Omega = \prod_{i=1}^{3g-3+n} dl_i\wedge d\theta_i
\eeq
and it can be extended to the compactified $\ovl{\cal M}_{g,n}$.
The Weil-Petersson volume of $\ovl{\cal M}_{g,n}$ is then defined as:
\beq
{\rm Vol}_{g,n}(L_1,\dots,L_n) = \int_{\ovl{\cal M}_{g,n}} \Omega
\eeq
where the $n$ external boundaries are restricted to have fixed geodesic lengths $L_i$'s and fixed angles (i.e. some marked points).

It can be proved \cite{Wolpert} that the Weil-Petersson metrics comes from the K\"{a}hler metrics:
\beq
2\pi^2 \kappa_1
\eeq
i.e. we have:
\bea
{\rm Vol}_{g,n}(L_1,\dots,L_n)
&=& \left<(2\pi^2\kappa_1 + {1\over 2}\sum_{i=1}^n L_i^2 \psi_i)^{d_{g,n}}\right> \cr
&=& 2^{-d_{g,n}}\,\sum_{d_0+\dots+d_n=3g-3+n} {2^{d_0}\over d_0!}\,\, \prod_{i=1}^n {L_i^{2d_i}\over d_i!} \,\, \left<(2\pi^2\kappa_1)^{d_0} \, \psi_1^{d_1} \dots \psi_n^{d_n}\right> .\cr
\eea

It can be made to coincide with \eq{Vngkappa}, provided that we choose $\td{t}_1=4\pi^2$ and $t_3=3$.
Doing the reverse transform using \eq{ttdKonts}, it corresponds to $-\ln{(1-f(z))} = 4\pi^2 z$, i.e. $f(z) = 1-\ee{-4\pi^2 z}$ and thus:
\beq
t_{2d+3} = {(2i\pi)^{2d}\over (2d+1)!}+2\delta_{d,0}
\eeq
i.e. it corresponds to the spectral curve
\beq\label{spcurveWP}
{\cal E}_{WP} = \left\{
\begin{array}{l}
x(z) = z^2  \cr
y(z) = {1\over 2\pi}\, \sin{(2\pi z)}
\end{array}
\right. .
\eeq
In other words, the Laplace transforms of the Weil-Petersson volumes ${\rm Vol}_{g,n}(L_1,\dots,L_n)$, are the $W_n^{(g)}$'s of the spectral curve ${\cal E}_{WP}$.

\bigskip

Taking the inverse Laplace transform of the recursion relations \eq{defWnginv} which define the $W_n^{(g)}$'s for the spectral curve \eq{spcurveWP}, we recover Mirzakhani's recursion \cite{Mirz1,Mirz2} relations for the Weil-Petersson volumes (see \cite{EOWP,volumes} for the proof).

\subsection{Application: Witten-Kontsevich theorem}\label{secwittenkontsevich}

Witten's conjecture \cite{Witten} was the assertion that the limit of the generating function of large discrete surfaces was indeed the generating function of intersection numbers for continuous Riemann surfaces.
In other words the double scaling limit of the $F_g$'s of maps, should coincide with the $F_g$'s of the Kontsevich integral.

We have seen that the $F_g$'s of maps are given by symplectic invariants, and thus their limits as $t\to t_c$, are given by the symplectic invariants of the blown up curve.
Thus the $F_g$'s of the double scaling limit of maps, are the $F_g$'s of the minimal $(p,2)$ model, i.e. the $F_g$'s of the spectral curve:
\beq
F_{g}({\cal E}_{\rm maps}) \sim (t-t_c)^{(2-2g)\mu}\, F_{g}({\cal E}_{(p,2)})
\eeq
where $\mu = {p+2\over p+1}$, and:
\beq
{\cal E}_{(p,2)} = \left\{
\begin{array}{l}
x(z) = z^2- 2u \cr
y(z) = {\displaystyle \sum_{k=0}^p} \ovl{t}_k z^k
\end{array}
\right. .
\eeq

On the other hand, the $F_g$'s of the Kontsevich integral, are also given by the symplectic invariants of a spectral curve, which is the equilibrium density of eigenvalues in the Kontsevich integral's Matrix Airy function. The spectral curve (see section \ref{seckontsevich}) is (again we assume for simplicity that $t_1=0$):
\beq
{\cal E}_{K} = \left\{
\begin{array}{l}
x(z) = z - {1\over N}\Tr {1\over 2\L(\L-z)} \cr
y(z) = z^2
\end{array}
\right. .
\eeq

We have seen (theorem \ref{thsymplinv}), that if two spectral curve are equivalent under symplectic transformations, then they have the same $F_g$'s.
In particular the $F_g$'s don't change if we change $x\to y$ and $y\to -x$, thus:
\beq
{\cal E}_k \sim \td{\cal E}_{K}
=
 \left\{
\begin{array}{l}
x(z) = z^2 \cr
y(z) = -z + {1\over N}\Tr {1\over 2\L(\L-z)} \cr
\end{array}
\right. .
\eeq
In that new formulation, there is only one branchpoint located at $z=0$, and since all quantities  computed are residues at this branchpoint, we may Taylor expand $y(z)$ in the vicinity of $z=0$ and thus:
\beq
{\cal E}_{K}
\sim
 \left\{
\begin{array}{l}
x(z) = z^2 \cr
y(z) = -z + {1\over 2} \sum_{k=0}^\infty z^k{1\over N}\Tr \L^{-k-2} \cr
\end{array}
\right. .
\eeq
If we choose the diagonal matrix $\L$ such that:
\beq
\ovl{t}_k = {1\over 2N}\Tr \L^{-k-2} - \delta_{k,1} ,
\eeq
we have:
\beq
{\cal E}_K \sim {\cal E}_{(p,2)}
\eeq
and therefore:
\beq
\encadremath{
F_g({\cal E}_K) = F_g({\cal E}_{(p,2)})
}\eeq
which proves Witten's conjecture.

In fact this conjecture was first proved by M. Kontsevich \cite{kontsevich}. Kontsevich's method consisted of two parts: first prove that the Airy matrix integral now known as Kontsevich integral was the generating function of intersection numbers, and then prove that both $F_{g,K}$ and $F_{g,(p,2)}$ obeyed the same set of differential equations, namely KdV hierarchy.
Witten's conjecture has received several proof since then, in particular Loijeenga's \cite{Looijenga}, or Okounkov-Pandaripande \cite{OP}.

Here we see that the spectral curve of the Kontsevich model and the spectral curve of the $(p,2)$ model are obtained from one another by the symplectic transformation $x\to y, y\to -x$.
That symplectic transformation (the $\pi/2$ rotation in the $(x,y)$ plane), leaves $F_g$ invariant, but it changes the $\om_{n}^{(g)}$'s.

%%%%%%%%%%%%%%%%%%%%%%%%%%%%%%%%%%%%%%%%%%%%%%%%%%%%%%%%%%%%%%%%%%%%%%%%%%%%%%%%%%%%%%%%%%%%%%%%%%%%%%%%%%

\section{Application: topological strings}\label{sectopostring}

The counting of surfaces with particular weights, or enumerative geometry, has been very important in physics since the arising
of string theories. Indeed, those theories consist in replacing the point-like particles of usual theories such as classical
mechanics (0 dimensional objects) by strings
(1 dimensional objects): a state is now given by a string state instead of a point state. A string evolving in time describes a surface in the space-time, the world-sheet,
instead of a line for a point evolving in space-time. Hence the usual path integral corresponding to sum over all possible stories from one initial state
to one final state is now a "sum" over all possible surfaces in space-time (target space), linking the initial strings to the
final ones. This could explain why these theories are related to the symplectic invariants since the latter already appear in many
problems of "surface enumeration". So far, no proof is available
but there exist many hints that the symplectic invariants of some specific spectral curve should be the partition functions of some particular string
theories: type IIB toplogical strings on some special target space.
Many checks have been made that this is indeed the case \cite{MMtor}, so that Bouchard, Klemm, Mari\~{n}o and Pasquetti
\cite{Bmodel} proposed to define the topological string partition function and observables as the symplectic invariants
and correlation functions, considering the spectral curve as the target space of the string theory.

Even if this conjecture is not proved yet, some new clues were recently given by Dijkgraaf and Vafa \cite{DV2}
who already conjectured
that some matrix models are dual to some particular topological string theories \cite{DV}.
Those new hints rely on the study of an effective field theory for the string theory: the Kodaira-Spencer theory.
We review those advances in the second part of the present section.

\subsection{Topological string theory}

\subsubsection{Introduction}

Type IIB topological string theories are obtained by twisting $N=2$ superconformal sigma model in 2 dimensions. The precise
description would lead us too far away from the main topic of this review, and the interested reader is
invited to consult \cite{MMbook,MMtopo,Vonk,Hori} for details as well as \cite{Bouchard} for the particular topic of toric
geometries. There exist two ways of twisting this theory leading
to two different models referred to as $A$ and $B$ models. We will be particularly interested in this paper in the $B$-model
and won't develop the $A$-model too far. Nevertheless, since the special geometries of the target space of the B-model
we are interested in, are inherited from the $A$-model, we will say a few words about the latter and especially the
possible geometries of its target space and of the objects one wants to compute in the next section.

First of all, let us mention that the topological string theories can be thought of, as theories of the maps from a Riemann
surface $\Sigma$ (the world-sheet) to a Calabi-Yau manifold $M$ (the target space) of arbitrary dimension. More precisely,
in the $A$ model side, the amplitudes to be computed are related to the Gromov-Witten invariants as follows. Let us
consider a worldsheet $\Sigma_k^{(g)}$ of genus $g$ with $k$ holes (or boundaries). One wants to count maps which
map the boundaries to a Lagrangian submanifold of $M$ denoted as the brane\footnote{The geometry of such submanifolds is studied in the next section.} $L$.
Such maps are characterized by two additional parameters prescribing how the boundaries are mapped to the brane $L$:
a bulk class $\beta \in H_2(M,L)$\footnote{One assumes for simplicity that $b_1(L)=1$.} and winding numbers $\omega_i \in \mathbb{Z}$,
$i=1,\dots, k$, telling how many times the boundaries wrap around the brane $L$. One sums up those "number of maps" called Gromov-Witten invariants
$N_{\beta,\omega}^{(g)}$ in generating functions:
\beq
F_{\omega}^{(g)} := \sum_{\beta \in H_2(M,L)} N_{\beta,\omega}^{(g)} e^{-\beta t}
\eeq
and one also considers the open string amplitudes
\beq
A_k^{(h)}(z_1, \dots , z_k):= \sum_{\omega \in \mathbb{Z}^k} F_{\omega}^{(g)} {\displaystyle \prod_{i=1}^k} z_i^{\omega_i}
\eeq
where the open string parameters $z_i$ are parameters of the moduli space of the brane $L$ as well as the closed string amplitudes:
\beq
{\cal{F}}^{(g)}:= F_{0}^{(g)}.
\eeq

Let us also precise the moduli spaces of both models, since they play an important role in the link between topological
string theories and the symplectic invariants. In the $A$ model side, the moduli are the K\"{a}lher parameters of the target
space $M$ whereas the $B$ model depends on the complex structure of $M$.

\subsubsection{Mirror symmetry, branes and toric geometries}

One of the most fascinating features of topological string theories is the existence of a duality linking the $A$ and $B$ models:
the mirror symmetry. This symmetry states the equivalence of the $A$ model on a target space $M$ and the $B$ model
on a mirror target space $\widetilde{M}$ obtained from $M$ by a mirror map which exchanges the K\"{a}lher structure
of $M$ with the complex structure of $\widetilde{M}$\footnote{The works on the subject of mirror symmetry are numerous
in physics and mathematics. Entering this subject would quickly lead us too far away from our main topic. For a nice review
of the subject, one can refer to the excellent book \cite{Hori}.}.

In the following, we will only be concerned with a special but interesting class of target spaces $M$: Toric Calabi-Yau threefolds
for the $A$ model and their image under mirror symmetry. We now precise the structure of those geometries as well as the
geometry of their Lagrangian submanifolds.

Let us start with a Toric Calabi-Yau on the $A$ model side. A Toric Calabi-Yau threefold $M$ can be built as a submanifold of $\mathbb C^{k+3}$ as follows.
Consider $k+3$ complex scalars $X_i= \left|X_i\right| e^{i \theta_i}$, $i=1,\dots,k+3$, transforming under the action of $U(1)^k$ as
\beq
X_i \to e^{i Q_i^\alpha \epsilon_\alpha} X_i
\eeq
for some integers $Q_i^\alpha$, $\alpha=1,\dots,k$.
One then considers the 3-dimensional submanifold of $\mathbb R_+^{k+3}$ obtained by constraining the $|X_i|^2$'s to satisfy
\beq\label{Aconstr}
\sum_{i=1}^{k+3} Q_i^\alpha \left|X_i\right|^2 = r_\alpha
\virg \alpha=1,\dots,k.
\eeq
The CY 3-fold $M$ is the bundle of Tori generated by the $\theta_i$'s modulo the action of $U(1)^k$, over this real submanifold.

The parameters $r_\alpha$ are the K\"{a}hler moduli of the Toric threefold $M$.
The Calabi-Yau condition is then:
\beq
\forall \alpha=1,\dots,k \virg \sum_{i=1}^{k+3} Q_i^\alpha =0 .
\eeq

Remark that one can see the coordinates $X_i$ as a $S_1$-fibration (coordinates $e^{i \theta_i}$) over $\mathbb{R}^+$ (coordinates $\left|X_i\right|$)
giving to $M$ a structure of $T^3$ fibration over the subspace of $\mathbb{R}_+^3$ defined by the constraints \eq{Aconstr}.
It is then interesting to note that this fibration has singular loci when one or several $|X_i|$ vanish. Indeed, the $S^1$ fiber
defined by the corresponding $\theta_i$ shrinks. These loci will be important in the folowing study of the type $A$ branes
and they are encoded in the so-called toric graph of the threefold $M$ \cite{MMtor,Bmodel,Bouchard}.

The boundaries of the worldsheet must be mapped to special Lagrangian submanifolds of $X$ called branes. For this purpose,
let us remind the notation
\beq
X_i=\left|X_i\right|e^{i \theta_i}.
\eeq
The canonical symplectic form on $M$ is then given by
\beq
\omega= {1 \over 2} \sum_{i=1}^3 d\left|X_i\right|^2 \wedge d\theta_i.
\eeq
One can cancel this form and obtain a special Lagrangian submanifold $L$ by fixing the $\theta_i$'s with the equation
\beq
\sum_{i=1}^3 \theta_i = 0 \quad  [\pi]
\eeq
as well as constraining the moduli of $X_i$ by
\beq\label{braneA}
\sum_{i=1}^{k+3} q_i^\alpha \left|X_i\right|^2 = c^\alpha
\eeq
for $\alpha=1,\dots,r$, where the $q_i^\alpha$ satisfy
\beq
\sum_{i=1}^{k+3} q_i^\alpha = 0.
\eeq
A special submanifold in $M$ is then given by the set of complex numbers $q_i^\alpha$ and $c^\alpha$.

Moreover, one can consider such manifolds $L$ passing through some singular locus of the manifold $M$. In this case, it splits
into two submanifolds $L^+$ and $L^-$. This is one of the latter submanifolds that we consider as $A$ brane, e.g. $L^+$\footnote{
One can chose $L^+$ or $L^-$  as this brane without changing anything in the following.}.

%%%%%%%%%%%%%%%%%%%%%%%%%%%%%%%%%%%%%%%%%%%%%%%%%%%%%%%%%%%%%%%%%%%%%%%%%%%%%%%%%%%%%%%%%%%%%%%%%%%%

Let us now describe the mirror geometry of this target space and of the branes in the $B$ model.

The mirror map transforming $M$ into $\widetilde{M}$ can be built as follows. $\widetilde{M}$ has homogenous coordinates
$\widetilde{X}_i:= e^{x_i} \in \mathbb{C}^*$ with $i=1, \dots, k+3$ whose moduli are constrained by
\beq
\left|\widetilde{X}_i\right| = e^{-\left|{X}_i\right|^2}.
\eeq
The mirror geometry $\widetilde{M}$ of $M$ is then given by
\beq
\om^+ \om^- = \sum_{i=1}^{k+3} \widetilde{X}_i
\eeq
for two complex scalars $\left(\om^+,\om^-\right) \in \mathbb{C}^2$ and non vanishing complex homogenous coordinates
$\widetilde{X}_i \in \mathbb{C}^*$ satisfying
\beq
\prod_{i=1}^{k+3} \widetilde{X}_i^{Q_i^\alpha} = e^{-t_\alpha} = q_\alpha
\eeq
for any $\alpha= 1, \dots, k$, where
\beq
t_\alpha:= r_\alpha + i \theta_\alpha
\eeq
are complexified K\"{a}hler parameters of the threefold $M$. The equation of the mirror geometry $\widetilde{M}$ reduces to
\beq
\widetilde{H}(\widetilde{X},\widetilde{Y}|t_\alpha)= \om^+ \om^- = H(\ee x,\ee y |t_\alpha)
\eeq
where $\widetilde{X} = e^{x}$ and $\widetilde{Y}= e^y$ are two non-vanishing coordinates chosen among the $\widetilde{X}_i$'s\footnote{The choice of such coordinates $\widetilde{X}_i$ and $\widetilde{X}_j$ depends on
the sector of the moduli space that we are studying. We explicit this choice in the next section.}. The holomorphic
volume form on $\widetilde{M}$ is then given by
\beq
\Omega = {d\om d\widetilde{X} d\widetilde{Y} \over \om \widetilde{X} \widetilde{Y}} = {d\om \over \om} dx dy.
\eeq

Under this mirror map, one can easily characterize the $B$ model branes, i.e. the image of the special Lagrangian submanifolds of $M$
under the mirror map \cite{AV}. The constraints \eq{braneA} are translated into constraints on the $\widetilde{X}_i$:
\beq
\prod_{i=1}^{k+3} \widetilde{X}_i^{q_i^\alpha} = e^{-c_\alpha}
\eeq
for $\alpha=1,\dots, r$.
Moreover, for $r=2$, if one considers the singular brane $L^+$, its image is a one dimensional complex submanifold described
by the algebraic equation
\beq
H(\ee x,\ee y)=0 = \widetilde{H}(\widetilde{X},\widetilde{Y})
\eeq
on $\mathbb{C}^*$
and can thus be obtained by fixing $\om^-=0$ and considering $\om^+$ as a parameter of this brane.
In the following, we will consider this equation as the spectral curve.

\subsubsection{Embedding of the spectral curve and open string parameters}

In the preceding section, we showed that the spectral curve corresponds to the moduli space of the open string boundaries,
i.e. of the B-branes. In particular, it was shown  that the description of this moduli space depends on a choice of local
coordinates. Let us make this statement more precise by studying the whole target space of the B model and its projection to
local patches.

Remember that the moduli space of the B-branes is a Riemann surface $\spcurve$ given by a set of coordinates $\widetilde{X}_i$ and $\om_{\pm}$ constrained by
\beq\label{Beq1}
\sum_{i=1}^{k+3} \widetilde{X}_i = \om_- = 0
\eeq
as well as the constraints
\beq\label{Beq2}
\prod_{i=1}^{k+3} \widetilde{X}_i^{q_i^\alpha} = e^{-t_\alpha} \virg \alpha=1,\dots,k
\eeq
and
\beq\label{Beq3}
\prod_{i=1}^{k+3} \widetilde{X}_i^{q_i^\alpha} = e^{-c_\alpha} \virg \alpha=1,\dots,r.
\eeq
In order to describe this moduli space, one has to choose as set of coordinates characterizing it. Indeed, one can describe
it as an embedding of the Riemann surface $\spcurve$ into $\mathbb{C}^*\times \mathbb{C}^*$ (resp. to $\mathbb{C}\times \mathbb{C}$) through the coordinates
$\widetilde{X}_i$ (resp. the coordinates $x_i$) , e.g. the choice of two coordinates
$\widetilde{X}_i$ and $\widetilde{X}_j$ (resp. $x_i$ and $x_j$) among the set $\left\{\widetilde{X}_a\right\}_{a=1}^{k+3}$ (resp. $\left\{x_a\right\}_{a=1}^{k+3}$) allowing to describe $\spcurve$ by an equation
\beq
\widetilde{H}_{i,j}(\widetilde{X}_i,\widetilde{X}_j)=H_{i,j}(\ee{x_i},\ee{x_j})=0
\eeq
obtained by elimination of all the other coordinates $\widetilde{X}_a$ in \eq{Beq1}, \eq{Beq2} and \eq{Beq3}.
In the following, one generically denotes this embedding of the spectral curve by the equation
\beq
H(\ee{x},\ee{y})=0=\widetilde{H}(\widetilde{X},\widetilde{Y}).
\eeq
If all these equations represent the
embedding of the same surface, they correspond to different description of the branes, i.e. different types of boundary
conditions for the worldsheet. The choice of such an embedding is not random: depending on the regime we are considering,
i.e. the sector of the moduli space we are studying, some embeddings are more appropriate (see for example the discussion in
section 2.2 of \cite{Bmodel}).

Remark that the reparameterization group of the spectral curve $\spcurve$ is
\beq
G_\spcurve = SL(2,\mathbb{Z}) \times \left(\begin{array}{cc}
0 & 1 \cr
1 & 0 \cr \end{array} \right)
\eeq
in terms of the variables $(\widetilde{X},\widetilde{Y})$ of $\widetilde{H}(\widetilde{X},\widetilde{Y})=0$.

\br
Note that these transformations preserve the symplectic form $\left|dx \wedge dy\right| = \left| {d\widetilde{X} \over \widetilde{X}} \wedge {d\widetilde{Y} \over \widetilde{Y}}\right|$.
This reminds of the symplectic transformations section \ref{secsymplinv} which do not change the symplectic invariants. In this
topological string setup, these transformations acting on the open string parameters should preserve the closed string
amplitudes.
\er

These transformations, i.e. changing the open string parameters, are important in the study of the open string amplitudes.
Indeed, the whole open string moduli space, or moduli space of branes, exhibits different phases. In each of these particular
regimes, one can use a specific embedding  to describe the brane moduli space in appropriate coordinates (see \cite{Hori,Bmodel}
for a review on the subject).
The usual methods of computation allow to know the open string amplitudes in some very particular regime. This means that one can
compute these amplitudes in a specific patch of $\widetilde{M}$ and not on the others. It is thus interesting to
be able to go from one patch to the others. These "phase transitions", corresponding for example to blow ups
of $\widetilde{M}$, are elements of $G_{\spcurve}$ encoding the transition from the open string parameters of one
embedding to the others.

The choice of an embedding, i.e. the choice of a coordinate $x$, does not only fix the location of a brane but also the
last remaining ambiguity known as the framing\footnote{See \cite{Bmodel} for explanations on this phenomenon.}.
Roughly speaking, the framing consists in a discrete ambiguity and can be fixed by choosing an integer $f$. This ambiguity
correspond to the elements of $G_\spcurve$:
\beq
(\widetilde{X},\widetilde{Y}) \to (\widetilde{X}\widetilde{Y}^f,\widetilde{Y})
\eeq
or, in $x$ and $y$ coordinates,
\beq
(x,y) \to (x+ f y , y)
\eeq
for integer $f$. Note that this is also a symplectic transformation considered in section \ref{secsymplectic}.

We have thus shown that fixing an embedding of the spectral curve $\spcurve$ in $\mathbb{C}\times \mathbb{C}$ (or $\mathbb{C}^*\times \mathbb{C}^*$),
one fixes the open string parameter space which can be seen as the coordinate $x$. Going from one patch in the parameter space
to another is obtained by changing embedding thanks to an element of $G_\spcurve$.

\subsubsection{Open/closed flat coordinates}\label{secopenflat}

As it was already mentioned in the preceding sections, the moduli space of the B model (resp. A model) is given  by the complex
(resp. K\"{a}hler) parameters of the target space $\widetilde{M}$ (resp. $M$).
Let us precise the flat coordinates describing these complex and K\"{a}hler structures which are mapped to each other
by the closed string mirror map.

These flat coordinates $T^\alpha, \alpha = 1,\dots ,\genus$ are given by the periods of the meromorphic one form $ydx = \ln \widetilde{Y} {d\widetilde{X} \over \widetilde{X}}$ on the spectral curve $\spcurve$:
\beq
T^\alpha = {1 \over 2 i \pi} \oint_{{\cal{A}}_\alpha} ydx
\eeq
where $({\cal{A}}_\alpha,{\cal{B}}_\alpha)$ is a canonical basis of one cycles on $\spcurve$.
This also ensures the existence of a holomorphic function $F(T_\alpha)$ such that the dual periods can be expressed as
\beq
{\partial F \over \partial T_\alpha} = {1 \over 2 i \pi} \oint_{{\cal{B}}_\alpha} ydx.
\eeq

What about the open flat coordinates? If the closed coordinates are given by closed integrals of $\l$ over the cycles, the
open flat coordinate is expected to be given by chain integrals
\beq
U = {1 \over 2 i \pi} \int_{\alpha_x} ydx
\eeq
where $\alpha_x$ is an open path over which $y$ jumps by $2 i \pi$.

Moreover, it is interesting to note that the open string disc amplitude $A_1^{(0)}(u)$ can be computed explicitly: it is
also a chain integral of the one form $\Theta$
\beq
A_1^{(0)}(x) = \int_{[x^*,x]} ydx.
\eeq

\br
Once again, it is interesting to note the similarities between the theory of symplectic invariants and B model. Let us summarize
this correspondence in a short array:
\begin{center}
\begin{tabular}{|c|c|}
\hline
{\bf Symplectic invariants} & {\bf B model} \cr
\hline
Spectral curve & Brane moduli space \cr
\hline
Symplectic transformations & phase transition \cr
\hline
filling fraction $\epsilon_i$ & closed flat coordinates $T_\alpha$ \cr
\hline
genus zero free energy $F^{(0)}$ & prepotential $F$ \cr
\hline
variations ${\partial F^{(0)} \over \partial \epsilon_i} = {1 \over 2 i \pi} \oint_{{\cal{B}}_i} ydx$ &
variations ${\partial F \over \partial T_\alpha} = {1 \over 2 i \pi} \oint_{{\cal{B}}_\alpha} ydx$ \cr
\hline
Genus 0 one point function $ydx$ & Disc amplitude $\int y dx$ \cr
\hline
\end{tabular}
\end{center}

\er

\subsubsection{Symplectic invariants formalism: a conjecture}

In \cite{Bmodel}, following some checks of \cite{MMtor} and the seminal paper \cite{DV}, Bouchard, Klemm, Mari\~{n}o
and Pasquetti proposed to define the open and closed string amplitudes of the $A$ model as the symplectic invariants
and correlation functions computed on the spectral curve of the mirror B model branes.

The conjecture of \cite{Bmodel} simply states that the open string amplitudes
$A_k^{(h)}(z_1, \dots , z_k)$ and closed string amplitudes ${\cal{F}}^{(h)}$ of the $A$ model
whose mirror background gives rise to the spectral curve $H(\ee{x},\ee{y})=0$ are given by correlation functions and symplectic invariants
built from the equation $H(\ee{x},\ee{y})=0$:
\beq\encadremath{
A_k^{(h)}(z_1, \dots , z_k)=\int \om_{k,string}^{(h)}(z_1, \dots , z_k)
}
\eeq
and
\beq\encadremath{
{\cal{F}}^{(h)}=F_h(t_\alpha) .
}\eeq

\br
In \cite{Bmodel}, the authors seem to slightly change the recursive rules defining the correlation functions and symplectic
invariants. However, this apparent transformation results from their choice to work with the coordinates
$\widetilde{X}= e^x$ and $\widetilde{Y}=e^y$ instead of $x$ and $y$ in order to start from an equation
\beq
\widetilde{H}(\widetilde{X},\widetilde{Y})=0
\eeq
which is algebraic. It appears that the algebraicity of the spectral curve is not essential and that one can directly work
with the coordinates $x$ and $y$, avoiding this change of coordinates.
\er

Let us emphasize a few important points. In order to get the $A$ model amplitudes, one first has to compute the correlation functions
and symplectic invariants from the $B$ model spectral curve and then plug in the mirror map to obtain the result in terms
of the $A$ model parameters. It should be underlined that the choice of coordinates $x$ and $y$ out of the $x_i$'s, i.e.
a choice of embedding of $\spcurve$ in $\mathbb{C}\times \mathbb{C}$, corresponds to a choice of brane in the $A$
model\footnote{That is to say, the choice of the location of the Brane as well as a choice of framing. This topic is well developed in \cite{Bmodel}.}. Thus, it is interesting to study the behavior of the amplitudes
when one moves in the brane moduli space, which corresponds to the space of parameterizations of the spectral curve. In particular,
the closed amplitudes should not depend on this choice of parametrization since only the boundaries of the worldsheet are
sensitive to the definition of the branes. This independence of the closed amplitudes on the embedding of the spectral curve
is indeed true and follows directly from the symplectic invariance of the $F^{(g)}$'s.

\subsubsection{Checks of the conjecture}

There have been many checks of this conjecture before the definition of the symplectic invariants and after.

First of all, most of this conjecture was inspired by the idea of Dijkgraaf and Vafa who conjectured that
the partition function of the type $B$ topological string on some special backgrounds is given by a random matrix
integral \cite{DV}. Now that the symplectic invariants extend the notion of random matrix partition function
to any spectral curve, it seemed natural to conjecture that these symplectic invariants do coincide with the partition
function of $B$ model topological string with more general backgrounds.

Another further property of the symplectic invariants points in the same direction. As it is reminded in section \ref{secholoano},
using the modular properties of the symplectic invariants, it was proved that one can promote the latter to modular invariants
whose non-holomorphic part is fixed by the same set of equations as the $B$ model topological string partition function
\cite{EMO}: the holomorphic anomaly equations of BCOV \cite{BCOV}. This means that the non-holomorphic part of these
functions coincide. To prove the conjecture, one thus have to prove it only for the holomorphic part.

Further studies were led by Mari\~{n}o and collaborators \cite{MMtor,Bmodel}: they checked for many explicit examples
of possible backgrounds for the $B$ model topological strings that the partition function and open string amplitudes indeed
coincide with the correlation functions and symplectic invariants computed on the associated spectral curve. Every single
check indeed works, giving more weight to this conjecture!

Another general check can be made by looking at the short summary made in the array in section \ref{secopenflat}.
It is also interesting to note that the disc and cylinder amplitudes can be computed independently from this conjecture for
any background in the B model: they satisfy the relation conjectured by \cite{Bmodel}.

Moreover, the computation of the sum over large partition with respect to the $q$-deformed Plancherel measure makes the
link with the topological vertex approach and proves the conjecture in a particular family of target spaces. The extension of this method could lead to a direct proof of this conjecture.

Finally, a last clue has been added recently by Dijkgraaf and Vafa \cite{DV2}, using an effective field theory conjectured
to be equivalent to $B$ model: the Kodaira-Spencer theory. This check is the subject of the next section.

\subsection{Kodaira-Spencer theory}

\subsubsection{Introduction: an effective field theory for the B-model}

The six dimensional Kodaira-Spence theory is the string field theory for the $B$ model on Calabi-Yau threefold. Consider the case
of non-compact Calabi-Yau threefold $\widetilde{M}$ defined by
\beq
H(x,y) = \om^+ \om^-
\eeq
whose holomorphic volume form is
\beq
\Omega = {d\om \over \om } dx dy.
\eeq
The Kodaira-Spencer theory is the quantization of the cohomologically trivial variations of the operator $\overline{\partial}$ on $\widetilde{M}$
with fixed complex structure.

The setup dual to the preceding section corresponds to the local surface
\beq
H(x,y) = 0.
\eeq
Since one can see that the periods of $\Omega$ can be reduced on this local surface to the integrals of the one form
$ydx$, the two dimensional reduction of the Kodaira-Spencer theory is defined by the pair $(\overline{\partial},ydx)$
on the spectral curve: it means that it is the study of the deformations of $\overline{\partial}$ keeping the cohomology
class of $ydx$ fixed. For this purpose, one looks at the variations of the this operator under the form
\beq
\overline{\partial} \to \overline{\partial} - {\overline{\partial} \phi \over ydx} \partial
\eeq
where $\phi$ is a scalar field satisfying
\beq
\overline{\partial} \partial \phi = 0 .
\eeq
Indeed, under this variation, the cohomology class of $ydx$ is not changed since it transforms as
\beq
ydx \to ydx + d \phi.
\eeq
Finally, we are left with a field theory on the spectral curve $\spcurve$ given by the action\footnote{For a very pedagogic
construction of this action, see \cite{DV2}.}:
\beq\label{actionKS}
S = \int_{\spcurve} \partial \phi \overline{\partial}\phi + { ydx \over \l} \overline{\partial}\phi
+ {\l \over ydx}  \overline{\partial}\phi \left(\partial \phi\right)^2
\eeq
where we rescale the differential $ydx$ by the string coupling constant $\l = {t \over N}$:
$ydx \to  {ydx \over \l}$.
Let us just explain the three different terms of this action. The first term is a simple kinetic term whereas the second term
corresponds to the coupling to a holomorphic background gauge field ${ydx\over \l}$. The most interesting term is the
third one: this cubic interaction encodes the perturbative corrections and is the fundamental ingredient of this action.

Let us now move to the observables of this theory. First of all, the partition function can be written as
\beq
{\cal{Z}}= e^{- {\cal{F}}}
\eeq
where the free energy ${\cal{F}}$ has a topological expansion in terms of the string coupling constant
\beq
{\cal{F}} = \sum_{g\geq 0} \l^{2g-2} F^{(g)}.
\eeq
One also defines the correlation functions
\beq
W_k(z_1,\dots,z_k|\l):=\left<\partial\phi(z_1) \dots \partial\phi(z_k)\right>_{c}
\eeq
where the subscript $c$ denotes the connected part. These correlation function also have a topological expansion
\beq
W_k(z_1,\dots,z_k|\l) = \sum_{g \geq 0} \l^{2g+k-2} W_k^{(g)}(z_1,\dots,z_k)
\eeq
coming from the interaction term
$e^{\l \int_\spcurve {\overline{\partial}\phi \left(\partial \phi\right)^2 \over ydx}}$.

\subsubsection{Recursive relations as Schwinger-Dyson equations}

One can remark that the integrant in this interaction can be written as a total derivative and does not give any contribution
except at the zeroes at the denominator, i.e. the zeroes of $ydx$. These zeroes are the zeroes of $y(z)$ and $dx(z)$.
However, one can show that only the zeroes $a_i$ of $dx$ do give non-vanishing contributions (see \cite{DV2}). Thus,
the interaction term can be written as follows:
\beq
\int_\spcurve {\overline{\partial}\phi \left(\partial \phi\right)^2 \over ydx}
= \sum_i \oint_{a_i} {\phi \, \partial\phi \, \partial\phi \over ydx} .
\eeq
This means that the interaction vertex is localized at the branch points.

In order to compute the correlation functions, one thus has to compute terms of the form
$\left< \partial\phi(z_1) \oint_{z \to a_i} {\phi(z) \, \partial\phi(z) \, \partial\phi(z) \over y(z)dx(z)} \dots \right>$.
A first step consists in the computation of the two point chiral operator $\left<\partial \phi(z) \partial \phi(z_1)\right>$
which is known to be the Bergmann kernel. From this point, one can easily compute the contraction of $\partial \phi(z_1)$
with $\phi(z)$
\beq
\left<\phi(z) \partial \phi(z_1)\right>_{twist} = {1 \over 2} \int_{\xi(z')= -\xi(z)}^{\xi(z)} B(z',z_1)
\eeq
where the subscript $twist$ refers to the fact that one constrains the scalar field $\phi$ to be an odd function
of a local variable $\xi(z)$ as $z$ approaches a branch point:
\beq
\phi(-\xi(z)) = - \phi(\xi(z)).
\eeq
In other terms, using the notations of section \ref{secsymplectic}, it implies, thanks to De l'H\^{o}pital's rule:
\beq
\lim_{z \to a_i} {\left<\phi(z) \partial \phi(z_1)\right>_{twist}\over ydx(z)} = \lim_{z \to a_i} {1 \over 2} {dE_z(z_1) \over (y(z)-y(\overline{z}))dx(z)}
= - \lim_{z \to a_i} K(z_1,z).
\eeq
Finally, taking into account the normal ordering of the cubic interaction term, the Schwinger-Dyson equations of this theory
give the recursion relation for the correlation functions
\beq
W_{n+1}^{(g)}(z_0,J)
= \sum_i \Res_{z\to a_i}\, K(z_0,z)\,\Big[
W_{n+2}^{(g-1)}(z,\bar{z},J)
+ \sum_{h=0}^g\sum'_{I\subset J} W_{1+|I|}^{(h)}(z,I) W_{1+n-|I|}^{(g-h)}(\bar{z},J\backslash I) \Big] .
\eeq
The correlators of the Kodaira-Spencer theory on $\spcurve$ are thus the correlation functions computed from the latter
curve.

What about the partition function? On the one hand, from the topological expansion, one easily finds that
\beq
\l {\partial {\cal{F}}\over \partial \l} = \sum_{g \ge 0} (2g-2) \l^{2g-2} F^{(g)}.
\eeq
On the other hand, let us reexpress the LHS directly in terms of the correlators of the theory.
Indeed, thanks to the expression of the action \eq{actionKS}, one gets
\beq
\l {\partial {\cal{Z}} \over \partial \l} = -{1 \over \l} \left<\int_\spcurve ydx \overline{\partial}\phi \right>
+ {\l^2 \over ydx} \left< \int_\spcurve \overline{\partial}\phi \left(\partial\phi\right)^2 \right> .
\eeq
In order to compute these terms, one proceed as in the case of the correlation functions by localizing these expressions
around the branch points. One can first remark that the second term vanishes since it corresponds to the interaction
operator with no field inserted. Let us thus compute the first term. Since the integrant can be written as a total derivative
$ydx \, \overline{\partial}\phi = d\left( ydx \, \phi \right)$, this term can be localized around the poles of the integrant
which are nothing but the branch points, i.e.
\beq
\l {\partial {\cal{Z}} \over \partial \l} ={1 \over \l} \sum_i \left< \oint_{a_i} ydx \, \phi \right>.
\eeq
Consider now a primitive $\Phi$ of $ydx$
\beq
d\Phi = ydx.
\eeq
Integrating by parts, it implies
\beq
\l {\partial {\cal{Z}} \over \partial \l} ={1 \over \l} \sum_i \oint_{a_i}  \Phi \left< \partial \phi \right>.
\eeq
In terms of the topological expansion, this equation coincides with the definition of the symplectic invariants
\beq
F^{(g)} = {1 \over 2-2g} \sum_i \oint_{a_i} \Phi(z) W_1^{(g)}(z).
\eeq
This means that the partition function of the Kodaira-Spencer theory is the tau function $\tau_N$ defined from the
symplectic invariants in section \ref{secsymplectic}.

%%%%%%%%%%%%%%%%%%%%%%%%%%%%%%%%%%%%%%%%%%%%%%%%%%%%%%%%%%%%%%%%%%%%%%%%%%%%%%%%%%%%%%%%%%%%%%%%%%%%%%%

\section{Conclusion}

In this review article, we have presented overview of the recent method introduced for solving matrix models loop equations, and its further extension to a more general context. We have defined the notion of symplectic invariants of a spectral curve, and we have studied its main applications, as in the present state of the art.
In some sense, starting from the spectral curve of a classical integrable system, we have proposed a way to reconstruct the full quantum integrable system.

The study of applications to enumerative geometry and integrable systems of those notions is probably only at its beginning, and in particular the consequences for topological string theory are still mostly to be understood...

\section*{Acknowledgments}
We would like to thank
M. Adler, M. Berg\`ere, M. Cafasso, P. Di Francesco, A. Kashani-Poor, M. Mari\~ no, P. Van Moerbeke
for useful and fruitful discussions on this subject and especially L. Chekhov and A. Prats Ferrer who have been very active in the development of these methods.
This work of is partly supported by the Enigma European network MRT-CT-2004-5652, by the ANR project G\'eom\'etrie et int\'egrabilit\'e en physique math\'ematique ANR-05-BLAN-0029-01,
by the European Science Foundation through the Misgam program and the work of B.E. is partly supported
by the Quebec government with the FQRNT. N.O. would like to thank the Universit\'{e} Catholique de Louvain where part of this work was done.

%%%%%%%%%%%%%%%%%%%%%%%%%%  Bibliography
%%%%%%%%%%%%%%%%%%%%%%%%%%%%%%%%%%

\end{document}